\documentclass[11pt,a4paper]{article}
\pdfoutput=1
\usepackage{jheppub}
\allowdisplaybreaks
\usepackage{graphicx,psfrag}% Include figure files
\usepackage{bm}% bold math
\usepackage{mathbbol,verbatim}
\usepackage{slashed}
\usepackage{graphics}
\usepackage{color,ulem}
\usepackage{makecell,multirow}
\graphicspath{{fig/}}

\allowdisplaybreaks

\definecolor{greeen}{rgb}{0.03,0.84,0.13}
\definecolor{test}{rgb}{0.03,0.74,0.33}
\definecolor{viol}{rgb}{0.44,0,0.94}
\definecolor{or}{rgb}{0.95,0.65,0}

\begin{document}

\title{Displaced vertex signatures of doubly charged scalars in the type-II seesaw and its left-right extensions}

\author{P. S. Bhupal Dev and}
\author{Yongchao Zhang}

\affiliation{Department of Physics and McDonnell Center for the Space Sciences,  Washington University, St. Louis, MO 63130, USA}

\emailAdd{bdev@wustl.edu}
\emailAdd{yongchao.zhang@physics.wustl.edu}

\date{\today}

\abstract{The type-II seesaw mechanism with an isospin-triplet scalar $\Delta_L$ provides one of the most compelling explanations for the observed smallness of neutrino masses. The triplet contains a doubly-charged component $H_L^{\pm\pm}$, which dominantly decays to either same-sign dileptons or to a pair of $W$ bosons, depending on the size of the triplet vacuum expectation value. However, there exists a range of Yukawa couplings $f_L$ of the triplet to the charged leptons, wherein a relatively light $H_L^{\pm\pm}$ tends to be long-lived, giving rise to distinct displaced-vertex signatures at the high-energy colliders. We find that the displaced vertex signals from the leptonic decays $H_L^{\pm\pm} \to \ell_\alpha^\pm \ell_\beta^\pm$ could probe a broad parameter space with $10^{-10} \lesssim |f_L| \lesssim 10^{-6}$ and 45.6 GeV $< M_{H_L^{\pm\pm}} \lesssim 200$ GeV at the high-luminosity LHC. Similar sensitivity can also be achieved at a future 1 TeV $e^+e^-$ collider. The mass reach can be extended to about 500 GeV at a future 100 TeV proton-proton collider. Similar conclusions apply for the right-handed triplet $H_R^{\pm\pm}$ in the TeV-scale left-right symmetric models, which provide a natural embedding of the type-II seesaw.
We show that the displaced vertex signals are largely complementary to the prompt same-sign dilepton pair searches at the LHC  and the low-energy, high-intensity/precision measurements, such as neutrinoless double beta decay, charged lepton flavor violation, electron and muon anomalous magnetic moments, muonium-antimuonium oscillation and M{\o}ller scattering. }
%, there also exists a , which is also long-lived when its Yukawa coupling $f_R$ is small. For a fixed right-handed scale $v_R = 5\sqrt2$ TeV, the displaced same-sign dilepton signatures are sensitive to a large parameter space with $10^{-10} \lesssim |f_R| \lesssim 10^{-6}$ and $m_Z/2 < M_{H_R^{\pm\pm}} \lesssim 200$ (400) GeV at the future high-luminosity LHC and ILC 1 TeV (FCC-hh).
%, which constrain relatively large Yukawa couplings of the doubly-charged scalars to charged leptons.


\maketitle

%\tableofcontents

\section{Introduction}

%After observation of the solar, atmospheric and reactor oscillations of the three active neutrinos, the tiny neutrino masses remain one of the biggest mysteries in the elementary particle physics, and they are also the only New Physics beyond the Standard Model (SM) of particles that have been confirmed in the terrestrial experiments.
To account for the tiny neutrino masses, as suggested by the neutrino oscillation experiments, the Standard Model (SM) has to be extended in the scalar, fermion and/or gauge sector; see Ref.~\cite{Mohapatra:2006gs} for a review. %~\cite{deGouvea:2013zba, Raidal:2008jk, Lindner:2016bgg}.
%There has been a plethora of global efforts in searching for the beyond SM particles in the high-energy colliders as well as in the low-energy high-intensity frontier, that might reveal the underlying theory for the neutrino masses as well as other phenomena like dark matter and baryon-anti-baryon asymmetry in the Universe~\cite{Aoki:2001rc, Tinsley:2004pe, Rubbia:2010fm, Edgecock:2013lga, Datta:1992qw, Ingelman:1993ve, Gluza:1996bz, Gluza:1997ts, Cvetic:1998vg, Porod:2000hv, Ali:2001gsa, Bigi:2001xb, Panella:2001wq, Berezhiani:2001rs, delAguila:2005ssc, delAguila:2005pin, Bray:2005wv, Han:2006ip, delAguila:2006bda, Atwood:2007zza, delAguila:2007qnc, Hektor:2007uu, Kersten:2007vk, Han:2007bk, Chen:2007dc, Chao:2007mz, Kadastik:2007yd, Huitu:2008gf, Perez:2008zc, Chao:2008mq, Perez:2008ha, delAguila:2008ir, Kovalenko:2009td, Geer:2009zz, Chen:2011hc, Ahriche:2014xra, Alva:2014gxa, Antusch:2015mia, Deppisch:2015qwa, Banerjee:2015gca, Kang:2015uoc, Degrande:2016aje, Antusch:2016vyf, Mitra:2016wpr, Antusch:2016ejd}.
%The seesaw mechanism is arguably the simplest paradigm for understanding the neutrino masses in a natural way.
By simply extending the SM scalar sector by an $SU(2)_L$-triplet scalar, the type-II seesaw~\cite{Magg:1980ut, Schechter:1980gr, Mohapatra:1980yp, Lazarides:1980nt, Konetschny:1977bn, Cheng:1980qt} is one of the most economical frameworks to generate the observed neutrino masses and mixing. In this paper we study the displaced vertex (DV) signatures from the doubly-charged scalars in the type-II seesaw and its left-right extensions, which could enrich the searches for new physics behind neutrino mass generation at the high-energy frontier, beyond the standard prompt decays currently being probed at the Large Hadron Collider (LHC). %\footnote{For DV signatures in type-I seesaw and its extensions, see e.g. Refs.~\cite{}.}

In the pure type-II seesaw, the doubly-charged component $H_L^{\pm\pm}$ of the $SU(2)_L$-triplet $\Delta_L$ couples to the SM charged leptons and the $W$ boson~\cite{Perez:2008ha, Melfo:2011nx, Kanemura:2014goa}. The strength of these interactions can be potentially suppressed by either the small Yukawa couplings $f_L$ or the small vacuum expectation value (VEV) of the neutral component of $\Delta_L$. Therefore, in a sizable parameter space, the doubly-charged scalar $H_L^{\pm\pm}$ tends to be long-lived at the hadron and lepton colliders, and the decay products of $H_L^{\pm\pm}$ form distinct displaced vertex (DV) signatures. In this paper we study only the simplest DV scenario, i.e. the displaced same-sign dilepton pairs $H_L^{\pm\pm} \to \ell_\alpha^\pm \ell_\beta^\pm$ (with $\alpha,\,\beta = e,\, \mu,\, \tau$ being the lepton flavor indices), though in principle we could also have the displaced multi-body final states through the (off-shell) $W$ bosons: $H_L^{\pm\pm} \to W^{\pm \, (\ast)} W^{\pm \, (\ast)} \to \bar{f} f^\prime \bar{f}^{\prime\prime} f^{\prime\prime\prime}$ (with the $f$'s being the SM fermions). For the sake of concreteness and illustration purpose, we estimate the DV prospects at the high-luminosity LHC (HL-LHC) with the center-of-mass energy of 14 TeV and an integrated luminosity of 3000 fb$^{-1}$~\cite{Apollinari:2017cqg}, as well as the proposed 100 TeV Future Circular Collider (FCC-hh) with a luminosity of 30 ab$^{-1}$~\cite{fcc-hh} and the International Linear Collider (ILC) with the center-of-mass energy of 1 TeV and a luminosity of 1 ab$^{-1}$~\cite{Baer:2013cma}. The sensitivities at other proposed facilities such as the Super Proton-Proton Collider (SPPC)~\cite{Tang:2015qga}, Circular Electron-Positron Collider (CEPC)~\cite{CEPC-SPPCStudyGroup:2015csa}, FCC-ee~\cite{Gomez-Ceballos:2013zzn} and Compact LInear Collider (CLIC)~\cite{Battaglia:2004mw} might be somewhat different, depending on the colliding energies and luminosities, but could be easily derived following our analysis.

The left-right symmetric model (LRSM)~\cite{LR1, LR2, LR3}, based on the gauge group $SU(3)_C \times SU(2)_L \times SU(1)_R \times U(1)_{B-L}$, provides a natural ultraviolet (UV)-completion of the type-II seesaw mechanism. In this case, as a ``partner'' of $H_L^{\pm\pm}$ under parity, there exists a right-handed (RH) doubly-charged scalar $H_R^{\pm\pm}$ originating from the $SU(2)_R$-triplet scalar $\Delta_R$, which couples predominantly to the RH charged leptons via the Yukawa couplings $f_R$ and the heavy $W_R$ boson via the RH gauge interaction~\cite{Gunion:1989in, Dev:2016dja}. For sufficiently heavy $W_R$ boson and small couplings $f_R$, the lifetime of $H_R^{\pm\pm}$ could also be large enough to give rise to DV signatures in $H_R^{\pm\pm} \to \ell_\alpha^\pm \ell_\beta^\pm$ at future colliders. The DV signatures from the off-shell $W_R$ bosons: $H_R^{\pm\pm} \to W_R^{\pm\, \ast} W_R^{\pm \, \ast} \to \bar{f} f^\prime \bar{f}^{\prime\prime} f^{\prime\prime\prime}$ (here the $f$'s stand for the SM fermions as well as the RH neutrinos (RHNs) in the LRSM) will not be covered in this paper, as a proper analysis of displaced jets and RH neutrinos from $W_R$ decay is more involved than the simplest case of displaced same-sign dilepton pairs.

As we will show below, for both the $H_L^{\pm\pm}$ in the pure type-II seesaw and $H_R^{\pm\pm}$ in the LRSM, the DV signatures from the decay of doubly-charged scalars are sensitive to relatively small Yukawa couplings, typically $f_{L,\,R} \lesssim 10^{-7}$. These are largely complementary to the searches of prompt same-sign dilepton pair signals~\cite{Perez:2008ha, Azuelos:2004mwa, Han:2007bk, delAguila:2008cj, Akeroyd:2009hb, Akeroyd:2010je, Akeroyd:2010ip, Alloul:2013raa, Chun:2013vma, delAguila:2013mia, Bambhaniya:2013wza, Mitra:2016wpr, Babu:2016rcr, Agrawal:2018pci, Dev:2016dja, Ghosh:2017pxl, Dev:2018upe, Borah:2018yxd} performed at LEP~\cite{Abbiendi:2001cr, Achard:2003mv, Abdallah:2002qj},  Tevatron~\cite{Aaltonen:2008ip, Acosta:2004uj, Abazov:2008ab, Abazov:2011xx} and LHC~\cite{Aaboud:2017qph, CMS:2017pet}, wherein the Yukawa couplings are assumed to be large such that the doubly-charged scalars decay promptly after production. If the Yukawa couplings $f_{L,R}$ happen to be very small, then we will not expect any prompt leptons at hadron and lepton colliders, and all these dilepton constraints are no longer applicable. Furthermore, with the lepton-flavor violating (LFV) couplings $(f_{L,\,R})_{\alpha\beta}$ ($\alpha \neq \beta$), both $H_L^{\pm\pm}$ and $H_R^{\pm\pm}$ could induce rare LFV processes like $\mu \to eee$ and $\mu \to e\gamma$~\cite{Pal:1983bf, Leontaris:1985qc, Swartz:1989qz, Mohapatra:1992uu, Cirigliano:2004mv, Cirigliano:2004tc, Akeroyd:2009nu, Tello:2010am, Chakrabortty:2012vp, Barry:2013xxa, Bambhaniya:2015ipg, Chakrabortty:2015zpm, Borah:2016iqd, Bonilla:2016fqd, Borgohain:2017akh, Crivellin:2018ahj, Lindner:2016bgg}, electron and muon anomalous magnetic moment~\cite{Leveille:1977rc, Moore:1984eg, Gunion:1989in, Lindner:2016bgg}, and muonium-antimuonium oscillation~\cite{Chang:1989uk, HM, Clark:2003tv, Swartz:1989qz}, which are all highly suppressed in the SM~\cite{PDG}. The current low-energy high-intensity experiments searching for these rare processes set severe constraints on the LFV couplings~\cite{Dev:2017ouk, Dev:2018upe, Dev:2018sel}, which could go down to $10^{-3}$ for a 100 GeV-scale doubly-charged scalar. The DV signatures probing much lower LFV couplings are complementary to these low-energy high-intensity experiments as well.

The rest of the paper is organized as follows: Section~\ref{sec:left} is devoted to the left-handed (LH) doubly-charged scalar $H_L^{\pm\pm}$ in the pure type-II seesaw, where after sketching the basic properties of $H_L^{\pm\pm}$ we collect all the LFV constraints, re-interpret the high-energy same-sign dilepton constraints, and estimate the DV prospects at HL-LHC, FCC-hh and ILC. In Section~\ref{sec:lrsm} we focus on the RH doubly-charged scalar $H_R^{\pm\pm}$ in parity-conserving LRSM, where the symmetry relation $f_L = f_R$ sets stringent constraints on the couplings $f_R$. Setting the RH scale $v_R = 5\sqrt2$ TeV, the DV prospects are found to be somewhat different from those of $H_L^{\pm\pm}$ in the type-II seesaw. The parity-violating case of LRSM follows in Section~\ref{sec:lrsm2}. Without parity in the Yukawa sector, i.e. $f_L \neq f_R$, all the elements of $f_R$ can be considered as free parameters. Taking as an explicit example the benchmark scenario of $(f_R)_{ee} \neq 0$ with all other elements vanishing, the DV prospects turn out to be quite similar to those in the parity conserving case. We conclude in Section~\ref{sec:conclusion}. The details of four-body decays of $H_{L (R)}^{\pm\pm} \to W_{(R)}^{\pm \, (\ast)} W_{(R)}^{\pm \, (\ast)} \to \bar{f} f^\prime \bar{f}^{\prime\prime} f^{\prime\prime\prime}$ can be found in Appendix~\ref{sec:decay}, and the LFV decay formulas are collected in Appendix~\ref{sec:appendix:LFV}.

%The MOLLER project~\cite{Benesch:2014bas, Moller} proposes such a global strategy which could in principle discover new physics signatures that might escape LHC detection.  We illustrate here this complementarity between the precision and energy frontiers by taking the scalar triplets as a case study.

%early limit~\cite{Swartz:1989qz}

\section{Left-handed doubly-charged scalar in type-II seesaw}
\label{sec:left}

In the type-II seesaw model~\cite{Magg:1980ut, Schechter:1980gr, Mohapatra:1980yp, Lazarides:1980nt, Konetschny:1977bn, Cheng:1980qt}, there exists a new complex scalar multiplet which transforms as a triplet under the SM $SU(2)_L$ gauge group. It can be written in terms of its components as
\begin{align}
\label{eqn:DeltaL}
\Delta_L \ = \  \left(\begin{array}{cc}
\delta_L^+/\sqrt{2} & \delta_L^{++} \\
\delta_L^0 & -\delta_L^+/\sqrt{2}
\end{array}\right) \, .
\end{align}
The most general scalar potential for the SM doublet $\phi = \left( \phi^+ ,\, \phi^0 \right)^{\sf T}$ and the triplet $\Delta_L$ reads~\cite{Arhrib:2011uy, DGOS, Chabab:2015nel}
\begin{eqnarray}
\label{eq:Vpd}
{\cal V}(\phi,\Delta_L) & \ = \ &
-\mu_\phi^2(\phi^\dag \phi)
+\mu^2_\Delta {\rm Tr}({\Delta}_L^\dag {\Delta}_L)
+\frac{\lambda}{2}(\phi^\dag \phi)^2
+ \frac{\lambda_1}{2}\left[{\rm Tr}({\Delta}_L^\dag {\Delta}_L)\right]^2\nonumber\\
&& +\frac{\lambda_2}{2}\left(\left[{\rm Tr}({\Delta}_L^\dag {\Delta}_L)\right]^2
-{\rm Tr}\left[({\Delta}_L^\dag {\Delta}_L)^2\right]\right) +\lambda_4(\phi^\dag \phi){\rm Tr}({\Delta}_L^\dag {\Delta}_L) \nonumber \\
&& +\lambda_5\phi^\dag[{\Delta}_L^\dag,{\Delta}_L]\phi
+\left(\frac{\lambda_6}{\sqrt 2}\phi^{\sf T}i\sigma_2{\Delta}_L^\dag \phi+{\rm H.c.}\right)\, ,
\end{eqnarray}
with all the couplings being real ($\lambda_6$ having the mass dimension) and $\sigma_2$ being the second Pauli matrix. A non-zero VEV for the doublet field $\langle \phi^0 \rangle = v_{\rm EW}/\sqrt2$ (with $v_{\rm EW} \simeq$ 246 GeV being the electroweak scale) induces a tadpole term for the scalar triplet field ${\Delta}_L$ via the $\lambda_6$ term in Eq.~(\ref{eq:Vpd}), thereby generating a non-zero VEV for its neutral component, $\langle \delta_L^0\rangle = v_L/\sqrt 2$, and breaking lepton number by two units, which is responsible for neutrino mass generation at tree-level.

As the VEV $v_L$ is in charge of the tiny neutrino masses, it is expected to be much smaller than the electroweak scale, or even close to the eV scale, depending on the corresponding Yukawa couplings. On the other hand, the electroweak precision data, and in particular, the $\rho$-parameter constraint requires that $v_L\lesssim 2$ GeV~\cite{delAguila:2008ks}. In the limit of $v_L \ll v_{\rm EW}$, after spontaneous symmetry breaking, the neutral component from the SM doublet has a mass $m_h^2 \simeq \lambda v_{\rm EW}^2$ and identified as the SM Higgs, whereas the neutral, singly-charged and doubly-charged components of the triplet $\Delta_L$ give rise to the additional physical scalars
\begin{eqnarray}
H \ \simeq \ \frac{{\rm Re} \, \delta_L^0}{\sqrt2} \,, \quad
A \ \simeq \ \frac{{\rm Im} \, \delta_L^0}{\sqrt2} \,, \quad
H^{\pm} \ \simeq \ \delta_L^\pm \,, \quad
H_L^{\pm\pm} \  = \ \delta_L^{\pm\pm} \,,
\end{eqnarray}
with ``${\rm Re}$'' and ``${\rm Im}$'' denoting respectively the real and imaginary parts. Their masses are respectively
\begin{eqnarray}
M^2_{H,\,A} &\ = \ & \mu_\Delta^2+\frac{1}{2}(\lambda_4-\lambda_5)v_{\rm EW}^2 ,
\label{eqn:mass1} \\
M^2_{H^\pm} &\ = \ & \mu_\Delta^2+\frac{1}{2}\lambda_4v_{\rm EW}^2 \,,
\label{eqn:mass2}\\
M^2_{H_L^{\pm\pm}} &\ = \ & \mu_\Delta^2+\frac{1}{2}(\lambda_4+\lambda_5)v_{\rm EW}^2 \,.
\label{eqn:mass3}
\end{eqnarray}
 The mass splitting of the triplet scalars is dictated by the quartic coupling $\lambda_5$ in Eq.~(\ref{eq:Vpd}) and tends to be small (compared to the triplet scalar masses), in particular when the electroweak precision data is taken into consideration~\cite{Chun:2012jw, Aoki:2012jj}.

The triplet $\Delta_L$ couples to the SM lepton doublet  $\psi_L =(\nu,\ell)_L^{\sf T}$ via the Yukawa interactions
\begin{eqnarray}
\label{eqn:lagrangian}
{\cal L}_Y \ = \
- \left(f_L \right)_{\alpha\beta} \psi_{L,\,\alpha}^{\sf T}Ci\sigma_2 {\Delta}_L \psi_{L,\,\beta} ~+~ {\rm H.c.},
\end{eqnarray}
where $\alpha,\,\beta=e,\mu,\tau$ denote the lepton flavors and $C$ is the charge conjugation matrix. Then the tiny neutrino mass matrix is obtained with the induced VEV $v_L$:
\begin{eqnarray}
\label{eq:neutrino}
m_\nu \ = \
\sqrt2 \, f_L v_L \ = \
U \widehat{m}_\nu U^{\sf T} \, .
\end{eqnarray}
The Yukawa coupling matrix $f_L$ is fixed by the active neutrino data, i.e. the observed neutrino mass squared differences and mixing angles, up to the unknown lightest neutrino mass $m_0$, the neutrino mass hierarchy and the Dirac and Majorana CP violating phases. In Eq.~(\ref{eq:neutrino}) $\widehat{m}_\nu = {\rm diag} \{ m_1,\, m_2,\, m_3 \}$ with $m_{1,2,3}$ the physical neutrino masses, and $U$ is the standard PMNS matrix, which can be parameterized as~\cite{PDG}
\begin{eqnarray}
\label{eqn:PMNS}
U \ = \ & \left(\begin{array}{ccc}
c_{12}c_{13} & s_{12}c_{13} & s_{13}e^{-i\delta_{\rm CP}}\\
-s_{12}c_{23}-c_{12}s_{23}s_{13}e^{i\delta_{\rm CP}} &
c_{12}c_{23}-s_{12}s_{23}s_{13}e^{i\delta_{\rm CP}} & s_{23}c_{13}\\
s_{12}s_{23}-c_{12}c_{23}s_{13}e^{i\delta_{\rm CP}} &
-c_{12}s_{23}-s_{12}c_{23}s_{13}e^{i\delta_{\rm CP}} & c_{23}c_{13}
\end{array}\right)\nonumber \\
& \qquad \qquad \times \ {\rm
  diag}\{1, e^{i\alpha_1/2},e^{i\alpha_2/2}\}\; ,
\end{eqnarray}
where $c_{ij}\equiv \cos\theta_{ij}$, $s_{ij}\equiv \sin\theta_{ij}$ with $\theta_{ij}$ the mixing angles, $\delta_{\rm CP}$ the Dirac CP phase and $\alpha_{1,2}$ the two Majorana phases. For the convenience of the calculations below, a recent global fit results~\cite{Esteban:2016qun, nufit} on the mass squared differences and mixing angles are collected in Table~\ref{tab:neutriodata}, including the central values and $1\sigma$ uncertainties for both the normal hierarchy (NH) and inverted hierarchy (IH) spectra of neutrino masses. We vary the CP phases within the whole range of $[0,2\pi]$ (unless otherwise specified). Note that the recent T2K~\cite{Abe:2017vif}  and  NO$\nu$A~\cite{NOvA:2018gge} results  indicate  a  mild preference  for non-zero $\delta_{\rm CP}$, but this has not been established at $5\sigma$ level yet.

\begin{table}[t]
  \centering
  \caption[]{Best-fit values and $1\sigma$ ranges of the neutrino mass squared difference and mixing parameters for both NH and IH of neutrino spectra from a recent global fit~\cite{Esteban:2016qun, nufit}. The Dirac CP violating phase $\delta_{\rm CP}$ and the Majorana phases $\alpha_{1,2}$ are considered to be unconstrained.}
  \label{tab:neutriodata}
  %\begin{tabular}[t]{C{3.2cm}|C{2.8cm}|C{3.5cm}|C{3.5cm}}
  \begin{tabular}[t]{ccc}
  \hline\hline
  parameters & NH & IH \\ \hline
  $\Delta m^2_{21}$ [$10^{-5}$ eV$^2$] & $7.40^{+0.21}_{-0.20}$ & $7.40^{+0.21}_{-0.20}$ \\
  $\Delta m^2_{32}$ [$10^{-3}$ eV$^2$] & $2.494^{+0.033}_{-0.031}$ & $-2.465^{+0.032}_{-0.031}$ \\ \hline

  $\sin^2\theta_{12}$ & $0.307^{+0.013}_{-0.012}$ & $0.307^{+0.013}_{-0.012}$ \\
  $\sin^2\theta_{23}$ & $0.538^{+0.033}_{-0.069}$ & $0.554^{+0.023}_{-0.033}$ \\
  $\sin^2\theta_{13}$ & $0.02206 \pm 0.00075$ & $0.02227 \pm 0.00074$ \\ \hline
  $\delta_{\rm CP}$ & $[0,\, 2\pi]$ & $[0,\, 2\pi]$ \\
  $\alpha_1$ & $[0,\, 2\pi]$ & $[0,\, 2\pi]$ \\
  $\alpha_2$ & $[0,\, 2\pi]$ & $[0,\, 2\pi]$ \\
  \hline\hline
  \end{tabular}
\end{table}

\subsection{Decay Length}

In the type-II seesaw, the doubly-charged scalar $H_{L}^{\pm\pm}$ has the following decay modes:
\begin{itemize}
\item $H_L^{\pm\pm} \to \ell_\alpha^\pm \ell_\beta^\pm$ which depends on the Yukawa couplings in Eq.~\eqref{eqn:lagrangian}. The partial width is given by
\begin{eqnarray}
\label{eqn:width1}
\Gamma (H_L^{\pm\pm} \to \ell_\alpha^\pm \ell_\beta^\pm) \ = \
\frac{M_{H_L^{\pm\pm}} \left|(m_\nu)_{\alpha\beta}\right|^2}{8\pi (1+ \delta_{\alpha\beta}) v_{L}^2} \,,
\end{eqnarray}
with $M_{H_L^{\pm\pm}}$ the mass of $H_L^{\pm\pm}$, $\delta_{\alpha\beta}$ the Kronecker $\delta$-function, and $(m_\nu)_{\alpha\beta}$ the neutrino mass matrix elements given by Eq.~\eqref{eq:neutrino}.

\item $H_L^{\pm\pm} \to W^\pm W^\pm$ (if kinematically allowed) which depends on the triplet VEV $v_L$. The partial width is given by
\begin{eqnarray}
\label{eqn:width2}
\Gamma (H_L^{\pm\pm} \to W^\pm W^\pm) \ = \
\frac{G_F^2 v_L^2 M_{H_L^{\pm\pm}}^3}{2\pi}
\sqrt{1-4x_W} (1-4x_W+12x_W^2) \,,
\end{eqnarray}
with $G_F$ the Fermi constant and $x_W \equiv m_W^2 / M_{H_L^{\pm\pm}}^2$. When $M_{H_L^{\pm\pm}} < 2m_W$, then at least one of the two $W$ bosons is off-shell and when both the $W$ bosons are off-shell, we have the four-body decay
\begin{eqnarray}
H_L^{\pm\pm} \to W^{\pm \,\ast} W^{\pm\,\ast} \to f \bar{f}^\prime f^{\prime\prime} \bar{f}^{\prime\prime\prime} \, ,
\end{eqnarray}
in which case the partial width calculation is a bit involved~\cite{Kanemura:2014goa} and is detailed in Appendix~\ref{sec:decay}.

In principle, $H_L^{\pm\pm}$ has other diboson decay modes as follows:
  \item $H_L^{\pm\pm} \to H^{\pm \, (\ast)} W^{\pm\, (\ast)}$ which depends on the triplet scalar mass splitting $M_{H_L^{\pm\pm}} - M_{H^\pm}$. Even if $M_{H_L^{\pm\pm}} > M_{H^\pm}$ which implies that $\lambda_5 >0$ in Eqs.~(\ref{eqn:mass1})-(\ref{eqn:mass3}), the mass splitting larger than 60 GeV is disfavored by current electroweak precision data~\cite{Aoki:2012jj}, thus the on-shell decay into $H^\pm W^\pm$ is not kinematically allowed. For $M_{H_L^{\pm\pm}} - M_{H^\pm} \lesssim 1$ GeV, the cascade decay width $\Gamma (H_L^{\pm\pm} \to H^{\pm} W^{\pm\, \ast})$ is smaller than that for the dilepton and $W$ boson pair channels given by Eqs.~(\ref{eqn:width1}) and (\ref{eqn:width2}) respectively~\cite{Kanemura:2014goa, Melfo:2011nx}.
  \item $H_L^{\pm\pm} \to H^{\pm\, \ast} H^{\pm\, \ast}$ which is subject to the trilinear scalar couplings in the potential~(\ref{eq:Vpd}). In light of the electroweak precision data~\cite{Aoki:2012jj}, both the singly-charged scalars $H^\pm$ in the final state are expected to be off-shell.
\end{itemize}

\begin{figure}[t!]
  \centering
  \includegraphics[width=0.49\textwidth]{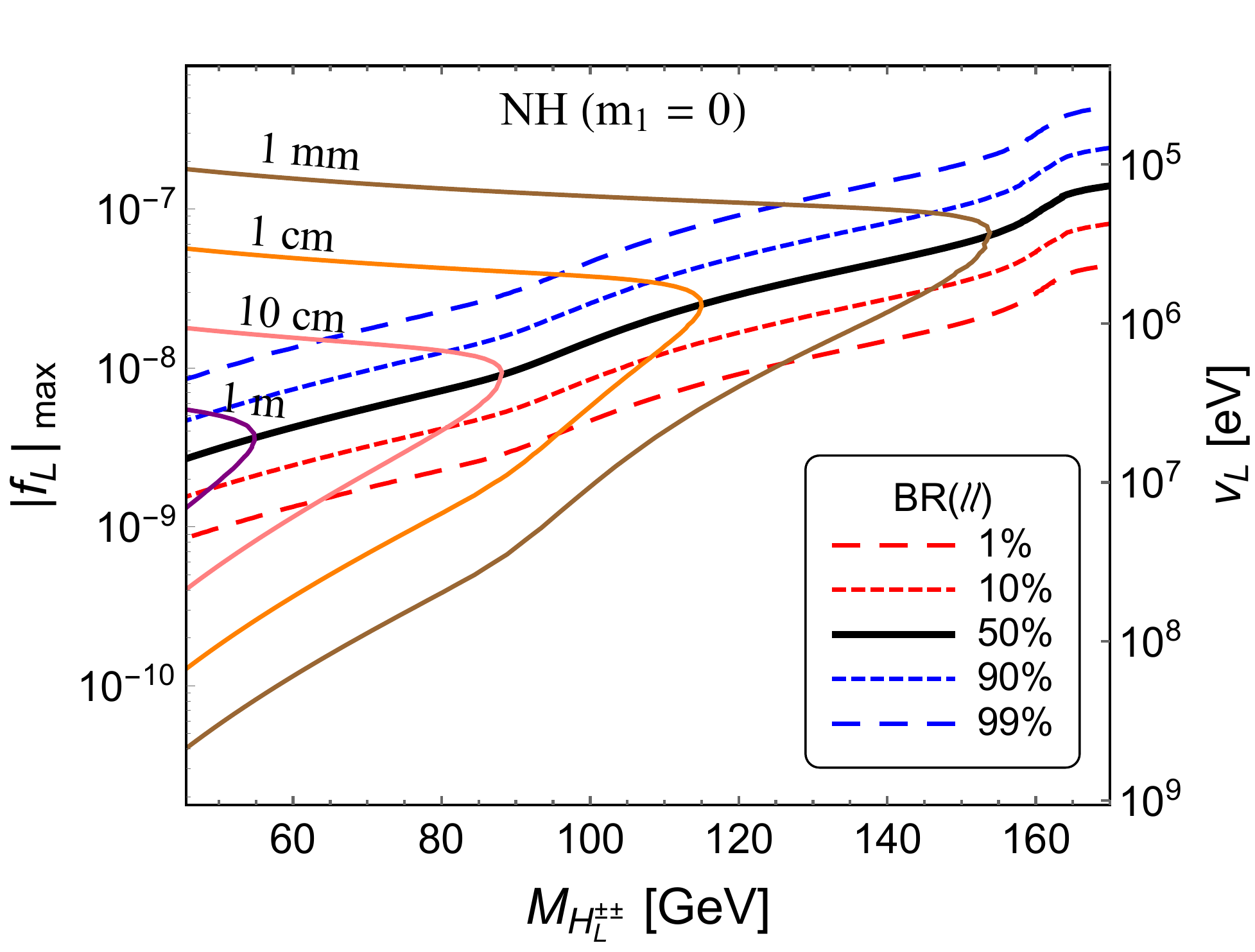} %\hspace{-10pt}
  \includegraphics[width=0.49\textwidth]{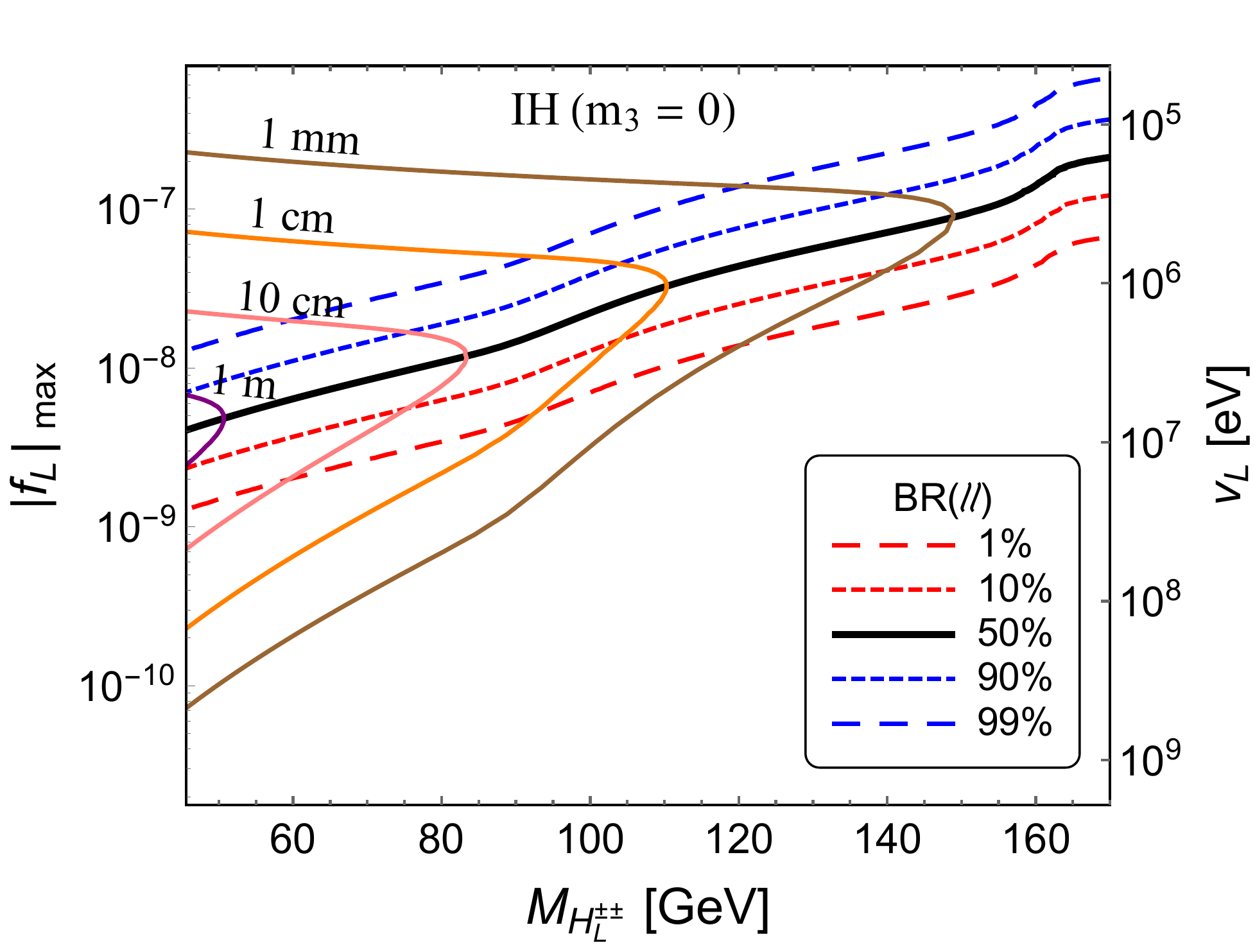}
  \caption{Contours of proper decay length $c\tau_0$ of 1 mm, 1 cm, 10 cm and 1 m of the LH doubly-charged scalar $H_L^{\pm\pm}$ in type-II seesaw, as functions of the doubly-charged scalar mass $M_{H_L^{\pm\pm}}$ and the largest Yukawa coupling $|(f_L)_{\rm max}|$. Also shown are the ${\rm BR} (H_L^{\pm\pm} \to \ell_\alpha^\pm \ell_\beta^\pm) = 1 - {\rm BR} (H_L^{\pm\pm} \to W^{\pm (\ast)} W^{\pm (\ast)}) = 1$\%, 10\%, 50\%, 90\%, 99\%. The left and right panels are respectively for the neutrino spectra of NH and IH with zero lightest neutrino mass.  The corresponding values of the VEV $v_L$ are also shown in the plots.}
  \label{fig:lifetime1}
\end{figure}

For simplicity, we neglect the $H^\pm W^\pm$ and $H^\pm H^\pm$ diboson channels, e.g. by choosing appropriate quartic couplings such that $M_{H_L^{\pm\pm}} < M_{H^\pm}$ and the trilinear coupling $H^{\pm\pm} H^\mp H^\mp$ is small. Then the total width is given by
\begin{eqnarray}
\label{eqn:widthtotal}
\Gamma_{\rm total} (H_L^{\pm\pm}) \ = \
\Gamma (H_L^{\pm\pm} \to \ell_\alpha \ell_\beta) +
\Gamma (H_L^{\pm\pm} \to W^{\pm \, (\ast)} W^{\pm \, (\ast)}) \,.
\end{eqnarray}
The dilepton width in Eq.~(\ref{eqn:width1}) is suppressed by the tiny neutrino masses when the VEV $v_L$ is comparatively large, while the $W$ pair width in Eq.~(\ref{eqn:width2}) is suppressed by the VEV $v_L$, which leads to a maximal total width at a VEV value of $v_L \sim 1$ MeV, depending on the doubly-charged scalar mass~\cite{Perez:2008ha}, as shown in Fig.~\ref{fig:lifetime1}.
For sufficiently light $H_L^{\pm\pm}$, roughly of order 100 GeV, the proper lifetime $c\tau_0 (H_L^{\pm\pm})$ is of order millimeter to meter and could thus generate DV signal at colliders. For the illustration purpose, the proper decay length $c\tau_0$ of 1 mm, 1 cm, 10 cm and 1 m are shown in Fig.~\ref{fig:lifetime1}, as functions of the Yukawa coupling $|f_L|$ and the doubly-charged scalar mass $M_{H_L^{\pm\pm}}$. The left and right panels are respectively for the NH and IH cases with the lightest neutrino mass $m_1 = 0$ (NH) and $m_3 = 0$ (IH). As all the elements of $f_L$ are strongly correlated by the neutrino mass and mixing data, as shown in Eq.~(\ref{eq:neutrino}), for concreteness we take the largest element $|f_L|_{\rm max}$ on the left $y$-axes. Also we take only the central values of the neutrino mixing angles and mass-squared differences in Table~\ref{tab:neutriodata} and choose the Dirac CP phase $\delta_{\rm CP} = 3\pi/2$, as suggested from the best-fit central value of the recent T2K~\cite{Abe:2017vif}  and  NO$\nu$A~\cite{NOvA:2018gge} data. The corresponding values of $v_L$ from Eq.~\eqref{eq:neutrino} are also shown on the right $y$-axes of the plots, with the relation
\begin{eqnarray}
v_L |f_L|_{\rm max} \ = \
\begin{cases}
0.027 \, {\rm eV}  \,, & \text{for NH with $m_1 = 0$} \,, \\
0.048 \, {\rm eV}  \,, & \text{for IH with $m_3 = 0$} \,.
\end{cases}
\end{eqnarray}
The contours of branching ratios (BRs) ${\rm BR} (H_L^{\pm\pm} \to \ell_\alpha^\pm \ell_\beta^\pm) = 1 - {\rm BR} (H_L^{\pm\pm} \to W^{\pm (\ast)} W^{\pm (\ast)}) = 1$\%, 10\%, 50\%, 90\%, 99\% are also depicted in Fig.~\ref{fig:lifetime1} as respectively the long-dashed red, short-dashed red, thick solid black, short-dashed blue and long-dashed blue lines. It is clear from Fig.~\ref{fig:lifetime1} that to have a proper decay length $c\tau_0$ of 1 mm to 1 m (in order to DV signatures), the LH doubly-charged scalar $H_L^{\pm\pm}$ in the type-II seesaw is required to have a mass from $m_Z/2$ (see Section~\ref{sec:dilepton} for the mass limit $M_{H_L^{\pm\pm}} \gtrsim m_Z/2$) to roughly 150 GeV, with Yukawa couplings $|f_L|_{\rm max} \sim 10^{-10}$ to $10^{-7}$ (or effectively $v_L = 10^{5}$ to $10^8$ eV). For larger values of lightest neutrino masses $m_{1,3}>0$, the neutrino mass elements in Eq.~(\ref{eqn:width1}) tend to be larger; however, as the (proper) lifetime of $H_L^{\pm\pm}$ is only sensitive to the total width, the lifetime contours, and as a result the DV sensitivities, do not change too much in Fig.~\ref{fig:lifetime1}.

Given the physical mass spectrum in the triplet sector, cf. Eqs.~\eqref{eqn:mass1}-\eqref{eqn:mass3}, we could also envision DV signatures due to the singly-charged Higgs bosons $H^\pm$ or the neutral CP-even (odd) Higgs bosons $H \ (A)$. As for the singly-charged one, the dominant decay modes are $H^\pm \to \ell^\pm \nu$ (depending on the Yukawa coupling $f_L$) and $H^\pm\to W^\pm Z, \ W^\pm h$ (depending on the VEV $v_L$). However, the Drell-Yan production cross section for $H^\pm$ pair at colliders is  smaller than that of $H^{\pm\pm}$ pair, due to the smaller electric charge, and for our choice $M_{H_L^{\pm\pm}} < M_{H^\pm}$, other production modes are also smaller for $H^\pm$~\cite{Perez:2008ha}. Therefore, the DV signatures of $H^\pm$ are expected to be sub-dominant compared to that of $H_L^{\pm\pm}$. As for the neutral CP-even scalar $H$, the dominant decays are into $\nu\bar{\nu}$ (depending on $f_L$) and $hh, \ ZZ, \ t\bar{t}, \ b\bar{b}$ (depending on $v_L$). Similarly, for the CP-odd scalar $A$, the dominant decays are into $\nu\bar{\nu}$ (depending on $f_L$) and $hZ, \ t\bar{t}, \ b\bar{b}$ (depending on $v_L$). Thus, in the parameter space of interest where $H_L^{\pm\pm}$ gives rise to DV dilepton decays, the neutral scalars would most likely lead to DV missing energy signal, which is not very promising. We postpone a detailed investigation of the possible DV prospects of singly-charged and neutral scalars in the type-II seesaw model at future colliders to a follow-up work.

\subsection{Low and high-energy constraints}
\label{sec:LFV}
In this section, we discuss various experimental constraints on the Yukawa couplings $(f_L)_{\alpha\beta}$ from both low- and high-energy observables.

\subsubsection{Lepton flavor violation}

For the LH doubly-charged scalar $H_L^{\pm\pm}$, the couplings $(f_L)_{\alpha\beta}$ with $\alpha\neq \beta$ could induce rare LFV decays such as $\ell_\alpha \to \ell_\beta \ell_\gamma \ell_\delta$, $\ell_\alpha \to \ell_\beta \gamma$~\cite{Pal:1983bf, Leontaris:1985qc, Swartz:1989qz, Mohapatra:1992uu, Cirigliano:2004mv, Cirigliano:2004tc, Akeroyd:2009nu, Tello:2010am, Chakrabortty:2012vp, Barry:2013xxa, Bambhaniya:2015ipg, Chakrabortty:2015zpm, Borah:2016iqd, Bonilla:2016fqd, Borgohain:2017akh, Crivellin:2018ahj, Lindner:2016bgg}, anomalous magnetic moments of electron and muon~\cite{Leveille:1977rc, Moore:1984eg, Gunion:1989in, Lindner:2016bgg}, and the muonium oscillation~\cite{Chang:1989uk, HM, Clark:2003tv, Swartz:1989qz} which are all highly suppressed in the SM~\cite{PDG}. The couplings $(f_L)_{e \ell}$ contribute also to the scattering $e^+e^- \to \ell^+\ell^-$ (with $\ell = e,\, \mu,\, \tau$) and are thus constrained by the LEP data~\cite{Abbiendi:2003pr, Achard:2003mv, Abdallah:2005ph}.\footnote{The coupling $(f_L)_{ee}$ of $H_L^{\pm\pm}$ to electrons also contributes to the M{\o}ller scattering $e^- e^-  \to e^- e^-$ and could be probed by the upcoming MOLLER experiment~\cite{Benesch:2014bas, Moller}, but in the pure type-II seesaw case, the MOLLER sensitivity is precluded by the LFV constraints~\cite{Dev:2018sel}.} All these limits on $|(f_L)_{\alpha\beta}|$ or $|f_L^\dagger f_L|$ are collected in the third column of Table~\ref{tab:limits} (see also Refs.~\cite{Dev:2017ouk, Dev:2018upe}). The flavor limits in this section apply also to the RH doubly-charged scalar $H_R^{\pm\pm}$ in the LRSM discussed in Sections~\ref{sec:lrsm} and \ref{sec:lrsm2}, thus the subscript ``L'' and ``R'' of the Yukawa couplings and the doubly-charged scalar mass are not shown explicitly in the third column of Table~\ref{tab:limits}, which are collectively dubbed as $f$ and $M_{\pm\pm}$. The relevant formulas and calculation details can be found in Appendix~\ref{sec:appendix:LFV}.

\begin{table}[!t]
  \centering
  \small
  \caption[]{Current experimental limits on the BRs of $\ell_\alpha \to \ell_\beta \ell_\gamma \ell_\delta$, $\ell_\alpha \to \ell_\beta \gamma$~\cite{PDG, Amhis:2016xyh}, anomalous electron~\cite{Hanneke:2008tm} and muon~\cite{Bennett:2006fi} magnetic moments, muonium oscillation~\cite{Willmann:1998gd}, and LEP $e^+e^- \to \ell^+\ell^-$ data~\cite{Abdallah:2005ph}, along with the corresponding constraints on the Yukawa couplings $|f|$ or $|f^\dagger f|$, in unit of $(M_{\pm\pm}/100 \, {\rm GeV})^2$ (third column). These limits apply to both $H_L^{\pm\pm}$ in the type-II seesaw and $H_R^{\pm\pm}$ in the LRSM (so the subscript ``L'' has been removed). The data in the last two columns are the resultant constraints on $(v_L M_{H_L^{\pm\pm}})$ in the type-II seesaw, in unit of (eV)(100 GeV), for both NH and IH with the lightest neutrino mass $m_0=0$ (0.05 eV).}
  \label{tab:limits}
  \begin{tabular}[t]{c|c|c|c|c}
  \hline\hline
  %process & current data & constraints [$\left(\frac{M^2_{\pm\pm}}{100 \, {\rm GeV}} \right)^2$] \\ \hline

  \multirow{4}{*}{Process } & \multirow{4}{*}{Experimental} &
  \multirow{4}{*}{\makecell{Constraint \\ $\times \left(\frac{M_{\pm\pm}}{100 \, {\rm GeV}} \right)^2$}} &
   \multicolumn{2}{c}{Lower limit on
   $\left( \frac{v_L}{{\rm eV}} \right)
   \left( \frac{M_{H_L^{\pm\pm}}}{100\,{\rm GeV}} \right)$ }  \\ \cline{4-5}
  & Bound && \makecell{NH \\ $m_1 = 0$ \\ ($m_1 = 0.05$ eV)} & \makecell{IH \\ $m_3 = 0$ \\ ($m_3 = 0.05$ eV)}  \\ \hline

  $\mu^- \to e^- e^+ e^-$ & $< 1.0 \times 10^{-12}$ &
  $|f_{ee}^\dagger f_{e\mu}|< 2.3 \times 10^{-7}$ &
  4.6 (36) & 23 (43) \\ \hline

  $\tau^- \to e^- e^+ e^-$ & $< 1.4 \times 10^{-8}$ &
  $|f_{ee}^\dagger f_{e\tau}| < 6.5 \times 10^{-5}$ &
  0.27 (2.2) & 1.4 (2.6) \\
  $\tau^- \to e^- \mu^+ \mu^-$ & $< 1.6 \times 10^{-8}$ &
  $|f_{e\mu}^\dagger f_{\mu\tau}| < 4.9 \times 10^{-5}$ &
  1.1 (1.0) & 1.2 (1.2) \\
  $\tau^- \to \mu^- e^+ \mu^-$ & $< 9.8 \times 10^{-9}$ &
  $|f_{e\tau}^\dagger f_{\mu\mu}| < 5.5 \times 10^{-5}$ &
  1.2 (2.6) & 1.1 (2.6)  \\

  $\tau^- \to \mu^- e^+ e^-$ & $< 1.1 \times 10^{-8}$ &
  $|f_{e\mu}^\dagger f_{e\tau}| < 4.1 \times 10^{-5}$ &
  0.69 (1.4) & 0.58 (1.4) \\
  $\tau^- \to e^- \mu^+ e^-$ & $< 8.4 \times 10^{-9}$ &
  $|f_{ee}^\dagger f_{\mu\tau}| < 5.1 \times 10^{-5}$ &
  0.57 (2.0) & 3.5 (2.8) \\
  $\tau^- \to \mu^- \mu^+ \mu^-$ & $< 1.2 \times 10^{-8}$ &
  $|f_{\mu\mu}^\dagger f_{\mu\tau}| < 6.1 \times 10^{-5}$ &
  2.2 (2.0) & 2.2 (2.4) \\ \hline

  $\mu^- \to e^- \gamma$ & $< 4.2 \times 10^{-13}$ &
  $|\sum_k f_{ek}^\dagger f_{\mu k}| < 2.7 \times 10^{-6}$ &
  6.9 (6.9) & 6.9 (6.9) \\
  $\tau^- \to e^- \gamma$ & $< 3.3 \times 10^{-8}$ &
  $|\sum_k f_{ek}^\dagger f_{\tau k}| < 1.8 \times 10^{-3}$ &
  0.27 (0.27) & 0.27 (0.27)  \\
  $\tau^- \to \mu^- \gamma$ & $< 4.4 \times 10^{-8}$ &
  $|\sum_k f_{\mu k}^\dagger f_{\tau k}| < 2.1 \times 10^{-3}$ &
  0.52 (0.52) & 0.54 (0.54) \\ \hline

  electron $g-2$ & $< 5.2 \times 10^{-13}$ &
  $\sum_k |f_{ek}|^2 < 1.2$ & $0.0058$ (0.033) & 0.032 (0.045) \\
  muon $g-2$ & $< 4.0 \times 10^{-9}$ &
  $\sum_k |f_{\mu k}|^2 < 0.17$ &
  0.06 (0.1) & 0.061 (0.11) \\ \hline

  \makecell{muonium \vspace{-3pt} \\  oscillation} & $<8.2 \times 10^{-11}$ &
  $|f_{ee}^\dagger f_{\mu\mu}| < 0.0012$ &
  0.13 (1.1) & 0.7 (1.3) \\ \hline

  $e^+e^- \to e^+e^-$ & $\Lambda_{\rm eff} > 5.2$ TeV &
  $|f_{ee}|^2 < 0.0012$ & 0.033 (0.98) & 1.0 (1.4) \\
  $e^+e^- \to \mu^+\mu^-$ & $\Lambda_{\rm eff} > 7.0$ TeV &
  $|f_{e\mu}|^2 < 6.4 \times 10^{-4}$ &  0.17 (0.36) & 0.15 (0.36) \\
  $e^+e^- \to \tau^+\tau^-$ & $\Lambda_{\rm eff} > 7.6$ TeV &
  $|f_{e\tau}|^2 < 5.4 \times 10^{-4}$ & 0.19 (0.39) & 0.16 (0.39) \\
 % \hline
 % MOLLER & $\delta A_{\rm PV} < 0.7$ ppb & $|f_{ee}|^2 <0.019$ &
 % 0.059 (1.8) & 1.8 (2.5) \\
  \hline\hline
  \end{tabular}
\end{table}

\begin{figure}[t!]
  \centering
  \includegraphics[height=0.37\textwidth]{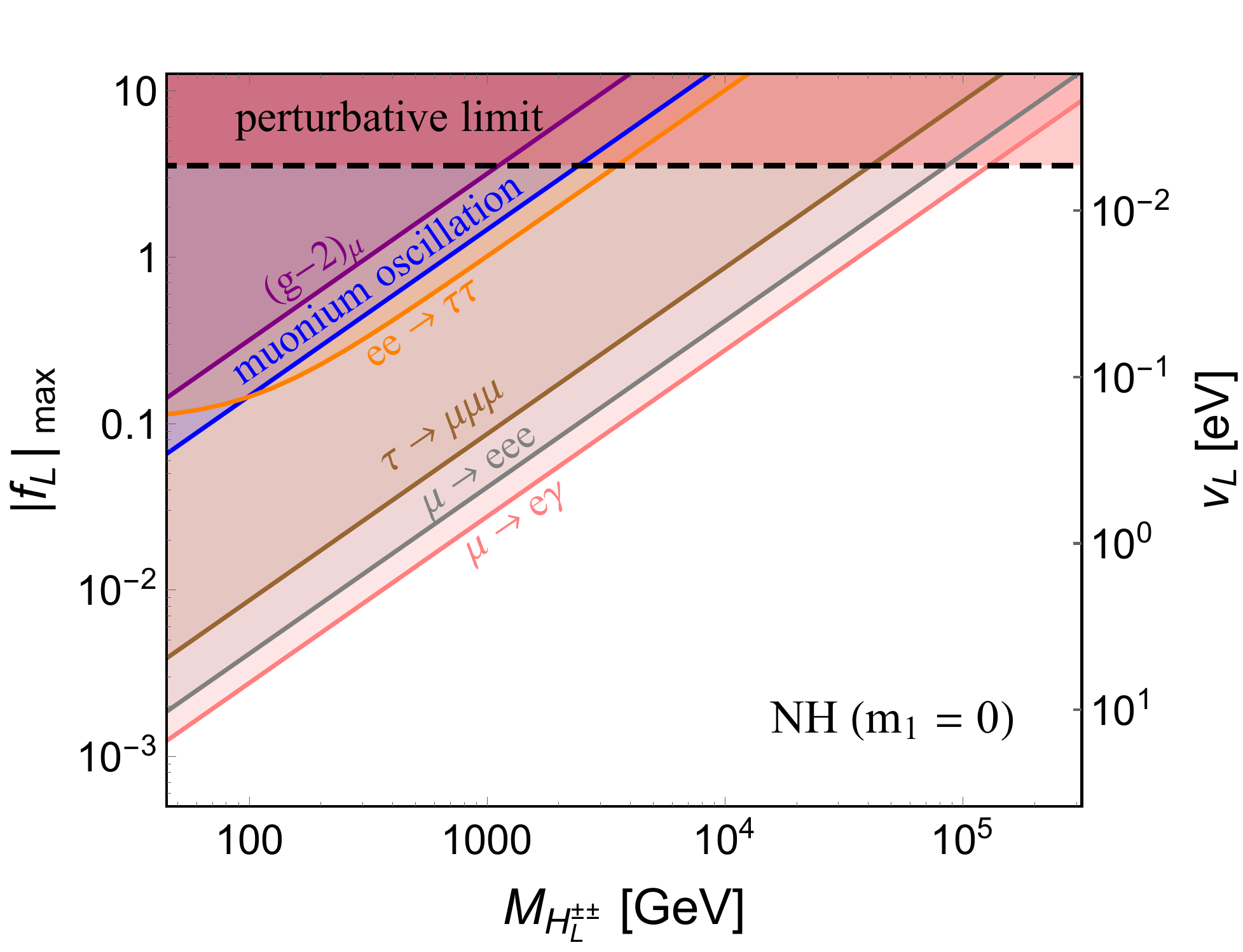}
  \includegraphics[height=0.37\textwidth]{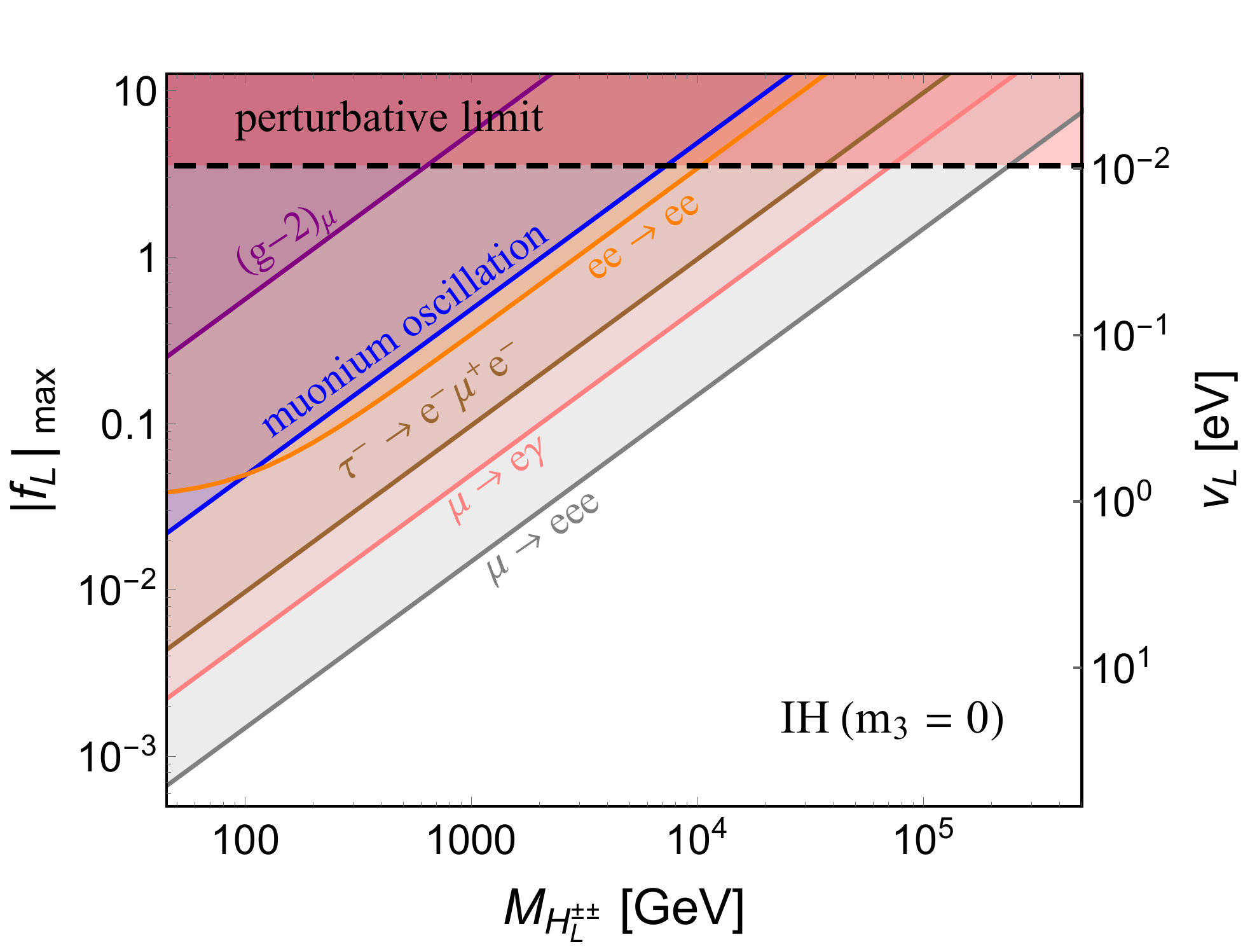} \\ \vspace{3pt}
  \includegraphics[height=0.37\textwidth]{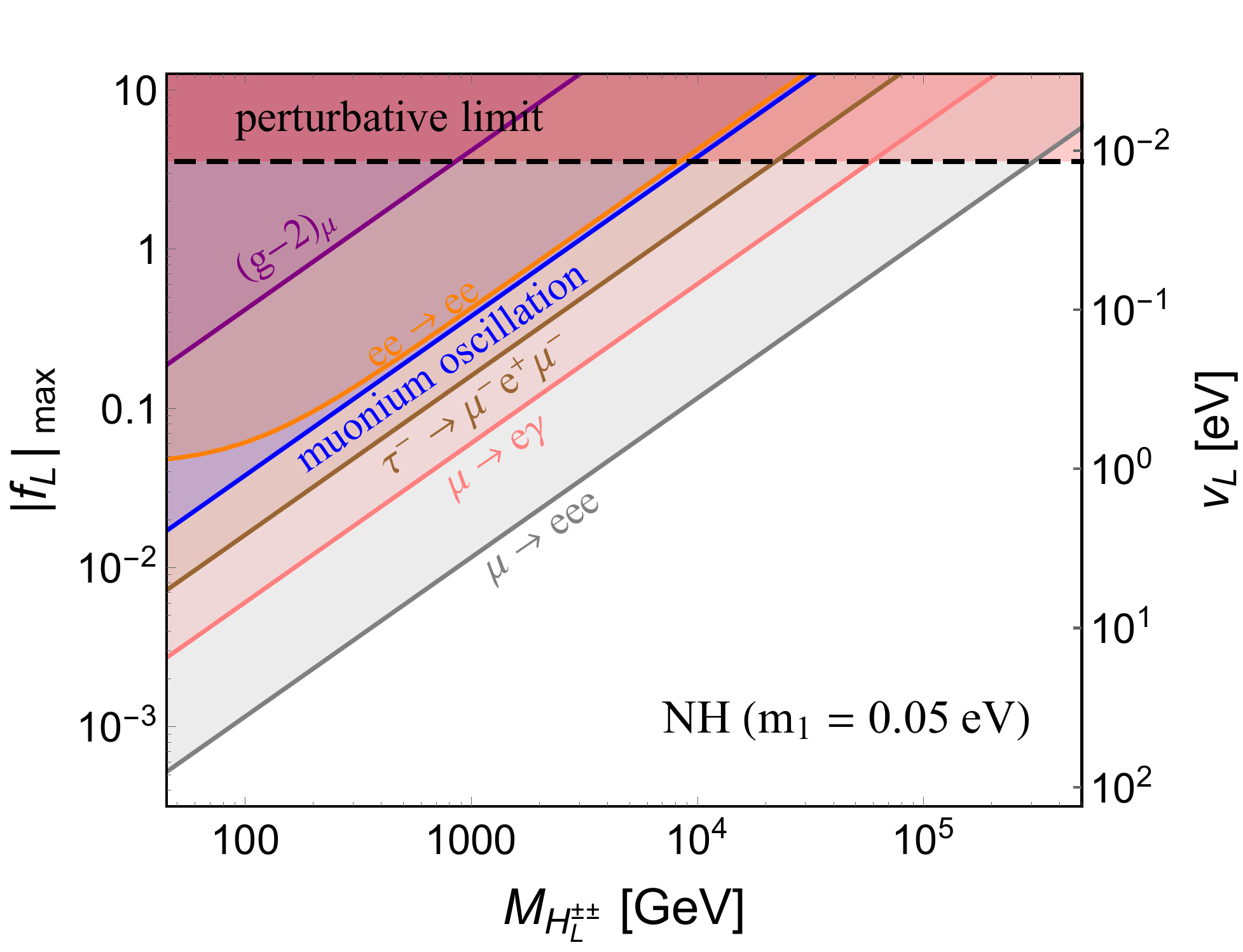}
  \includegraphics[height=0.37\textwidth]{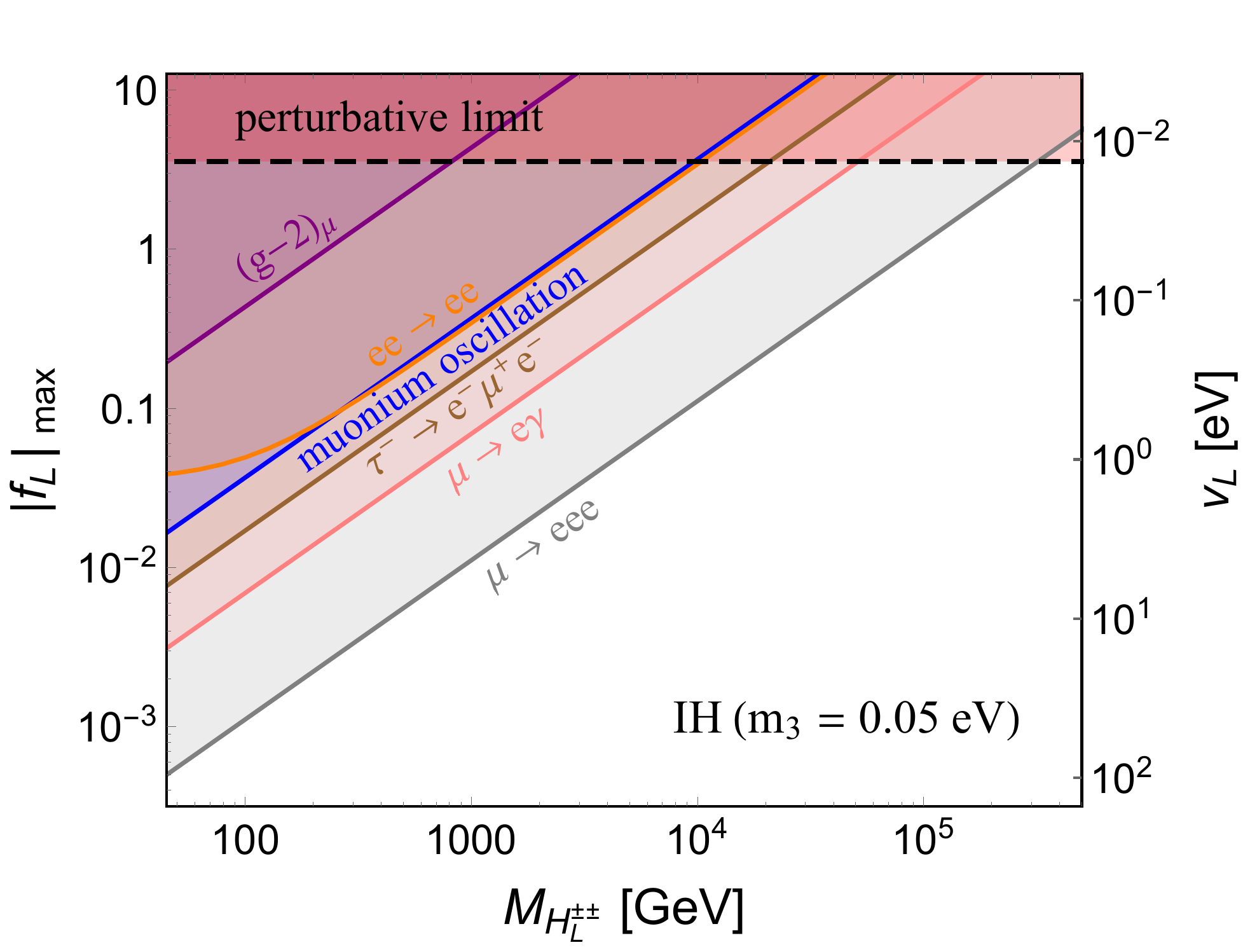}
  \caption{LFV constraints on the doubly-charged scalar mass $M_{H_L^{\pm\pm}}$ and the largest Yukawa coupling $|f_L|_{\rm max}$ in the type-II seesaw, for the NH (left) and IH (right) with the lightest neutrino mass $m_{0} = 0$ (upper) or 0.05 eV (lower). All the shaded regions are excluded by either the low-energy LFV constraints on BR($\ell_\alpha \to \ell_\beta \ell_\gamma \ell_\delta)$, BR($\ell_\alpha \to \ell_\beta \gamma$)~\cite{PDG}, anomalous muon $g-2$, muonium oscillation, or the LEP $e^+e^- \to \ell^+\ell^-$ data~\cite{Abdallah:2005ph}. More constraints can be found in Table~\ref{tab:limits}. The horizontal black line represents the perturbative limit of $|f_L|_{\rm max} < \sqrt{4\pi}$. }
  \label{fig:flavor1}
\end{figure}

In light of the neutrino mass relation in Eq.~(\ref{eq:neutrino}), all these limits on the couplings $|(f_L)_{\alpha\beta}|$ or $|f_L^\dagger f_L|$ can be traded for constraints on the VEV $v_L$ in the type-II seesaw, up to the unknown lightest neutrino mass $m_0$, the Dirac CP phase $\delta_{\rm CP}$ and the neutrino mass hierarchy~\cite{Dev:2017ouk}. To be concrete, we utilize only the central values of the neutrino data in Table~\ref{fig:lifetime1}, with the Dirac CP violating phase $\delta_{\rm CP} = 3\pi/2$. To set limits on $v_L$, we consider both the NH and IH spectra, and adopt two benchmark values of $m_0= 0$ and 0.05 eV in each case. All the corresponding constraints on the product $v_L M_{H_L^{\pm\pm}}$ are collected in the last two columns of Table~\ref{tab:limits}, in unit of $(\rm eV) \, (100 \, \rm GeV)$. The limits for these four benchmark scenarios (NH and IH, $m_0=0$ and 0.05 eV) are graphically depicted in Fig.~\ref{fig:flavor1} in the plane of $ M_{H_L^{\pm\pm}}$ and $|f_L|_{\rm max}$, along with the corresponding values of $v_L$. All the shaded regions are excluded by current data.

Two comments are in order:
\begin{itemize}
\item As the doubly-charged scalar mass is much larger than the charged lepton masses or the energy scale of the low-energy experiments, the limits on $|f_L^\dagger f_L| / M^2_{H_L^{\pm\pm}}$ are almost constants, corresponding to an effective cutoff scale $\Lambda_{\rm eff} \simeq M_{H_L^{\pm\pm}} / |f_L|$. One exception is the limit from LEP $e^+e^- \to \ell^+\ell^-$ data~\cite{Dev:2017ftk}: When the $H_L^{\pm\pm}$ mass is smaller than the center-of-mass energy at LEP, i.e. $M_{H_L^{\pm\pm}} \lesssim 100$ GeV, the propagator is dominated by the kinetic term, viz.
\begin{eqnarray}
\frac{1}{q^2 - M_{H_L^{\pm\pm}}^2} \ \to \  \frac{1}{q^2} \,.
\end{eqnarray}
Therefore, the LEP limits in Fig.~\ref{fig:flavor1} get to some extent weaker for lighter $H_L^{\pm\pm}$ and do not depend on $M_{H_L^{\pm\pm}}$ in the limit of $|q| \gg M_{H_L^{\pm\pm}}$.

\item In the NH case with a massless neutrino, i.e. $m_1 = 0$, the neutrino mass matrix element $(m_\nu)_{ee}$ is suppressed either by the solar neutrino mass squared difference $\Delta m_{12}^2$ (compared to $\Delta m_{23}^2$) or the reactor neutrino mixing angle $\sin\theta_{13}$
\begin{eqnarray}
(m_\nu)_{ee}^{\rm NH} \ \sim \
m_2 s_{12}^2 c_{13}^2 + m_3 s_{13}^2 \ \simeq \
\sqrt{\Delta m^2_{12}} s_{12}^2 +
\sqrt{|\Delta m^2_{23}|} s_{13}^2 \quad (m_1 =0) \,,
\end{eqnarray}
where we have neglected all the Dirac and Majorana phases and used the fact that $\sin\theta_{13} \ll 1$. This is significantly smaller than that in the IH case, where
\begin{eqnarray}
(m_\nu)_{ee}^{\rm IH} \ \sim \
m_1 c_{12}^2 c_{13}^2 + m_2 s_{12}^2 c_{13}^2  \ \simeq \
\sqrt{|\Delta m^2_{23}|}  \quad (m_3 =0) \,.
\end{eqnarray}
Thus in the fourth and fifth columns of Table~\ref{tab:limits}, the limits involving the coupling $f_{ee}$, like $\mu \to eee$, for the case of NH ($m_1 = 0$) is weaker than that for IH ($m_3 = 0$). When the three active neutrinos becomes heavier, for instance in the scenarios of NH with $m_1 = 0.05$ eV and IH with $m_{3} = 0.05$ eV in Table~\ref{tab:limits}, the matrix elements $(m_\nu)_{\alpha\beta}$ tend to be larger (though some of them would get smaller due to mild cancellation in the summation $\sum_i m_i U_{\alpha i} U_{\beta i}$), and most of the limits in the parentheses of Table~\ref{tab:limits} are somewhat stronger than the cases with a massless neutrino.
\end{itemize}

\subsubsection{Neutrinoless double beta decay}

\begin{figure}[!t]
  \centering
  \includegraphics[width=0.3\textwidth]{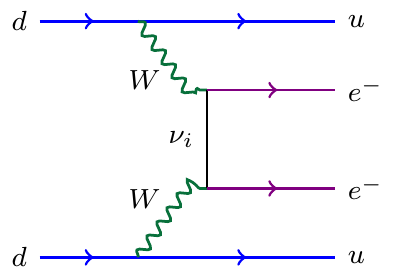}
  \includegraphics[width=0.3\textwidth]{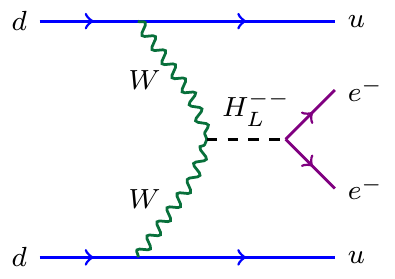}
  \caption{Feynman diagram for the parton-level $0\nu\beta\beta$ induced by the active neutrinos $\nu_i$ (left) and the LH doubly-charged scalar $H_L^{\pm\pm}$ (right), which correspond to the amplitudes $\eta_\nu$ and $\eta_{\rm DCS}^L$ in Eq.~(\ref{eqn:0nubetabeta}), respectively.}
  \label{fig:diagram1}
\end{figure}

Due to its direct interaction with the SM $W$ boson and the electrons, the LH doubly-charged scalar $H_{L}^{\pm\pm}$ in type-II seesaw contributes to $0\nu\beta\beta$, in addition to the canonical light neutrino contribution. The parton-level Feynman diagrams of $0\nu\beta\beta$ from the light neutrinos $\nu_i$ and $H_L^{\pm\pm}$ are presented in Fig.~\ref{fig:diagram1}. The corresponding half lifetime of $0\nu\beta\beta$ can be factorized as~\cite{Barry:2013xxa}
\begin{eqnarray}
\label{eqn:0nubetabeta}
\left[ T^{0\nu}_{1/2} \right]^{-1} \ = \
G \, \left| {\cal M}_\nu \left( \eta_\nu + \eta_{\rm DCS}^{L} \right) \right|^2 \,,
\end{eqnarray}
with $G$ the phase space factor, ${\cal M}_\nu$ the nuclear matrix element (NME) for the light neutrino contribution, and the dimensionless term $\eta_\nu = (m_\nu)_{ee} / m_e$ is the amplitude of the canonical light neutrino contribution, with the effective electron neutrino mass
\begin{eqnarray}
(m_\nu)_{ee}  \ = \
\sum_i U_{ei}^2 m_{i}  \ = \
m_1 c_{12}^2 c_{13}^2 +
m_2 s_{12}^2 c_{13}^2 e^{i\alpha_1} +
m_3 s_{13}^2 e^{i\alpha_2}
\end{eqnarray}
encoding the Majorana phases $\alpha_{1,2}$. The second term in Eq.~(\ref{eqn:0nubetabeta}) denotes the amplitude that is mediated by $H_L^{\pm\pm}$, and is proportional to the coupling of $H_L^{\pm\pm}$ to the SM $W$ boson (the VEV $v_L$) and the elements of the Yukawa coupling matrix $f_L$:
\begin{eqnarray}
\eta_{\rm DCS}^L \ = \
\frac{U_{ei}^2 m_i m_e}{M_{H_{L}^{\pm\pm}}^{2}} \ = \
\frac{m_e^2}{M_{H_{L}^{\pm\pm}}^{2}} \eta_\nu \,.
\end{eqnarray}
Compared to the canonical $\eta_\nu$ term, the extra $H_L^{\pm\pm}$ contribution is highly suppressed by the doubly-charged scalar mass. Therefore we can not set any limits on $H_L^{\pm\pm}$ in the pure type-II seesaw.

\subsubsection{High-energy collider constraints}
\label{sec:dilepton}

The doubly-charged scalar $H_L^{\pm\pm}$ couples directly to the SM $Z$ boson, with the coupling proportional to the factor $(1 - 2\sin^2\theta_w)$, where $\sin\theta_w$ the weak mixing angle. For $M_{H_L^{\pm\pm}} < m_Z/2$, it contributes to the total width of $Z$ via the decay
\begin{eqnarray}
\label{eqn:Zwidth}
\Gamma (Z \to H_L^{++} H_L^{--})  \ = \
\frac{G_F \, m_Z^3 \, (1-2\sin^2\theta_w)^2}{6\sqrt2\pi}
\left( 1 - \frac{4 M^2_{H_L^{\pm\pm}}}{m_Z^2} \right)^{3/2} \,,
\end{eqnarray}
and is thus stringently constrained by the high precision $Z$-pole data~\cite{PDG}. This puts a {\it lower} bound on $M_{H_L^{\pm\pm}} > m_Z/2 \simeq 45.6$ GeV, irrespective of how the doubly-charged scalar decays or whether it is long-lived or not~\cite{Kanemura:2014goa}.

\begin{figure}[t!]
  \centering
  \includegraphics[width=0.48\textwidth]{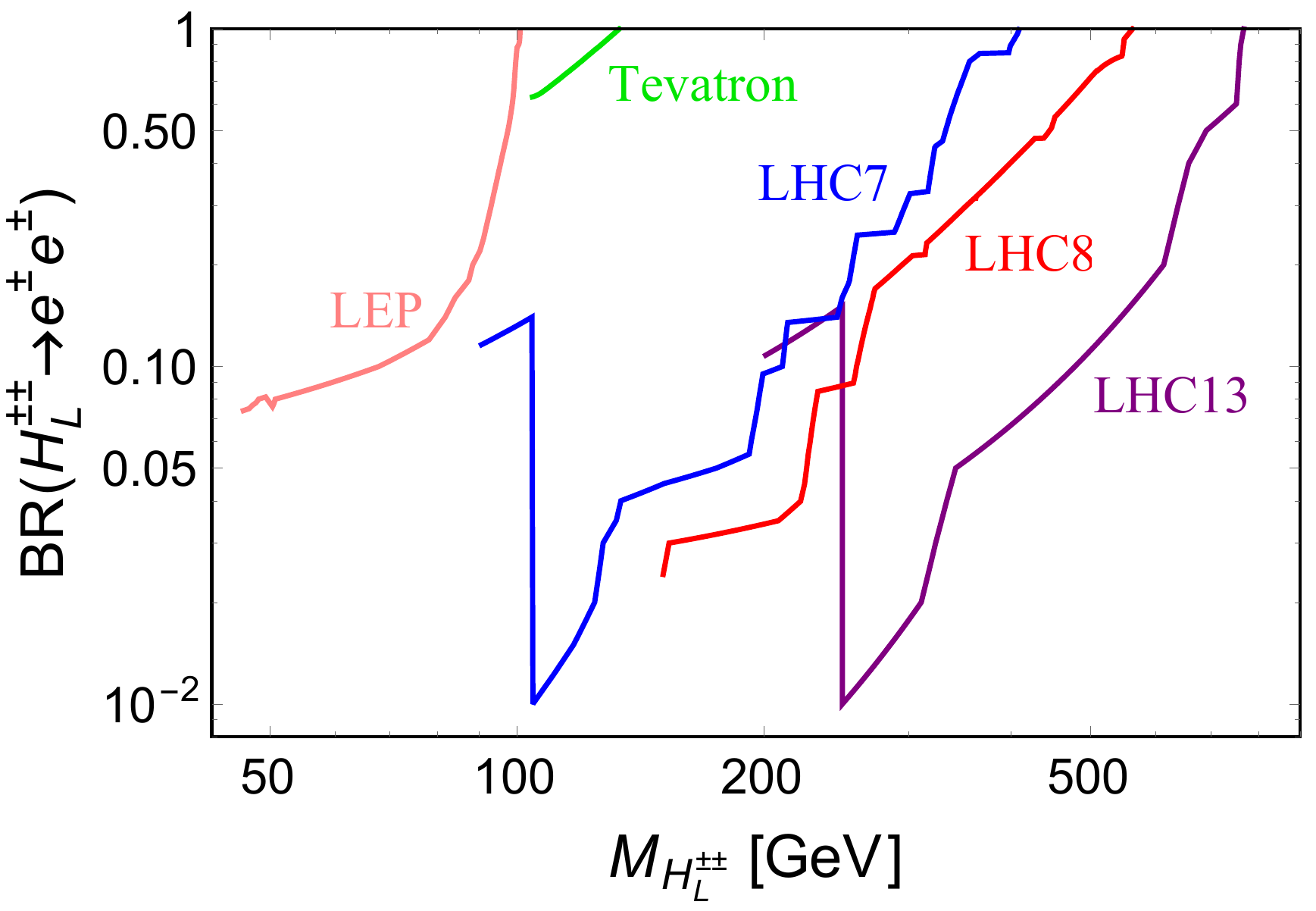}
  \includegraphics[width=0.48\textwidth]{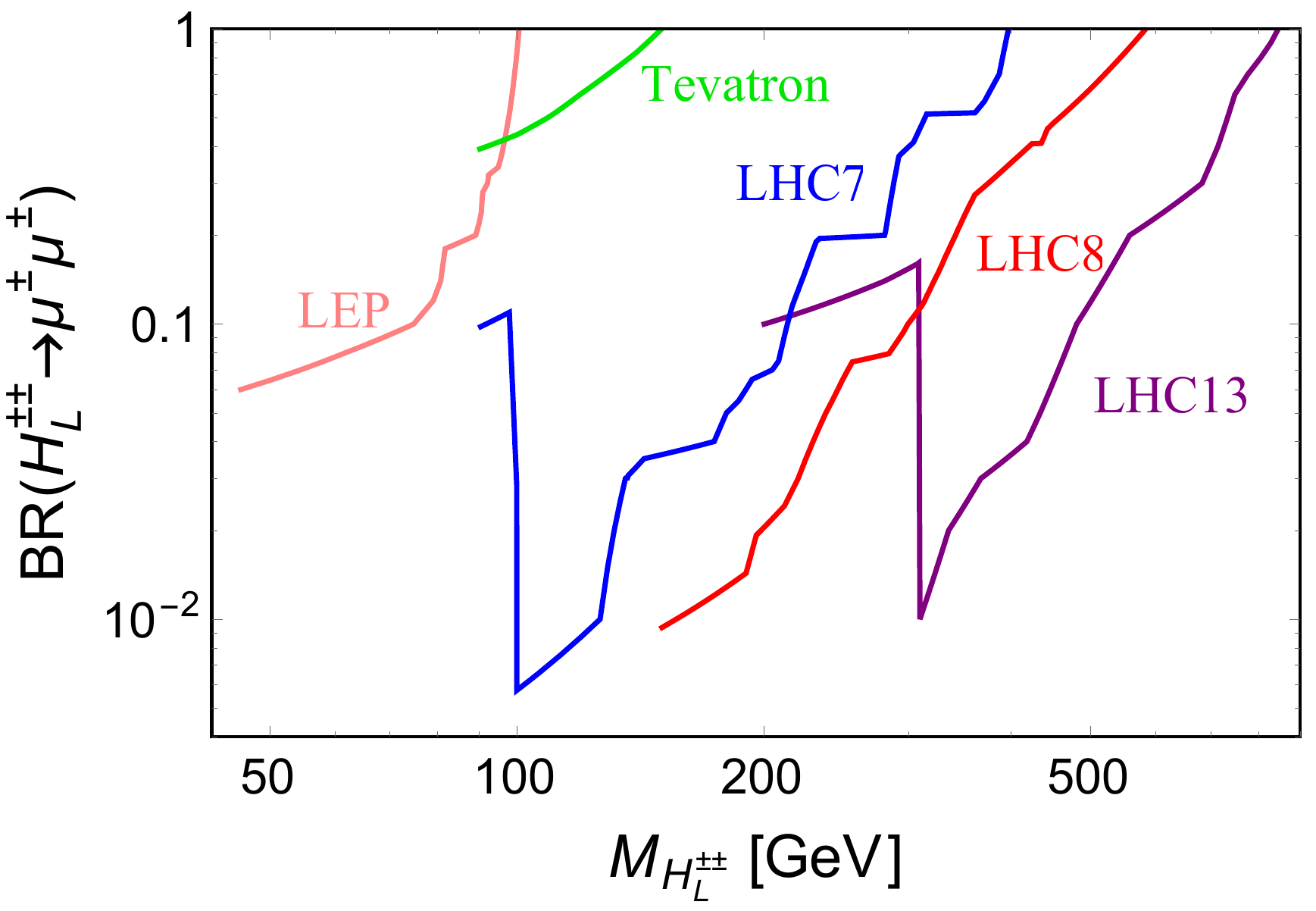}
  \includegraphics[width=0.48\textwidth]{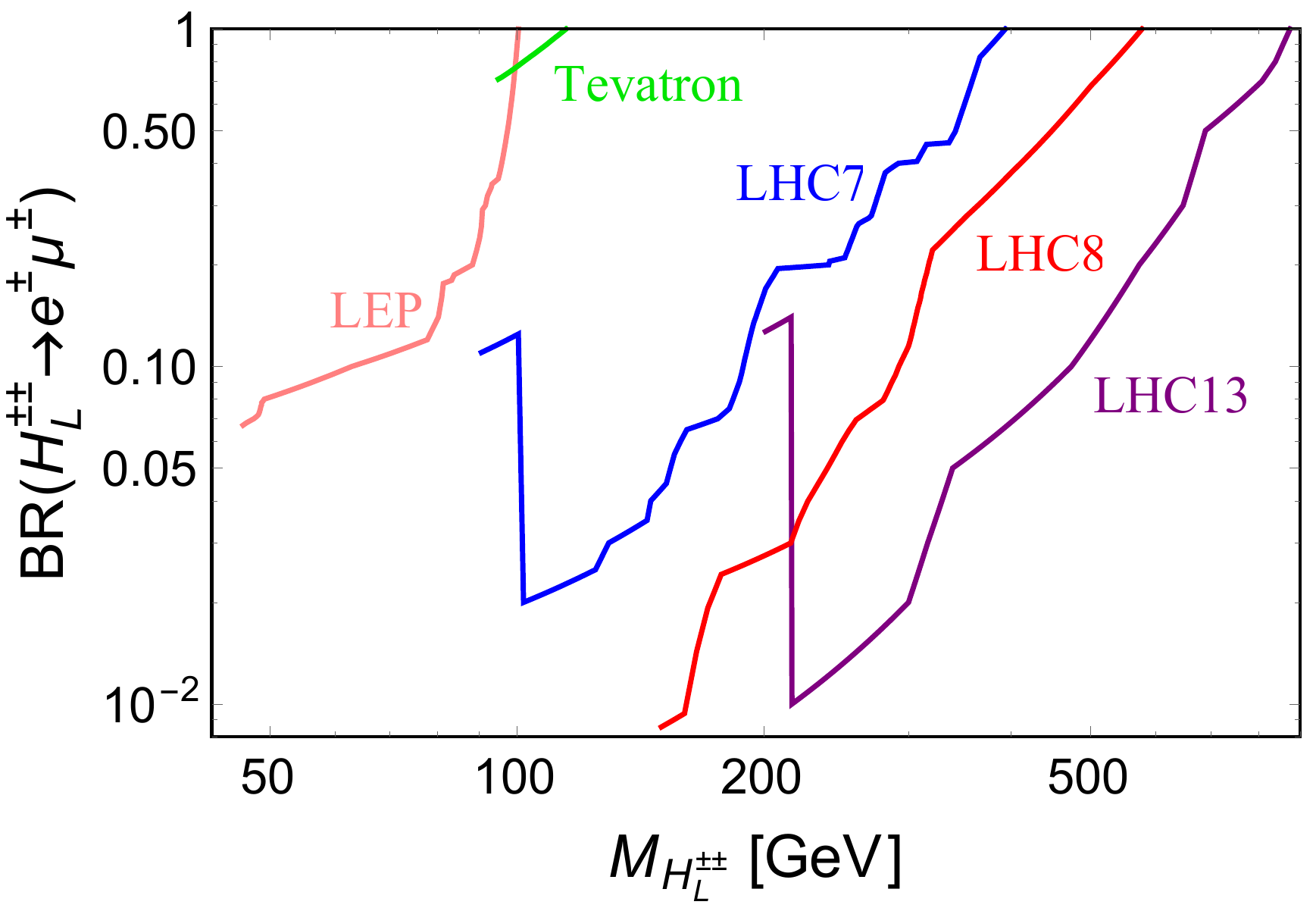}
  \includegraphics[width=0.48\textwidth]{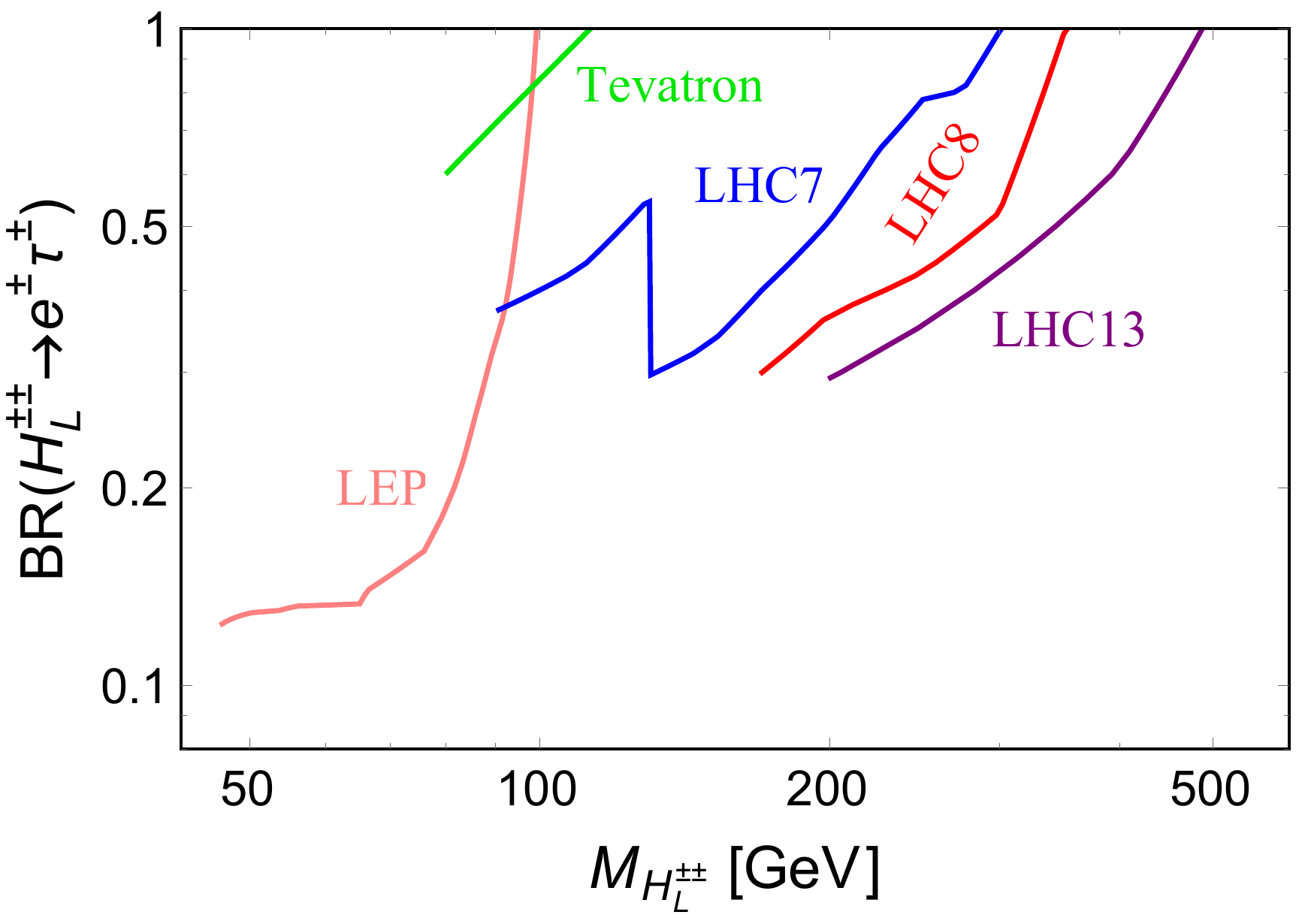}
  \includegraphics[width=0.48\textwidth]{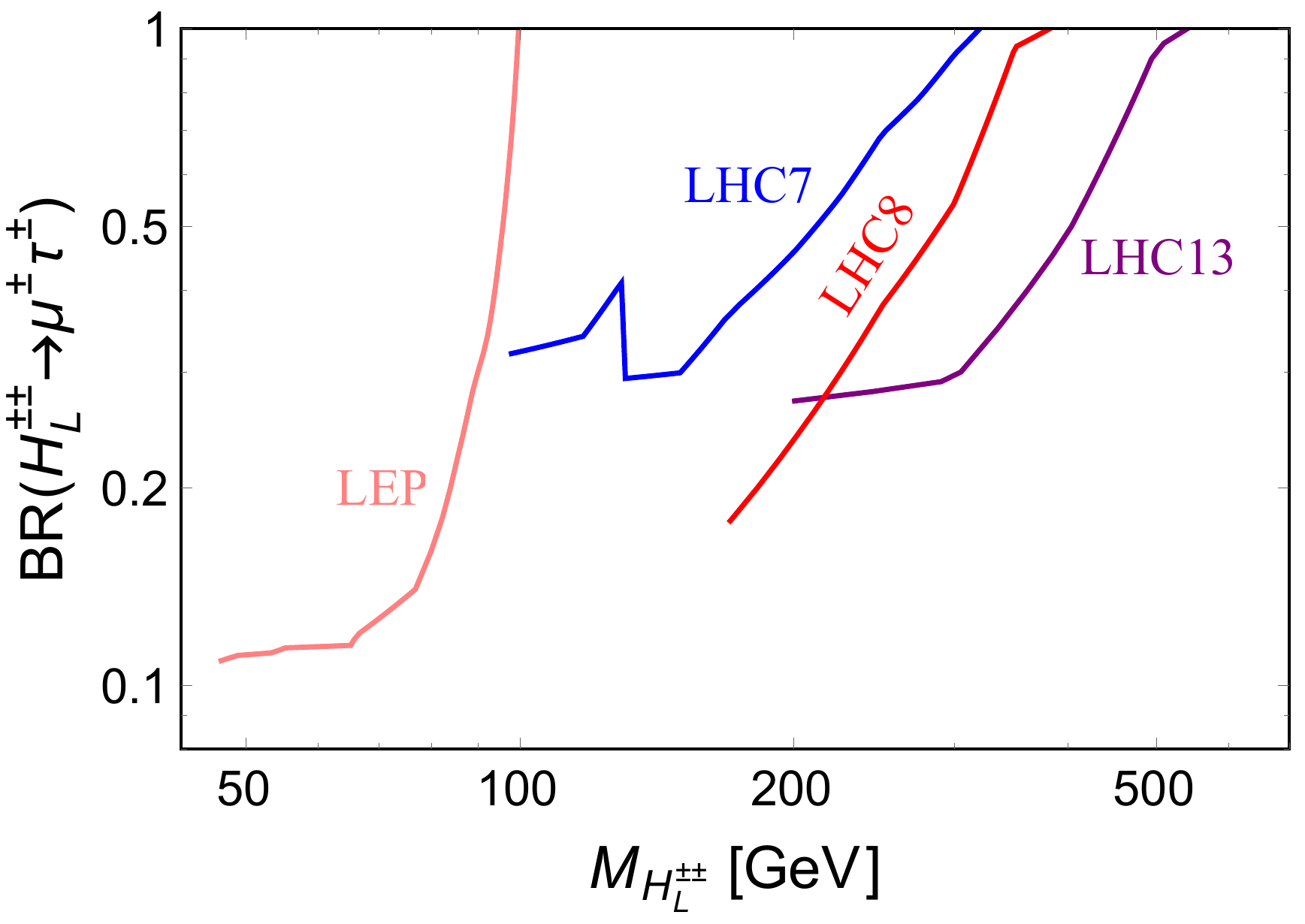}
  \includegraphics[width=0.48\textwidth]{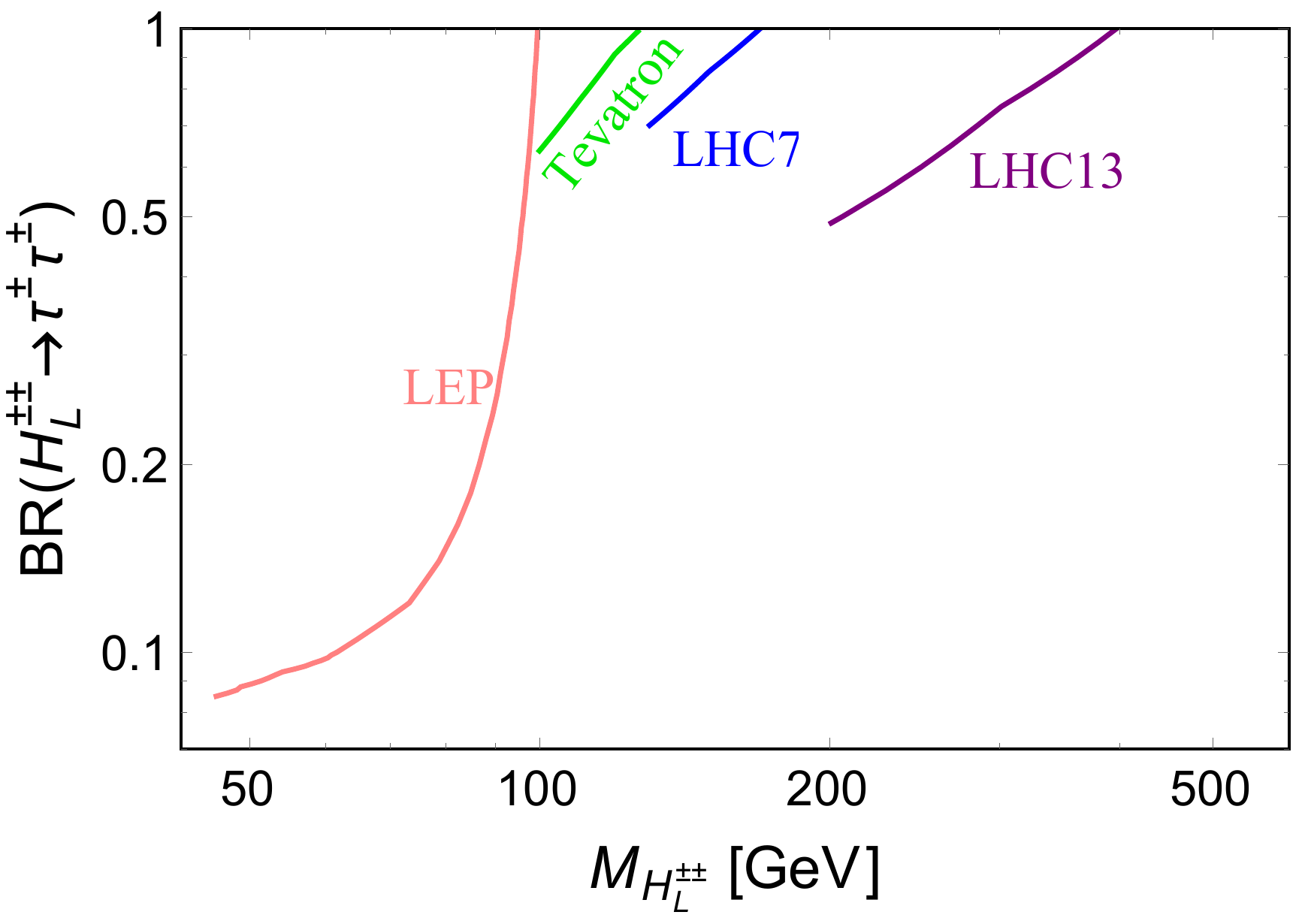}
  \caption{90\% CL lower limits on $M_{H_L^{\pm\pm}}$ in the type-II seesaw as functions of ${\rm BR} (H_L^{\pm\pm} \to \ell_\alpha^\pm \ell_\beta^\pm)$, in all the six flavor combinations of $\ell_\alpha \ell_\beta = ee$ (upper left), $\mu\mu$ (upper right), $e\mu$ (middle left), $e\tau$ (middle right), $\mu\tau$ (lower left) and $\tau\tau$ (lower right) using the data from LEP~\cite{Abbiendi:2001cr, Achard:2003mv, Abdallah:2002qj}, Tevatron~\cite{Aaltonen:2008ip, Acosta:2004uj, Abazov:2008ab, Abazov:2011xx} and LHC 7 TeV~\cite{ATLAS:2011rha, CMS:2011sqa}, 8 TeV~\cite{ATLAS:2014kca, CMS:2016cpz} and 13 TeV~\cite{Aaboud:2017qph, CMS:2017pet}.}
  \label{fig:dilepton:left:1}
\end{figure}

Given the gauge interactions to the SM photon and $Z$ bosons, the doubly-charged scalar can be pair produced from the electron-positron and quark annihilation processes:
\begin{eqnarray}
\label{eqn:DY}
e^+ e^-,\, q \bar{q} \ \to \ \gamma^\ast/Z^\ast \ \to \  H^{++} H^{--} \,.
\end{eqnarray}
This leads to the strikingly clean signal of same-sign dilepton pair from prompt decay of each doubly-charged scalar: $H_L^{\pm\pm} \to \ell_\alpha^\pm \ell_\beta^\pm$, with potentially LFV signatures (for $\alpha\neq \beta$), which is almost SM background free. Direct same-sign dilepton pair searches of this kind have been performed at LEP~\cite{Abbiendi:2001cr, Achard:2003mv, Abdallah:2002qj}, Tevatron~\cite{Aaltonen:2008ip, Acosta:2004uj, Abazov:2008ab, Abazov:2011xx} and LHC 7 TeV~\cite{ATLAS:2011rha, CMS:2011sqa}, 8 TeV~\cite{ATLAS:2014kca, CMS:2016cpz} and 13 TeV~\cite{Aaboud:2017qph, CMS:2017pet}. All these limits are presented in Fig.~\ref{fig:dilepton:left:1}, as functions of $M_{H_L^{\pm\pm}}$ and the BRs into six distinct flavor combinations $\ell_\alpha \ell_\beta = ee,\, \mu\mu,\, e\mu,\, e\tau,\, \mu\tau,\, \tau\tau$.\footnote{Including the photon fusion process $\gamma\gamma \to H^{++} H^{--}$, these limits could be slightly strengthened~\cite{Babu:2016rcr}.} With more data taken at LHC 13 TeV and future 14 TeV and high-luminosity stages, the doubly-charged scalars could be probed up to about 1 TeV~\cite{Dev:2016dja, Mitra:2016wpr}.  Future 100 TeV hadron colliders like SPPC~\cite{Tang:2015qga} or FCC-hh~\cite{fcc-hh} could push the mass reach to beyond 5 TeV~\cite{Arkani-Hamed:2015vfh, Dev:2016dja, Contino:2016spe}.

Limited by the center-of-mass energy, the LEP data could only probe $H_L^{\pm\pm}$ up to the masses of $\sim$100 GeV in the pair production mode. Furthermore, in the data analysis of Refs.~\cite{Abbiendi:2001cr, Abdallah:2002qj} the Yukawa couplings are assumed to be larger than $10^{-7}$, otherwise the reconstruction efficiency of the charged leptons would be affected by the non-prompt decays of $H_L^{\pm\pm}$. The doubly-charged scalar has also been searched in the single production mode, via the process $e^+ e^- \to e^\mp e^\mp H^{\pm\pm}$~\cite{Abbiendi:2003pr}. Analogous searches have also been performed at the lepton-hadron collider HERA~\cite{Aktas:2006nu} in the process $e^+ p \to \ell^- p H^{++}$. The single production is dictated by the Yukawa interaction $(f_L)_{ee}$ but not the gauge couplings, thus these experimental data can be used to  directly constrain the Yukawa couplings, but not the BRs like ${\rm BR} (H^{\pm\pm} \to e^\pm \mu^\pm)$~\cite{Dev:2018upe}. Therefore, the limits on the Yukawa couplings derived in Ref.~\cite{Abbiendi:2003pr} are not shown in the plots of BR constraints in Fig.~\ref{fig:dilepton:left:1}.

Benefiting from the higher energy and larger luminosity, the $H_L^{\pm\pm}$ mass limits from the LHC data are much more stringent, up to $\sim$ 800 GeV in the $ee$, $e\mu$ and $\mu\mu$ channels and $\sim 500$ GeV if the $\tau$ lepton is involved. The BRs are probed up to $\sim 10^{-2}$ for lighter $H_L^{\pm\pm}$ in the $e$ and $\mu$ channels, whereas in the channels involving the $\tau$ flavor, the limits are much weaker, at most up to the level of 0.2. In hadron collisions, $H_L^{\pm\pm}$ could also be produced in association with the singly-charged scalar $H^\pm$, i.e. $pp \to W^{\pm\, \ast} \to H^\mp H_L^{\pm\pm}$. This depends however on the mass splitting $M_{H^\pm} - M_{H_L^{\pm\pm}}$~\cite{Chun:2012jw} and the decay of $H^\pm$~\cite{Perez:2008ha}, which involves the couplings in the scalar potential (\ref{eq:Vpd}).  For simplicity, the associated production channel is not considered in this paper, though the corresponding production cross section $\sigma (pp \to H^\mp H_L^{\pm\pm})$ tends to be larger than that of the Drell-Yan pair production cross section $\sigma (pp \to H_L^{++}H_L^{--})$.

\begin{figure}[t!]
  \centering
  \includegraphics[width=0.48\textwidth]{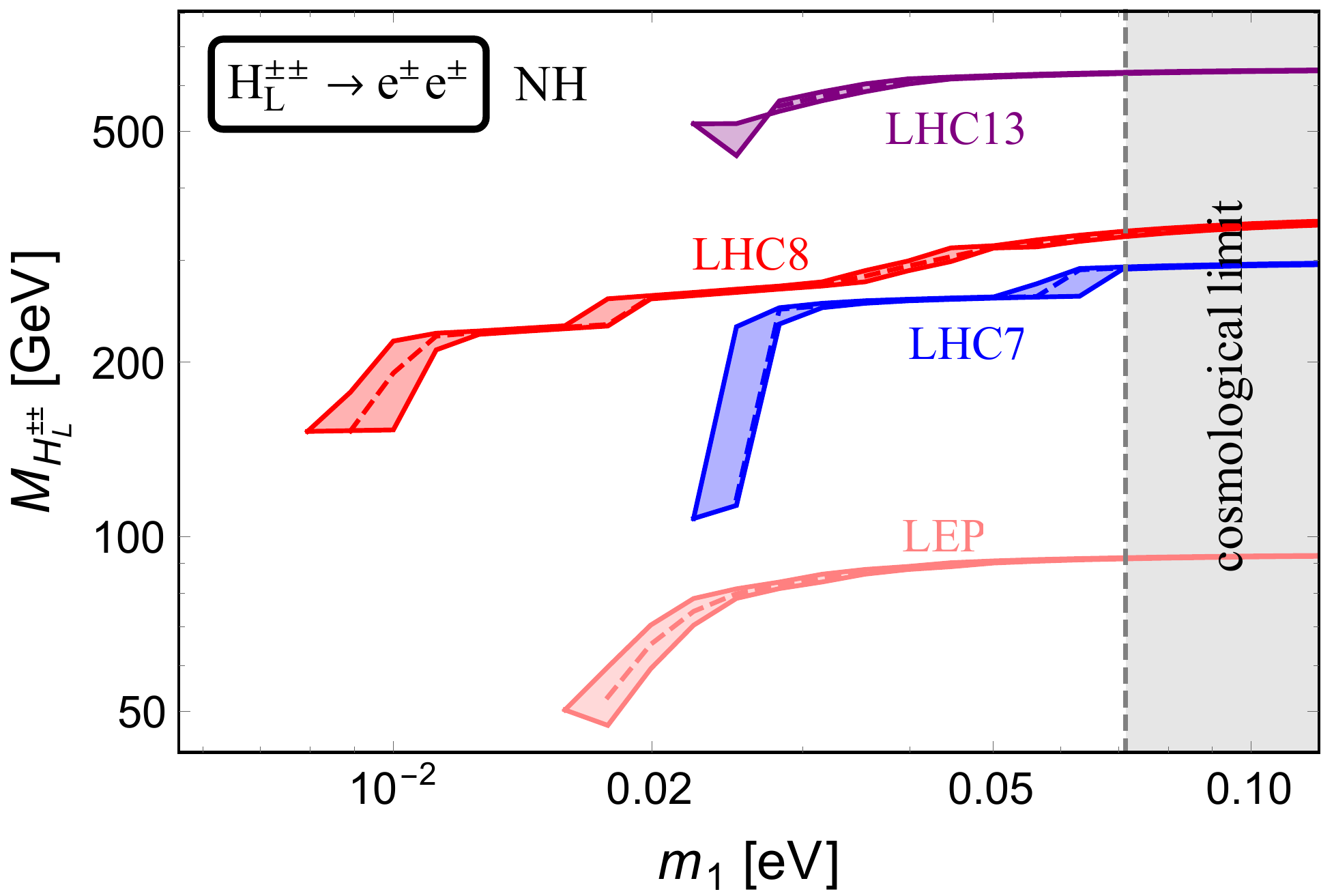}
  \includegraphics[width=0.48\textwidth]{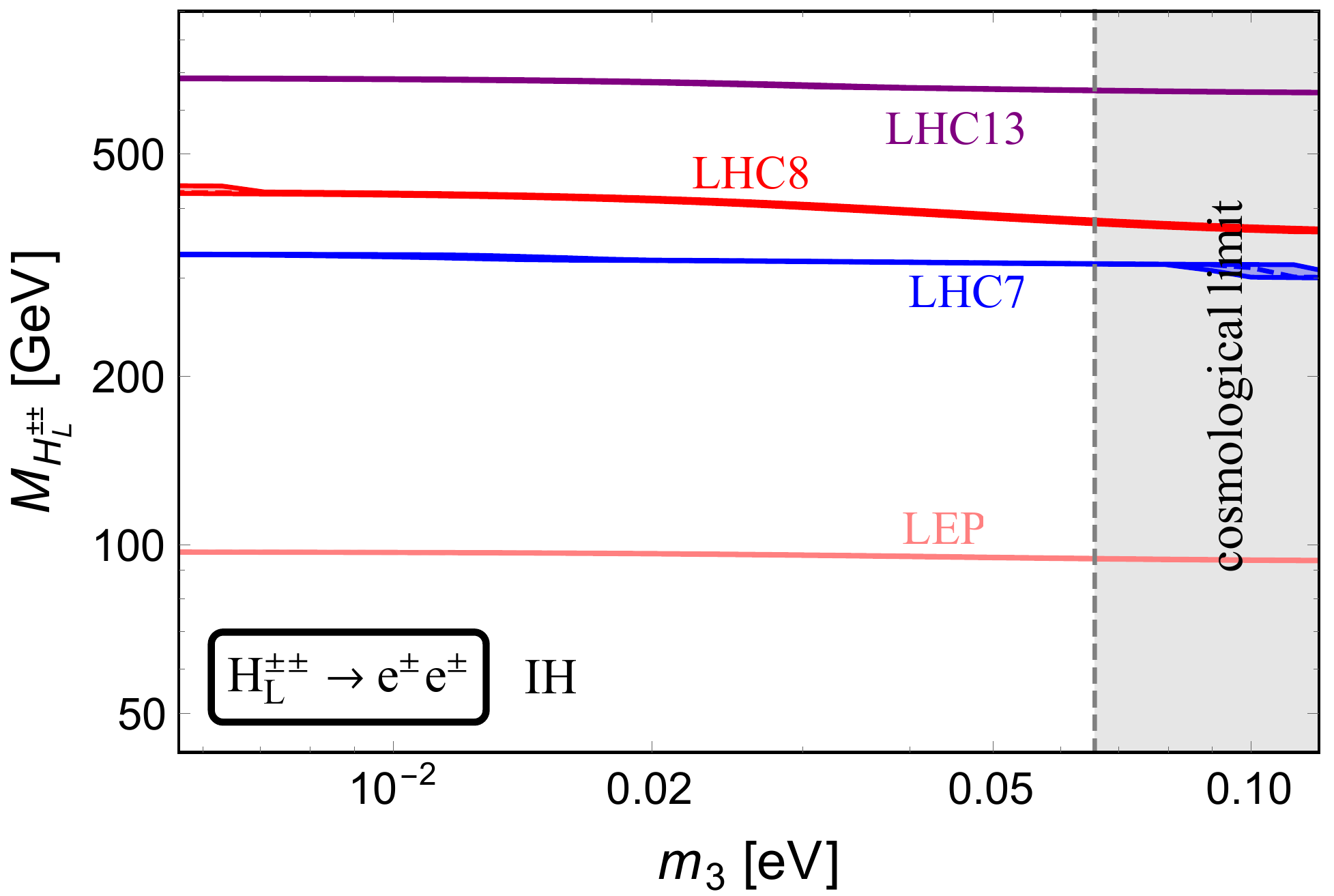} \vspace{3pt} \\
  \includegraphics[width=0.48\textwidth]{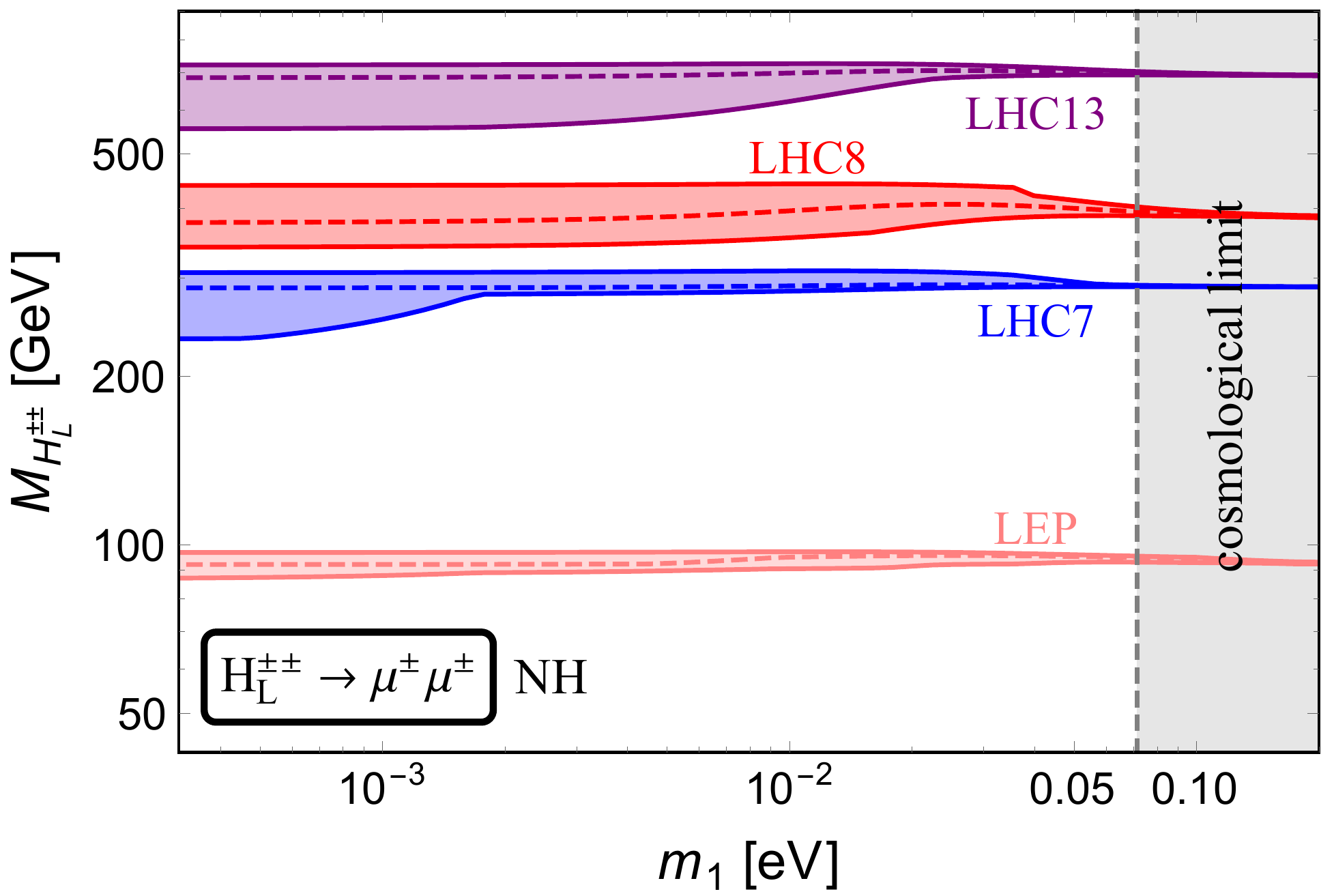}
  \includegraphics[width=0.48\textwidth]{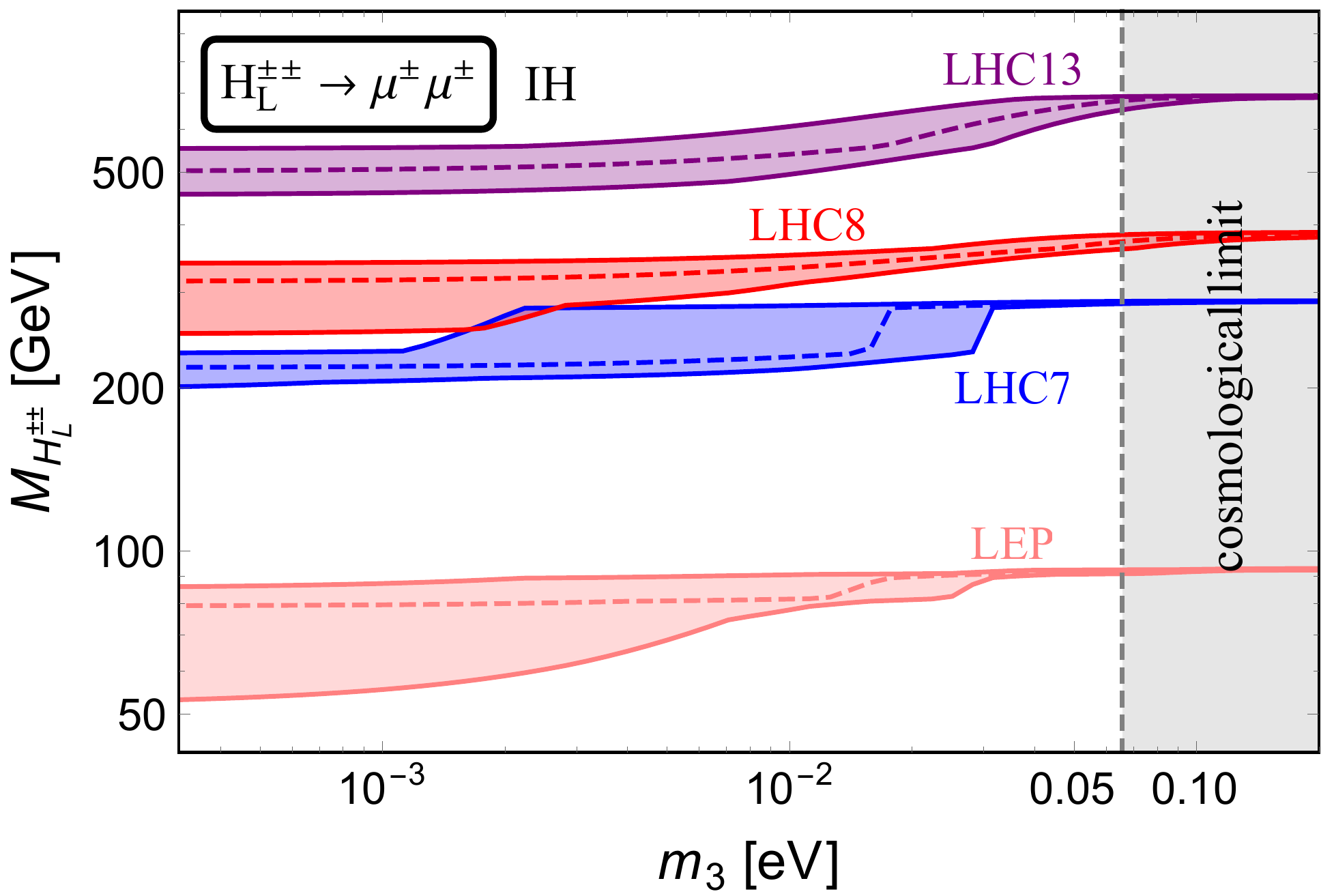} \vspace{3pt} \\
  \includegraphics[width=0.48\textwidth]{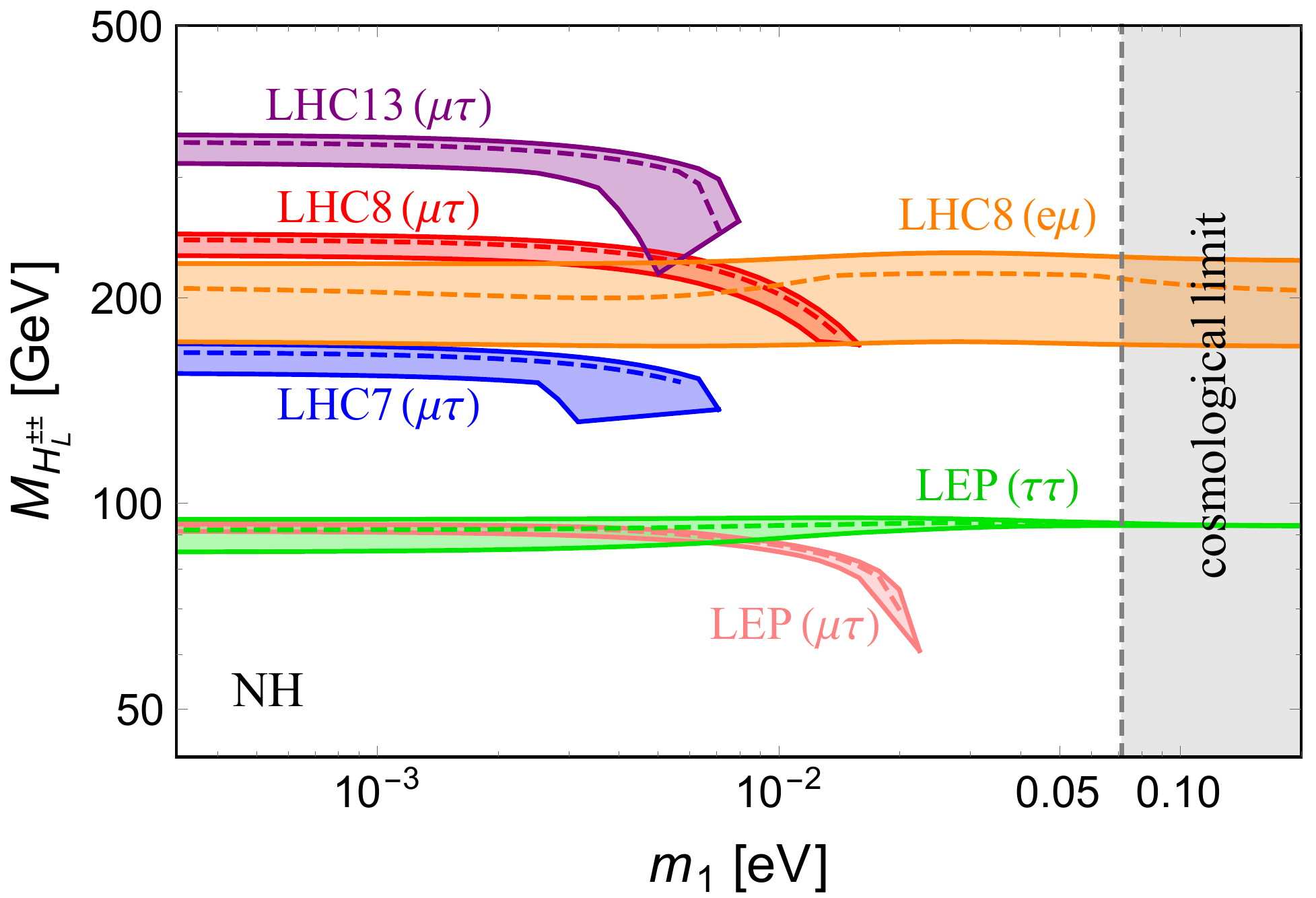}
  \includegraphics[width=0.48\textwidth]{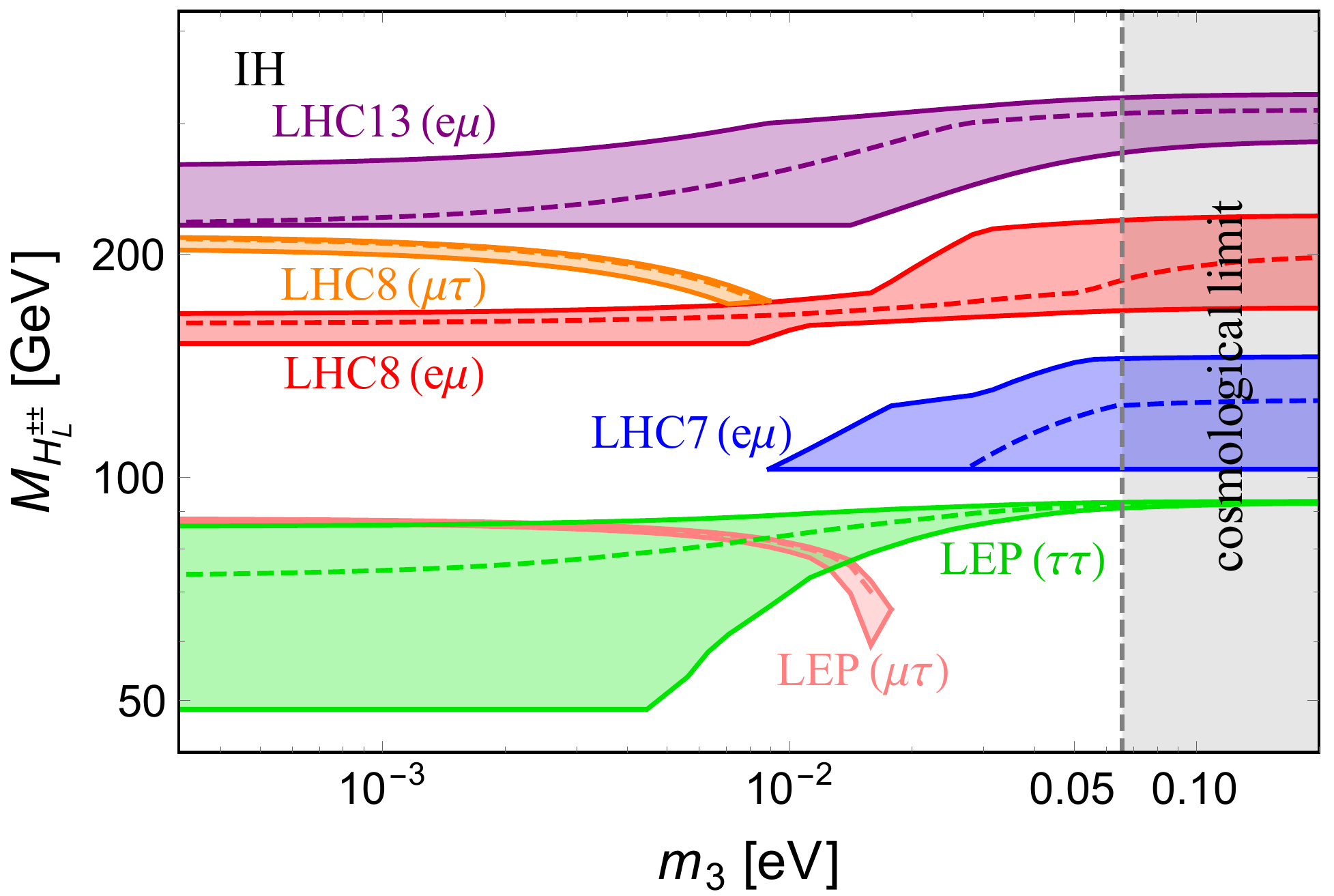}
  \caption{Lower limits on $M_{H_L^{\pm\pm}}$ in the type-II seesaw, as functions of the lightest neutrino mass  for NH (left) and IH (right), for different flavor combinations $H_{L}^{\pm\pm} \to \ell_\alpha^\pm \ell_\beta^\pm$, derived from the data in Fig.~\ref{fig:dilepton:left:1}, in the small $v_L$ limit [cf.~Eq.~\eqref{eqn:BR}]. The dashed curves corresponds to the central values of neutrino data in Table~\ref{tab:neutriodata}, and the colorful shaded bands are due to the $3\sigma$ uncertainties. The gray shaded region is excluded by the cosmological constraint on the sum of light neutrino masses $\sum_i m_{i} < 0.23$ eV~\cite{Ade:2015xua}.}
  \label{fig:dilepton:left:2}
\end{figure}

As shown in Eqs.~(\ref{eqn:width1}) and (\ref{eqn:width2}) and in Fig.~\ref{fig:lifetime1}, the triplet VEV $v_L$ plays a crucial rule in determining the BRs of $H_L^{\pm\pm}$ into same-sign leptons and $W$ boson pairs. In the limit of small $v_L$, i.e. $v_L\lesssim 0.1$ MeV~\cite{Kanemura:2014goa}, the $W$ pair channel is suppressed  and $H_L^{\pm\pm}$ decays predominantly into the same-sign dileptons. Since the Yukawa coupling $f_L$ is related to the neutrino mass matrix $m_\nu$ via Eq.~(\ref{eq:neutrino}), the leptonic BRs can be readily obtained from Eq.~(\ref{eqn:width1}) in terms of the neutrino masses:
\begin{eqnarray}
\label{eqn:BR}
{\rm BR} (H_L^{\pm\pm} \to \ell_\alpha^\pm \ell_\beta^\pm) \ = \
\frac{|(m_\nu)_{\alpha\beta}|^2}{\sum_{\alpha \leqslant \beta} (1+\delta_{\alpha\beta}) |(m_\nu)_{\alpha\beta}|^2} \quad
\text{(small $v_L$ limit)} \,.
\end{eqnarray}
Then the constraints in Fig.~\ref{fig:dilepton:left:1} can be translated into limits on $M_{H_L^{\pm\pm}}$ as functions of the unknown lightest neutrino mass $m_0$, which are all collected in Fig.~\ref{fig:dilepton:left:2} for both NH (left) and IH (right) cases. Though the neutrino mass squared difference and mixing data in Table~\ref{tab:neutriodata} are rather precise, when the uncertainties are taken into account, some of the decay BRs in Eq.~(\ref{eqn:BR}) like ${\rm BR} (H_L^{\pm\pm} \to \mu^\pm \mu^\pm)$ might vary significantly. Therefore, we consider the central values of the neutrino oscillation parameters as shown in Table~\ref{tab:neutriodata}, as well as their $3\sigma$ uncertainties, and take the whole range of $[0,\,2\pi]$ for the Dirac CP phase $\delta_{\rm CP}$. In Fig.~\ref{fig:dilepton:left:2} the dashed curves correspond to the central values of neutrino data, while the shaded bands are due to the $3\sigma$ uncertainties. The gray shaded region in these plots is excluded by the cosmological limit on the sum of light neutrino masses $\sum_i m_{i} < 0.23$ eV~\cite{Ade:2015xua}.

We see from Fig.~\ref{fig:dilepton:left:2} that for both NH and IH the dilepton limits are the most stringent in the $ee$ (upper panels) and $\mu\mu$ (middle panels) channels, whereas those involving $\tau$ lepton are much less constraining, mainly limited by the $\tau$ lepton reconstruction efficiency at colliders. Similarly, the $e\mu$ channel is suppressed by the solar mixing angle ($\sin^2\theta_{12}$) when compared to the $ee$ and $\mu\mu$ decay modes.
{For the NH case, when the lightest neutrino mass gets small, say $m_1 \lesssim 0.01$ eV [cf. Eq.~\eqref{eqn:BR}], the branching fraction ${\rm BR} (H_L^{\pm\pm} \to e^\pm e^\pm)$ is so small that it goes out of the range of the LEP and LHC data (see the upper left panel in Fig.~\ref{fig:dilepton:left:1}). Thus there is no dilepton limit for $m_1 \lesssim 0.01$ eV in the $ee$ channel for the NH case, as shown in the upper left panel of Fig.~\ref{fig:dilepton:left:2}. Same thing happens for the $e\mu$ limits in the IH case (lower right panel).  By the same token, there is no limit in the $\mu\tau$ channel for $m_0 \gtrsim 0.01$  in both the NH and IH cases (cf. the two lower panels in Fig.~\ref{fig:dilepton:left:2}), as the dilepton limits in this channel is comparatively weaker than those without the tau lepton.}
%In addition, although the neutrino data in Table~\ref{tab:neutriodata} seems to be high precision, the mass limits  in Fig.~\ref{fig:dilepton:left:2} are still subject to  the experimental uncertainties in some of the channels, such as the $\mu\mu$ channel when the lightest neutrino mass $m_0$ is small.
The constraints from the Tevatron data in Fig.~\ref{fig:dilepton:left:1} are much weaker and are not shown in Fig.~\ref{fig:dilepton:left:2}.

\begin{figure}[t!]
  \centering
  \includegraphics[width=0.48\textwidth]{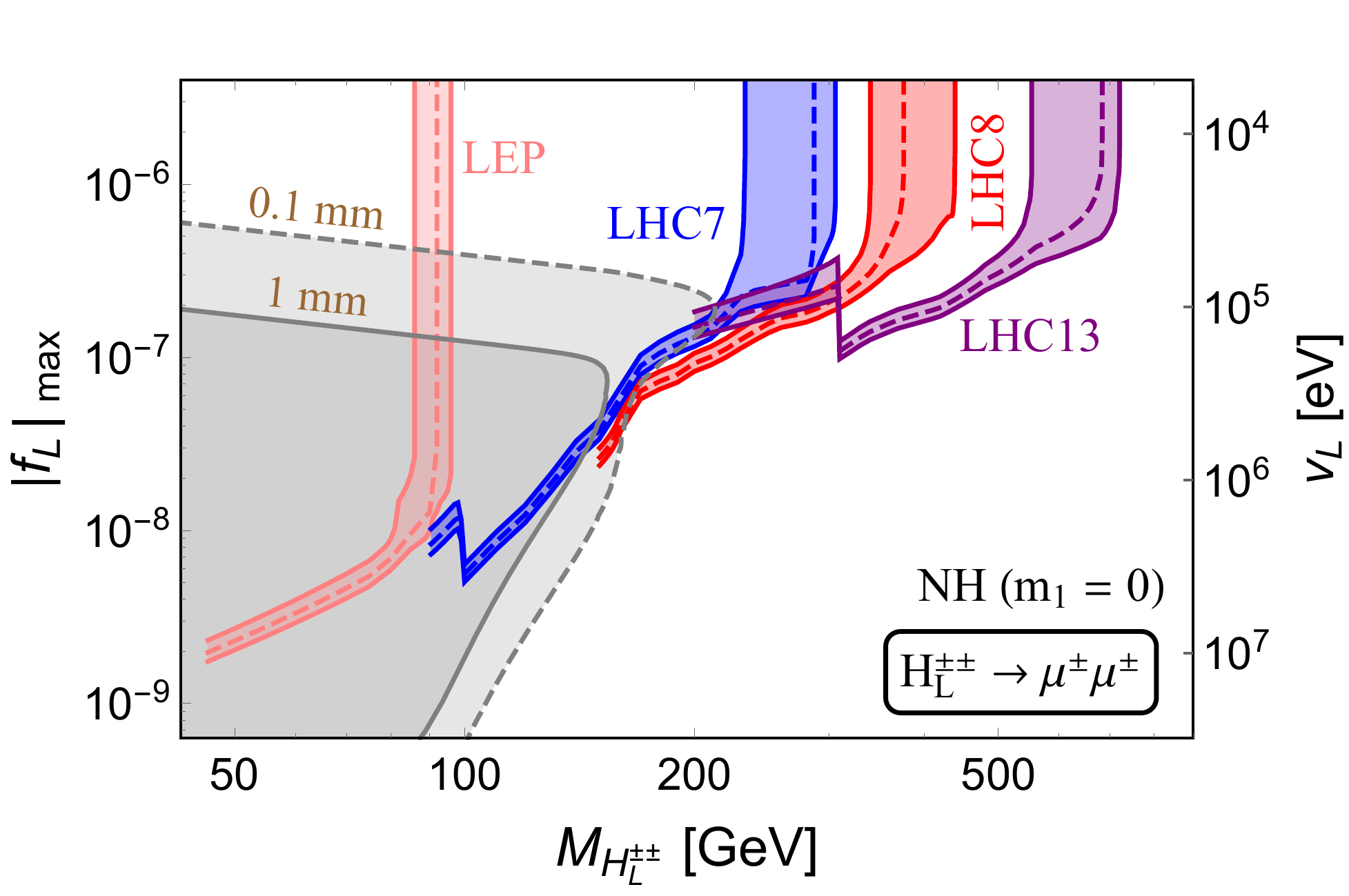}
  \includegraphics[width=0.48\textwidth]{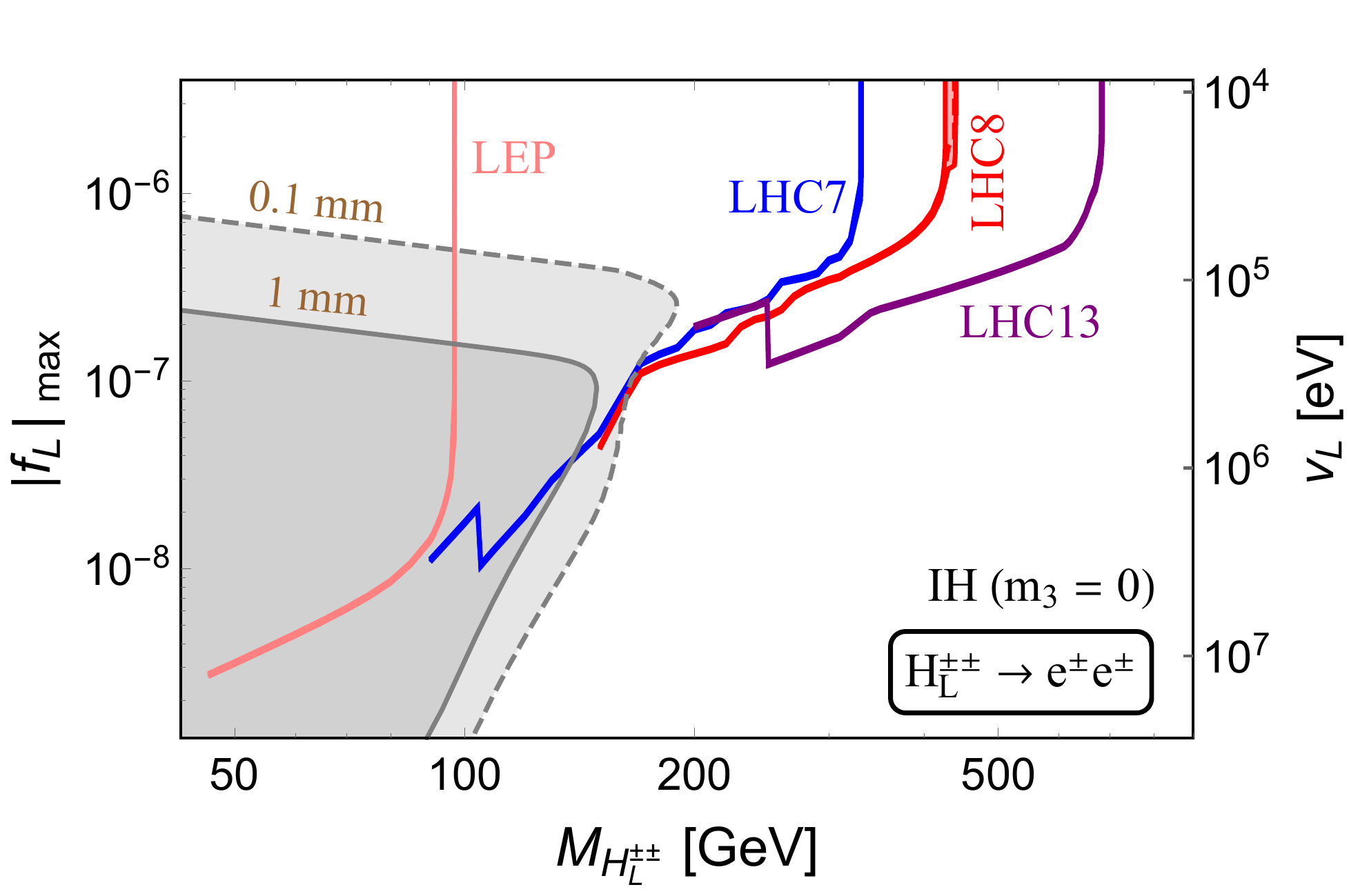} \vspace{-7pt} \\
  \includegraphics[width=0.48\textwidth]{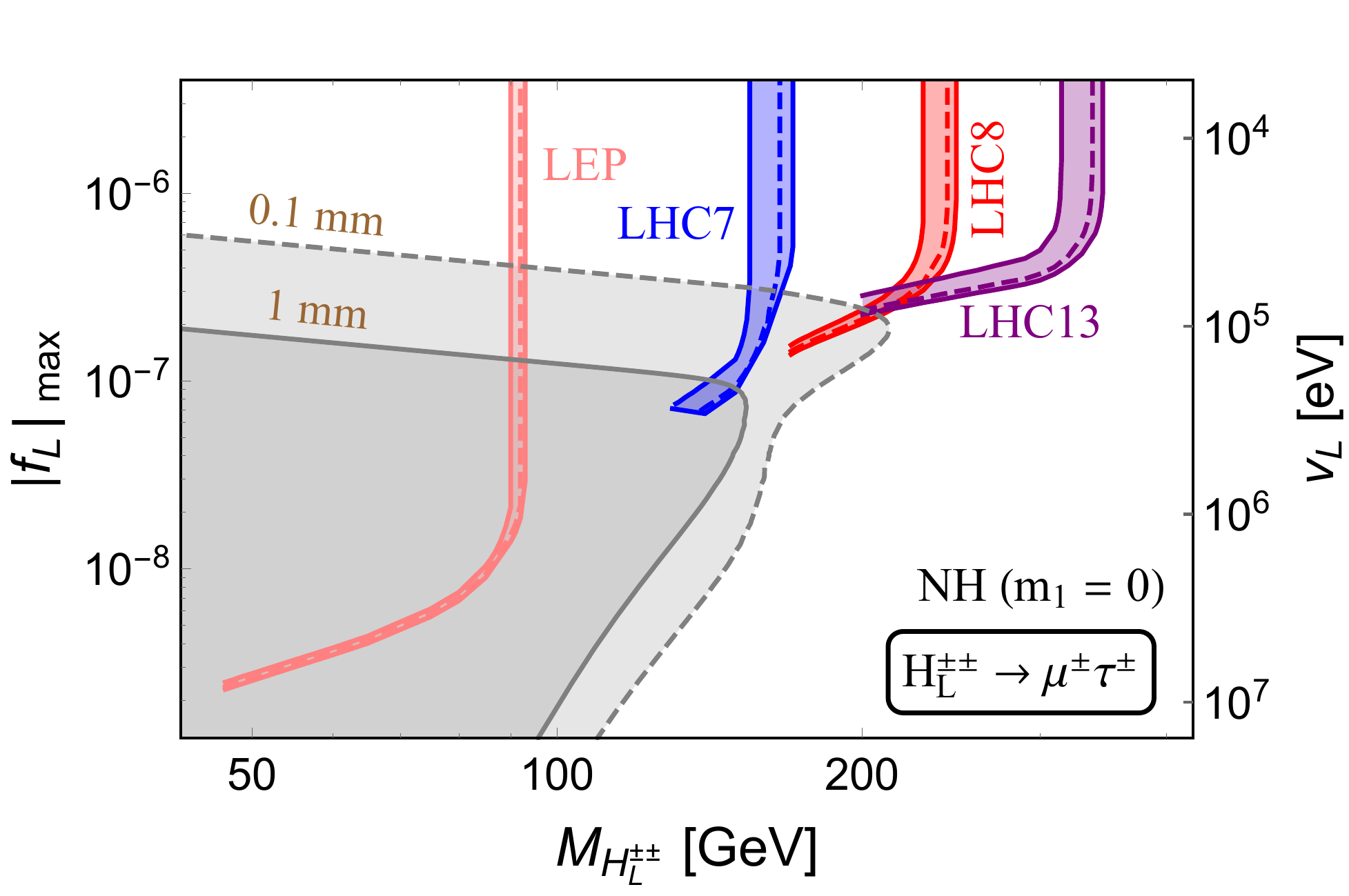}
  \includegraphics[width=0.48\textwidth]{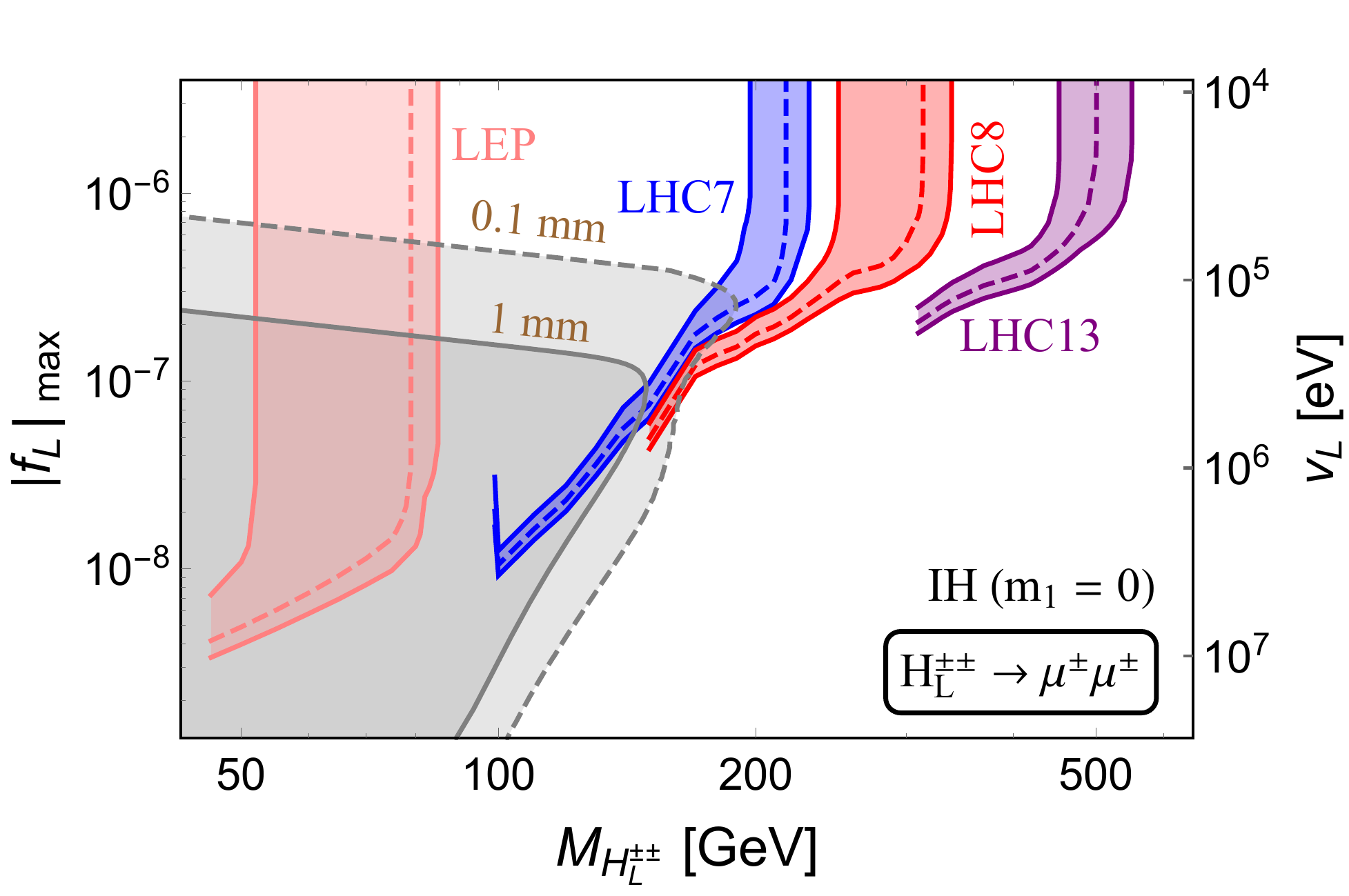} \vspace{-7pt} \\
  \includegraphics[width=0.48\textwidth]{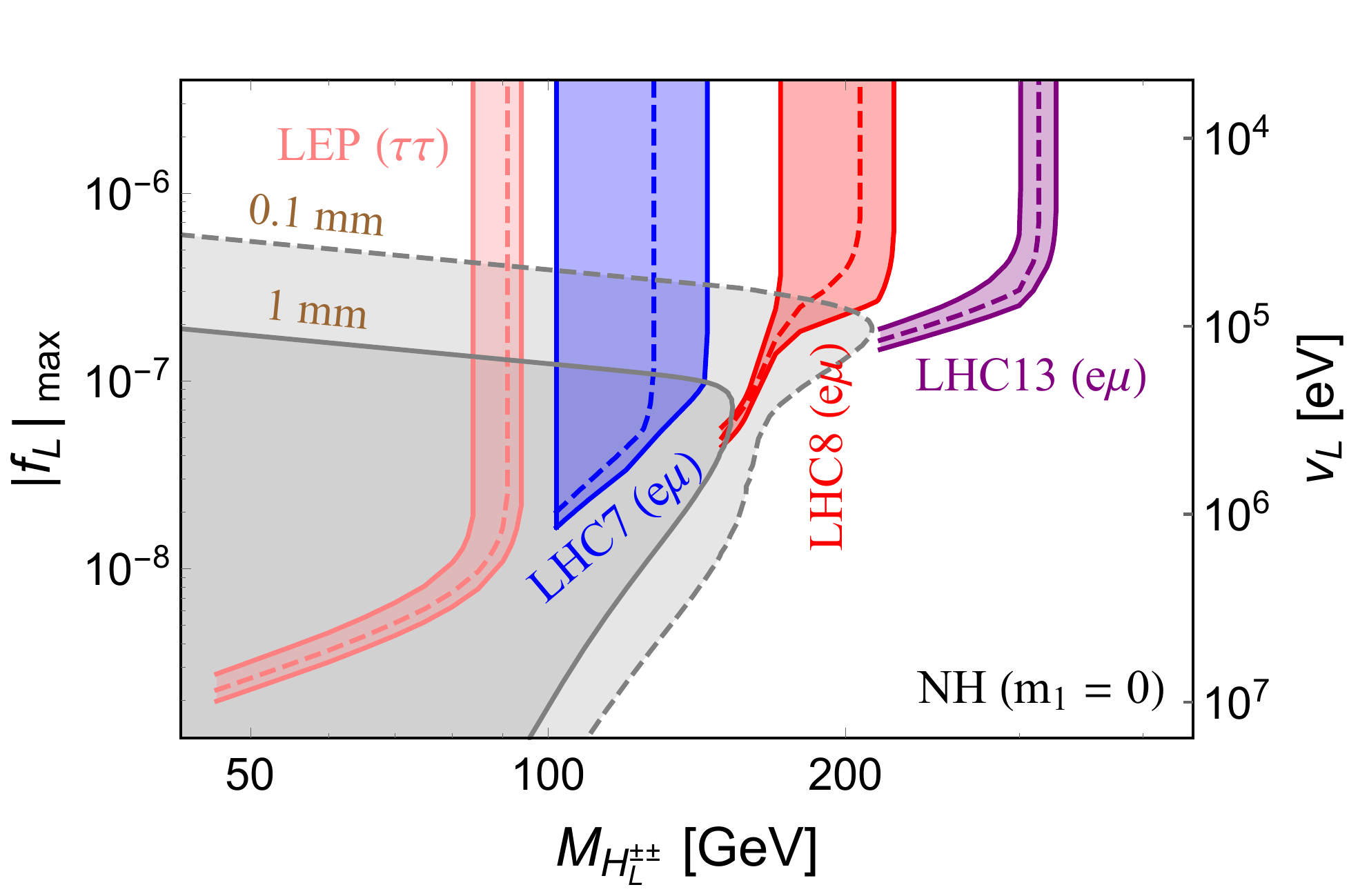}
  \includegraphics[width=0.48\textwidth]{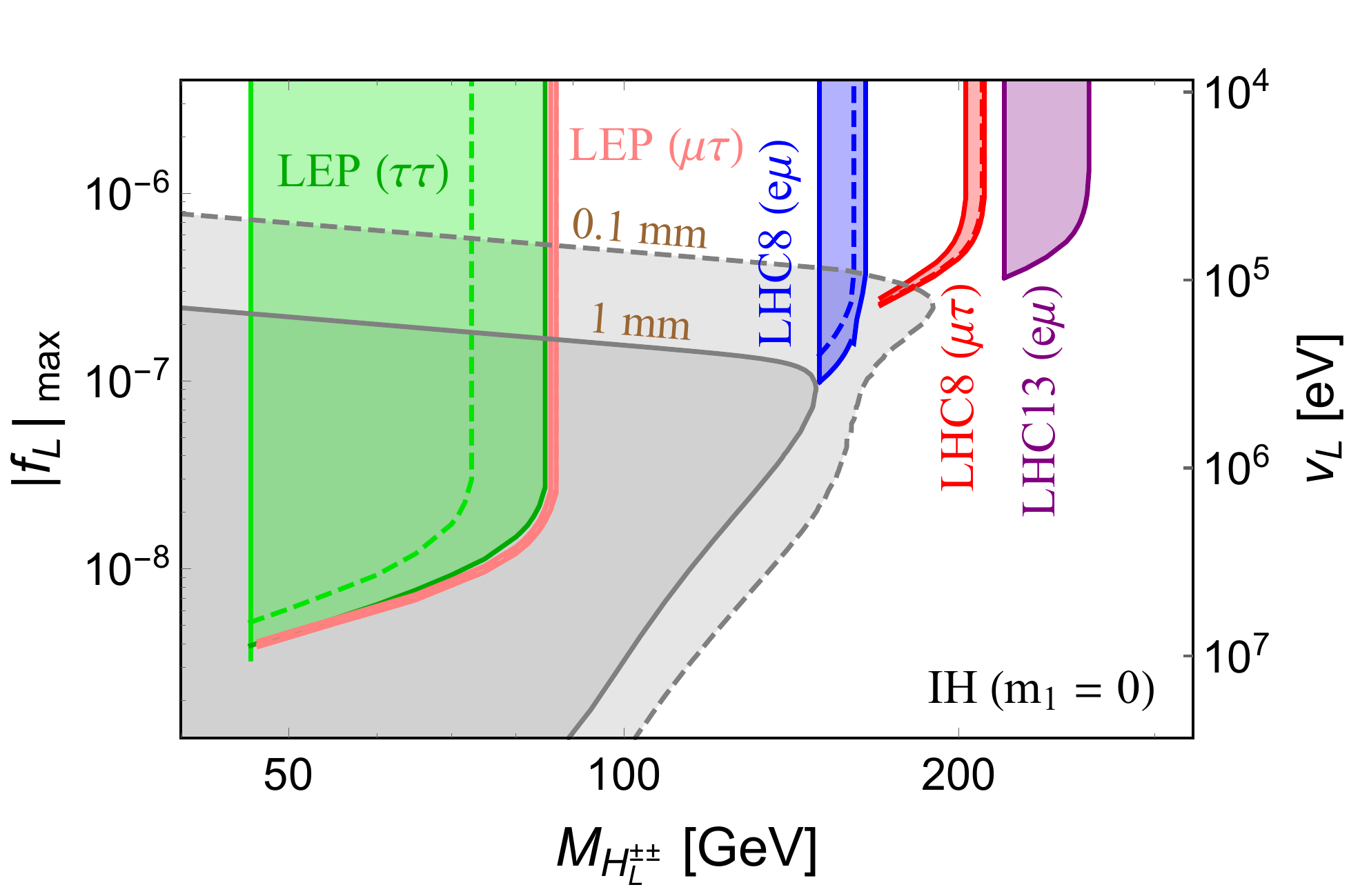}
  \caption{Lower limits on $M_{H_L^{\pm\pm}}$ in the type-II seesaw from the data in Fig.~\ref{fig:dilepton:left:1}, as functions of the value of largest Yukawa coupling $|f_L|_{\rm max}$. The left panels are for the NH case with $m_1 = 0$, in the $\mu\mu$ (upper left), $\mu\tau$ (middle left), $e\mu$ and $\tau\tau$ channels (lower left). The right panels are the limits for the IH case with $m_3=0$ in the $ee$ (upper right), $\mu\mu$ (middle right), $e\mu$, $\mu\tau$ and $\tau\tau$ channels (lower right). The dashed curves correspond to the central values of neutrino data in Table~\ref{tab:neutriodata}, and the colorful bands are due to the $3\sigma$ uncertainties. The corresponding lower limits on the VEV $v_L$ are also shown in these plots. The darker and lighter gray regions correspond to the proper decay lengths $c\tau_0> 1$ mm and 0.1 mm respectively; within these regions the prompt dilepton limits are not applicable. }
  \label{fig:dilepton:left:3}
\end{figure}

\begin{figure}[t!]
  \centering
  \includegraphics[width=0.48\textwidth]{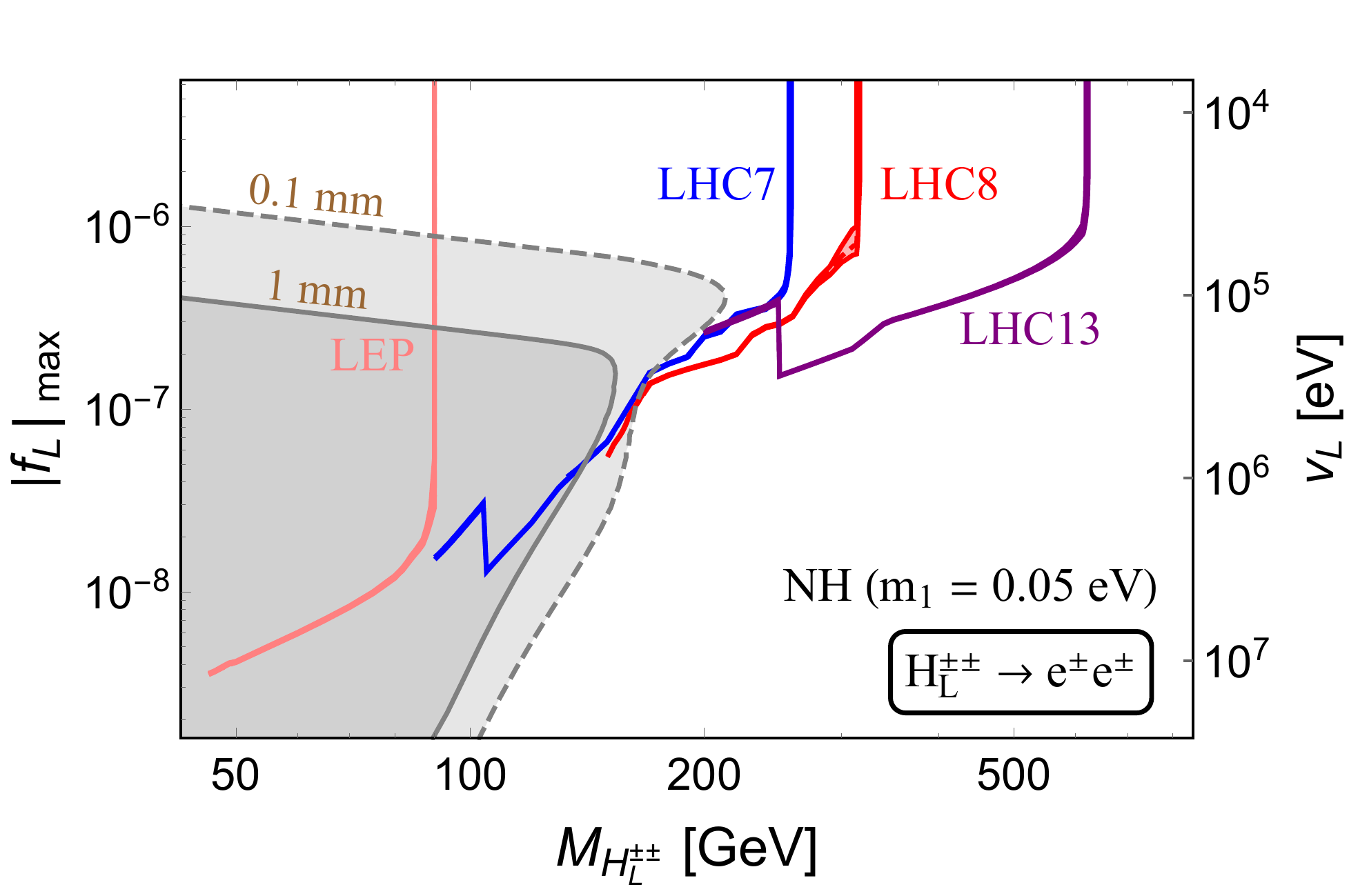}
  \includegraphics[width=0.48\textwidth]{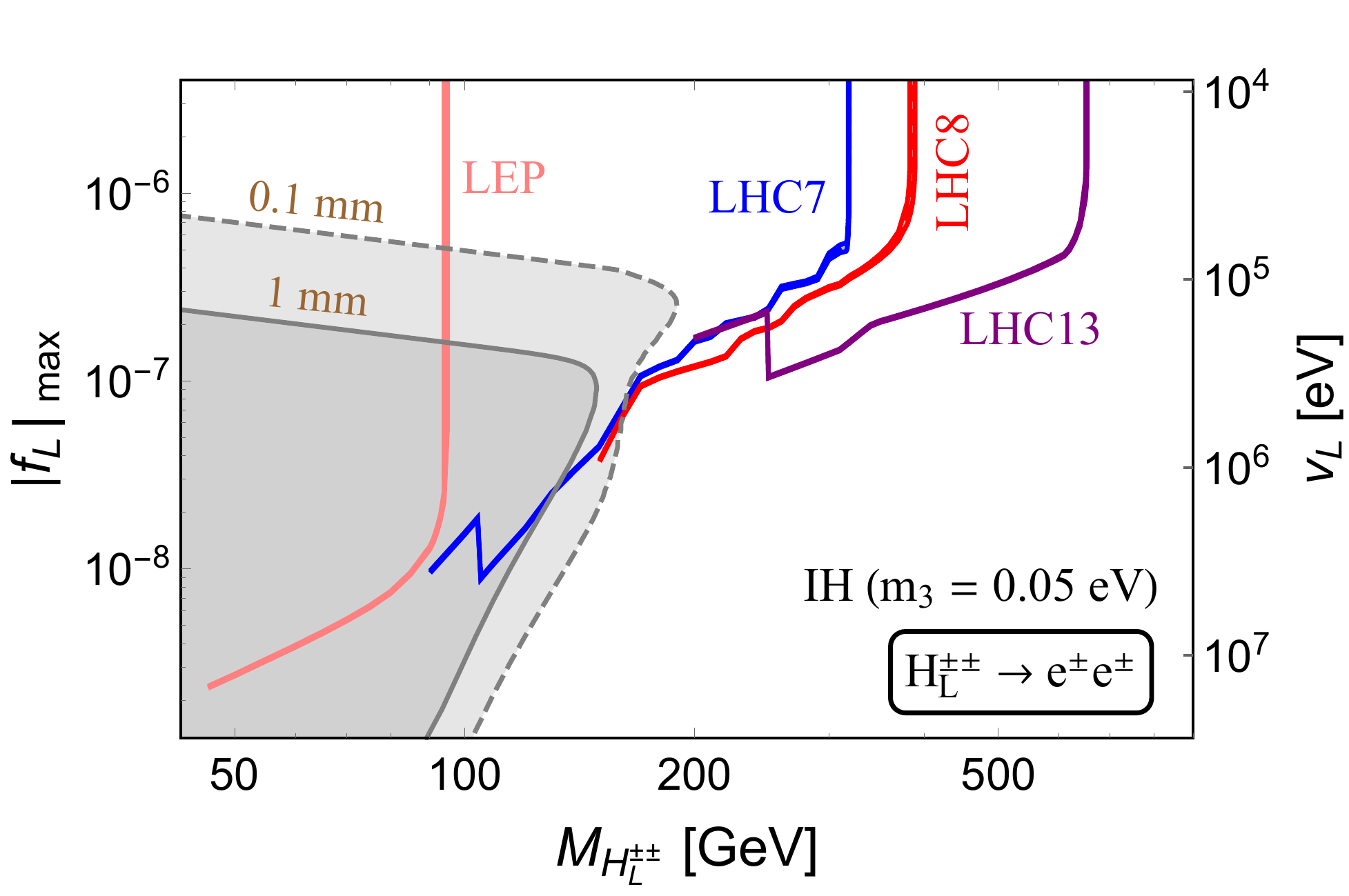} \vspace{-7pt} \\
  \includegraphics[width=0.48\textwidth]{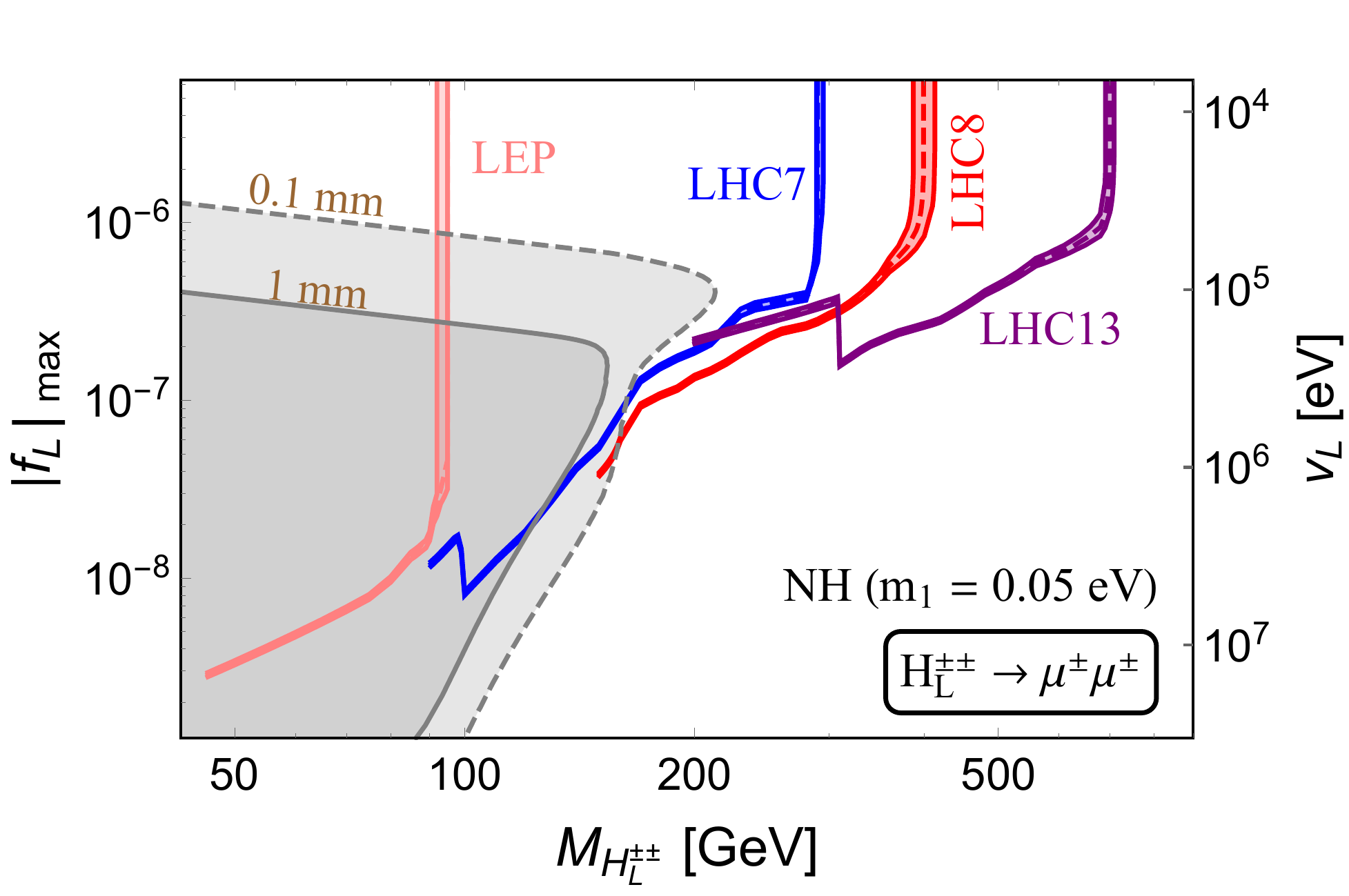}
  \includegraphics[width=0.48\textwidth]{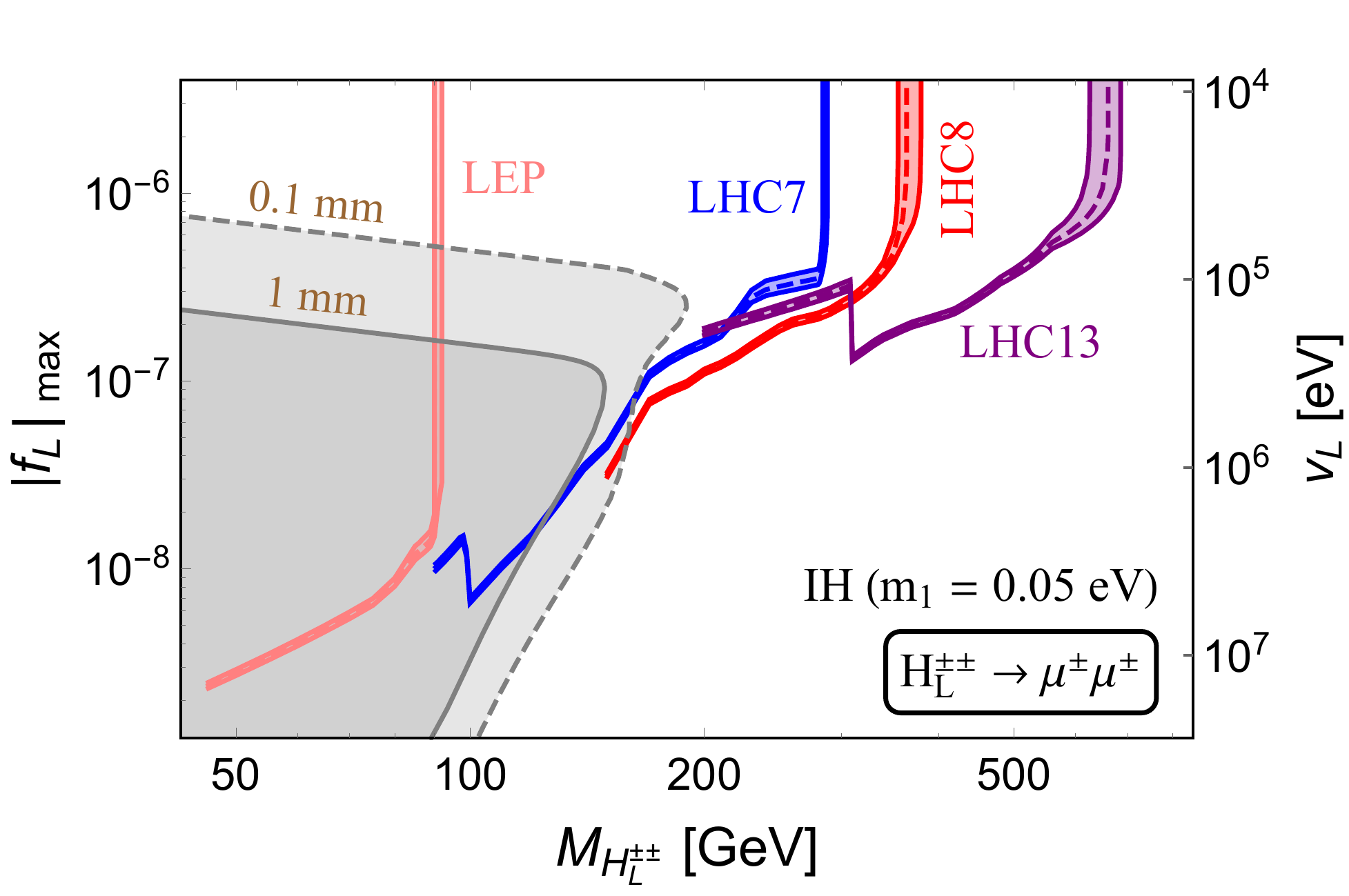} \vspace{-7pt} \\
  \includegraphics[width=0.48\textwidth]{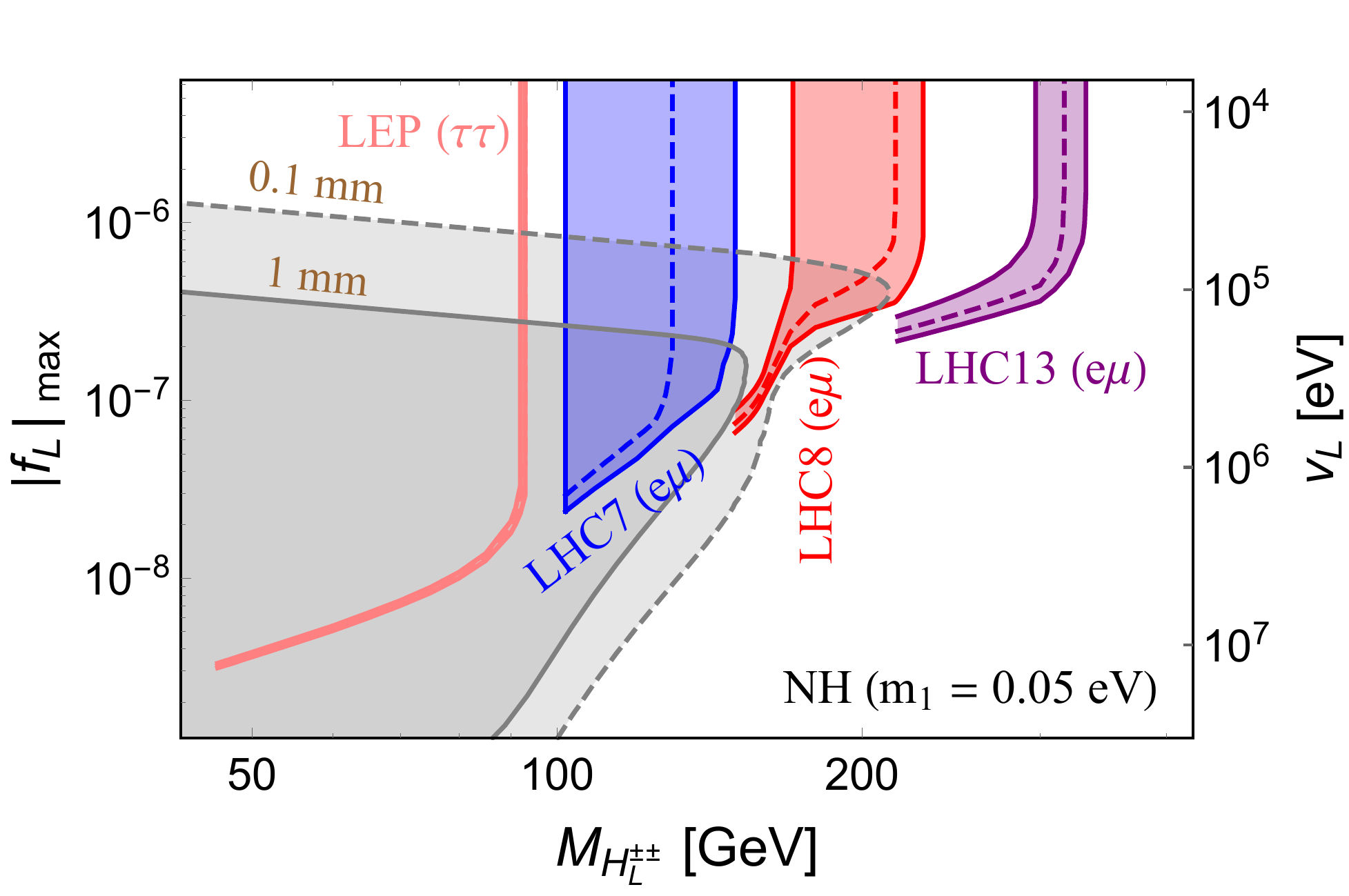}
  \includegraphics[width=0.48\textwidth]{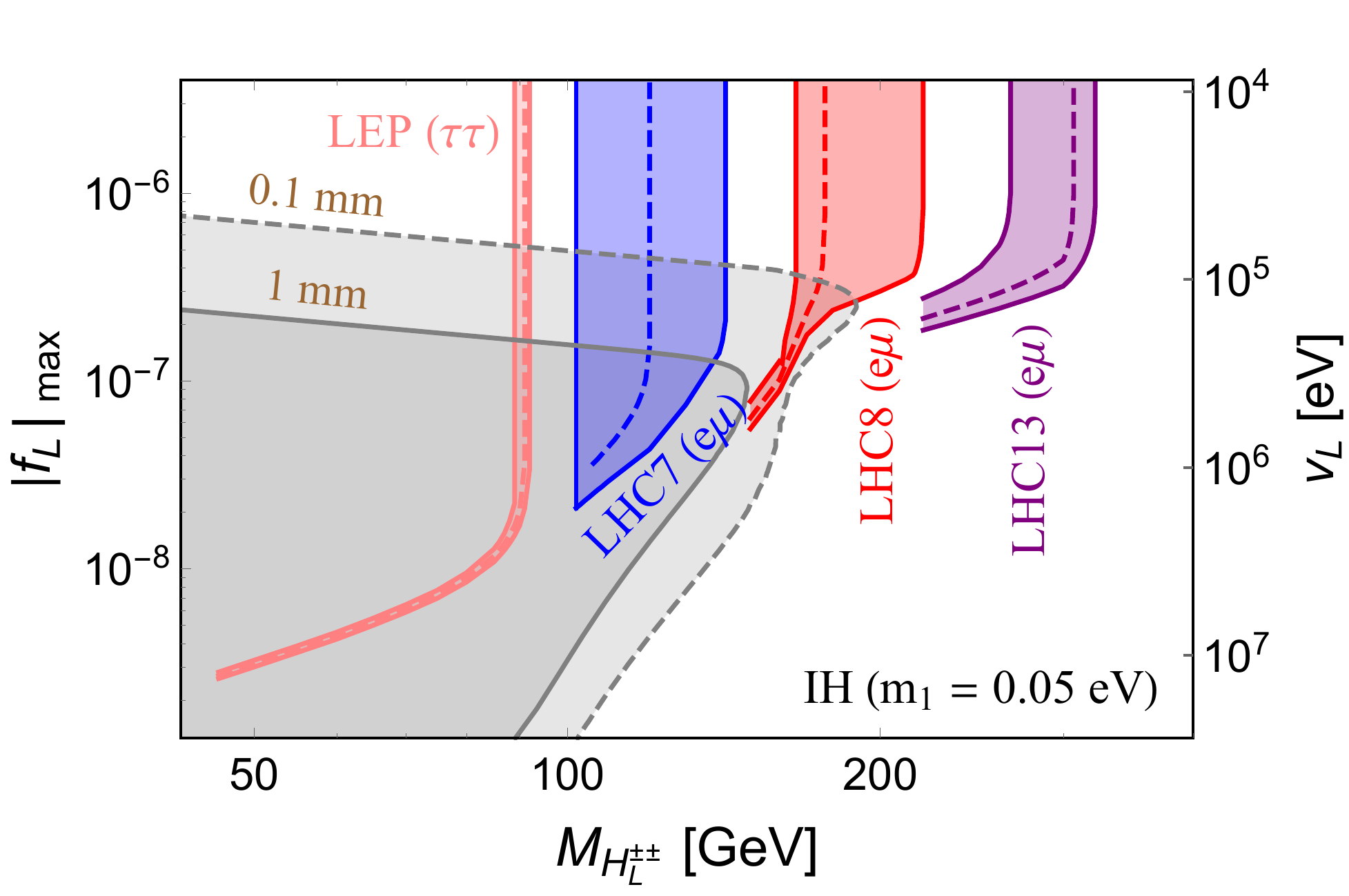}
  \caption{Same as in Fig.~\ref{fig:dilepton:left:3}, but for the NH case with $m_1 = 0.05$ eV (left) and IH with $m_3 = 0.05$ eV (right). The upper, middle and lower panels are respectively in the channels of $ee$, $\mu\mu$, $e\mu$ and $\tau\tau$ for both NH and IH scenarios.}
  \label{fig:dilepton:left:4}
\end{figure}

When the $W$ boson channel becomes important i.e. $\Gamma (H_L^{\pm\pm} \to W^{\pm\,(\ast)} W^{\pm\,(\ast)}) \gtrsim \Gamma (H_L^{\pm\pm} \to \ell_\alpha^\pm \ell_\beta^\pm)$, the dependence of $\Gamma (H_L^{\pm\pm} \to \ell_\alpha^\pm \ell_\beta^\pm)$ on the VEV $v_L$, or equivalently the dependence on the magnitudes of Yukawa couplings $(f_L)_{\alpha\beta}$, has to be taken into consideration. In this case, the leptonic branching fractions could be obtained from Eq.~(\ref{eqn:widthtotal}). For illustration purpose, the NH and IH scenarios with the lightest neutrino mass $m_{0} = 0$ are shown respectively in the left and right panels of  Fig.~\ref{fig:dilepton:left:3}, and with $m_{0} = 0.05$ eV are presented in Fig.~\ref{fig:dilepton:left:4}. As in Fig.~\ref{fig:dilepton:left:2}, the dashed curves in Figs.~\ref{fig:dilepton:left:3} and \ref{fig:dilepton:left:4} correspond to the central values of neutrino data in Table~\ref{tab:neutriodata}, and the ``widths'' of the curves are due to the $3\sigma$ uncertainties. All the regions above the curves are excluded by the same-sign dilepton data from LEP and LHC just aforementioned, which set upper bounds on the Yukawa coupling $|f_L|$ (or lower bounds on the VEV $v_L$). To be concise, the limits in Fig.~\ref{fig:dilepton:left:3} and \ref{fig:dilepton:left:4} are all expressed as functions of the largest Yukawa coupling element $|f_L|_{\rm max}$, which is the $\mu\mu$ ($ee$) element in the case of NH (IH).

As shown in Fig.~\ref{fig:lifetime1}, for a light $H_L^{\pm\pm} \lesssim 150$ GeV, when the Yukawa coupling is small, say $\lesssim 10^{-7}$, the decay length of $H_L^{\pm\pm}$ is sizable at the high-energy colliders, and the prompt dilepton limits can not be used to set limits on the mass of $H_L^{\pm\pm}$ and the Yukawa couplings $f_L$, because the prompt lepton efficiencies are significantly affected~\cite{Abbiendi:2001cr, Abdallah:2002qj, Aad:2014yea}. This is because of two reasons: (i) the algorithm for reconstructing particle tracks in the electromagnetic calorimeter or the muon spectrometer has a loose requirement of extrapolation to the interaction point, and (ii) the opening angle between the two leptons decreases as the boost increases.  To this end, we show in Figs.~\ref{fig:dilepton:left:3} and \ref{fig:dilepton:left:4} the regions of proper decay length $c\tau_0 > 1$ mm and $0.1$ mm in the darker and lighter gray color, which can be considered respectively as the aggressive and conservative estimates of the regions, within which the prompt dilepton limits are not applicable. {In the analysis of the LHC data~\cite{ATLAS:2011rha, CMS:2011sqa, ATLAS:2014kca, CMS:2016cpz, Aaboud:2017qph, CMS:2017pet}, the doubly-charged scalars are assumed to decay promptly, with a lifetime $c\tau < 10 \, \mu{\rm m}$, corresponding to a coupling $f \sim 10^{-6}$ for a doubly-charged scalar with mass of 200 GeV. For smaller couplings, a sizable fraction of $H_L^{\pm\pm}$ tends to be non-prompt, and the LHC sensitivities would be significantly weakened and even not applicable. For simplicity we just exclude the LHC limits inside the shaded gray regions in Figs.~\ref{fig:dilepton:left:3} and \ref{fig:dilepton:left:4} to make sure that $H_L^{\pm\pm}$ decay promptly at the LHC.}

It is clear that in all the four benchmark scenarios considered above, the higher-energy data tend to be more sensitive to large couplings and larger $H_L^{\pm\pm}$ mass.
%although in Fig.~\ref{fig:dilepton:left:1} the LEP data seem to be much more stringent.
This could be easily understood by looking closer at the two partial widths in Eqs.~(\ref{eqn:width1}) and (\ref{eqn:width2}): the width in the $W$ channel is proportional to $G_F M_{H_L^{\pm\pm}}^3$ while in the leptonic channel the width is proportional to $M_{H_L^{\pm\pm}}$. Thus when $H_L^{\pm\pm}$ becomes heavier, the diboson channel is comparatively enhanced, and the dilepton channel needs a larger Yukawa coupling to compensate for the suppression. {On the other hand, the production cross section times branching fractions $\sigma (pp,\, e^+ e^- \to H_L^{++} H_L^{--}) \times {\rm BR} (H_L^{++} \to \ell_\alpha^+ \ell_\beta^+) \times {\rm BR} (H_L^{--} \to \ell_\gamma^- \ell_\delta^-)$ becomes smaller when the Yukawa couplings are smaller, with a sizable fraction of $H_L^{\pm\pm}$ decaying into same-sign $W$ pairs. Thus the dilepton limits get to some extent weaker and $H_L^{\pm\pm}$ could go to smaller mass values than those shown in Fig.~\ref{fig:dilepton:left:2}, which are valid only in the small $v_L$ (or large $f_L$) limit. The limits shown in Fig.~\ref{fig:dilepton:left:2} can be recovered by just drawing horizontal lines at large $f_L\sim 10^{-6}$ in Figs.~\ref{fig:dilepton:left:3} and \ref{fig:dilepton:left:4}. }
%However, for a light $H_L^{\pm\pm}$ with mass $m_Z/2 < M_{H_L^{\pm\pm}} \lesssim 100$ GeV and couplings $f_L \lesssim 10^{-7}$, the proper lifetime $c\tau_0 (H_L^{\pm\pm}) \gtrsim 1$ cm, as seen in Fig.~\ref{fig:lifetime1}, and the prompt charged leptons in the LEP data are no longer applicable. Therefore, following the experimental analysis~\cite{Abbiendi:2001cr, Abdallah:2002qj}, we set a lower cut on the Yukawa couplings $|f_L|_{\rm max} > 10^{-7}$. As shown in Fig.~\ref{fig:dilepton:left:3} and \ref{fig:dilepton:left:4}, only the Yukawa couplings larger than $10^{-7}$ are excluded by the LEP data.

In the case of NH ($m_1 = 0$), the doubly-charged scalar decays predominantly into the muon and tau leptons, and the dilepton limits are mostly from the $\mu\mu$ and $\mu\tau$ channels, as shown in the upper left and middle left panels of Fig.~\ref{fig:dilepton:left:3}. In contrast, for all the other three scenarios, i.e. the NH with $m_1 = 0.05$ eV (left panels in Fig.~\ref{fig:dilepton:left:4}), the IH with $m_3 = 0$ (right panels in Fig.~\ref{fig:dilepton:left:3}) and $m_3 = 0.05$ eV (right panels in Fig.~\ref{fig:dilepton:left:4}), the most important constraints are from $ee$ and $\mu\mu$ channels. As stated above, the $e\tau$, $\mu\tau$ and $\tau\tau$ channels are limited by the $\tau$ lepton reconstruction efficiency, while the $e\mu$ channel is comparatively suppressed by $\sin^2\theta_{12}$.

The same-sign dilepton search results at LHC 8 TeV~\cite{ATLAS:2014kca} were also interpreted as constraints on the (fiducial) cross section $\sigma (pp\to H_L^{++}H_L^{--} \to W^{+ \, \ast} W^{+ \, \ast}W^{- \, \ast} W^{- \, \ast} \to \ell_\alpha^\pm \ell_\beta^\pm\ell_\gamma^\mp \ell_\delta^\mp+\slashed{E}_T)$~\cite{Kanemura:2014goa, Kanemura:2014ipa}. As a consequence of the small branching ${\rm BR} (W \to \ell\nu)$, the diboson limits turn out to be much weaker than the ``direct'' dilepton limits in Figs.~\ref{fig:dilepton:left:2}-\ref{fig:dilepton:left:4}, and could only exclude a narrow mass range of $m_Z/2 \lesssim M_{L}^{\pm\pm} < 84$ GeV. From Fig.~\ref{fig:lifetime1}, the dominance of the diboson decay mode (or suppression of the dilepton mode) implies that in the above mass range, $|f_L| < 10^{-8}$ in both the NH and IH cases, which is similar to the LEP limits in Fig.~\ref{fig:dilepton:left:3} and \ref{fig:dilepton:left:4}. In this region $H_L^{\pm\pm}$ anyway tends to be long-lived (cf. Fig.~\ref{fig:lifetime1}), and therefore, the prompt diboson limits derived in Refs.~\cite{Kanemura:2014goa, Kanemura:2014ipa} are not applicable.

\begin{figure}[t!]
  \centering
  \includegraphics[width=0.55\textwidth]{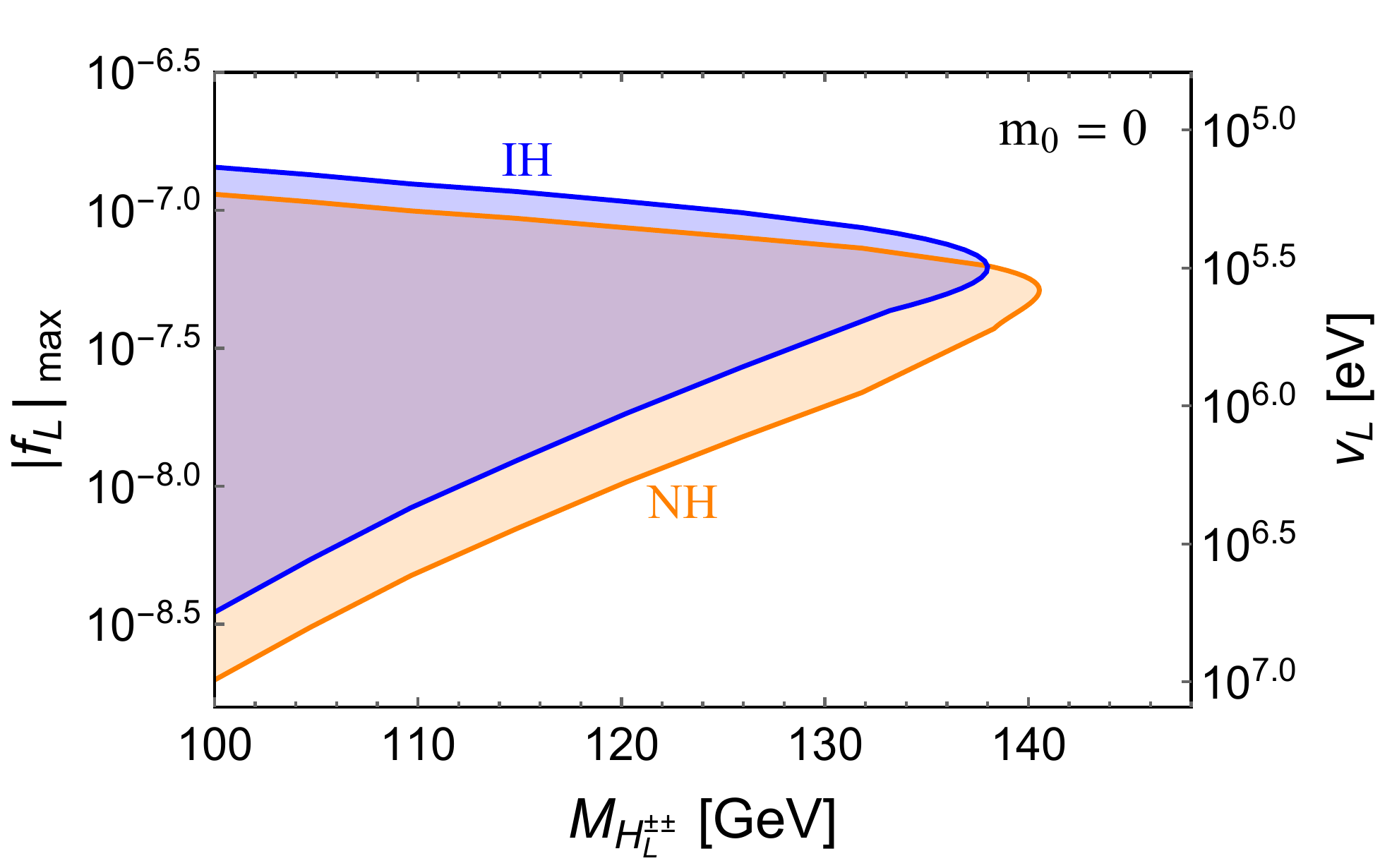}
  \caption{Limits on $M_{H_L^{\pm\pm}}$ in the type-II seesaw and its largest Yukawa coupling $|f_L|_{\rm max}$ from searches of doubly-charged HSCPs by the CMS group~\cite{CMS:2016ybj}. The orange and blue regions are excluded respectively for the NH and IH cases with the lightest neutrino mass $m_0 = 0$. The corresponding limits on the VEV $v_L$ are also shown in this plot. }
  \label{fig:HSCP}
\end{figure}

\subsubsection{Heavy stable charged particle search}
For sufficiently small total width, the lifetime of $H_L^{\pm\pm}$ is sizable, even comparable to the detector sizes, as shown in Fig.~\ref{fig:lifetime1}. If $H_L^{\pm\pm}$ decays outside either the inner silicon tracker or the whole detector, it would be recorded at the detector as a heavy stable charged particle (HSCP) and leave a trail behind as it passes through the detector. The doubly-charged HSCP has been searched for by the CMS group~\cite{CMS:2016ybj}. Both the ionization energy loss in the tracker and the time-of-flight can be used to set limits on the HSCPs. Conservatively, we use only the ``tracker-only'' analysis in~\cite{CMS:2016ybj} to constrain $H_L^{\pm\pm}$ in the type-II seesaw, as it could hardly fly out of the whole detector if its mass is larger than the lowest value of 100 GeV for the HSCP mass used in the analysis of Ref.~\cite{CMS:2016ybj}. Requiring that the decay length $43 \, {\rm mm} < bc\tau_0 (H_L^{\pm\pm}) < 1100$ mm~\cite{Chatrchyan:2008aa} ($b$ being the Lorentz boost factor), and rescaling the theoretical production cross section in Ref.~\cite{CMS:2016ybj} for the center-of-mass energy of $\sqrt{s} = 13$ TeV, we obtain the shaded orange and blue regions in Fig.~\ref{fig:HSCP} as the excluded regions respectively for the NH and IH cases, with the lightest neutrino mass $m_0 = 0$. This corresponds to the Yukawa coupling range $10^{-8.5} \lesssim |f_L| \lesssim 10^{-7}$, depending on the doubly-charged scalar mass within the narrow range $100 \, {\rm GeV} < M_{H_L^{\pm\pm}} \lesssim 140$ GeV. For heavier active neutrinos, as long as they are within the cosmological constraints~\cite{Ade:2015xua}, the total width of $H_L^{\pm\pm}$ and the exclusion regions in Fig.~\ref{fig:HSCP} would not change too much. With better particle identification using the time-of-flight measurement at the upgraded LHC detectors, the HSCP search limits could in principle be improved by up to an order of magnitude~\cite{Cerri:2018rkm}.

\subsection{Displaced vertex prospects}
\label{sec:DV1}

The decay of $H_L^{\pm\pm}$ in the dilepton and diboson channels are suppressed respectively by the small Yukawa couplings $|f_{L}|$ and the small VEV $v_L$, and the widths are proportional respectively to $v_L^{-2}$ and $v_L^{2}$, as seen in Eq.~(\ref{eqn:width1}) and (\ref{eqn:width2}). The total width of $H_L^{\pm\pm}$ reaches at a minimum when $v_L \sim 1$ MeV (depending also on the mass of $H_L^{\pm\pm}$) and the proper lifetime could go up to 1 m, as aforementioned and shown in Fig.~\ref{fig:lifetime1}. This would lead to DV signals from the decay of $H_L^{\pm\pm} \to \ell_\alpha^\pm \ell_\beta^\pm$ at the LHC and future higher energy hadron colliders like FCC-hh~\cite{fcc-hh} and SPPC~\cite{Tang:2015qga}, as well as future lepton colliders such as CEPC~\cite{CEPC-SPPCStudyGroup:2015csa}, ILC~\cite{Baer:2013cma}, FCC-ee~\cite{Gomez-Ceballos:2013zzn} and CLIC~\cite{Battaglia:2004mw}. As a strikingly clean signature beyond the SM, this kind of fully reconstructible DV signal from the doubly-charged scalar is largely complementary to the prompt same-sign dilepton pair searches at the high energy colliders: the prompt decays apply to relatively large couplings, while the DVs are sensitive to smaller couplings. In addition, if $H_{L}^{\pm\pm}$ is produced from the gauge interactions, then the prompt decays can only be used to constrain the branching fractions, as shown in Fig.~\ref{fig:dilepton:left:1}; for sufficiently small Yukawa couplings $|f_L|$, the decay products from the DVs can, in principle, be used to measure the lifetime $c\tau_0 (H_L^{\pm\pm})$, and even fix all the couplings $f_L$ involved in the decay of $H_L^{\pm\pm}$.

{Requiring at least one pair of displaced same-sign dileptons is to be reconstructed at colliders, the dominant SM backgrounds are from the low-mass Drell-Yan processes $pp \to e^+ e^-,\, \mu^+ \mu^-$, with the charges of the electron or muon misidentified (and the electron misidentified as a muon or vice versa)~\cite{Aaboud:2018jbr} (see also Refs.~\cite{Englert:2016ktc, Alcaide:2017dcx}). However, these contributions are most substantial for small values of dilepton mass $M_{H_L^{\pm\pm}} \simeq m_{\ell\ell^\prime}$, and the dileptons from Drell-Yan processes tend to be back-to-back at colliders, which could be easily distinguished from the four-body process $pp \to H_L^{++} H_L^{--} \to \ell_\alpha^\pm \ell_\beta^\pm \ell_\gamma^\mp \ell_\delta^\mp$. Thus the backgrounds are expected to be smaller than the assumed number of events (10 and 100) below. A jet might also be mis-identified as a lepton, with an energy-dependent fake rate  $\lesssim 2 \times 10^{-3}$ for the lepton energy $p_T (\ell) \gtrsim M_{H_L^{\pm\pm}}/2 \simeq M_Z/4$~\cite{ATLAS:2013hta}. To mimic the two leptons from the same vertex, we need two jets both misidentified, and  the rate is even smaller. For simplicity, we have neglected the SM backgrounds for all the prospects below. Following a recent ATLAS analysis for displaced dilepton searches, which includes a SM background estimation for same-charge displaced dimuon vertices~\cite{Aaboud:2018jbr}, we found that the backgrounds would not have substantial effects on our estimates of the signal sensitivities.}

\begin{figure}[t!]
  \centering
  \includegraphics[width=0.49\textwidth]{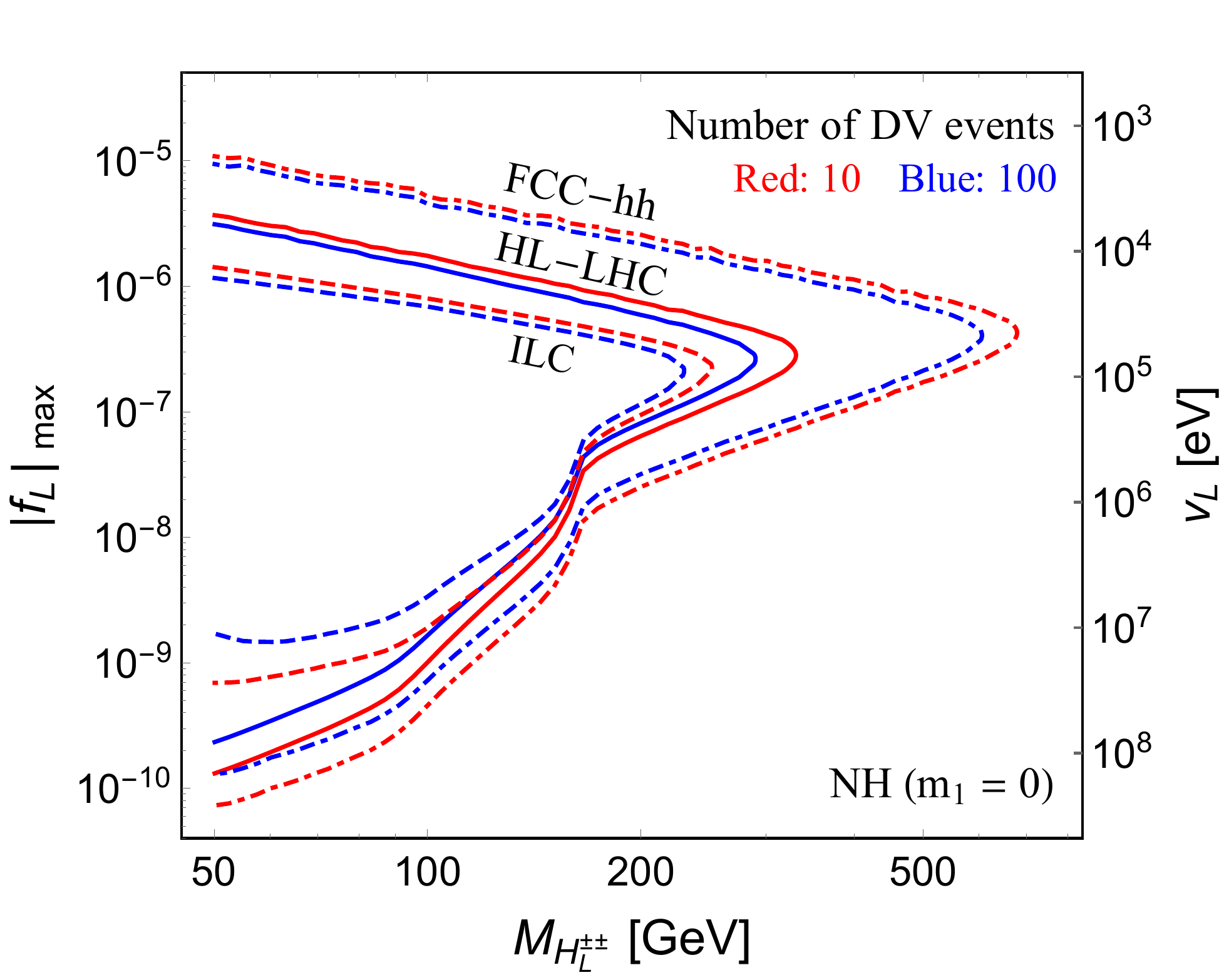}
  \includegraphics[width=0.49\textwidth]{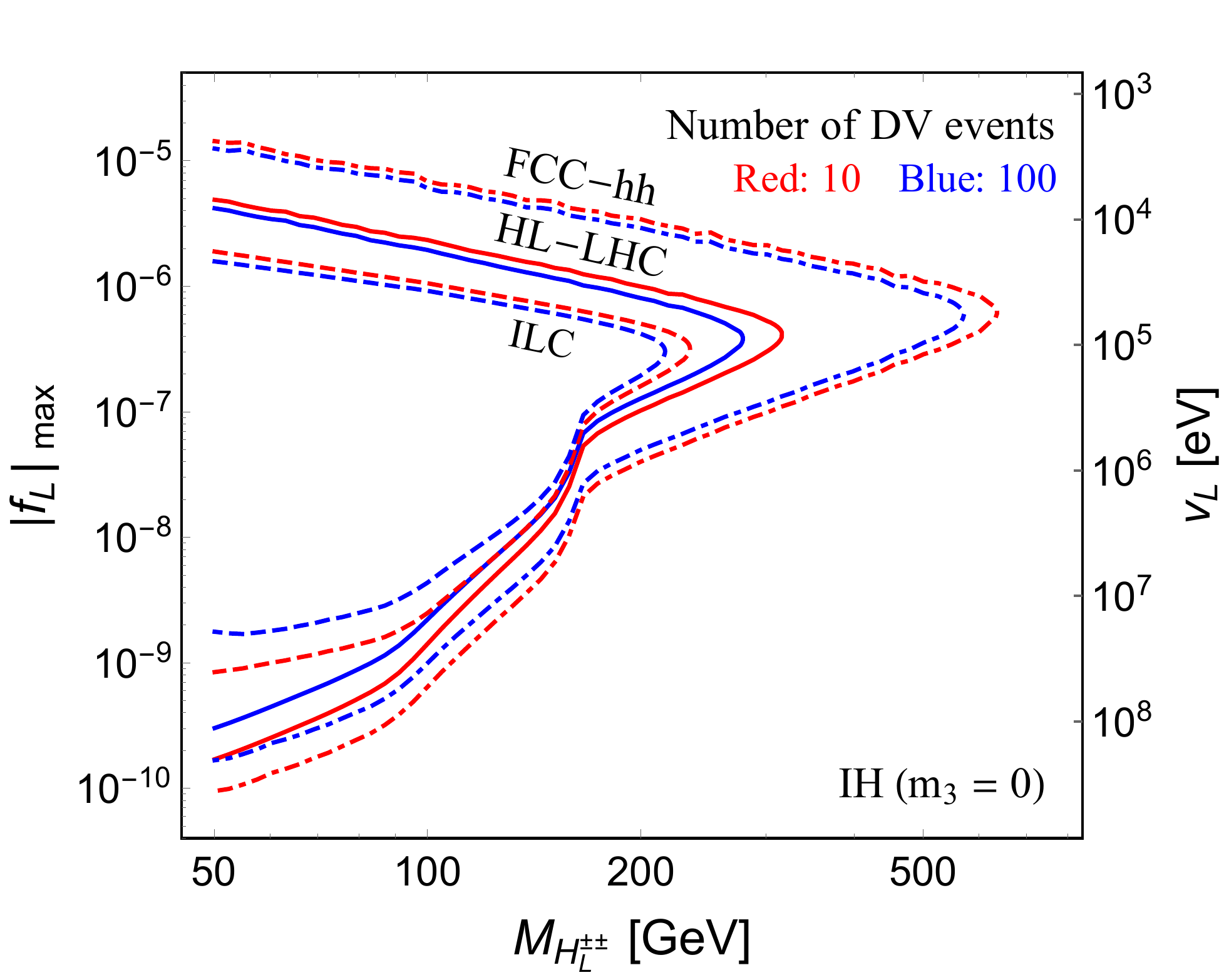}
  \caption{Prospects of DVs from the decay $H_L^{\pm\pm} \to e^\pm e^\pm,\, e^\pm\mu^\pm,\, \mu^\pm\mu^\pm$ in type-II seesaw, at HL-LHC 14 TeV and an integrated luminosity of 3000 fb$^{-1}$ (solid contours), the 100 TeV collider FCC-hh with a luminosity of 30 ab$^{-1}$ (dot-dashed contours) and ILC 1 TeV with 1 ab$^{-1}$ (dashed contours). The red and blue contours respectively correspond to 10 and 100 DV events, as functions of $M_{H_L^{\pm\pm}}$ and the largest Yukawa coupling $|f_L|_{\rm max}$, for the neutrino spectra of NH (left) and IH (right) with lightest neutrino mass $m_0 = 0$. The corresponding values of the VEV $v_L$ are also shown in these plots. }
  \label{fig:DV:left}
\end{figure}

Requiring that the decay length $1\, {\rm mm} < b c \tau_0 (H_L^{\pm\pm}) < 1$ m,  we have estimated the numbers of DV events at the HL-LHC at 14 TeV with an integrated luminosity of 3000 fb$^{-1}$ and the ILC 1 TeV with 1 ab$^{-1}$ luminosity, which are shown respectively as the solid and dashed contours in the plots of Fig.~\ref{fig:DV:left}.  The prospects at future 100 TeV collider FCC-hh are presented as the dot-dashed lines in Fig.~\ref{fig:DV:left}, with a higher luminosity of 30 ab$^{-1}$ and the decay length of $1\, {\rm mm} < b c \tau_0 (H_L^{\pm\pm}) < 3$ m. Here we have considered only the Drell-Yan production [cf.~Eq.~(\ref{eqn:DY})] at both the hadron and lepton colliders, and count only the leptonic decays $H_L^{\pm\pm} \to e^\pm e^\pm,\, e^\pm \mu^\pm,\, \mu^\pm \mu^\pm$, which are the most promising channels with almost no SM backgrounds. The $K$-factors for the higher-order QCD corrections at HL-LHC depends on the doubly-charged scalar mass and could be even larger at the 100 TeV collider; for simplicity, we take a universal conservative $K$-factor of 1.2~\cite{Muhlleitner:2003me} for both HL-LHC and FCC-hh. The higher-order electroweak corrections at lepton colliders like ILC are comparatively less important and are  neglected here. In this sense, the DV prospects presented throughout this work are rather conservative; when the higher-order corrections are fully taken into consideration, the DV sensitivities might actually be enhanced to some extent.

At HL-LHC and FCC-hh we take the nominal cuts on the displaced leptons $p_T (\ell) > 25$ GeV and $|\eta(\ell)| < 2.5$ and $\Delta\phi(\ell\ell') > 0.4$, implemented by using {\tt CalcHEP}~\cite{Belyaev:2012qa}; at ILC we set an lower momentum cut $p_T (\ell) > 10$ GeV and keep other cuts the same as above. For simplicity, we have assumed na\"{i}vely the efficiency factor to be one for all these different decay channels of $ee$, $e\mu$ and $\mu\mu$. To be concrete, we adopt the central values of neutrino data in Table~\ref{tab:neutriodata} and assume the lightest neutrino mass $m_0 = 0$ for both the NH (left) and IH (right) cases. As in Fig.~\ref{fig:HSCP}, the effects of larger neutrino masses on these DV prospects are minimal. The photon fusion process is not important for a relatively light doubly-charged scalar~\cite{Babu:2016rcr} and is not considered here. However, at lepton colliders, the laser photon fusion could largely enhance  the production cross sections~\cite{Dev:2018upe}, and hence, the DV prospects.

At the HL-LHC, with an integrated luminosity of $3000 \, {\rm fb}^{-1}$, a large parameter space in the type-II seesaw can be probed in the searches of displaced same-sign dilepton pairs, spanning over $m_Z/2 < M_{H_L^{\pm\pm}} \lesssim 250$ GeV for the scalar mass and $10^{-10} \lesssim |f_L| \lesssim 10^{-5.5}$ for the Yukawa couplings, which corresponds to a VEV of $10^4 \, {\rm eV} \lesssim v_L \lesssim 10^{8} \, {\rm eV}$, as depicted in Fig.~\ref{fig:DV:left}. With a higher energy and larger luminosity at future 100 TeV collider FCC-hh, $H_L^{\pm\pm}$ is likely to be more boosted and a much larger parameter space can be reached, up to $M_{H_L^{\pm\pm}} \sim 500$ GeV and broader $f_L$ ranges. At the ILC, the center-of-mass energy is lower than at LHC, and the production cross section of $H_L^{\pm\pm}$ is smaller, thus in Fig.~\ref{fig:DV:left} the mass reach at ILC is weaker than that at HL-LHC and FCC-hh. Comparing the contours in the left and right panels of Fig.~\ref{fig:DV:left}, we see that the DV signals have only a weak dependence on the neutrino data, as the most  relevant quantity is the total width of $H_L^{\pm\pm}$ in Eq.~(\ref{eqn:widthtotal}). For larger neutrino masses with $m_0 > 0$, as long as they are within the cosmological bound $\sum_i m_i < 0.23$ eV~\cite{Ade:2015xua}, the total width of $H_L^{\pm\pm}$ and the contours in Fig.~\ref{fig:DV:left} would not change too much.

%Note that the DV signal, more specifically the production cross section and number of DV events, has a weak dependence on the active neutrino data, e.g. the lightest neutrino mass and the Dirac CP violating phase $\delta_{\rm CP}$.
%The lepton flavors in the DV events, i.e. the probability of the DV events being electron or muon or even the tau lepton, are also dictated totally by the neutrino mass and mixing data in the type-II seesaw.
%thus the type-II seesaw can be tested at LHC and future hadron colliders in the searches of DV same-sign lepton pairs, which is more sensitive to a relatively light doubly-charged scalar and largely complementary to the prompt dilepton searches and the low-energy high-precision experiments like Moller~\cite{Benesch:2014bas}.

It is worth noting that the diboson decay $H_L^{\pm\pm} \to W^{\pm (\ast)} W^{\pm (\ast)}$ could also induce DVs at high energy colliders, and the searches of the displaced $W$ decay products are largely complementary to the dilepton DV signals discussed above, in the sense that they are sensitive to different ranges of the VEV $v_L$ (or equivalently different ranges of Yukawa couplings $f_L$), as implied by the BR contours in Fig.~\ref{fig:lifetime1}. For pair produced $H_L^{\pm\pm}$ in the Drell-Yan process, we have in total four (off-shell) $W$ boson, i.e. $pp \to H^{++} H^{--} \to 4W^{(*)}$. With the $W$ boson decaying either hadronically or leptonically, we can have different sorts of DV signals involving a large number of jets ($j$) or charged leptons ($\ell$) and neutrinos ($\slashed{E}_T$), such as $8j$, $6j \ell\nu$, $4j 2\ell 2\nu$, $2j 3\ell 3\nu$ and $4\ell 4 \nu$. The data analysis with multiple jets and $\slashed{E}_T$ is more challenging than the pure, all visible leptonic channels above, and the prospects are expected to be weaker.

%This has also not been studied by the experimental collaborations and worth proposing to them in the context of comparison with MOLLER.

%In addition, for smaller Yukawa couplings, if the decay of doubly-charged Higgs to the dilepton pair is still dominant (i.e. if the $WW$ mode is suppressed due to small triplet VEV),

%\subsection{Complementarity of the high-energy and high-precision experiments}

\begin{figure}[t!]
  \includegraphics[height=0.38\textwidth]{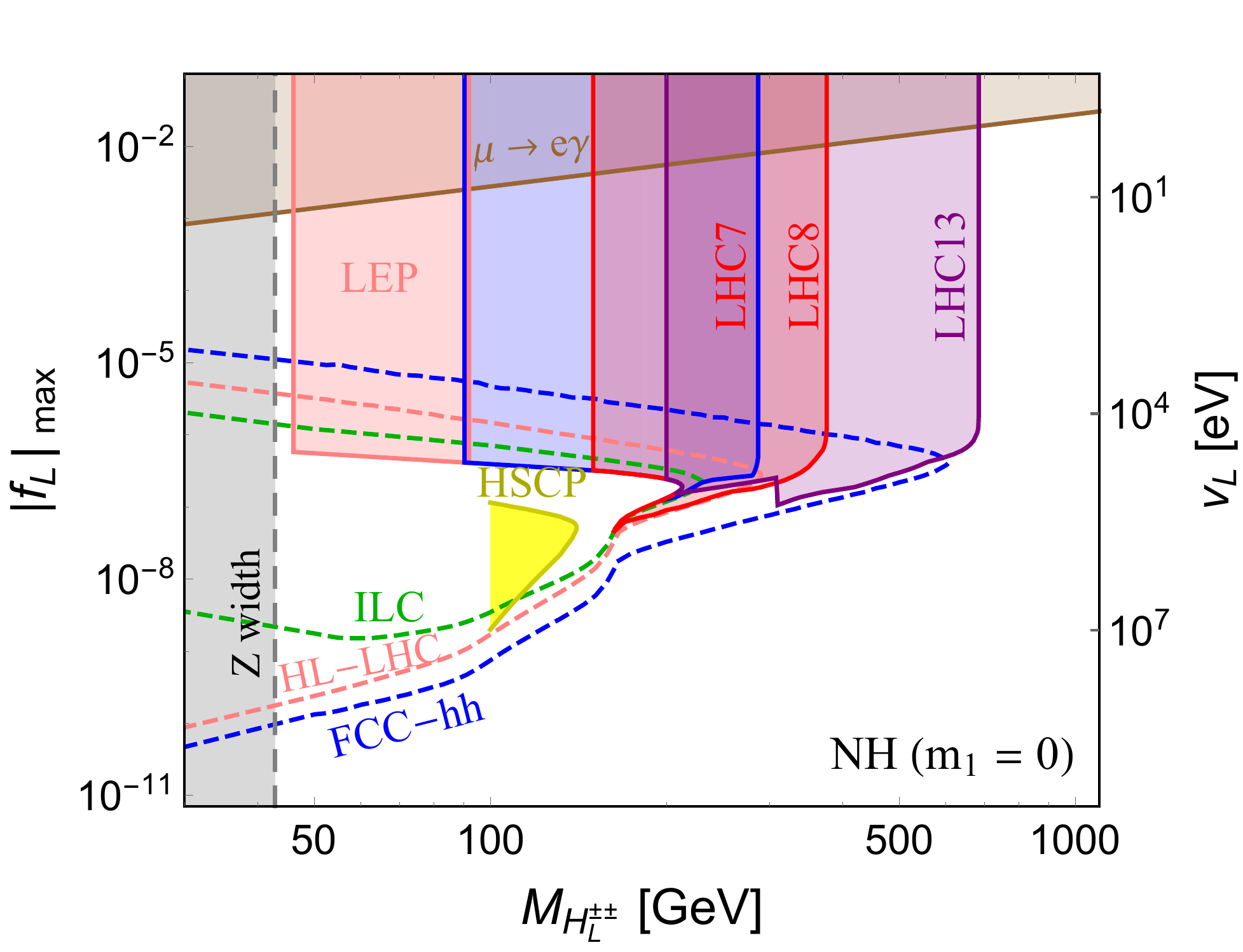}
  \includegraphics[height=0.38\textwidth]{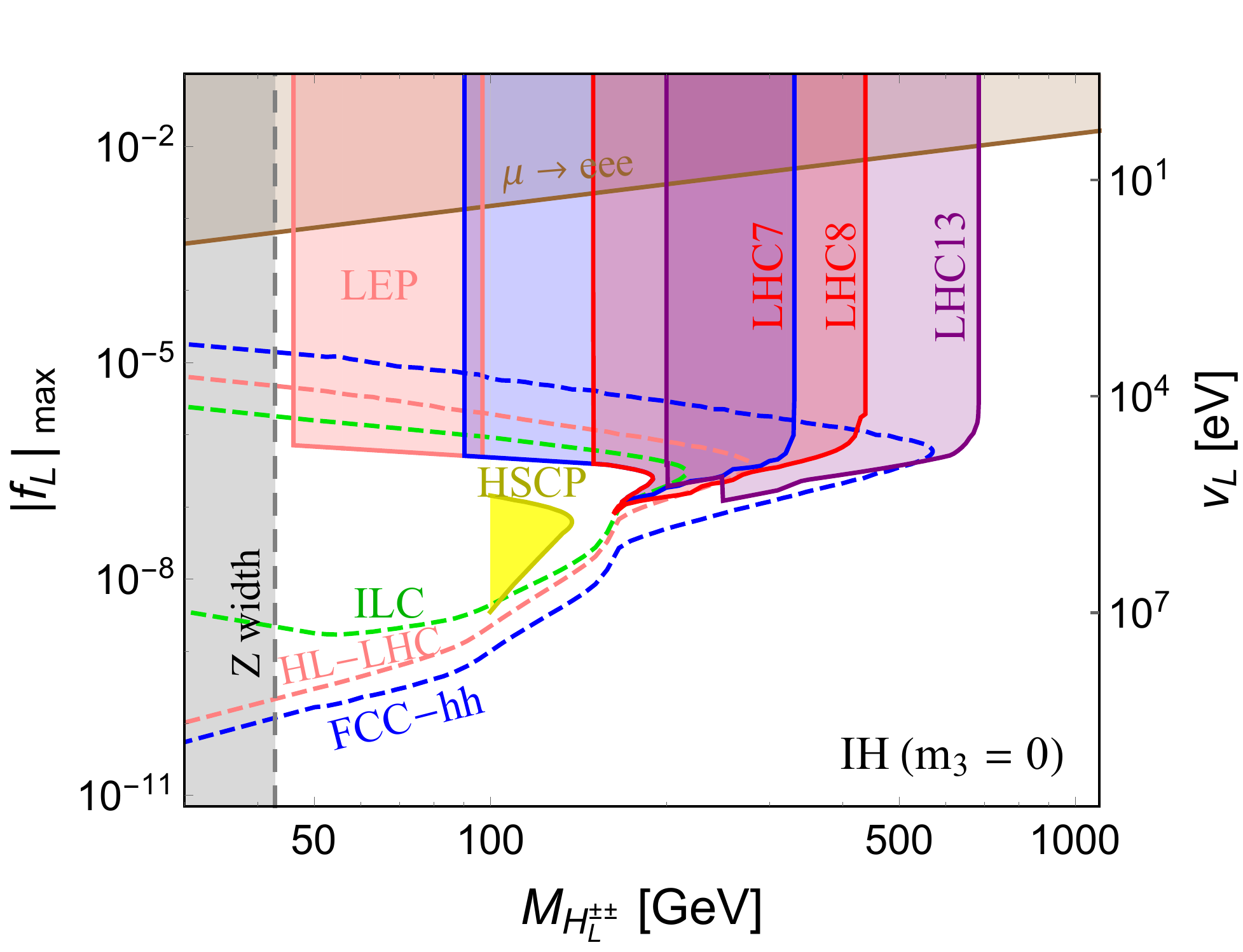}
  \caption{Summary of the most important constraints and sensitivities on the mass $M_{H_L^{\pm\pm}}$ and the Yukawa coupling $|f_L|_{\rm max}$ in type-II seesaw, extracted from Figs.~\ref{fig:flavor1}, \ref{fig:dilepton:left:3}, \ref{fig:HSCP} and \ref{fig:DV:left}. The corresponding values of the VEV $v_L$ are also shown in these plots. The DV prospects at ILC 1 TeV with a luminosity of 1 ab$^{-1}$ (dashed green), HL-LHC 14 TeV and 3000 fb$^{-1}$ (dashed pink) and FCC-hh 100 TeV and 30 ab$^{-1}$ (dashed blue) are shown assuming at least 100 signal events. The left and right panels are respectively for the NH and IH cases, both with the lightest neutrino mass $m_0 = 0$. All the shaded regions are excluded, which are derived from a combination of the LFV $\mu \to e\gamma$ ($\mu \to eee$) limit (brown), the dilepton constraints from LEP (pink), LHC 7 TeV (blue), 8 TeV (red) and 13 TeV (purple), limit on $M_{H_L^{\pm\pm}}$ from $Z$ boson width (gray), and the CMS HSCP limit (bright yellow). For all the dilepton limits at LEP and LHC, we have left out the regions with $c\tau_0 (H_L^{\pm\pm}) > 0.1$ mm (cf. Fig.~\ref{fig:dilepton:left:3}).
  }
  \label{fig:complementarity}
\end{figure}

In the low-energy high-intensity experiments, $H_L^{\pm\pm}$ could only be produced off-shell, and the high precision measurements can be used to set limits on the effective cutoff scales $\Lambda_{\rm eff} \sim M_{H_L^{\pm\pm}} / |f_L|$, as shown in Table~\ref{tab:limits} and Fig.~\ref{fig:flavor1}. At the high-energy colliders, the doubly-charged scalar $H_L^{\pm\pm}$ can be produced on-shell, and the prompt decays and DV signals are respectively sensitive to relatively large and small Yukawa couplings $|f_L|$. These high-energy and high-intensity experiments are largely complementary to each other; this can be clearly seen in the summary plot Fig.~\ref{fig:complementarity}. Here we have collectively presented the lower limit of $M_{H_L^{\pm\pm}} \lesssim m_Z/2$ from $Z$ boson width (gray), the most stringent LFV constraints in Table~\ref{tab:limits} and Fig.~\ref{fig:flavor1} from $\mu \to e\gamma$ for NH and $\mu \to eee$ for IH (brown), the same-sign dilepton pair constraints in Fig.~\ref{fig:dilepton:left:3} from LEP (pink), LHC 7 TeV (blue), 8 TeV (red) and 13 TeV (purple), and the HSCP searches by CMS in Fig.~\ref{fig:HSCP} (bright yellow). The dashed green, pink and blue curves in Fig.~\ref{fig:complementarity} correspond respectively to the DV prospects at ILC 1 TeV, HL-LHC and 100 TeV in Fig.~\ref{fig:DV:left}, all assuming 100 signal events. The left and right panels are respectively for the NH and IH cases with the lightest neutrino mass $m_0 = 0$. For all the dilepton limits, we have left out the regions with $c\tau_0 (H_L^{\pm\pm}) > 0.1$ mm as in Fig.~\ref{fig:dilepton:left:3}, as in this region, the doubly-charged scalar $H_L^{\pm\pm}$ is very likely to be long-lived and the prompt same-sign dilepton limits are not applicable, or at least weakened~\cite{Abbiendi:2001cr, Abdallah:2002qj}. At future high-energy hadron and lepton colliders, a light $H_L^{\pm\pm}$ will be highly boosted, thus there is a small region with $|f_L| \sim 10^{-6}$ to $10^{-5}$ where the DV sensitivity regions in Fig.~\ref{fig:complementarity} overlap with the current limits from LEP and LHC, where $H_L^{\pm\pm}$ is less boosted.

\section{Right-handed doubly-charged scalar in the LRSM}
\label{sec:lrsm}

The LRSM~\cite{LR1, LR2, LR3}, which provides a natural embedding of the type-II seesaw, contains two $SU(2)$ triplets -- $\Delta_L$ and $\Delta_R$ -- that transform nontrivially under $SU(2)_L$ and $SU(2)_R$, respectively. In the limit of small mixing between all the components of $\Delta_L$ and $\Delta_R$, the LH triplet $\Delta_L$ can be identified as that in the type-II seesaw in Eq.~(\ref{eqn:DeltaL}). The RH triplet \begin{align}
\label{eqn:DeltaR}
\Delta_R \ = \  \left(\begin{array}{cc}
\delta_R^+/\sqrt{2} & \delta_R^{++} \\
\delta_R^0 & -\delta_R^+/\sqrt{2}
\end{array}\right)
\end{align}
is the counterpart of $\Delta_L$ under parity, and it couples to the RH lepton doublets $\psi_R =(N,\ell_R)^{\sf T}$ via the Yukawa interactions, analogous to Eq.~\eqref{eqn:lagrangian} for the LH sector:
\begin{eqnarray}
\label{eqn:lagrangian2}
{\cal L}_Y \ = \
- \left(f_R \right)_{\alpha\beta} \psi_{R,\,\alpha}^{\sf T}Ci\sigma_2 {\Delta}_R \psi_{R,\,\beta} ~+~ {\rm H.c.},
\end{eqnarray}
with $N_\alpha$ the heavy RHNs, and $\alpha,\,\beta = e,\,\mu,\,\tau$  the lepton flavor indices. The parity symmetry dictates the Yukawa couplings $f_L = f_R$ in Eqs.~\eqref{eqn:lagrangian} and \eqref{eqn:lagrangian2}. A non-zero VEV of the neutral RH component $\langle \delta_{R} \rangle = v_R/\sqrt2$ gives rise to the Majorana masses for the heavy RHNs, $M_N = \sqrt2 f_R v_R$.  In the LRSM the tiny active neutrino mass receive, in principle, contributions from both type-I~\cite{seesaw1, seesaw2, seesaw3, seesaw4, seesaw5} and type-II~\cite{Magg:1980ut, Schechter:1980gr, Mohapatra:1980yp, Lazarides:1980nt, Konetschny:1977bn, Cheng:1980qt} seesaw mechanisms:
\begin{eqnarray}
m_\nu \ \simeq \ - m_D M_N^{-1} m_D^{\sf T} + \sqrt2 f_L v_L  \,,
\end{eqnarray}
with $m_D$ the Dirac mass matrix. For simplicity we assume here the type-I seesaw contribution is small, or in other words the LRSM is in the type-II dominance regime for neutrino mass generation, and the heavy and light neutrino masses are related via $m_\nu / M_N \simeq v_L / v_R$~\cite{Tello:2010am, Barry:2013xxa, Awasthi:2015ota, Pritimita:2016fgr, Dev:2013vxa}. In this case, the RHN masses are proportional to that of the active neutrinos, rescaled by the VEV ratio $v_R / v_L$, and the RHN mixing matrix $U_R$ is identical to the LH PMNS matrix $U$ in Eq.~(\ref{eqn:PMNS}).

\subsection{Decay Length}
\label{sec:lrsm:basic}

\begin{figure}[t!]
  \centering
  \includegraphics[width=0.48\textwidth]{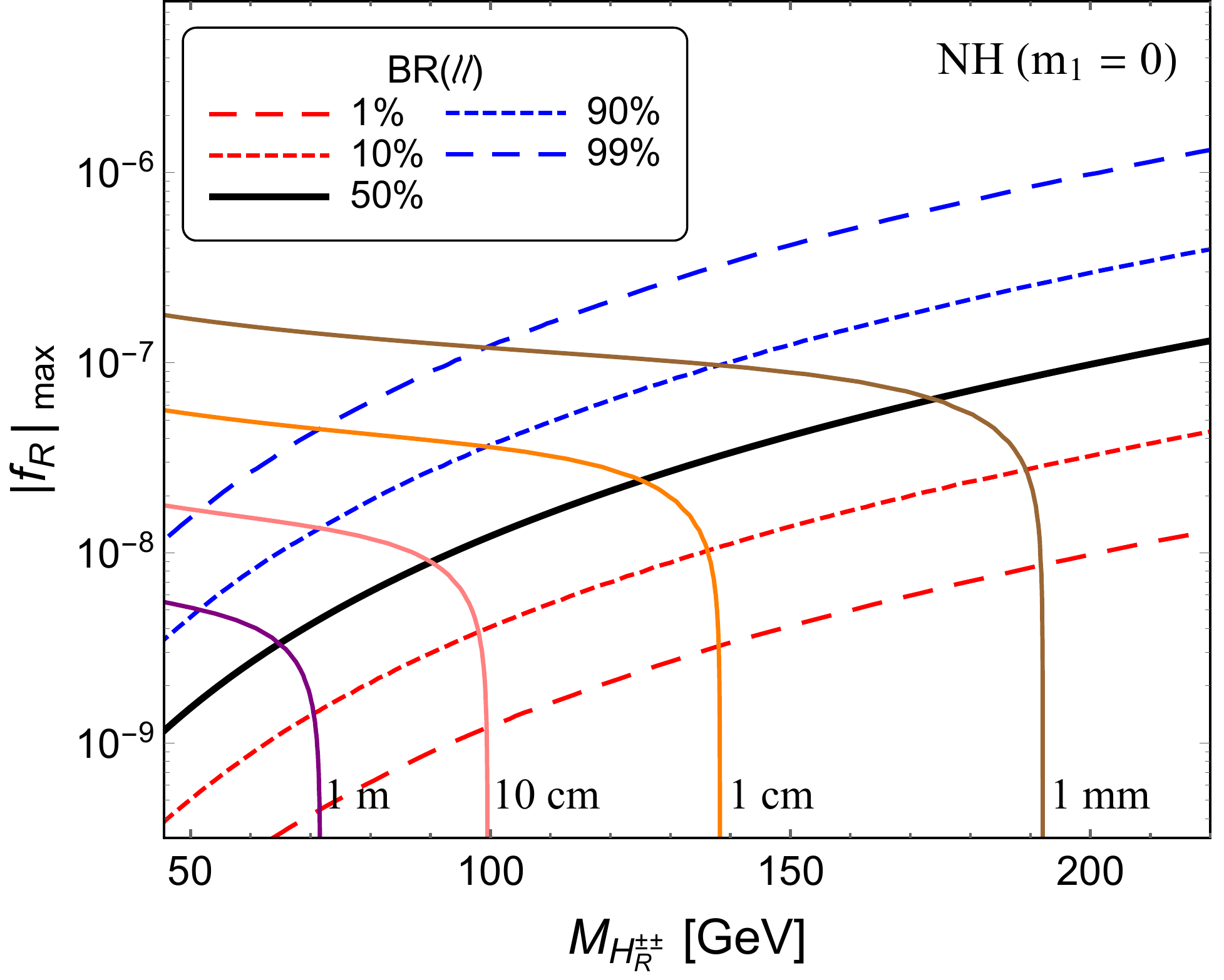}
  \includegraphics[width=0.48\textwidth]{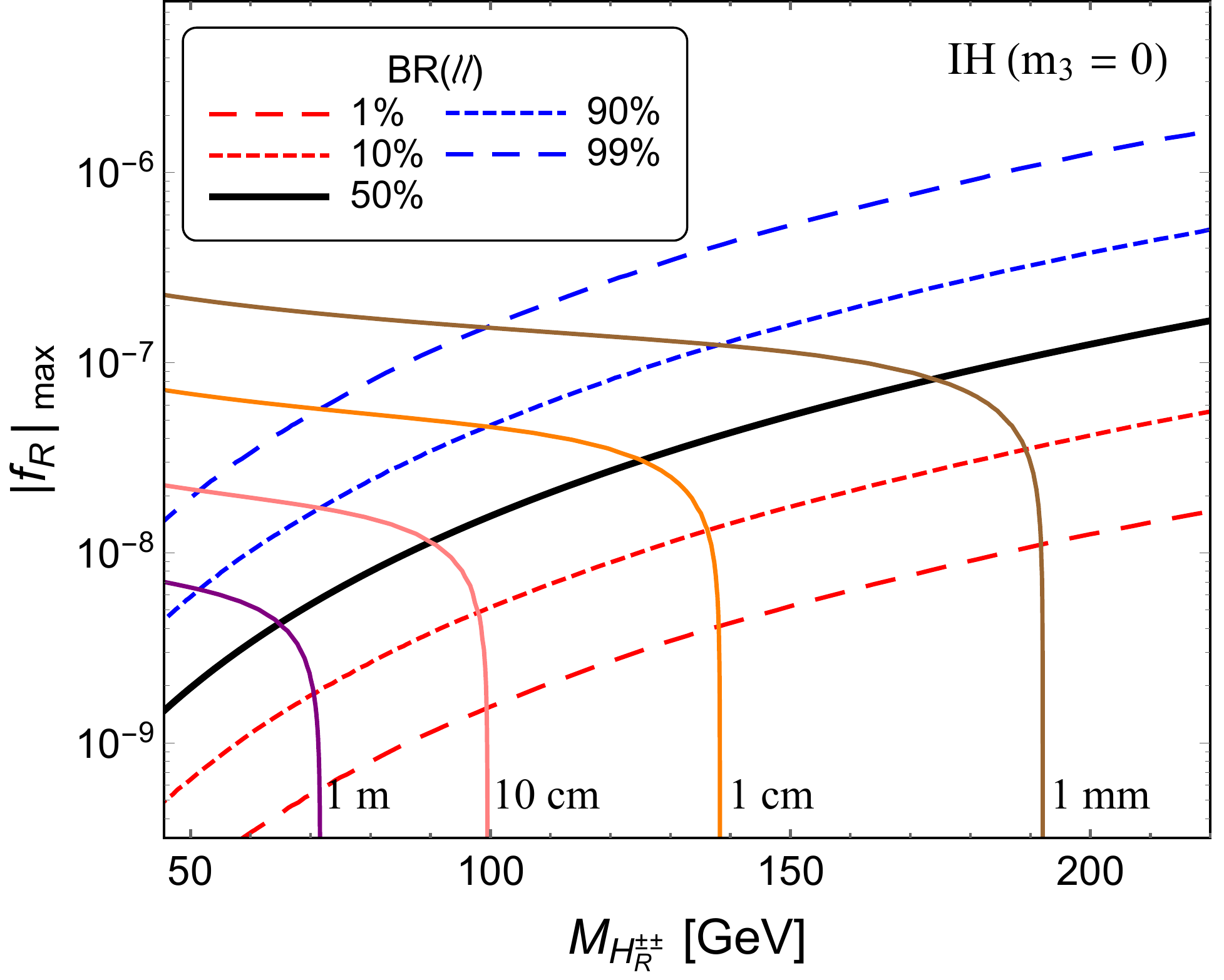}
  \caption{Contours of proper decay length $c\tau_0 = 1$ mm, 1 cm, 10 cm and 1 m of $H_R^{\pm\pm}$ in the LRSM, as functions of its mass $M_{H_R^{\pm\pm}}$ and the largest Yukawa coupling $|f_R|_{\rm max}$. Also shown are the ${\rm BR} (H_R^{\pm\pm} \to \ell_i^\pm \ell_j^\pm) = 1 - {\rm BR} (H_R^{\pm\pm} \to W_R^{\pm \ast} W_R^{\pm \ast}) = 1$\%, 10\%, 50\%, 90\%, 99\%. The left (right) panel is for NH (IH) and the lightest neutrino mass is taken to be zero. We have chosen the RH scale $v_R = 5\sqrt2$ TeV and the gauge coupling $g_R = g_L$.}
  \label{fig:lifetime2}
\end{figure}

The RH doubly-charged scalar $H_R^{\pm\pm}$ decays predominantly to a pair of same-sign charged RH leptons: $H_R^{\pm\pm} \to \ell_\alpha^\pm \ell_\beta^\pm$, and a pair of same-sign (off-shell) heavy $W_R$ bosons: $H_R^{\pm\pm} \to W_R^{\pm (\ast)} W_R^{\pm (\ast)}$.\footnote{The singly-charged scalar from $\Delta_R$ is eaten by the heavy $W_R$ boson after symmetry breaking, so there is no cascade decay to singly charged scalars unlike in the $\Delta_L$ case~\cite{Dev:2016dja}. Moreover, the heavy neutral CP-even and odd scalars from the bidoublet are required to be at least 10-20 TeV from the flavor changing neutral current (FCNC) constraints~\cite{Zhang:2007da, Bertolini:2014sua}. So in the RH scalar sector, the only other long-lived candidate, apart from the doubly-charged scalar, is the real part of the neutral component of the triplet Re($\Delta_R^0$), which has been studied in Refs.~\cite{Dev:2016vle, Dev:2017dui, Dev:2017ozg}.} The widths for the leptonic and bosonic channels are quite similar to those for the $H_L^{\pm\pm}$ in Eq.~(\ref{eqn:width1}) and (\ref{eqn:width2}):
\begin{eqnarray}
\label{eqn:width3}
\Gamma (H_R^{\pm\pm} \to \ell_\alpha^\pm \ell_\beta^\pm) & \ = \ &
\frac{M_{H_R^{\pm\pm}}}{8\pi(1+\delta_{\alpha\beta})}  |(f_R)_{\alpha\beta}|^2 \,, \\
\label{eqn:width4}
\Gamma (H_R^{\pm\pm} \to W_R^\pm W_R^\pm) & \ = \ &
\frac{M_{H_R^{\pm\pm}}^3}{16\pi \, v_R^2}
\sqrt{1-4x_{W_R}} (1-4x_{W_R}+12x_{W_R}^2) \,,
\end{eqnarray}
with $x_{W_R} \equiv m_{W_R}^2 / M_{H_R^{\pm\pm}}^2$. The current FCNC constraints from $K$ and $B$ meson oscillation data require that the $W_R$ boson is beyond roughly 3 TeV~\cite{Zhang:2007da, Bertolini:2014sua} for the gauge coupling $g_R=g_L$. The 13 TeV LHC searches yield a similar mass bound from the same-sign dilepton channel $pp\to W_R\to N\ell\to \ell^\pm\ell^\pm jj$~\cite{Keung:1983uu}, depending on the RHN mass~\cite{Sirunyan:2018pom}. Thus in the diboson decay mode of a TeV-scale (or lighter) doubly-charged scalar, both $W_R$'s can only be off-shell, which decay further into the SM fermions and heavy RHNs (if lighter than $H_R^{\pm\pm}$). The decay width of $H_R^{\pm\pm} \to W_R^{\pm \ast} W_R^{\pm \ast} \to f \bar{f}^\prime f^{\prime\prime} \bar{f}^{\prime\prime\prime}$ (where the fermions $f$ run over all the SM quarks, charged leptons and heavy RHNs) can be found in Appendix~\ref{sec:decay}. In the type-II seesaw dominance of LRSM, dictated by the parity symmetry $f_L = f_R$, the couplings $f_R$ are also related to the active neutrino masses and mixing angles, as $f_L$ is in the pure type-II seesaw. The BRs ${\rm BR} (H_R^{\pm\pm} \to \ell_\alpha^\pm \ell_\beta^\pm) = 1 - {\rm BR} (H_R^{\pm\pm} \to W_R^{\pm \ast} W_R^{\pm \ast})$ depend also on the heavy $W_R$ mass and the $v_R$ scale, cf.~Eq.~(\ref{eqn:width4}). To be concrete, we set $v_R = 5 \sqrt2$ TeV and the gauge coupling $g_R = g_L$ (unless otherwise specified) which leads to $M_{W_R} = g_R v_R /\sqrt2 \simeq 3.3$ TeV, consistent with the LHC and low-energy constraints. Some representative ${\rm BR} (H_R^{\pm\pm} \to \ell_\alpha^\pm \ell_\beta^\pm) = 1$\%, 10\%, 50\%, 90\%, 99\% are shown in Fig.~\ref{fig:lifetime2}, with the left and right panels respectively for the NH and IH of neutrino spectrum with the lightest neutrino mass $m_0 = 0$. As in the type-II seesaw, the total width and decay lifetime of $H_R^{\pm\pm}$ are not very sensitive to the lightest neutrino mass $m_0$.

The $W_R$ channel of $H_R^{\pm\pm}$ decay is heavily suppressed by the large $W_R$ mass, thus for sufficiently light $H_R^{\pm\pm}$, if the Yukawa couplings $f_R$ are small, $H_R^{\pm\pm}$ could be long-lived, as shown in Fig.~\ref{fig:lifetime2}. For an RH doubly-charged scalar mass $m_Z/2 < M_{H_R^{\pm\pm}} \lesssim 200$ GeV, the proper decay length could reach from 1 mm up to 1 meter if $|f_R| \lesssim 10^{-7}$. Unlike the LH case in Fig.~\ref{fig:lifetime1}, for fixed $M_{W_R}$ and $v_R$, the width $\Gamma (H_R^{\pm\pm} \to W_R^\pm W_R^\pm)$ depends only on the doubly-charged scalar mass $M_{H_R^{\pm\pm}}$, thus for sufficiently small $|f_R|$ the partial width $\Gamma (H_R^{\pm\pm} \to \ell_i^\pm \ell_j^\pm) \lesssim \Gamma (H_R^{\pm\pm} \to W_R^{\pm \ast} W_R^{\pm \ast})$ and the lifetime contours in Fig.~\ref{fig:lifetime2} tend to be flat in the downward direction.

All the relevant production channels of $H_R^{\pm\pm}$ at hadron and lepton colliders can be found in Refs.~\cite{Dev:2016dja} and \cite{Dev:2018upe} respectively. It could be produced at hadron colliders in the scalar portal from couplings with the SM Higgs and other heavy scalars in the LRSM, or in the gauge portal from interacting with the SM photon and $Z$ boson (and the heavy $Z_R$ boson). The pair production of $H_R^{\pm\pm}$ in the Drell-Yan process turns out to be much larger than that in the SM Higgs portal, as the latter is suppressed by the loop-induced effective $hgg$ coupling ($g$ here being the gluon)~\cite{Dev:2016dja}. The associated production of $H_R^{\pm\pm}$ with the $W_R$ boson is suppressed by the $W_R$ mass and can be neglected for a doubly-charged scalar with mass $M_{H_R^{\pm\pm}} \lesssim 700$ GeV for $g_R = g_L$. Similarly, at lepton colliders, the dominant pair-production channel is either Drell-Yan or photon fusion, depending on the doubly-charged scalar mass~\cite{Dev:2018upe}. For the sake of DV searches at future hadron and lepton colliders, we consider in this paper only the Drell-Yan production of $H_R^{\pm\pm}$ in the LRSM.

\subsection{Low and high-energy constraints}
\label{sec:limits:LRSM}

Similar to the $H_L^{\pm\pm}$ case in Eq.~(\ref{eqn:Zwidth}), the $H_R^{\pm\pm}$ also contributes to the $Z$ boson width and is constrained to have mass $M_{H_R^{\pm\pm}}>m_Z/2$ from  the precision $Z$-pole data, irrespective of how it decays or whether it is long-lived. Note that as a singlet under the SM gauge group $SU(2)_L$, the coupling of $H_R^{\pm\pm}$ to the SM $Z$ boson is only due to the hypercharge, proportional to $-2\sin^2\theta_w$, and does not depend on the RH gauge coupling $g_R$~\cite{Dev:2016dja}. The LFV constraints on $H_R^{\pm\pm}$ are the same as those on $H_L^{\pm\pm}$ in Table~\ref{tab:limits} (third column) and Fig.~\ref{fig:flavor1}.

\subsubsection{High-energy collider constraints}

\begin{figure}[t!]
  \centering
  \includegraphics[width=0.48\textwidth]{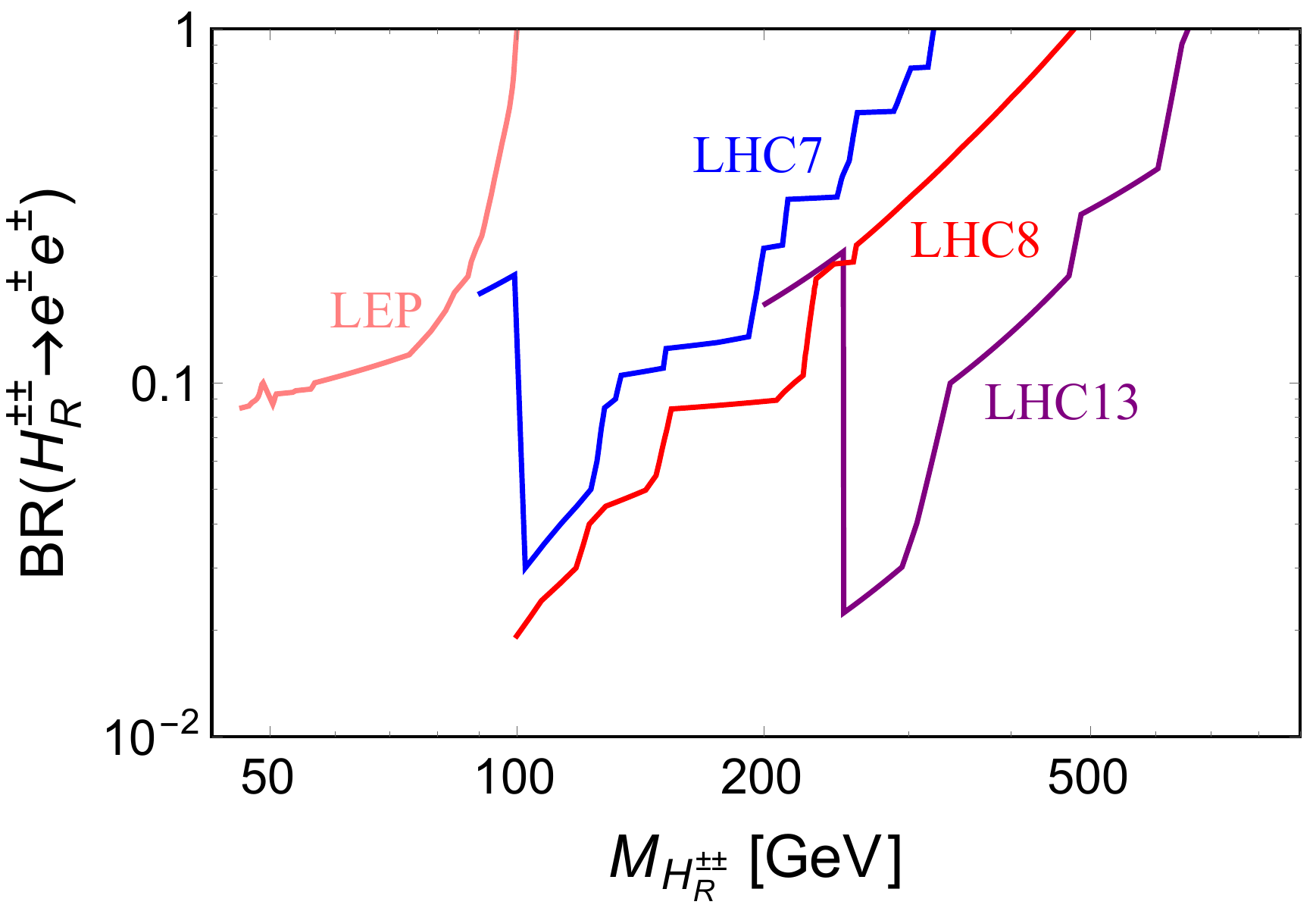}
  \includegraphics[width=0.48\textwidth]{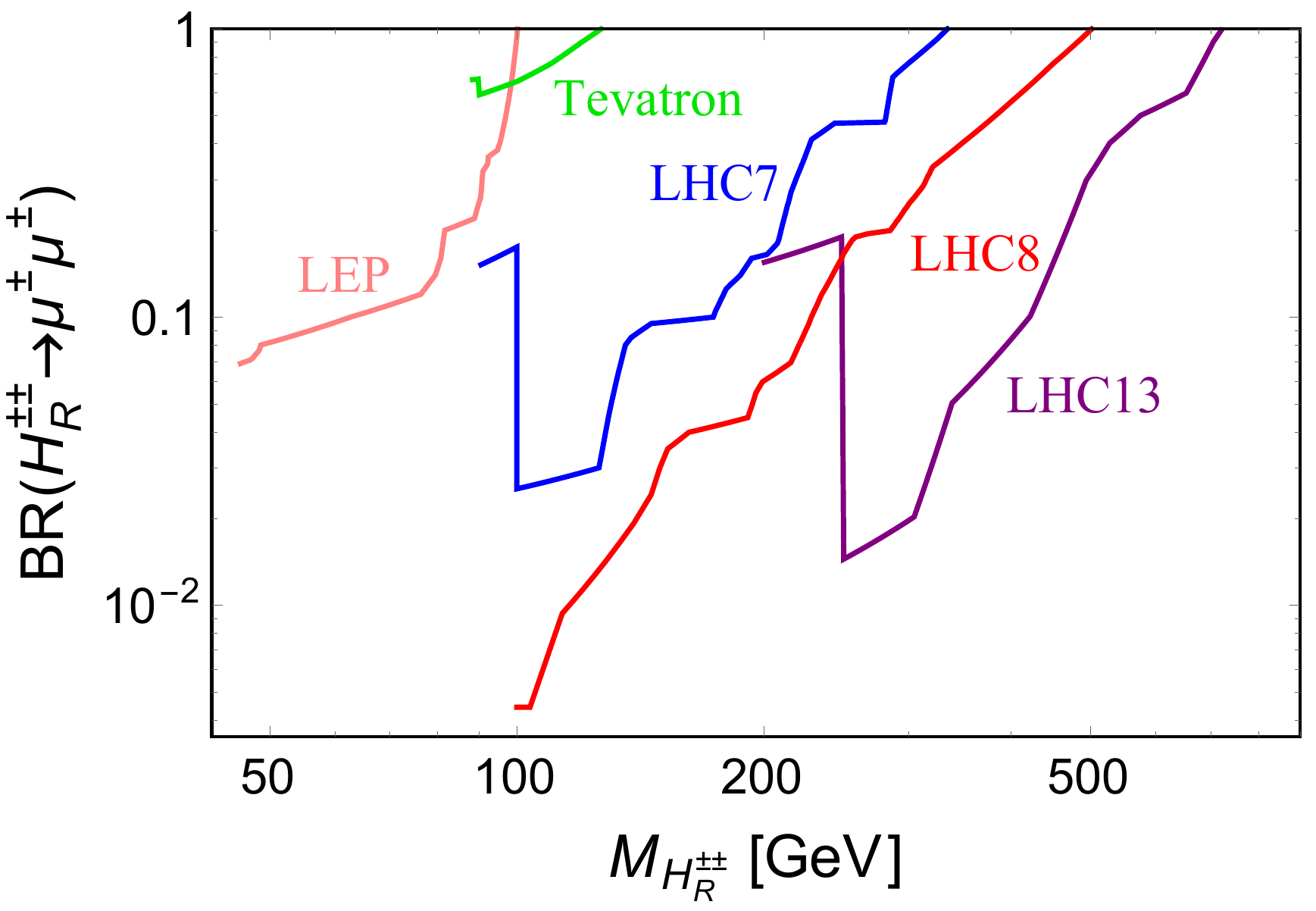}
  \includegraphics[width=0.485\textwidth]{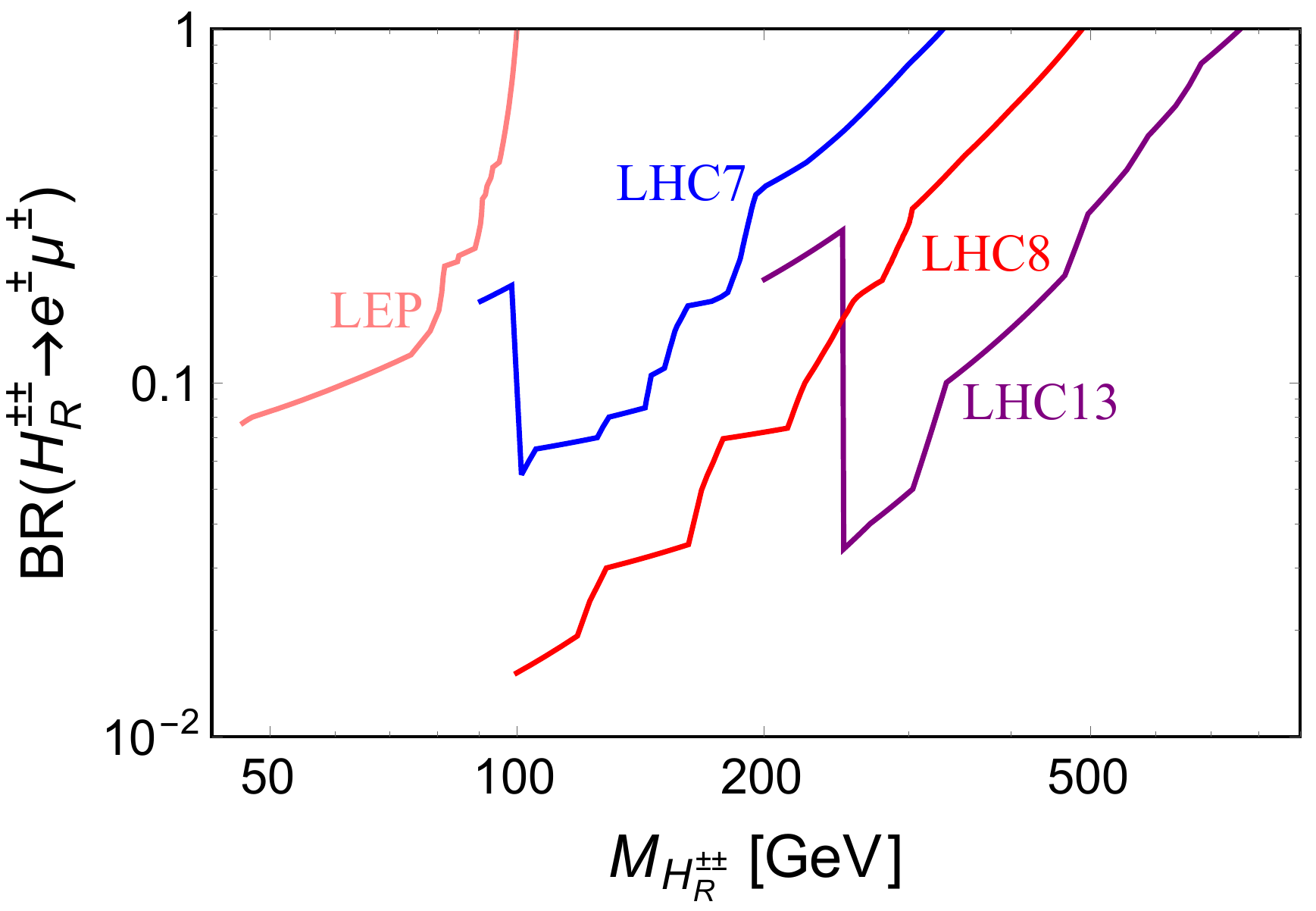}
  \includegraphics[width=0.475\textwidth]{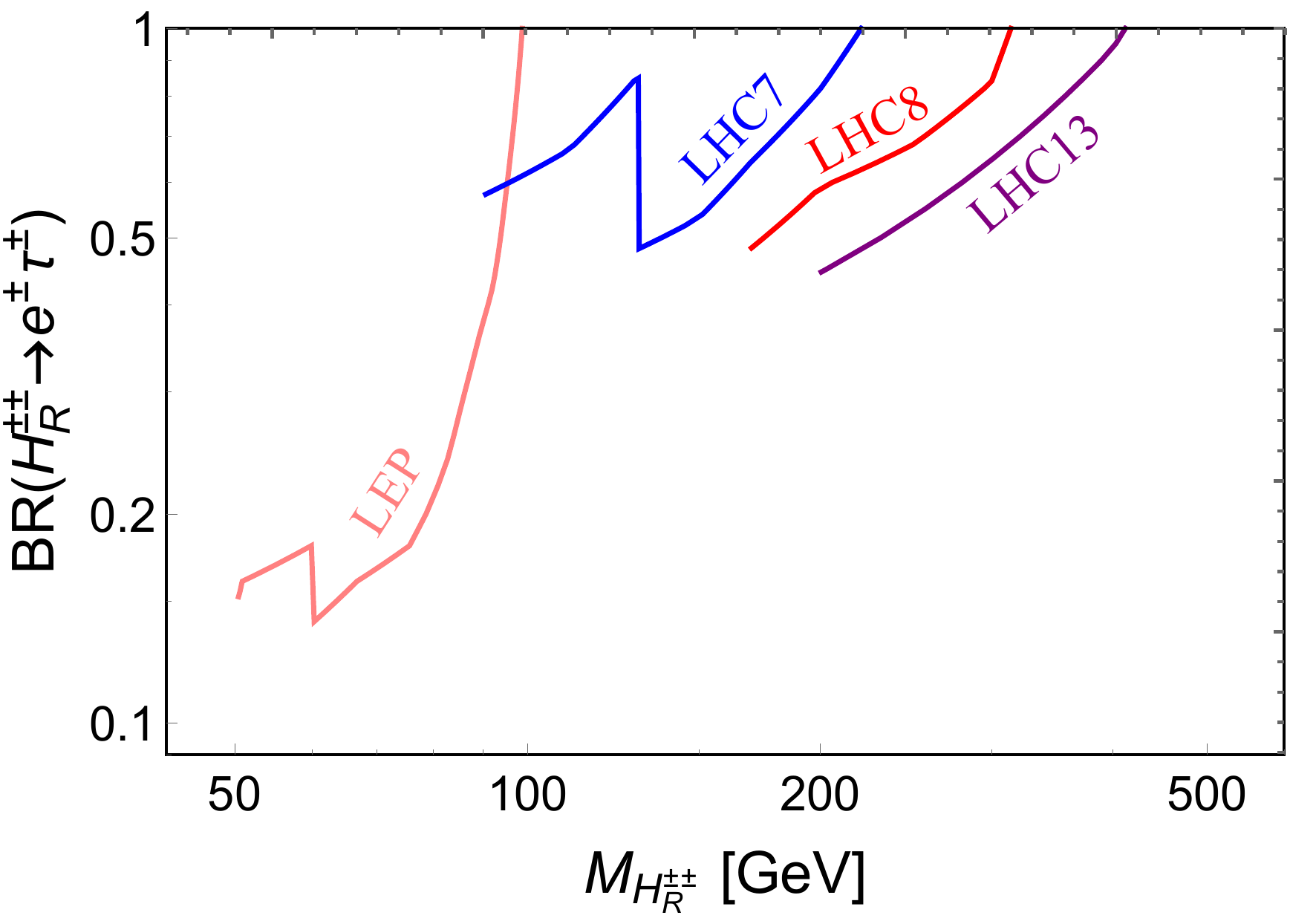}
  \includegraphics[width=0.48\textwidth]{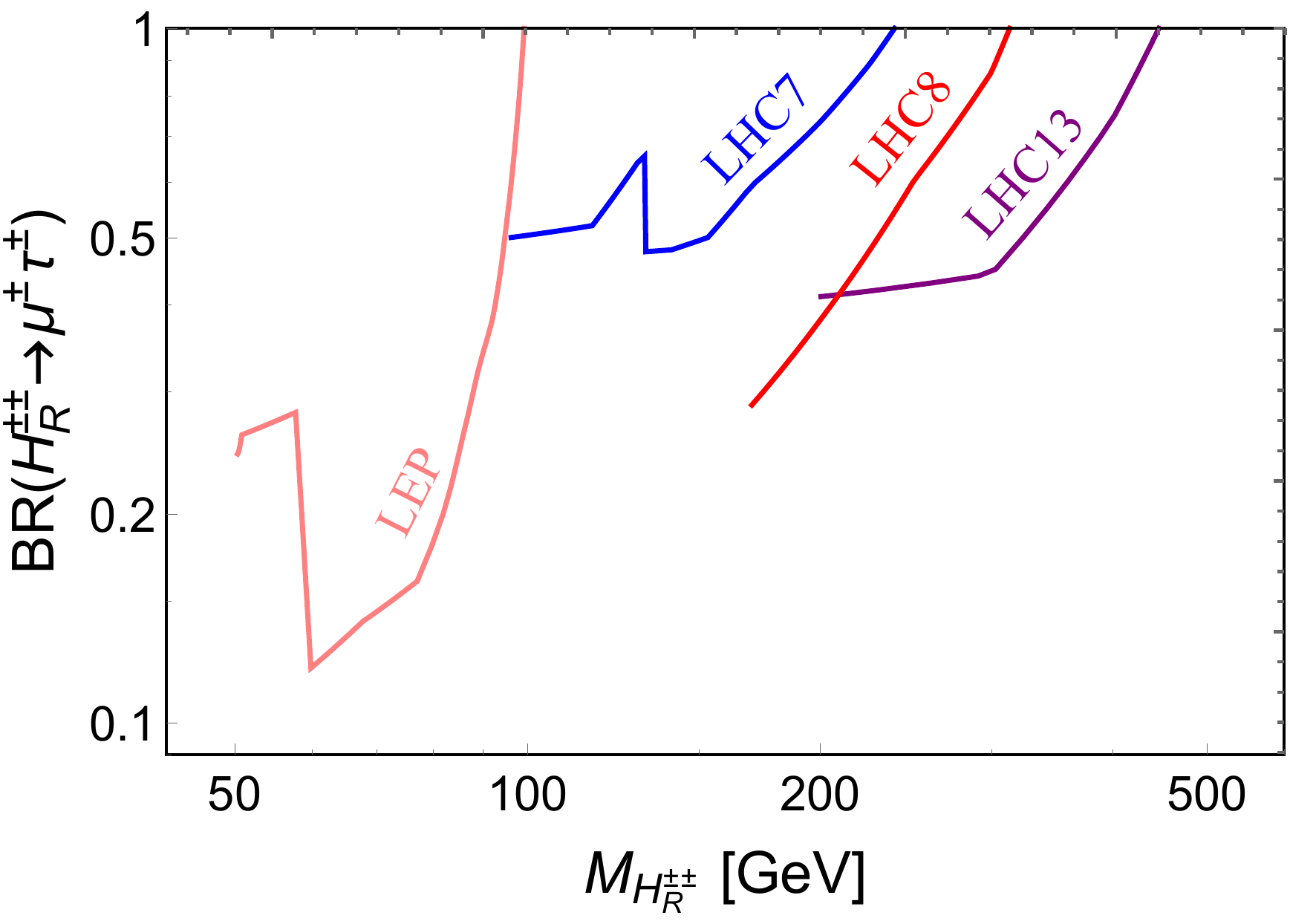}
  \includegraphics[width=0.48\textwidth]{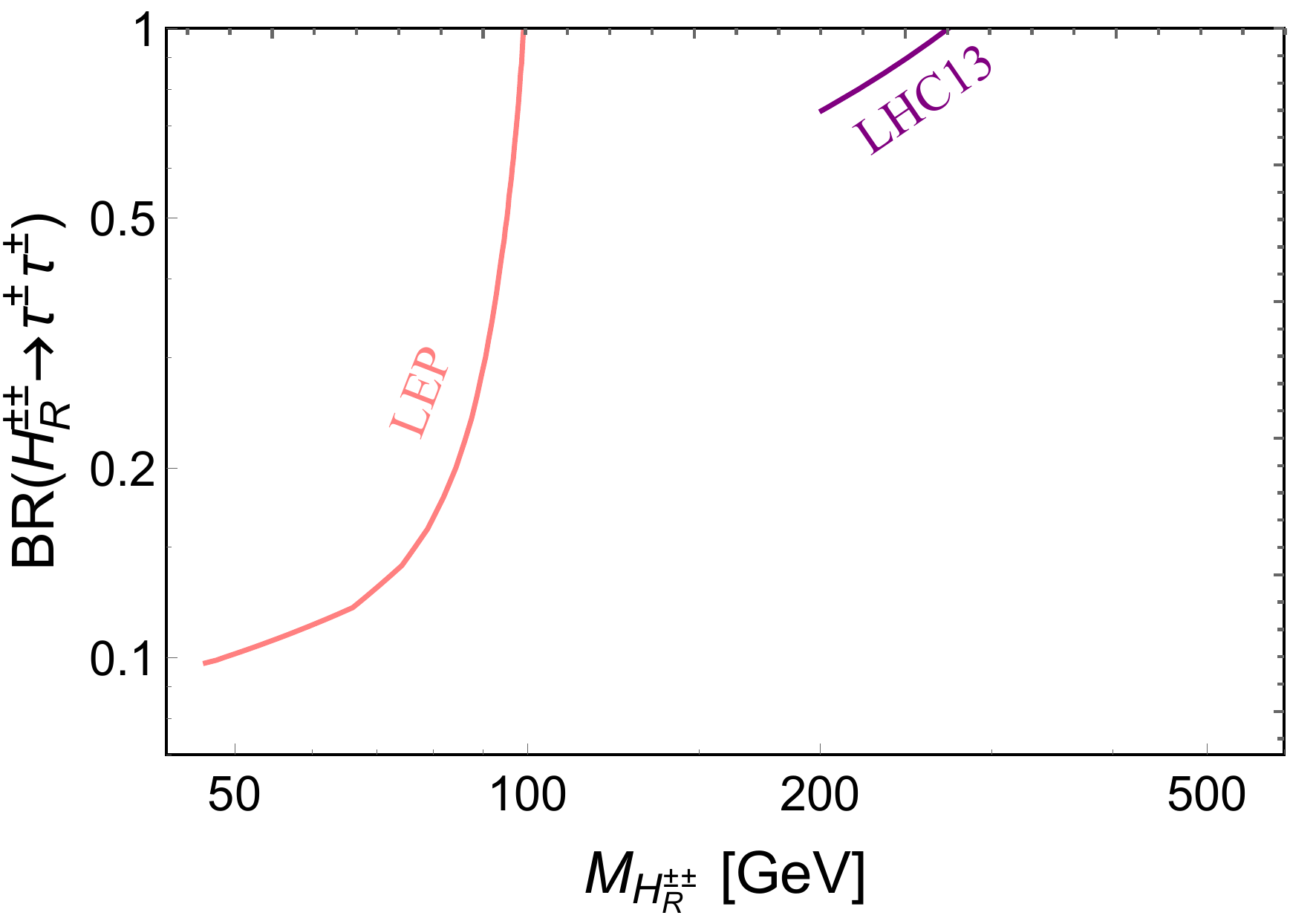}
  \caption{90\% CL lower limits on $M_{H_R^{\pm\pm}}$ in the LRSM as functions of ${\rm BR} (H_R^{\pm\pm} \to \ell_\alpha^\pm \ell_\beta^\pm)$, in all the six flavor combinations of $\ell_\alpha \ell_\beta = ee$ (upper left), $\mu\mu$ (upper right), $e\mu$ (middle left), $e\tau$ (middle right), $\mu\tau$ (lower left) and $\tau\tau$ (lower right). We have used the data from LEP~\cite{Abbiendi:2001cr, Achard:2003mv, Abdallah:2002qj}, Tevatron~\cite{Aaltonen:2008ip, Acosta:2004uj, Abazov:2008ab, Abazov:2011xx} and LHC 7 TeV~\cite{ATLAS:2011rha, CMS:2011sqa}, 8 TeV~\cite{ATLAS:2014kca, CMS:2016cpz} and 13 TeV~\cite{Aaboud:2017qph, CMS:2017pet}. }
  \label{fig:dilepton:right:1}
\end{figure}

As the coupling of $H_R^{\pm\pm}$ to the SM $Z$ boson is proportional to $-2\sin^2\theta_w$, smaller than that of $H_L^{\pm\pm}$ which is $(1-2\sin^2\theta_w)$, the Drell-Yan production cross sections of $H_R^{\pm\pm}$ at lepton and hadron colliders are thus significantly smaller than that of $H_{L}^{\pm\pm}$, roughly 1.3 times smaller at LEP and 2.4 times smaller at Tevatron and LHC. The same-sign dilepton searches of doubly-charged scalars in Section~\ref{sec:dilepton} apply also to the $H_R^{\pm\pm}$ case, i.e. those in LEP~\cite{Abbiendi:2001cr, Achard:2003mv, Abdallah:2002qj}, Tevatron~\cite{Aaltonen:2008ip, Acosta:2004uj, Abazov:2008ab, Abazov:2011xx} and LHC running at 7 TeV~\cite{ATLAS:2011rha, CMS:2011sqa}, 8 TeV~\cite{ATLAS:2014kca, CMS:2016cpz} and 13 TeV~\cite{Aaboud:2017qph, CMS:2017pet}. In some of the data analysis, the doubly-charged scalar is assumed to be an LH triplet; the production cross sections therein have to be rescaled accordingly, with the theoretical predictions multiplied by a factor of 1/1.3 at LEP and 1/2.4 at Tevatron and LHC. All the same-sign dilepton limits on $H_R^{\pm\pm}$ are collected in Fig.~\ref{fig:dilepton:right:1}, in the six different flavor channels: $ee$ (upper left), $\mu\mu$ (upper right), $e\mu$ (middle left), $e\tau$ (middle right), $\mu\tau$ (lower left) and $\tau\tau$ (lower right). As a result of the smaller couplings of $H_R^{\pm\pm}$ to the $Z$ boson, the dilepton limits are to some extent weaker than those for $H_L^{\pm\pm}$ in Fig.~\ref{fig:dilepton:left:1}.

\begin{figure}[t!]
  \centering
  \includegraphics[width=0.48\textwidth]{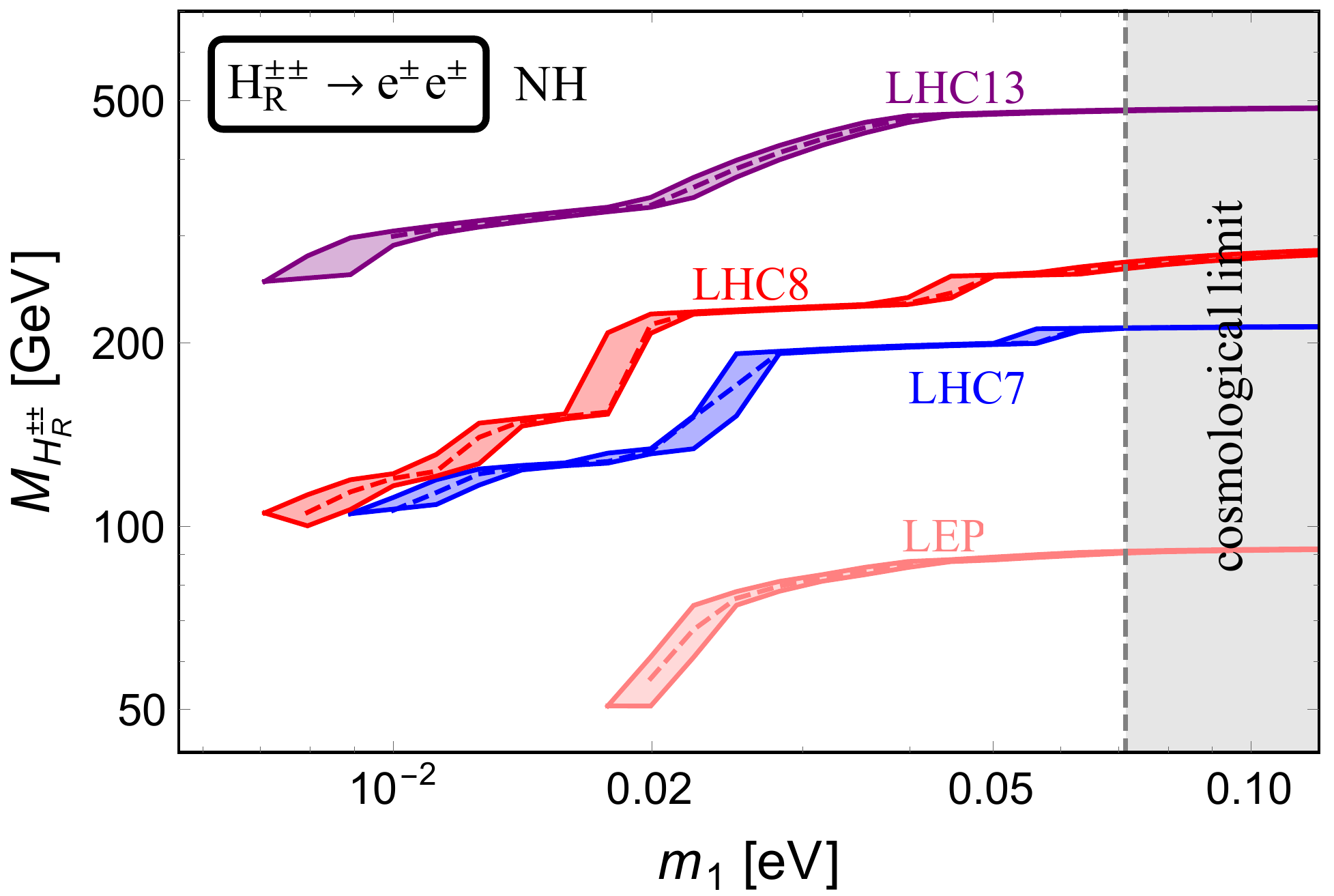}
  \includegraphics[width=0.48\textwidth]{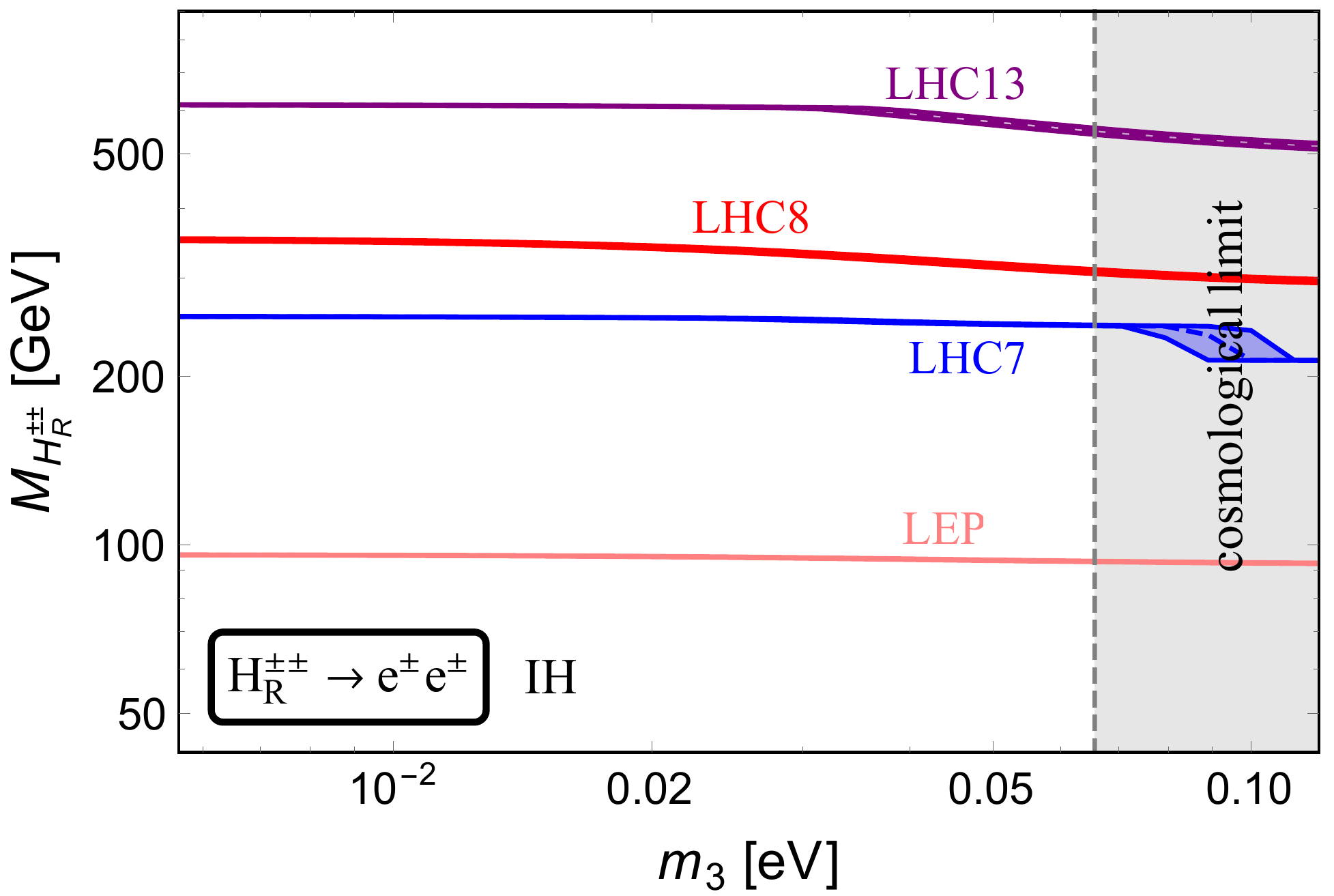} \vspace{3pt} \\
  \includegraphics[width=0.48\textwidth]{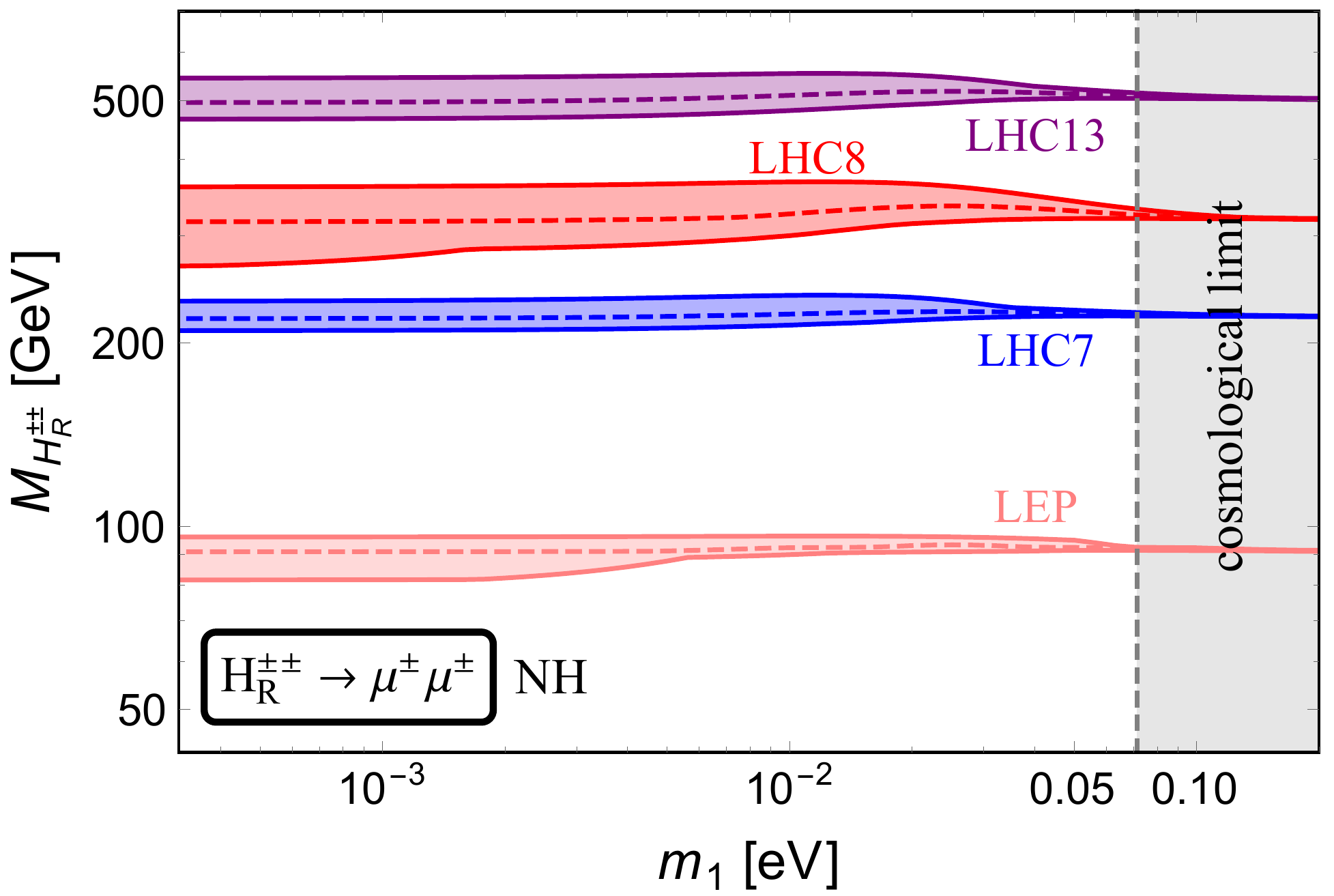}
  \includegraphics[width=0.48\textwidth]{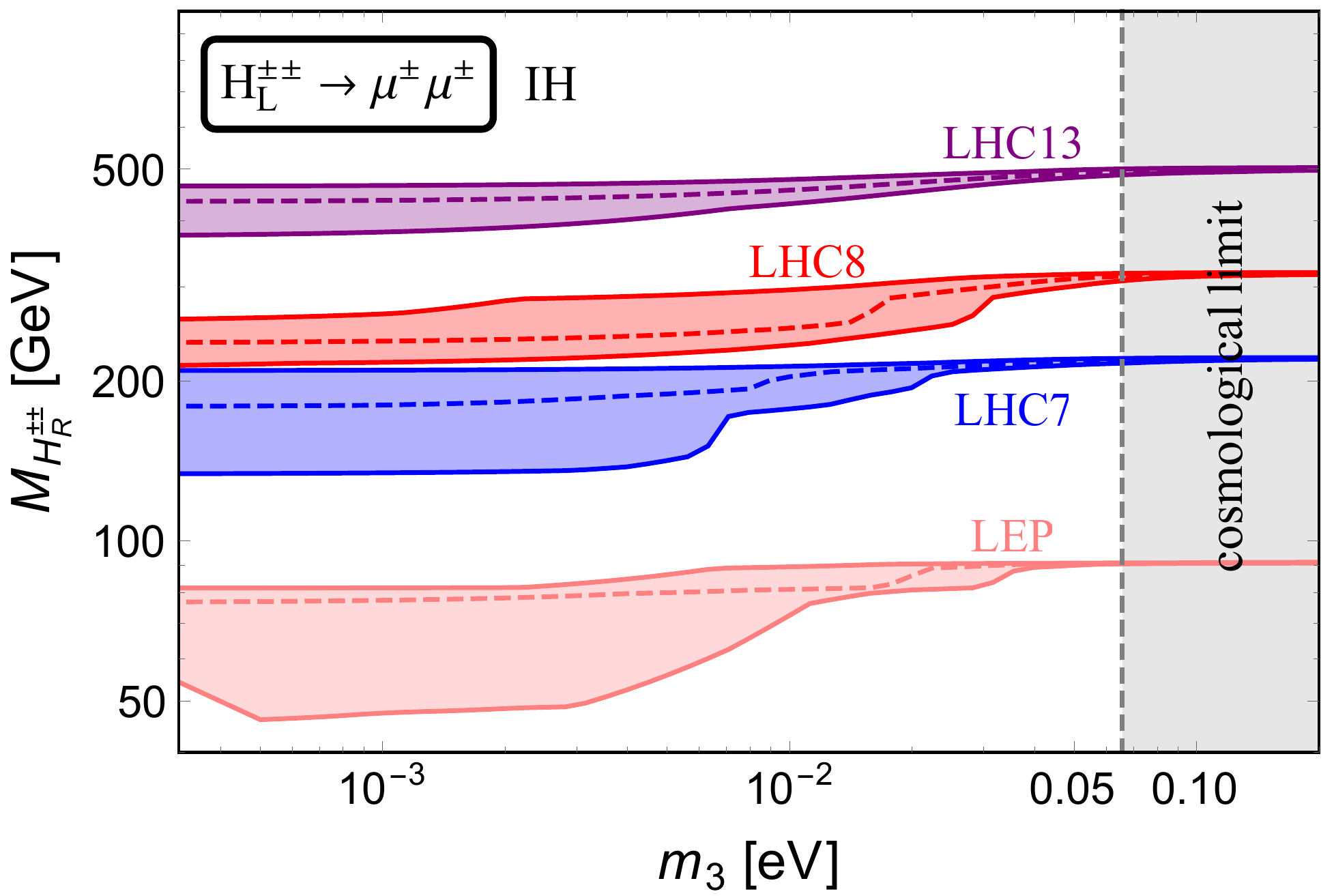} \vspace{3pt} \\
  \includegraphics[width=0.48\textwidth]{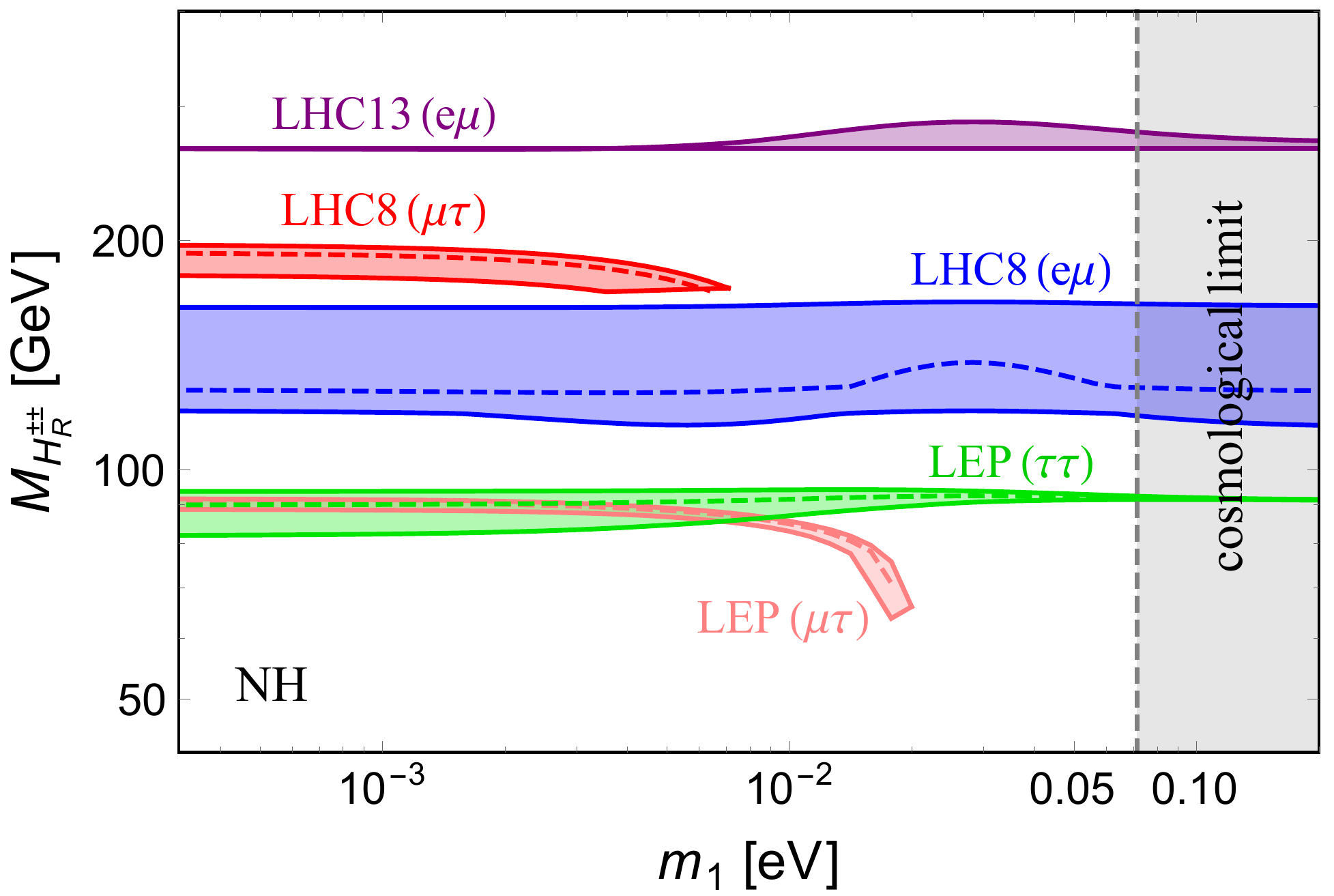}
  \includegraphics[width=0.48\textwidth]{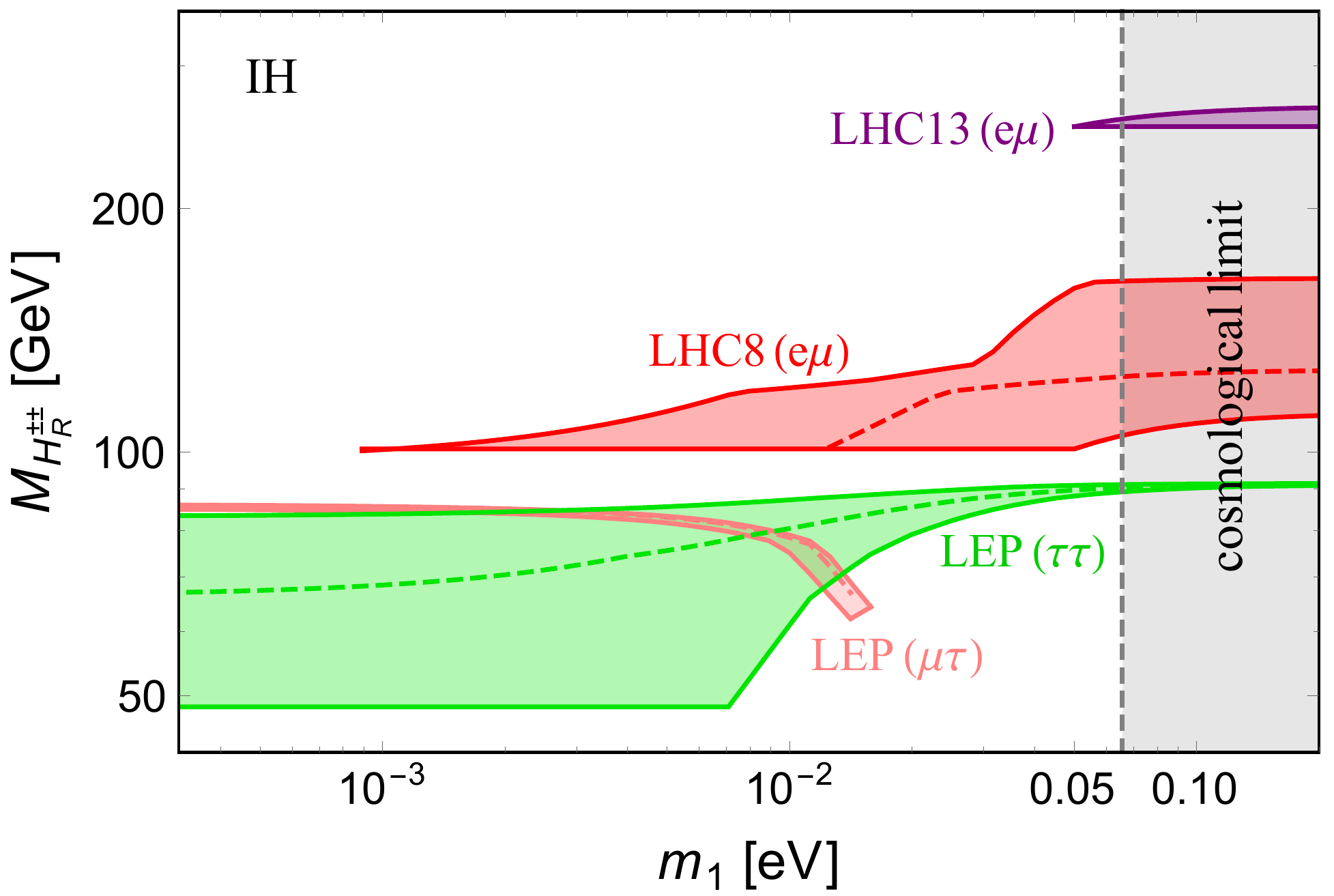}
  \caption{Same-sign dilepton lower limits on $M_{H_R^{\pm\pm}}$ in the LRSM from the data in Fig.~\ref{fig:dilepton:right:1} for different flavor combinations $H_R^{\pm\pm} \to \ell_\alpha^\pm \ell_\beta^\pm$, as functions of the lightest neutrino mass for NH (left) and IH (right). The dashed curves correspond to the central values of the neutrino data in Table~\ref{tab:neutriodata}, and the colorful bands are due to the $3\sigma$ uncertainties. The gray shaded region is excluded by the cosmological constraint on the sum of light neutrino masses $\sum_i m_{i} < 0.23$ eV~\cite{Ade:2015xua}.}
  \label{fig:dilepton:right:2}
\end{figure}

With the parity relation $f_L = f_R$, the Yukawa coupling matrix $f_R$ is also related to the active neutrino data in Table~\ref{tab:neutriodata}, as in the pure type-II seesaw. Analogous to Fig.~\ref{fig:dilepton:left:2}, the dilepton limits also depend  on the lightest neutrino mass $m_0$, which are collected in Fig.~\ref{fig:dilepton:right:2} for both the NH and IH neutrino spectra, in the limit of $\Gamma (H_R^{\pm\pm} \to \ell_\alpha^\pm \ell_\beta^\pm) \gg (H_R^{\pm\pm} \to W_R^{\pm \ast} W_R^{\pm \ast})$.

\begin{figure}[t!]
  \centering
  \includegraphics[width=0.48\textwidth]{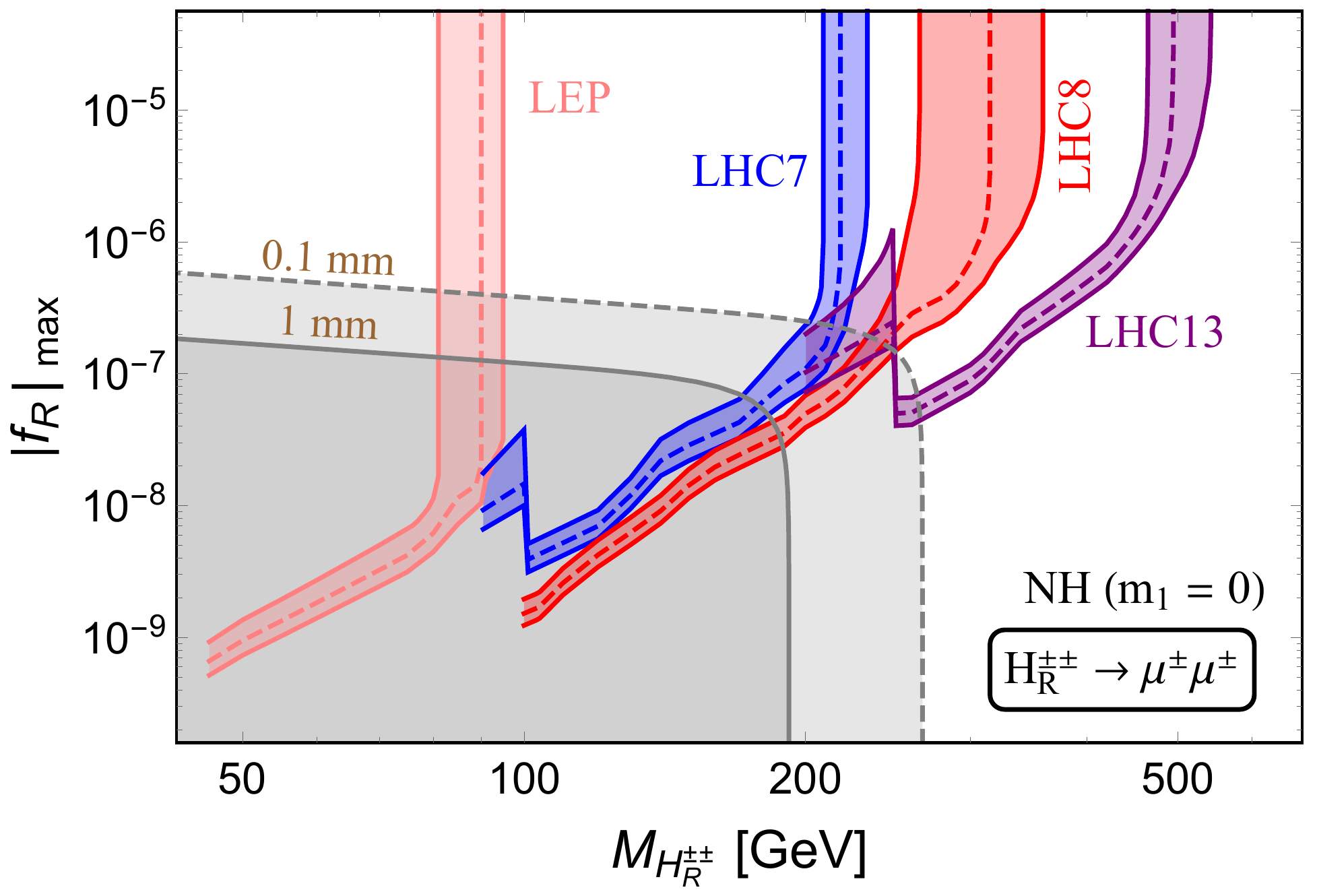}
  \includegraphics[width=0.48\textwidth]{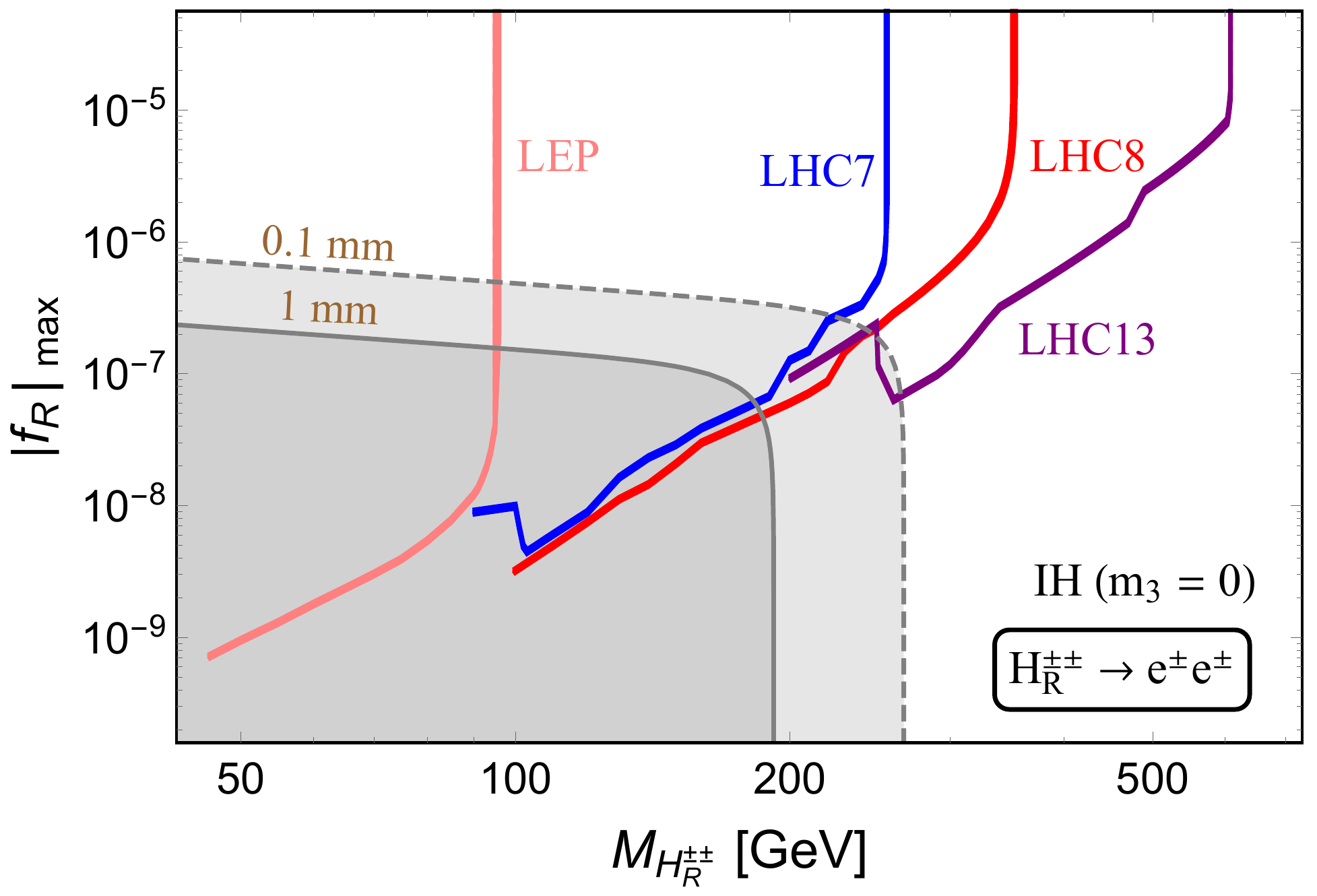} \vspace{3pt} \\
  \includegraphics[width=0.48\textwidth]{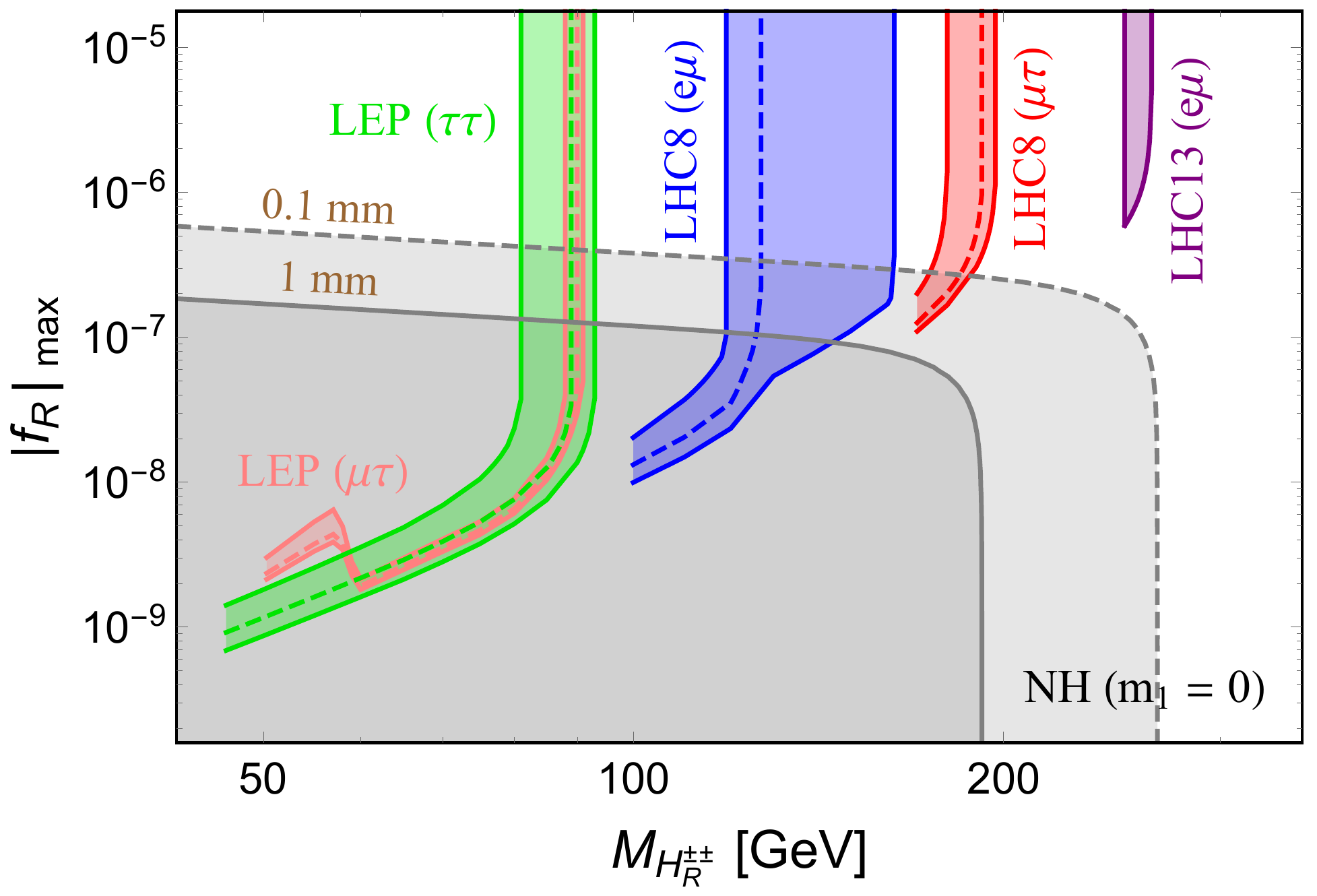}
  \includegraphics[width=0.48\textwidth]{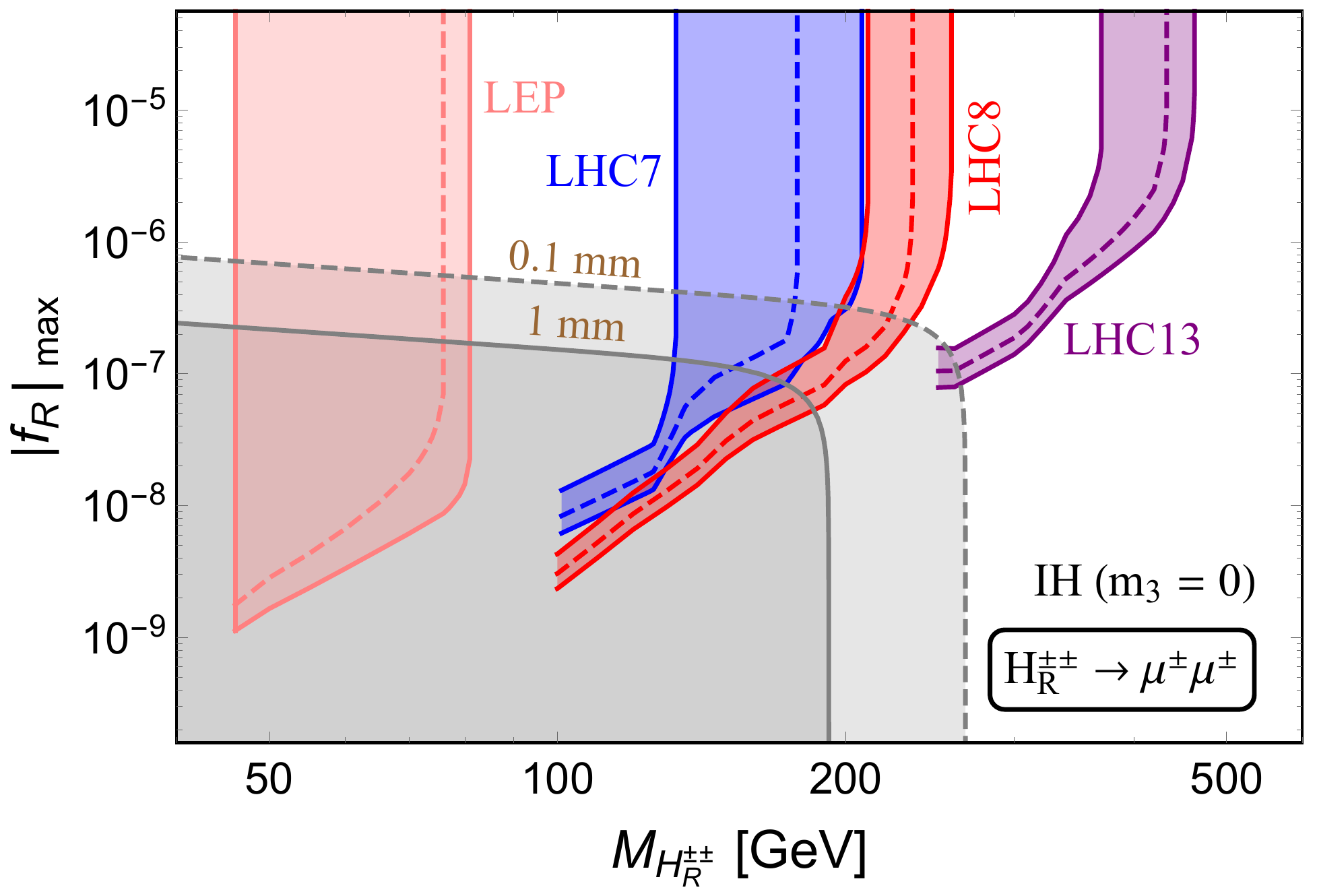} \vspace{3pt} \\
  \includegraphics[width=0.48\textwidth]{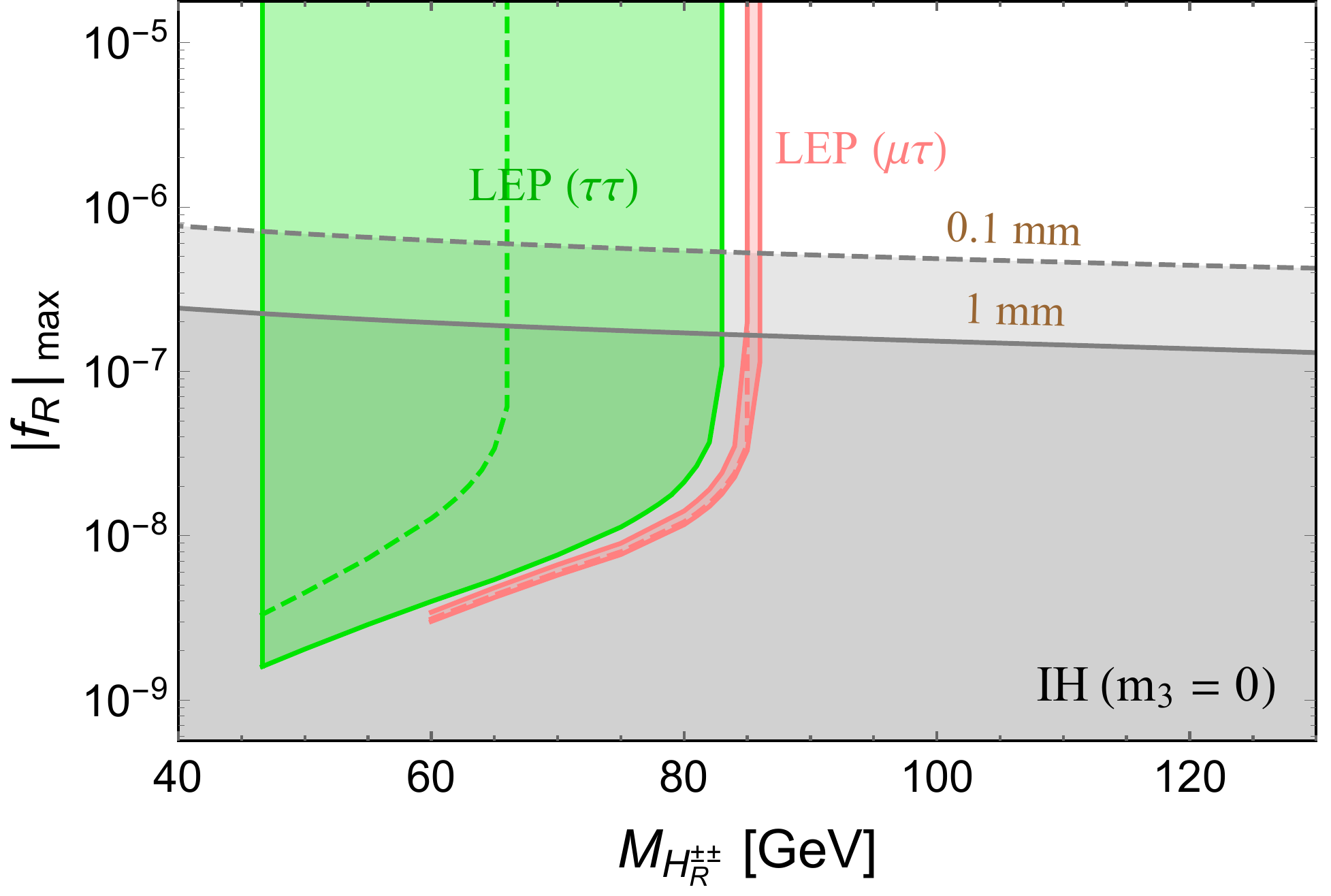}
  \caption{Same-sign dilepton lower limits on $M_{H_R^{\pm\pm}}$ in the LRSM from the data in Fig.~\ref{fig:dilepton:right:1}, as functions of the value of largest Yukawa coupling $|f_R|_{\rm max}$ for the lightest neutrino mass $m_0=0$. The upper and middle left panels are for the NH case, in the $\mu\mu$ channel and $e\mu$, $\mu\tau$ and $\tau\tau$ channels. The upper and middle right panels are the limits for the IH case in the $ee$ and $\mu\mu$ channels. The bottom panel are for the IH case in the $\mu\tau$ and $\tau\tau$ channels. The dashed curves correspond to the central values of the neutrino data in Table~\ref{tab:neutriodata}, and the colorful bands are due to the $3\sigma$ uncertainties. The darker and lighter gray regions correspond respectively to the proper decay length $c\tau_0 (H_R^{\pm\pm}) > 1$ mm and 0.1 mm; within these regions the prompt dilepton limits are not applicable.}
  \label{fig:dilepton:right:3}
\end{figure}

\begin{figure}[t!]
  \centering
  \includegraphics[width=0.48\textwidth]{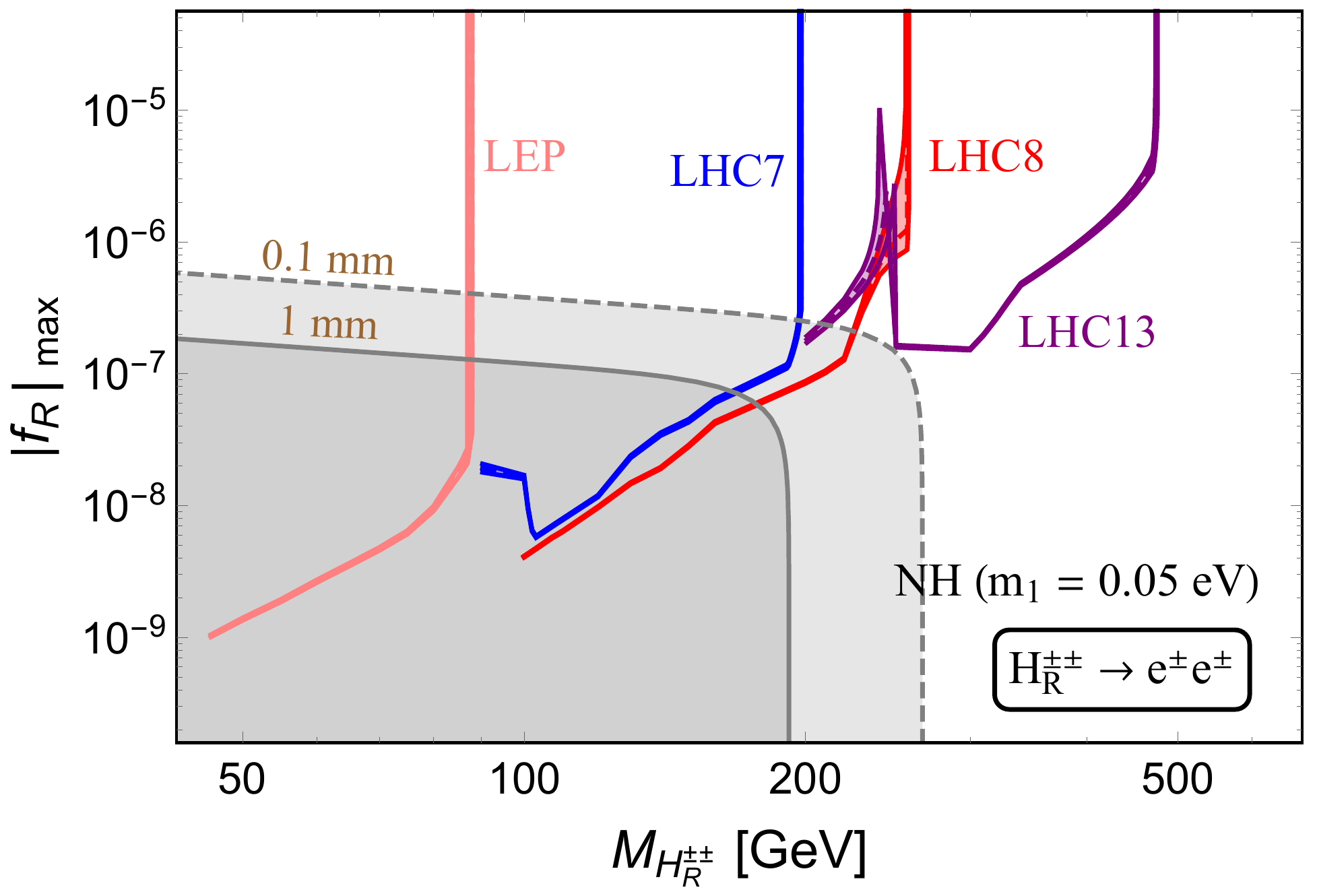}
  \includegraphics[width=0.48\textwidth]{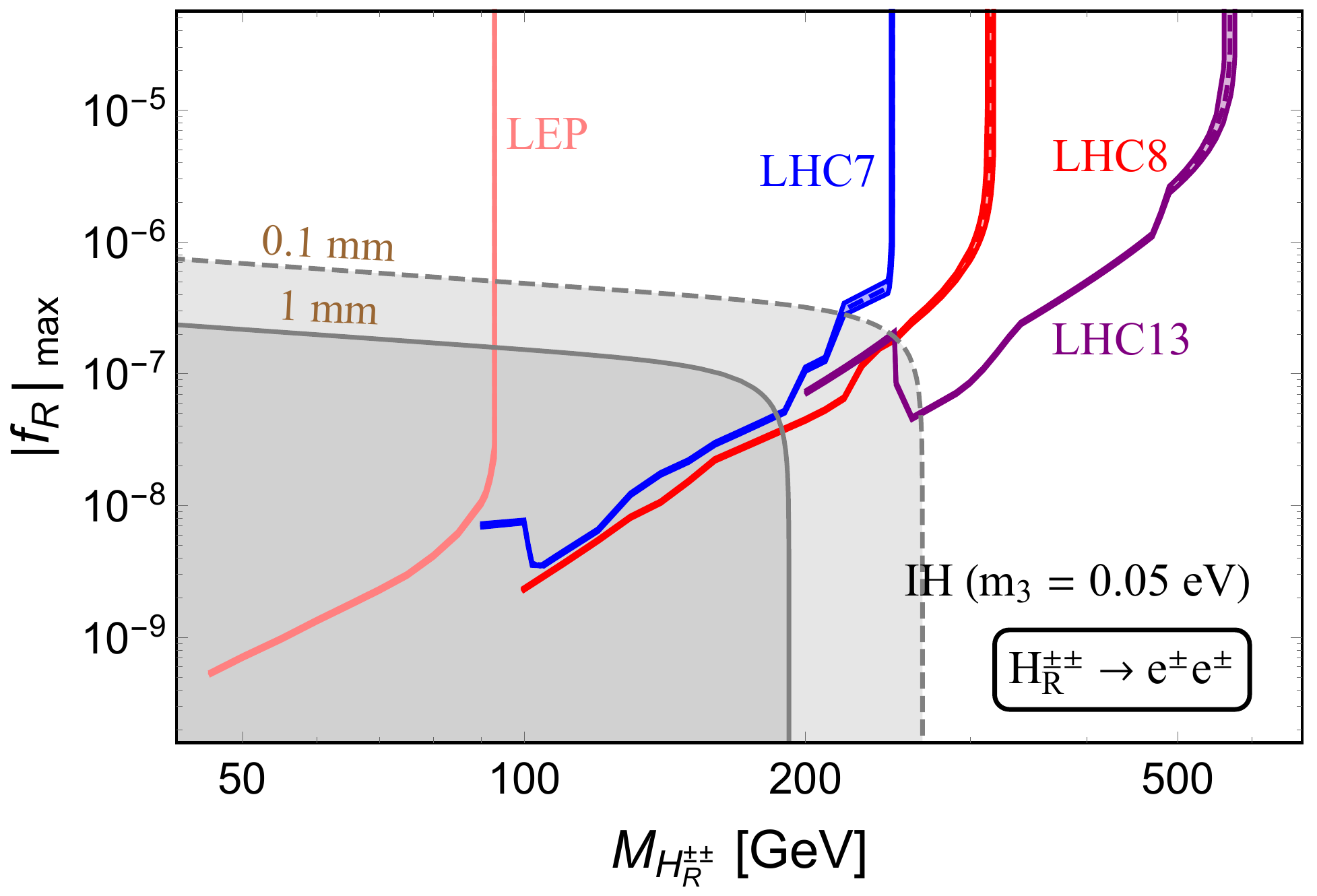} \vspace{3pt} \\
  \includegraphics[width=0.48\textwidth]{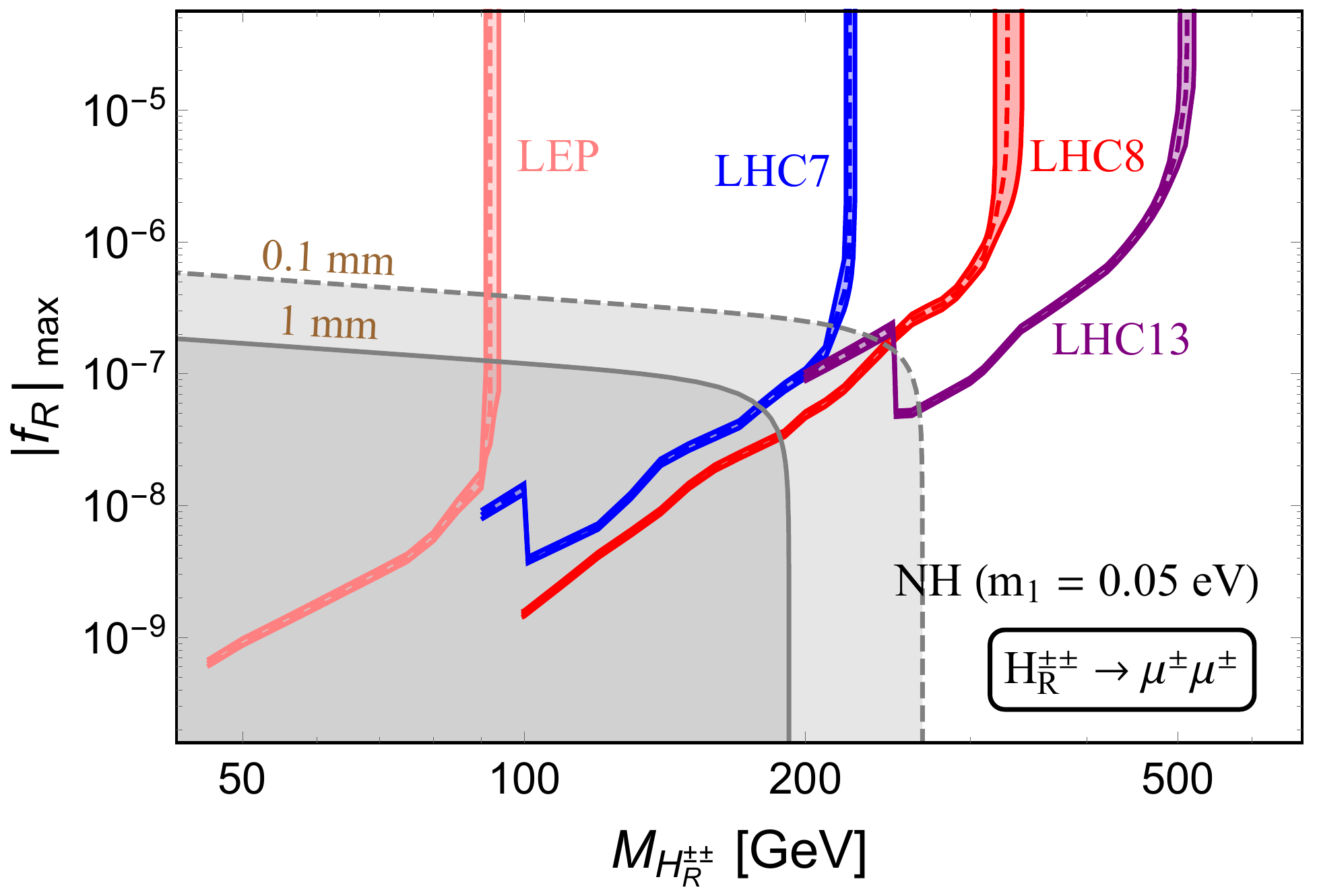}
  \includegraphics[width=0.48\textwidth]{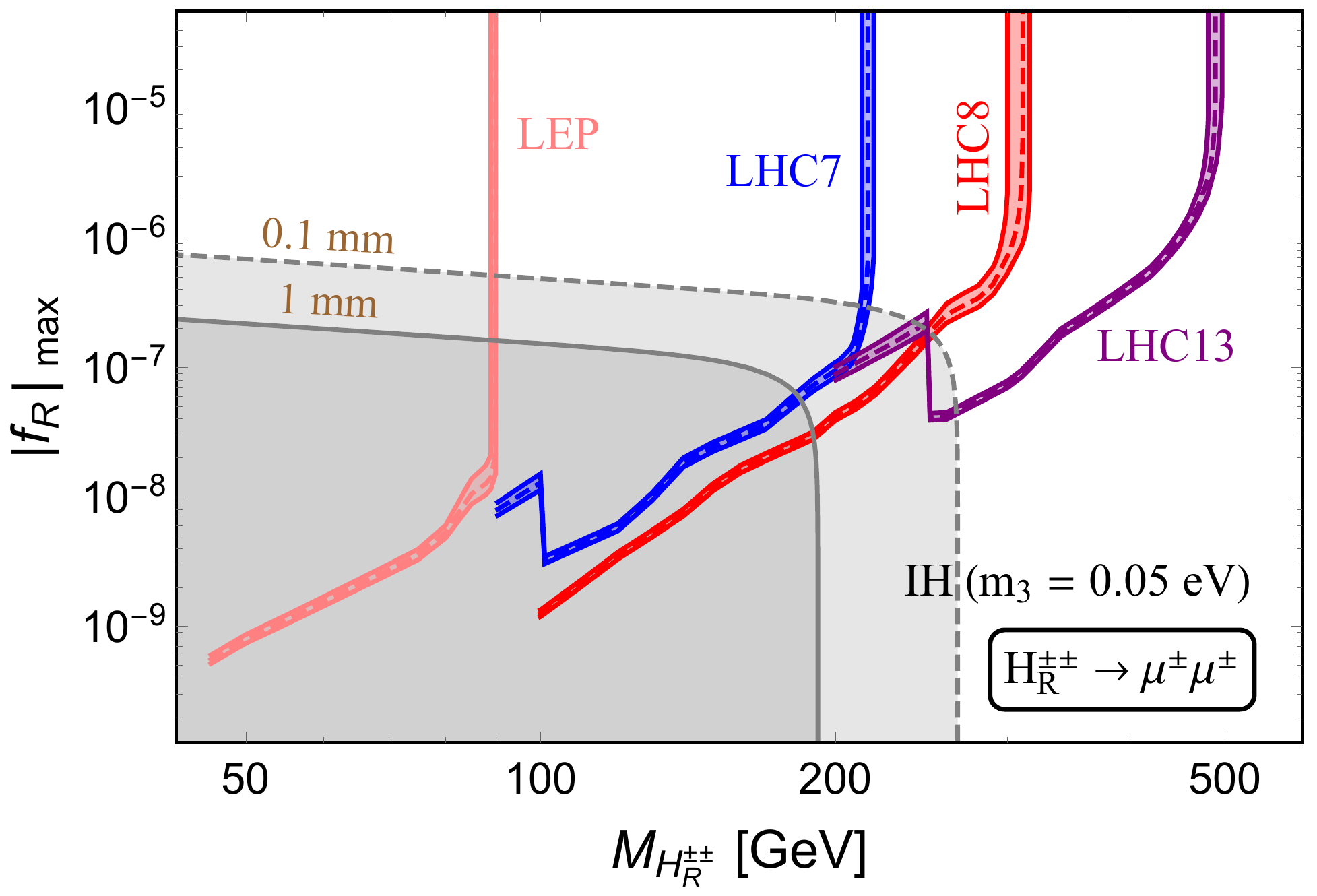} \vspace{3pt} \\
  \includegraphics[width=0.48\textwidth]{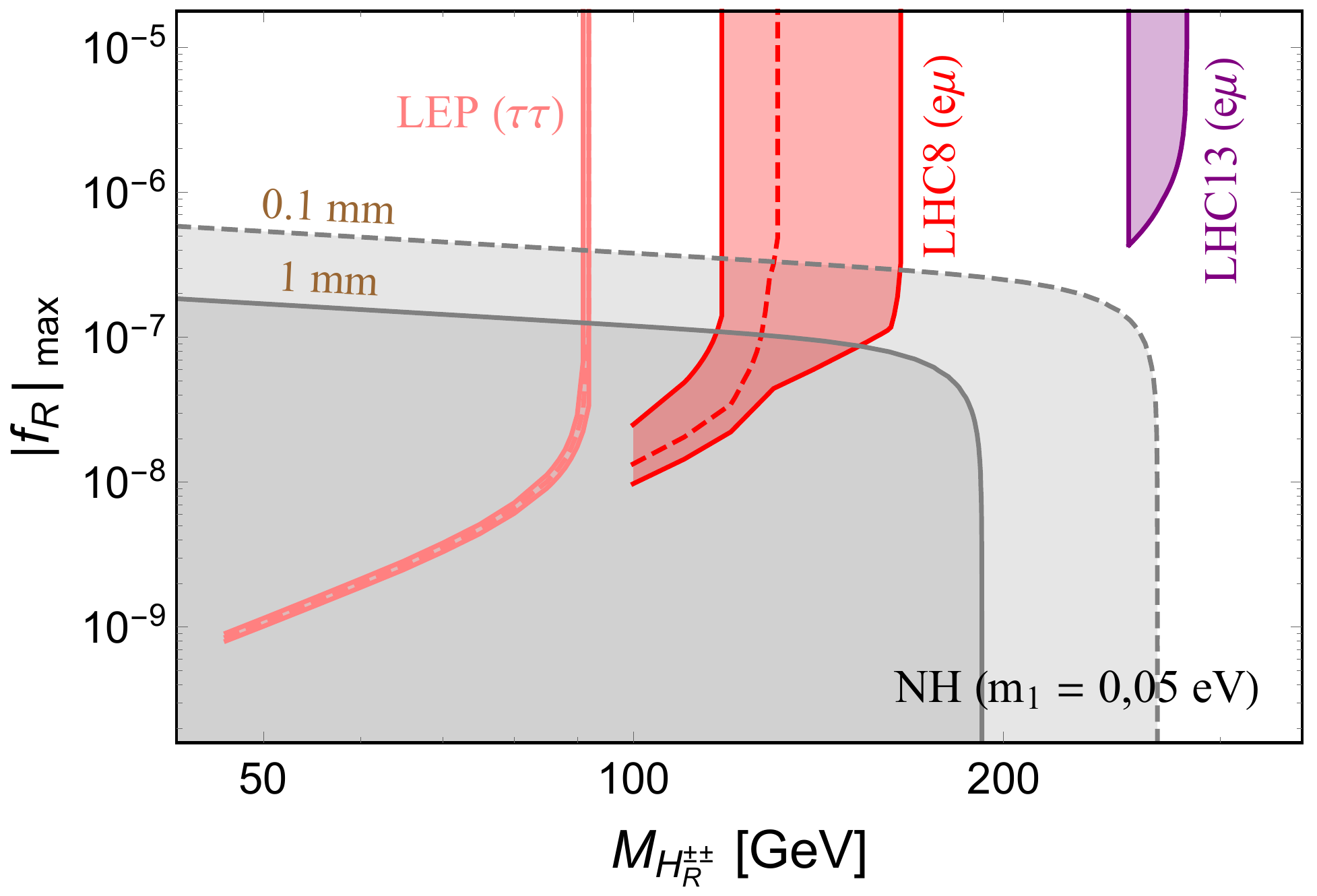}
  \includegraphics[width=0.48\textwidth]{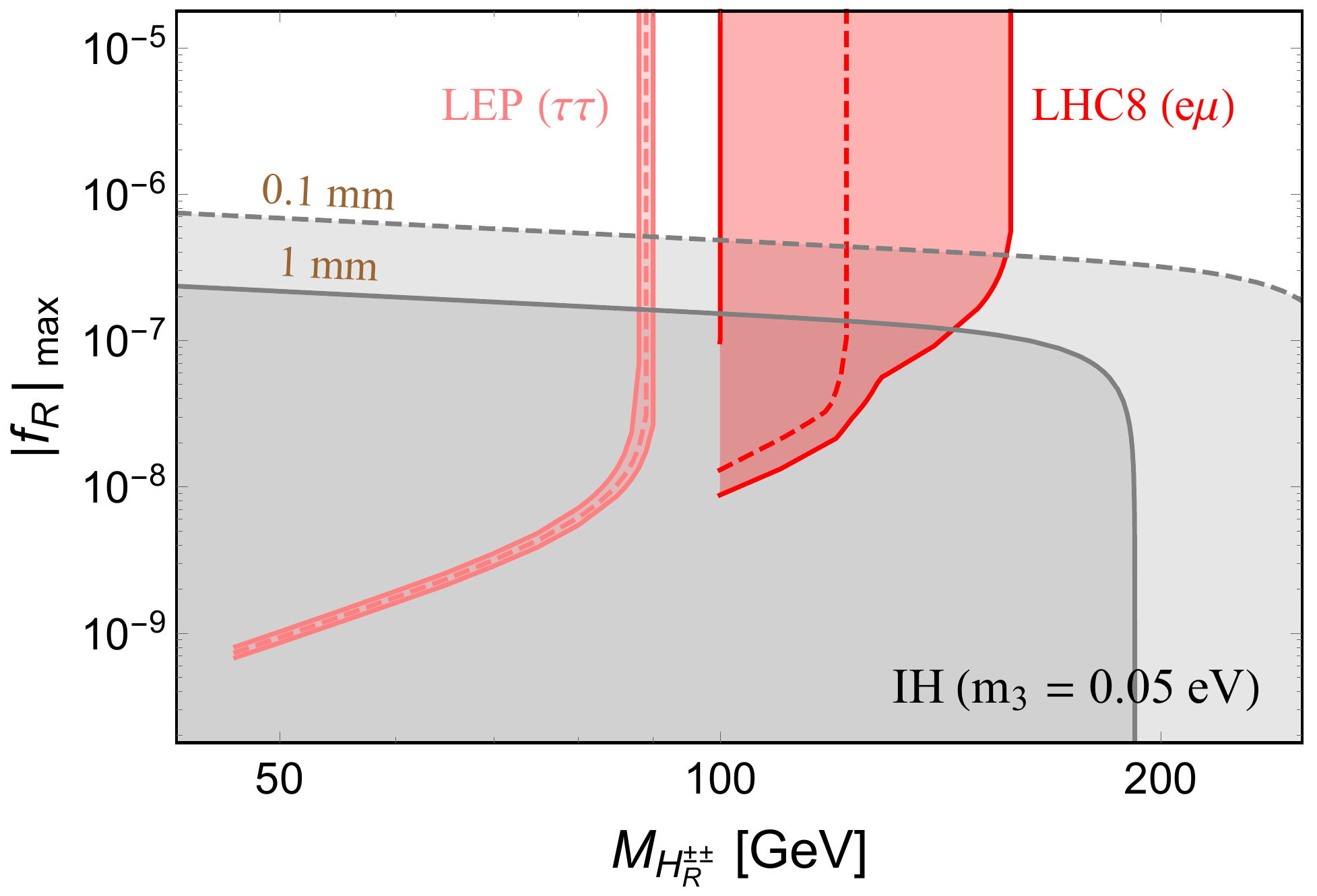}
  \caption{Same as in Fig.~\ref{fig:dilepton:right:3}, but for the lightest neutrino mass $m_0=0.05$ eV. The upper, middle and lower panels are respectively in the channels of ee, $e\mu$, $e\mu$ and $\tau\tau$ for both the NH and IH scenarios.}
  \label{fig:dilepton:right:4}
\end{figure}

In the LRSM, for sufficiently small Yukawa couplings $|f_R|$, a sizable portion of $H_R^{\pm\pm}$ decays into the heavy $W_R$ boson pairs. Then the same-sign dilepton limits from LEP, Tevatron and LHC can be interpreted as constraints on the mass of $H_R^{\pm\pm}$ and the largest coupling $|f_R|_{\rm max}$. Following Figs.~\ref{fig:dilepton:left:3} and \ref{fig:dilepton:left:4}, the NH and IH cases with the lightest neutrino mass $m_0 = 0$ are presented in Fig.~\ref{fig:dilepton:right:3}, and those with $m_0 = 0.05$ eV are shown in Fig.~\ref{fig:dilepton:right:4}. Again we have chosen the RH scale $v_R  = 5\sqrt2$ TeV and the gauge coupling $g_R = g_L$ in all these plots. For relatively light $H_R^{\pm\pm}$ with mass $M_{H_R^{\pm\pm}} \lesssim 200$ GeV and the coupling $|f_R| \lesssim 10^{-7}$, the decay lifetime $c\tau_0 (H_R^{\pm\pm})$ is noticeable at LEP and LHC (cf. Fig.~\ref{fig:lifetime2}), and therefore, in the regions with $c\tau_0 (H_R^{\pm\pm}) > 1$ mm (conservative) and 0.1 mm (aggressive), shaded respectively in darker and lighter gray in Figs.~\ref{fig:dilepton:right:3} and \ref{fig:dilepton:right:4}, the prompt same-sign dilepton limits are not applicable.

\begin{figure}[t!]
  \centering
  \includegraphics[width=0.48\textwidth]{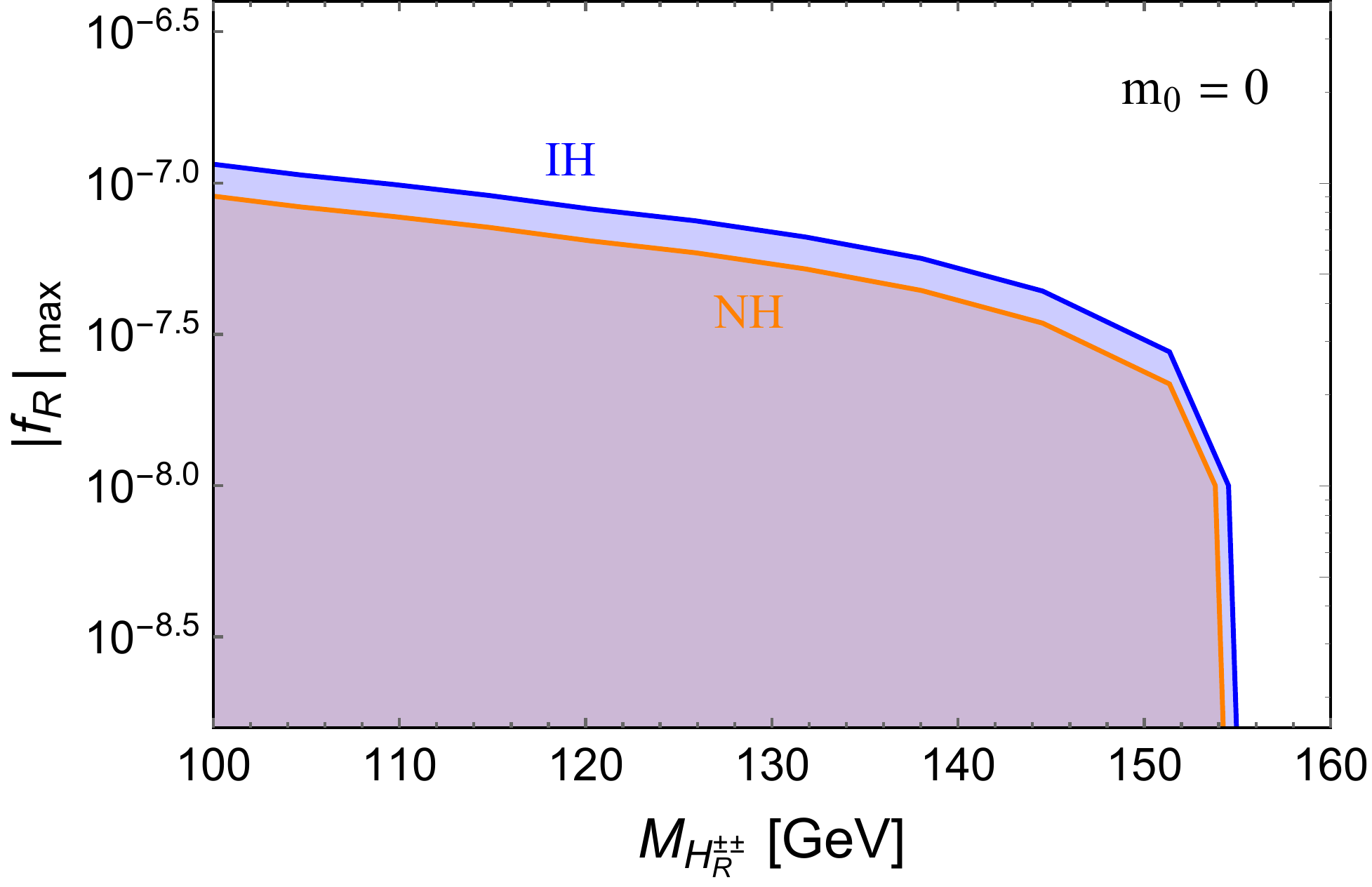}
  \caption{Limits on $M_{H_R^{\pm\pm}}$ in the LRSM and its largest Yukawa coupling $|f_R|_{\rm max}$ from searches of doubly-charged HSCPs by the CMS group~\cite{CMS:2016ybj}. The orange and blue regions are excluded respectively for the NH and IH cases with the lightest neutrino mass $m_0 = 0$.}
  \label{fig:HSCP2}
\end{figure}

As in Fig.~\ref{fig:HSCP} for the $H_L^{\pm\pm}$ in the type-II seesaw, the HSCP limits from Ref.~\cite{CMS:2016ybj} can be applied to the $H_R^{\pm\pm}$ case. To be conservative, we again use only the  ``tracker-only'' data in Ref.~\cite{CMS:2016ybj} to constrain the couplings of $H_R^{\pm\pm}$, with the decay length range of $43 \, {\rm mm} < bc\tau_0 (H_L^{\pm\pm}) < 1100$ mm~\cite{Chatrchyan:2008aa}. Setting again the RH scale $v_R = 5\sqrt2$ TeV, $g_R = g_L$ and rescaling the production cross section in Ref.~\cite{CMS:2016ybj} to that of $H_R^{\pm\pm}$ at $\sqrt{s} = 13$ TeV, the shaded orange and blue regions in Fig.~\ref{fig:HSCP2} are excluded respectively for the NH and IH cases, with the lightest neutrino mass $m_0 = 0$. This corresponds to the Yukawa coupling range $|f_L| \lesssim 10^{-7}$ for the mass range $100 \, {\rm GeV} < M_{H_R^{\pm\pm}} \lesssim 155$ GeV. Again, as in Fig.~\ref{fig:HSCP}, the HSCP exclusion regions in Fig.~\ref{fig:HSCP2} are not sensitive to the lightest neutrino mass $m_0$ for both the NH and IH scenarios.

\subsubsection{Neutrinoless double beta decay}
\label{sec:lrsm:DBD}

%By interacting with the $W_R$ boson, $H_R^{\pm\pm}$ contributes to $0\nu\beta\beta$, which is suppressed by heavy $W_R$ mass, or effectively by the right-handed scale $v_R$.

%To set limits by the limits from KamLAND-Zen~\cite{KamLAND-Zen:2016pfg} and GERDA Phase I~\cite{Agostini:2013mzu}, one has to compare the doubly-charged scalar mediated amplitudes with the canonical light neutrino diagrams, and make sure the former dominates.

\begin{figure}[!t]
  \centering
  \includegraphics[width=0.3\textwidth]{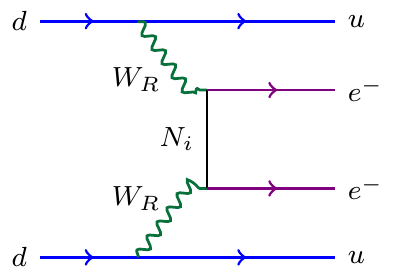}
  \includegraphics[width=0.3\textwidth]{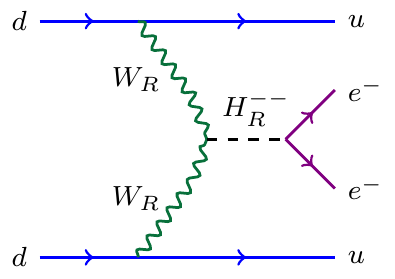}
  \caption{Additional Feynman diagram in the LRSM (in addition to Fig.~\ref{fig:diagram1}) for the parton-level $0\nu\beta\beta$, induced respectively by the heavy RHNs $N_i$ (left) and the RH doubly-charged scalar $H_R^{\pm\pm}$ (right), which correspond respectively to the amplitudes $\eta_N$ and $\eta_{\rm DCS}^R$ in Eq.~(\ref{eqn:0nubetabeta2}).}
  \label{fig:diagram2}
\end{figure}

In the LRSM, with the heavy $W_R$ and $H_R^{\pm\pm}$ bosons, the RHNs $N_i$ and the RH interactions, there are additional diagrams contributing to $0\nu\beta\beta$, cf. Fig.~\ref{fig:diagram2}, as  compared to the pure type-II seesaw case, cf. Fig.~\ref{fig:diagram1}, which could be important, depending on the heavy particle masses and the couplings and mixings involved. In the type-II dominance, neglecting the heavy-light neutrino mixing (responsible for the type-I seesaw) and the small $W - W_R$ mixing, the $0\nu\beta\beta$ half-life is given by~\cite{Chakrabortty:2012mh, Borah:2015ufa, Deppisch:2017vne, Ge:2015yqa, Cirigliano:2004tc, Barry:2013xxa, Mohapatra:1981pm, Hirsch:1996qw, Bambhaniya:2015ipg}\footnote{If the $W - W_R$ mixing is sizable, then the NME for this contribution is enhanced by chiral symmetry, and in principle can compete with the light and heavy neutrino contributions in Eq.~(\ref{eqn:0nubetabeta2}). See Ref.~\cite{Prezeau:2003xn} for more details.}
\begin{eqnarray}
\label{eqn:0nubetabeta2}
\left[ T^{0\nu}_{1/2} \right]^{-1} \ = \
G \, \left| {\cal M}_\nu \left( \eta_\nu + \eta_{\rm DCS}^{L} \right) + {\cal M}_N \left( \eta_N + \eta_{\rm DCS}^{R} \right)\right|^2 \,,
\end{eqnarray}
where the first term on the RHS is the LH contribution and same as in Eq.~(\ref{eqn:0nubetabeta}), whereas the second term is the RH contribution with ${\cal M}_N$ being the corresponding NME and
\begin{eqnarray}
\label{eqn:etaN}
\eta_N & \ = \ & m_p
\left( \frac{g_R}{g_L} \right)^4
\left( \frac{m_W}{M_{W_R}} \right)^4
\sum_i \frac{U_{ei}^2}{M_{N_i}}  \ = \
\frac{m_p}{4}
\left( \frac{v_{\rm EW}}{v_{R}} \right)^4
\sum_i \frac{U_{ei}^2}{M_{N_i}} \,, \\
\label{eqn:etaDCSR}
\eta_{\rm DCS}^R & \ = \ & m_p
\left( \frac{g_R}{g_L} \right)^4
\left( \frac{m_W}{M_{W_R}} \right)^4
\sum_i \frac{U_{ei}^2 M_{N_i}}{M_{H_R^{\pm\pm}}^2}  \ = \
\frac{m_p}{4}
\left( \frac{v_{\rm EW}}{v_{R}} \right)^4
\sum_i \frac{U_{ei}^2 M_{N_i}}{M_{H_R^{\pm\pm}}^2}
\end{eqnarray}
where $m_p$ is the proton mass, $M_{N_i}$ the mass eigenvalues for the three heavy RHNs, and we have applied the fact that the mixing matrix of RHNs, $U_R = U$ under parity. Note that there is essentially no gauge dependence on the gauge coupling $g_R$, as the $W_R$ boson couples to the fermions and $H_R^{\pm\pm}$ with the strength $g_R^2$, which is completely canceled out by the $g_R$ dependence in the $W_R$ propagator~\cite{Dev:2018sel}. The $H_R^{\pm\pm}$ contribution in Eq.~(\ref{eqn:etaDCSR}) is effectively suppressed by the RH scale $v_R$. Using the left-right symmetry $f_L = f_R$ which implies $M_{N_i} = (v_R / v_L) m_{i}$, we get
\begin{eqnarray}
\sum_i U_{ei}^2 M_{N_i} \ = \
\frac{v_R}{v_L} \sum_i U_{ei}^2 m_{i} \ = \
\frac{v_R}{v_L} \, (m_\nu)_{ee} \,,
\end{eqnarray}
and therefore,
\begin{eqnarray}
\eta_{\rm DCS}^R  & \ = \ &
%\left[ \frac{m_e m_p}{M_{H_R^{\pm\pm}}^2}
%\left( \frac{g_R}{g_L} \right)^4
%\left( \frac{m_W}{M_{W_R}} \right)^4
%\left( \frac{v_L}{v_R} \right)^{-1} \right] \nonumber \\
%& \ = \ &
\frac{m_e m_p}{4M_{H_R^{\pm\pm}}^2}
\left( \frac{v_{\rm EW}}{v_{R}} \right)^4
\left( \frac{v_R}{v_L} \right)  \eta_\nu \ \equiv R_{\rm DCS}\eta_\nu \,.
\end{eqnarray}
To have a large ${\eta}^R_{\rm DCS}$, the VEV ratio $v_L / v_R$ has to be small, say $v_L \sim {\rm eV}$ and $v_R \sim$ few TeV, such that the Yukawa couplings $f_{R} = f_L \simeq m_\nu / v_L$ are sizable or equivalently the RHN masses $M_{N_i}$ are large in Eq.~(\ref{eqn:etaDCSR}). This is a natural scenario one needs in the type-II seesaw dominance of LRSM to generate the tiny neutrino masses.

Comparing the two RH terms in Eq.~(\ref{eqn:0nubetabeta2}), we find that to have a large $H_R^{\pm\pm}$ contribution to $0\nu\beta\beta$ we require the mass ratio $M_{N_i}^2 / M_{H_R^{\pm\pm}}^2$ to be large, such that the $\eta_N$ term is  suppressed by the RHN masses, as compared to ${\eta}^R_{\rm DCS}$. To be specific, we define the ratio
\begin{eqnarray}
{\cal R}_{\rm DCS}^\prime \ = \
\frac{\eta_{\rm DCS}^R}{\eta_N} \ = \
\sum_i \frac{U_{ei}^2 M_{N_i}}{M_{H_R^{\pm\pm}}^2} \bigg/
\sum_i \frac{U_{ei}^2}{M_{N_i}}  \,.
\end{eqnarray}
When both ratios
\begin{eqnarray}
({\cal M}_N / {\cal M}_\nu) {\cal R}_{\rm DCS} \ > \ 1 \;\; {\rm and} \;\;
|{\cal R}_{\rm DCS}^\prime| \ > \ 1
\end{eqnarray}
the $0\nu\beta\beta$ process in the LRSM is dominated by the $H_R^{\pm\pm}$ contribution and we can set meaningful limits on $M_{H_R^{\pm\pm}}$ from the null results in current $0\nu\beta\beta$ searches, such as EXO-200~\cite{Albert:2017owj}, KamLAND-Zen~\cite{KamLAND-Zen:2016pfg}, GERDA~\cite{Agostini:2018tnm}, MAJORANA DEMONSTRATOR~\cite{Aalseth:2017btx}, CUORE~\cite{Alduino:2017ehq} and NEMO-3~\cite{Arnold:2018tmo}. In particular, $({\cal M}_N / {\cal M}_\nu) {\cal R}_{\rm DCS}$ does not depend on any of the neutrino data in Table~\ref{tab:neutriodata} but is only subject to the uncertainties of the NMEs ${\cal M}_{\nu}$ and ${\cal M}_N$. In addition, if the RHNs are too heavy (or equivalently the ratio $(v_L/v_R)$ is small for fixed value of $v_R$) and the doubly-charged scalar $H_R^{\pm\pm}$ is sufficiently light in Eq.~(\ref{eqn:etaDCSR}), then the half life $T^{0\nu}_{1/2}$ might be too small to be allowed by the current limits.

\begin{figure}[t!]
  \centering
  \includegraphics[width=0.48\textwidth]{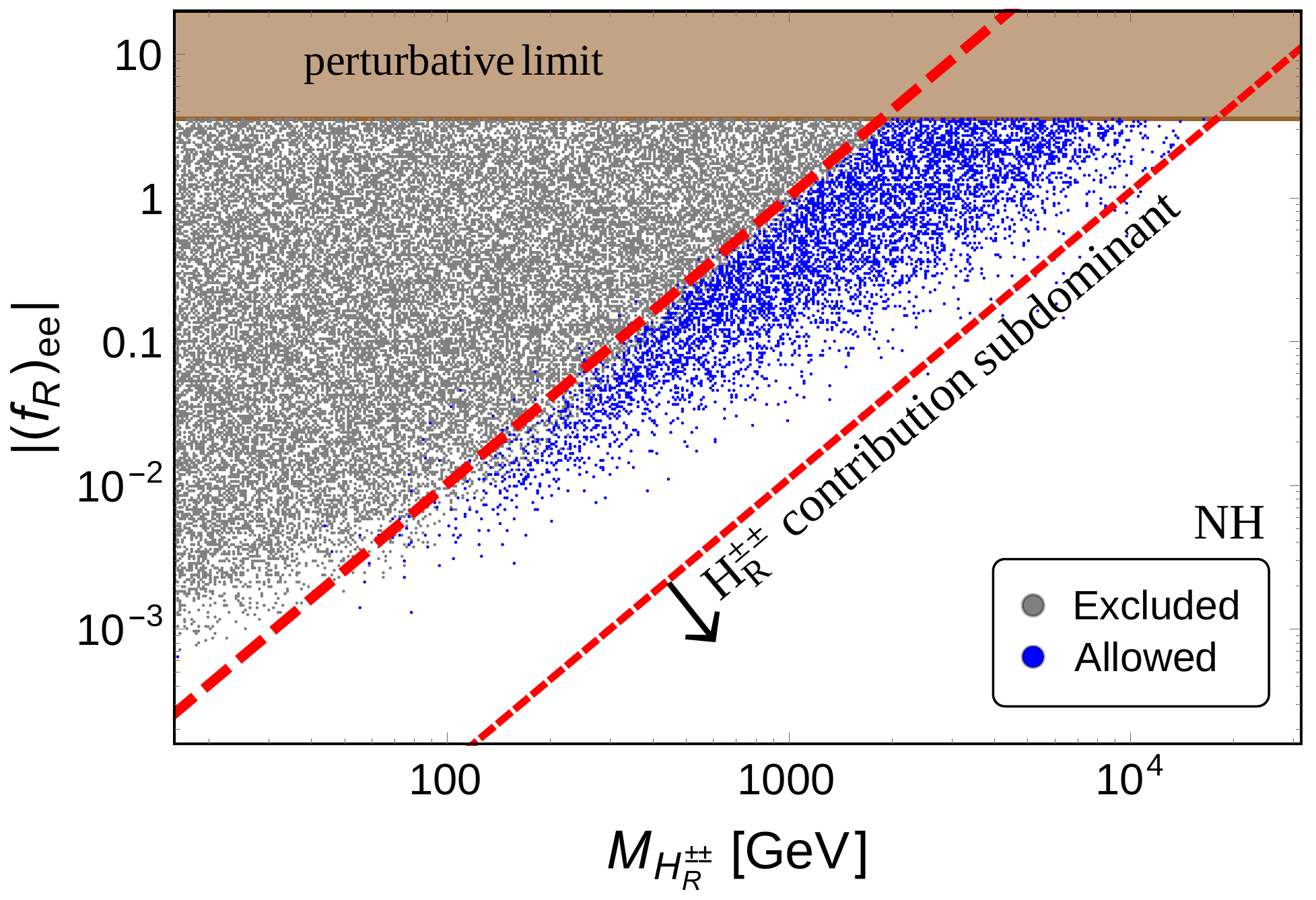}
  \includegraphics[width=0.48\textwidth]{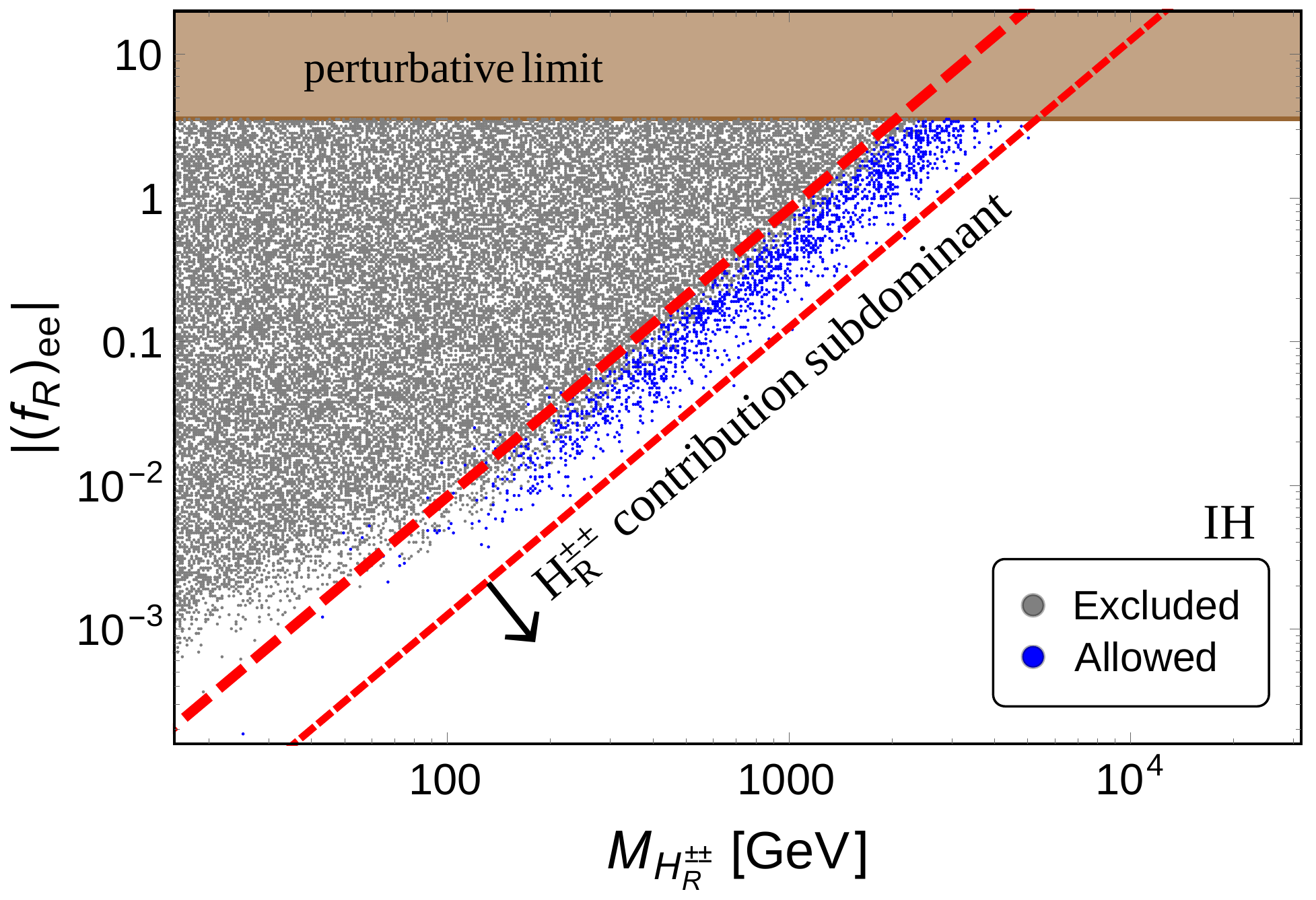}
  \caption{$0\nu\beta\beta$ constraints on $H_R^{\pm\pm}$ mass and its Yukawa coupling to electrons $|(f_R)_{ee}|$ for the neutrino spectra of NH (left) and IH (right). All the gray points are excluded by the KamLAND-Zen~\cite{KamLAND-Zen:2016pfg} and GERDA~\cite{Agostini:2018tnm} limits, while those in blue are allowed. The brown bands are excluded by the perturbativity requirement of $|(f_R)_{ee}| < \sqrt{4\pi}$. Below the short-dashed red lines, the $H_R^{\pm\pm}$ contribution to $0\nu\beta\beta$ is sub-dominant to other terms in Eq.~(\ref{eqn:0nubetabeta2}).}
  \label{fig:0nubetabeta}
\end{figure}

To set limits on $H_R^{\pm\pm}$ from $0\nu\beta\beta$, we use the most stringent half-life limits of $1.07 \times 10^{26}$ yrs for $^{136}$Xe from KamLAND-Zen~\cite{KamLAND-Zen:2016pfg} and $8.0 \times 10^{25}$ yrs for $^{76}$Ge from GERDA~\cite{Agostini:2018tnm}, both at the 90\% CL. We assume the RH scale $v_R = 5\sqrt2 $ TeV as above, and adopt the NMEs
\begin{align}
{\cal M}_\nu&: & [2.58,\, 6.64] &\text{ for $^{76}$Ge, } &
[1.57,\, 3.85] &\text{ for $^{136}$Xe} \,, \\
{\cal M}_N&: & [233,\, 412] &\text{ for $^{76}$Ge, } &
[164,\, 172] &\text{ for $^{136}$Xe} \,,
\end{align}
and the phase space factor $G = 5.77 \times 10^{-15}$ yr$^{-1}$ for $^{76}$Ge and $3.56 \times 10^{-14}$ yr$^{-1}$ for $^{136}$Xe from Ref.~\cite{Meroni:2012qf}. We vary the neutrino oscillation parameters in Table~\ref{tab:neutriodata} within their $3\sigma$ ranges and the lightest neutrino mass $m_0 \in [0,\, 0.05]$ eV. Our results are shown in Fig.~\ref{fig:0nubetabeta} for both NH (left) and IH (right) cases. All the gray points are excluded by either KamLAND-Zen~\cite{KamLAND-Zen:2016pfg} or  GERDA~\cite{Agostini:2018tnm} limit, which implies an upper bound on the combination
\begin{eqnarray}
\label{eqn:DBDlimit}
\frac{|(f_R)_{ee}|} {M_{H_R^{\pm\pm}}^2} <
\begin{cases}
1.0 \times 10^{-6} \, {\rm GeV}^{-2} & \text{for NH , } \\
8.3 \times 10^{-7} \, {\rm GeV}^{-2} & \text{for IH ,}
\end{cases}
\end{eqnarray}
as shown by the long-dashed red lines in Fig.~\ref{fig:0nubetabeta}.
Note that the dependence on the doubly-charged scalar mass and the Yukawa coupling is different from the LFV limits in Table~\ref{tab:neutriodata}. For heavier $H_R^{\pm\pm}$ and/or smaller coupling $|(f_R)_{ee}|$, the contribution of $H_R^{\pm\pm}$ is suppressed [cf.~Eq.~(\ref{eqn:etaDCSR})] and the $0\nu\beta\beta$ decay is dominated by the light and heavy neutrino terms in Eq.~(\ref{eqn:0nubetabeta2}). In this case, the KamLAND-Zen and GERDA limits are no longer applicable to $H_R^{\pm\pm}$, which is indicated by  the short-dashed red lines in Fig.~\ref{fig:0nubetabeta}.

%To have effective limits on the right-handed doubly-charged scalar from $0\nu\beta\beta$, we need to compare further the two terms

\subsection{Displaced vertex prospects}
\label{sec:lrsm:DV}

\begin{figure}[t!]
  \centering
  \includegraphics[width=0.48\textwidth]{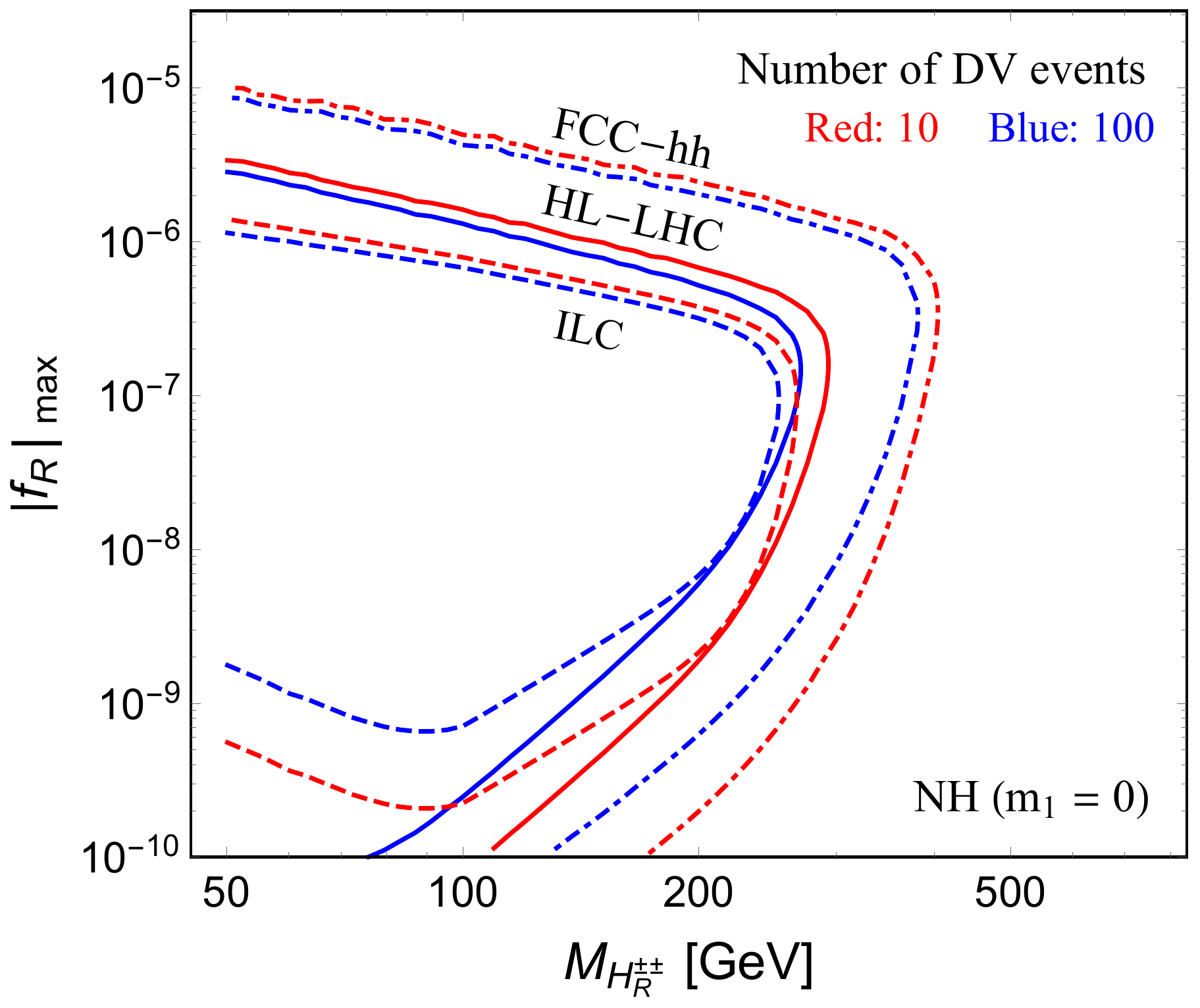}
  \includegraphics[width=0.48\textwidth]{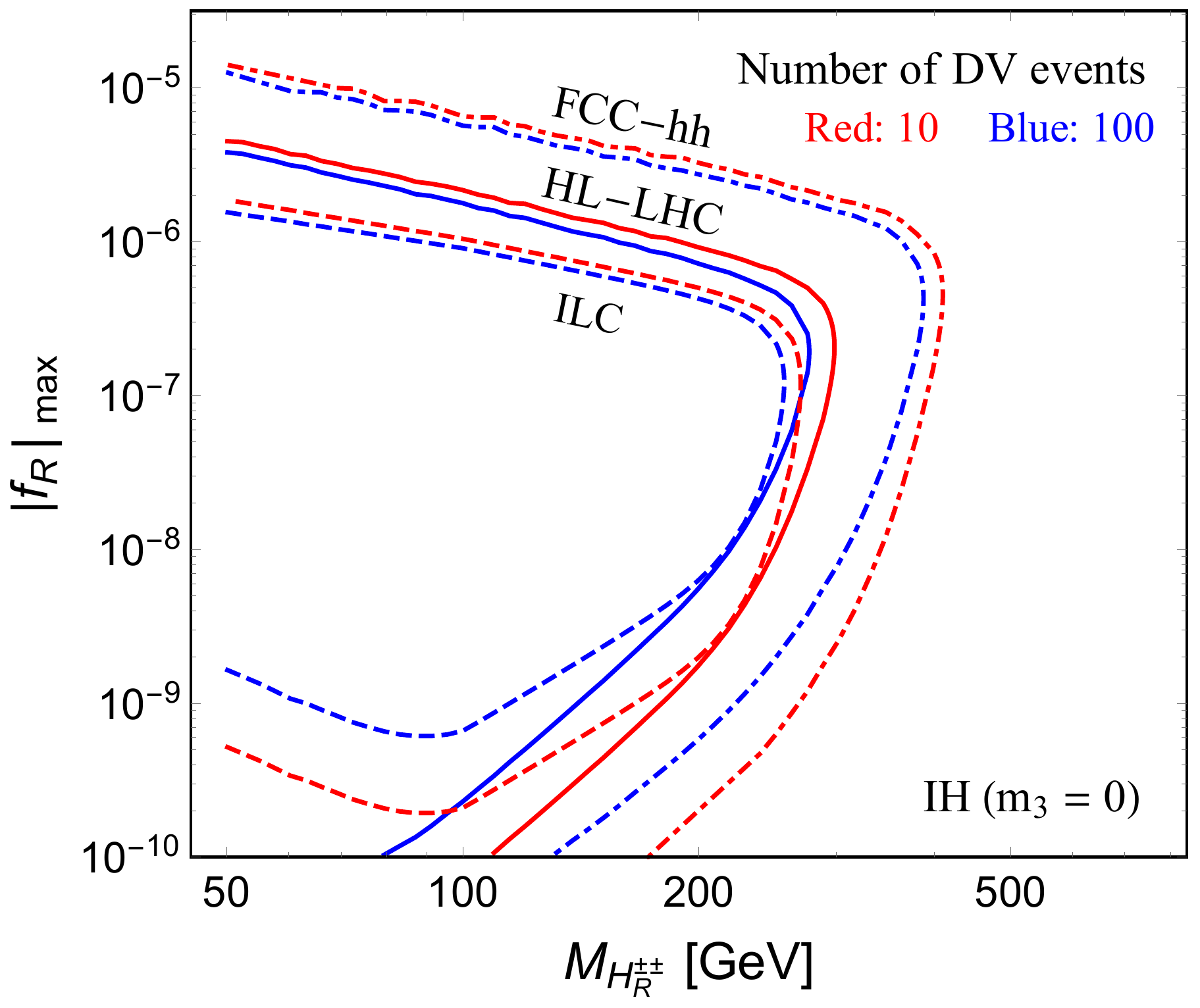}
  \caption{DV prospects from the decay of $H_R^{\pm\pm} \to e^\pm e^\pm,\, e^\pm\mu^\pm,\, \mu^\pm\mu^\pm$ in the LRSM, at HL-LHC 14 TeV with an integrated luminosity of 3000 fb$^{-1}$ (solid contours), the 100 TeV collider FCC-hh with a luminosity of 30 ab$^{-1}$ (dot-dashed contours) and ILC 1 TeV with 1 ab$^{-1}$ (dashed contours). The red and blue contours respectively correspond to 10 and 100 DV events, as functions of $M_{H_R^{\pm\pm}}$ and the largest Yukawa coupling $|f_R|_{\rm max}$, for the neutrino spectra of NH (left) and IH (right) with lightest neutrino mass $m_0 = 0$. }
  \label{fig:DV:right}
\end{figure}

As shown in Fig.~\ref{fig:lifetime2}, for sufficiently light $H_R^{\pm\pm}$ and sufficiently small Yukawa coupling $|f_R|$, the decay length $c\tau_0 (H_R^{\pm\pm})$ could reach up to 1 m. Requiring again a decay length of $1 \, {\rm mm} < bc\tau_0 (H_R^{\pm\pm}) < 1$ (3) m at the LHC and ILC (future 100 TeV collider FCC-hh) and adopting the same setups as in Section~\ref{sec:DV1}, we predict the number of DV events for $H_R^{\pm\pm} \to e^\pm e^\pm,\, e^\pm \mu^\pm,\, \mu^\pm \mu^\pm$ at the HL-LHC 14 TeV with a luminosity of 3000 fb$^{-1}$, ILC 1 TeV with a luminosity of 1 ab$^{-1}$ and FCC-hh 100
TeV with a luminosity of 30 ab$^{-1}$, which are depicted as the solid, dashed and dot-dashed curves in Fig.~\ref{fig:DV:right} respectively. The red and blue curves are the contours for respectively  10 and 100 DV events, for both NH with $m_1 = 0$ (left) and IH with $m_3 = 0$ (right). For extremely small couplings $f_R \lesssim 10^{-10}$, with a fixed RH scale $v_R = 5\sqrt2$ TeV, the RHNs are expected to be lighter than the keV scale, i.e. $M_{N} = \sqrt2 f_R v_R \lesssim $ keV, and are tightly constrained by the cosmological data~\cite{Vincent:2014rja, Abazajian:2017tcc}. Thus we impose a lower cut of $10^{-10}$ on the Yukawa coupling $|f_R|$ in Fig.~\ref{fig:DV:right}.
As shown in Fig.~\ref{fig:DV:right}, we can have at least 100 DV events at the HL-LHC for a broad parameter space with $10^{-10} \lesssim |f_R| \lesssim 10^{-5.5}$ and $m_Z/2 < H_R^{\pm\pm} \lesssim 200$ GeV. A larger mass and coupling range can be probed at  future 100 TeV colliders. Limited by the smaller center-of-mass energy, the sensitivities at ILC 1 TeV is relatively weaker. As in the pure type-II seesaw case, the total width of $H_R^{\pm\pm}$ and the DV prospects are not so sensitive to the lightest neutrino mass $m_0$ for either NH or IH case.

\begin{figure}[t!]
  \includegraphics[height=0.38\textwidth]{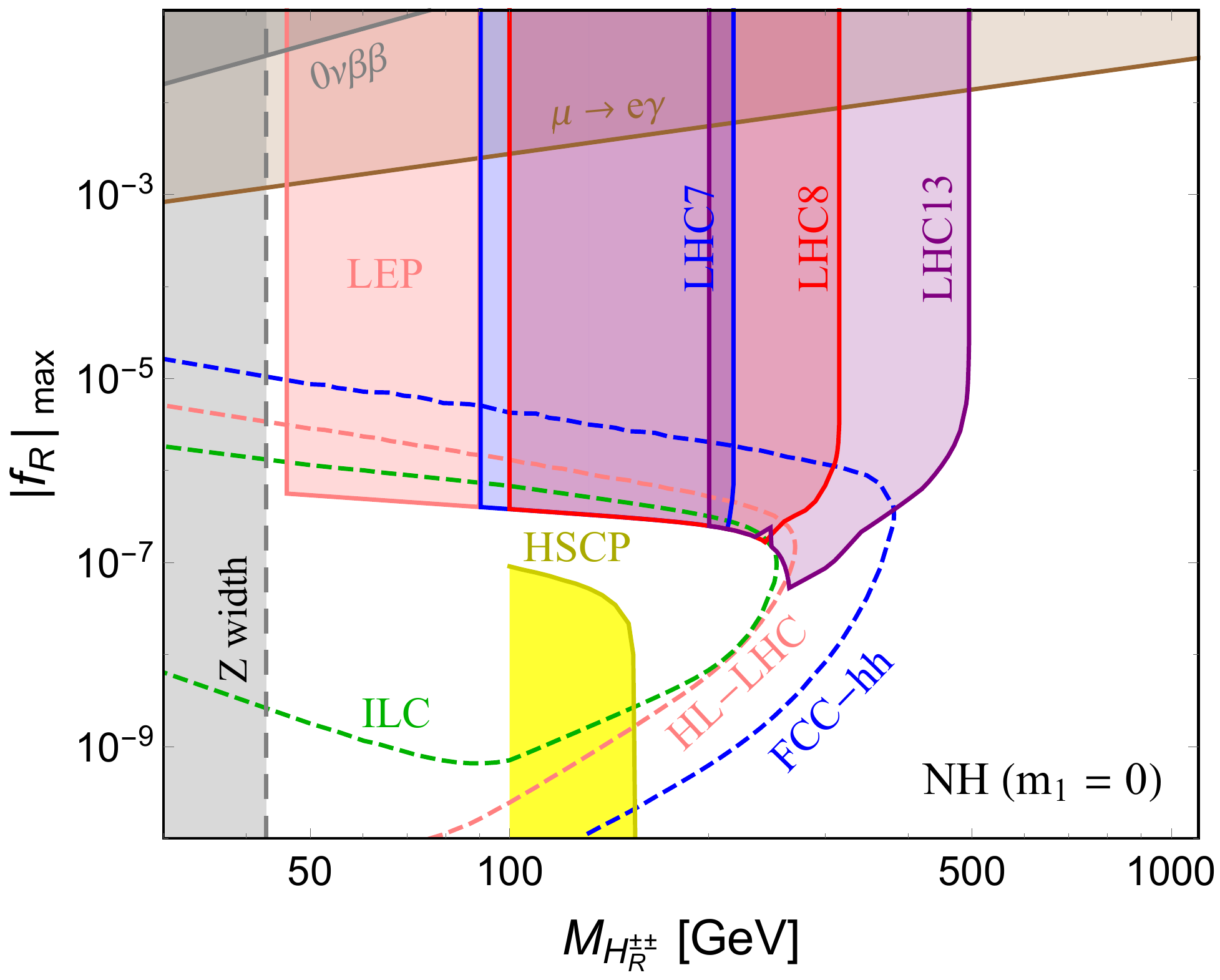}
  \includegraphics[height=0.38\textwidth]{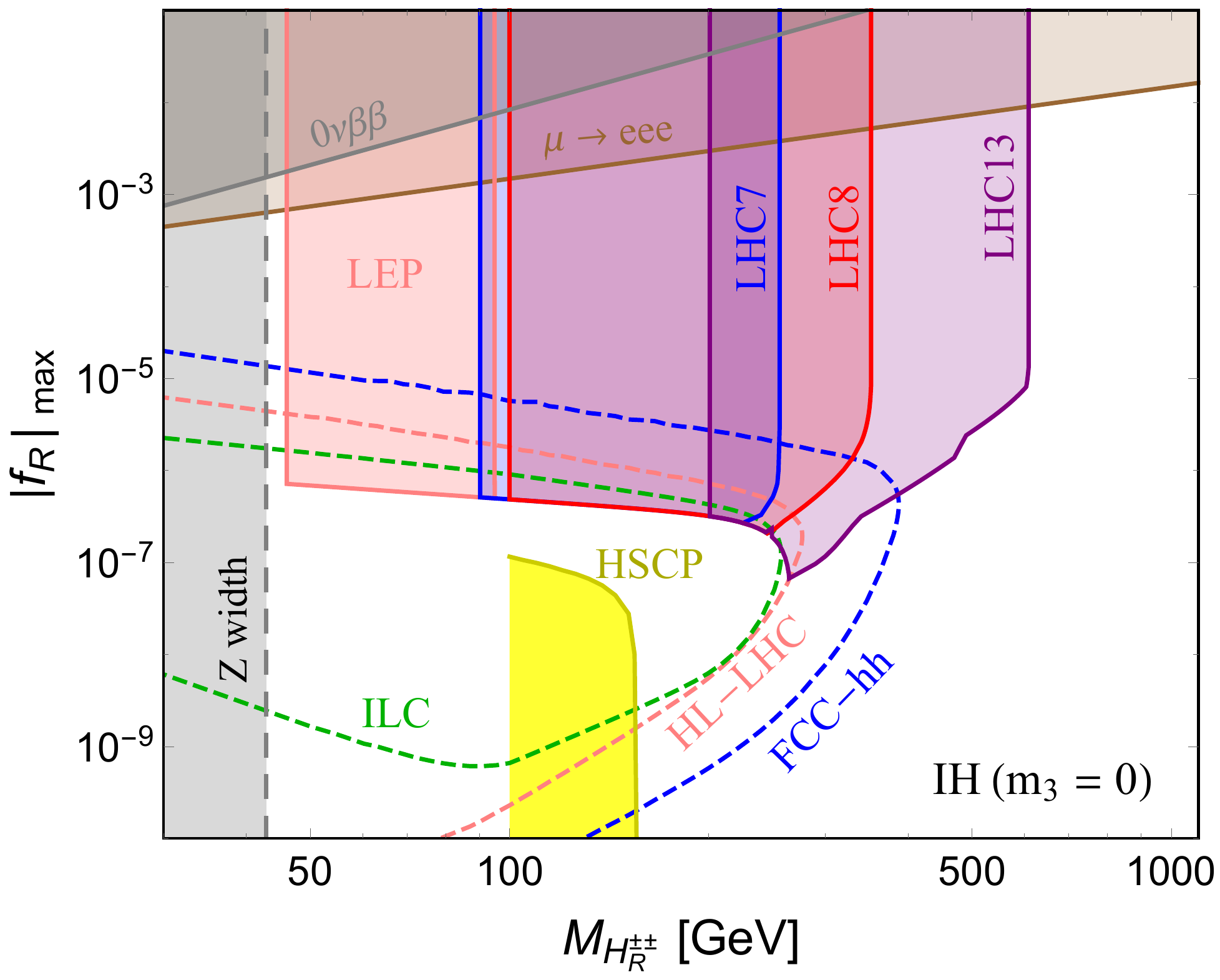}
  \caption{Summary of the most important constraints and sensitivities on $M_{H_R^{\pm\pm}}$ and the Yukawa coupling $|f_R|_{\rm max}$ in the LRSM, extracted from Figs.~\ref{fig:flavor1}, \ref{fig:dilepton:right:3}, \ref{fig:HSCP2}, \ref{fig:0nubetabeta} and \ref{fig:DV:right}. The DV prospects at ILC 1 TeV with a luminosity of 1 ab$^{-1}$ (dashed green), HL-LHC 14 TeV with 3000 fb$^{-1}$ (dashed pink) and FCC-hh 100 TeV and 30 ab$^{-1}$ (dashed blue) are shown assuming at least 100 signal events. The left and right panels are respectively for
  the NH and IH cases with the lightest neutrino mass $m_0 = 0$. All the shaded regions are excluded, which are derived from the $\mu \to e\gamma$ ($\mu \to eee$) limit (brown), the dilepton constraints from LEP (pink), LHC 7 TeV (blue), 8 TeV (red) and 13 TeV (purple), limit on $M_{H_R^{\pm\pm}}$ from $Z$ boson width, the CMS HSCP limit (bright yellow), and the $0\nu\beta\beta$ constraints (gray).   For all the dilepton limits at LEP and LHC, we have left out the regions with $c\tau_0 (H_R^{\pm\pm}) > 0.1$ mm (cf. Fig.~\ref{fig:dilepton:right:3}). }
  \label{fig:complementarity2}
\end{figure}

As in the case of $H_L^{\pm\pm}$ in type-II seesaw, the searches of DV same-sign dilepton signals from $H_R^{\pm\pm}$ are sensitive to relatively small Yukawa couplings $|f_R|$, and are largely complementary to the low and high-energy constraints in Section~\ref{sec:limits:LRSM}. Similar to Fig.~\ref{fig:complementarity}, we have collected the most important constraints and DV prospects for $H_R^{\pm\pm}$ in Fig.~\ref{fig:complementarity2}, which includes the most stringent LFV limit from $\mu \to e\gamma$ ($\mu \to eee$) in Table~\ref{tab:limits} and Fig.~\ref{fig:flavor1} (brown), the $0\nu\beta\beta$ constraints in Fig.~\ref{fig:0nubetabeta} (gray), the prompt dilepton constraints at LEP (pink), LHC 7 TeV (blue), 8 TeV (red) and 13 TeV (purple) in Fig.~\ref{fig:dilepton:right:3}, the HSCP limits in Fig.~\ref{fig:HSCP2} (bright yellow). The DV prospects with at least 100 events are shown for the HL-LHC (dashed pink), ILC 1 TeV (dashed green) and 100 TeV FCC-hh  (dashed blue). The left and right panels are respectively for the NH and IH cases with the lightest neutrino mass $m_0 = 0$. We have spared the regions with $c\tau_0 (H_L^{\pm\pm}) > 0.1$ mm from all the prompt dilepton search limits from LEP and LHC (cf. Fig.~\ref{fig:dilepton:right:3}).
In Fig.~\ref{fig:complementarity2} we have re-interpreted the $0\nu\beta\beta$ limits in Eq.~(\ref{eqn:DBDlimit}) as functions of the largest Yukawa coupling $|f_R|_{\rm max}$. In the case of NH with $m_1 = 0$, the element $|(m_\nu)_{ee}|$ is roughly 17 times smaller than the largest element $|(m_\nu)_{\mu\mu}|$, thus the $0\nu\beta\beta$ limit in the left panel of Fig.~\ref{fig:complementarity2} is much weaker than the IH case in the right panel wherein $|(m_\nu)_{ee}|$ is the largest element.

\section{Right-handed doubly-charged scalar in the LRSM with parity violation}
\label{sec:lrsm2}

If parity is not completely restored in the LRSM at the TeV scale, the Yukawa couplings $f_{L}$ and $f_R$ might be unequal. In addition, the minimization conditions of the scalar potential require that $v_L \sim v_{\rm EW}^2/v_R$~\cite{Zhang:2007da}. This implies that for TeV scale $v_R$ and heavy RHNs we have $v_L \sim {\cal O}$(MeV), which gives an unacceptably large type-II seesaw contribution $\sim f_L v_L$ to the light neutrino masses if $f_L = f_R \sim {\cal O}(1)$. Without large cancellation of the type-I and type-II contributions, one natural solution is to eliminate the left-handed triplet $\Delta_L$ from the low-energy scale, e.g. in a LRSM with $D$-parity breaking~\cite{CMP}. Then the neutrino masses are generated via the type-I seesaw $m_\nu \simeq - m_D M_N^{-1} m_D^{\sf T}$~\cite{seesaw1, seesaw2, seesaw3, seesaw4, seesaw5}.

Without the parity symmetry, the couplings $f_{R}$ of $H_R^{\pm\pm}$ to the charged leptons are no longer directly related to the low-energy neutrino oscillation data, and all these elements can be considered as free parameters, though they are intimately connected to the heavy RHN masses through $M_N = \sqrt2 f_R v_R$. For illustration purpose and comparison to the parity-conserving case in Section~\ref{sec:lrsm}, in this section we study a benchmark scenario with only one coupling $(f_R)_{ee}$ sizable in the Yukawa sector, and all other elements $(f_R)_{\alpha\beta}$ ($\alpha\beta \neq ee$) negligible.\footnote{For other textures of $(f_R)_{\alpha\beta}$, the low-energy LFV limits, the high-energy prompt dilepton limits and the DV sensitivities might differ, depending on the specific flavor content.} The total width is then
\begin{eqnarray}
\label{eqn:widthtotal3}
\Gamma_{\rm total} (H_R^{\pm\pm}) & \ \simeq \ &
\Gamma (H_R^{\pm\pm} \to e^\pm e^\pm) +
\Gamma (H_R^{\pm\pm} \to W_R^{\pm\, (\ast)} W_R^{\pm\, (\ast)}) \,,
\text{ (parity-violating)} \nonumber \\ &&
\end{eqnarray}
which can be readily evaluated as in Section~\ref{sec:lrsm:basic}.

\subsection{Low and high-energy constraints}
\label{sec:collider}
In our case here with only one element $(f_R)_{ee}$, most of the LFV constraints in Table~\ref{tab:limits}, such as those from $\mu \to eee$ and $\mu \to e\gamma$, can not be used to constrain the coupling $(f_R)_{ee}$, as they depend also on other entries of the $f_R$ matrix like $(f_R)_{e\mu}$. Therefore, we focus on the collider and $0\nu\beta\beta$ constraints that are directly applicable to $(f_R)_{ee}$, irrespective of other Yukawa elements.

\subsubsection{Collider constraints}

The heavy $H_R^{\pm\pm}$ in the $t$-channel could mediate the Bhabha scattering $e^+ e^- \to e^+ e^-$ and interfere with the SM diagrams. This alters both the total cross section and the differential distributions. If the Yukawa coupling $(f_R)_{ee}$ is of order one, $H_R^{\pm\pm}$ could be probed up to the TeV scale~\cite{Abbiendi:2003pr, Achard:2003mv}. By Fierz transformations, the coupling $(f_R)_{ee}$ of $H_R^{\pm\pm}$ contributes to the effective contact four-fermion interaction
\begin{eqnarray}
\frac{1}{\Lambda_{\rm eff}^2}
(\bar{e}_R \gamma_\mu e_R)
(\bar{e}_R \gamma^\mu e_R) \,,
\end{eqnarray}
where  $\Lambda_{\rm eff} \simeq M_{H_R^{\pm\pm}}/|(f_R)_{ee}|$ corresponds to the effective cutoff scale. This is constrained by the LEP $e^+e^- \to e^+e^-$ data in Ref.~\cite{Abdallah:2005ph}, which turns out to be more stringent than those in Refs.~\cite{Abbiendi:2003pr, Achard:2003mv}, and requires that $\Lambda_{\rm eff} > 5.2$ TeV.

By mediating the M{\o}ller scattering $e^- e^- \to e^- e^-$, the coupling $(f_R)_{ee}$ of $H_R^{\pm\pm}$ will also be constrained by  the upcoming MOLLER data, which could probe the effective scale $\Lambda = M_{H_R^{\pm\pm}}/|(f_R)_{ee}| \simeq 5.3$ TeV, slightly stronger than the current limit from LEP data above~\cite{Dev:2018sel}.

%The LEP $ee \to ee$ limit on the doubly-charged scalar mass $M_{H_R^{\pm\pm}}$ and the coupling $|(f_R)_{ee}|$ is shown in Fig.~\ref{fig:final} as the orange curve.

\begin{figure}[t!]
  \centering
  \includegraphics[width=0.48\textwidth]{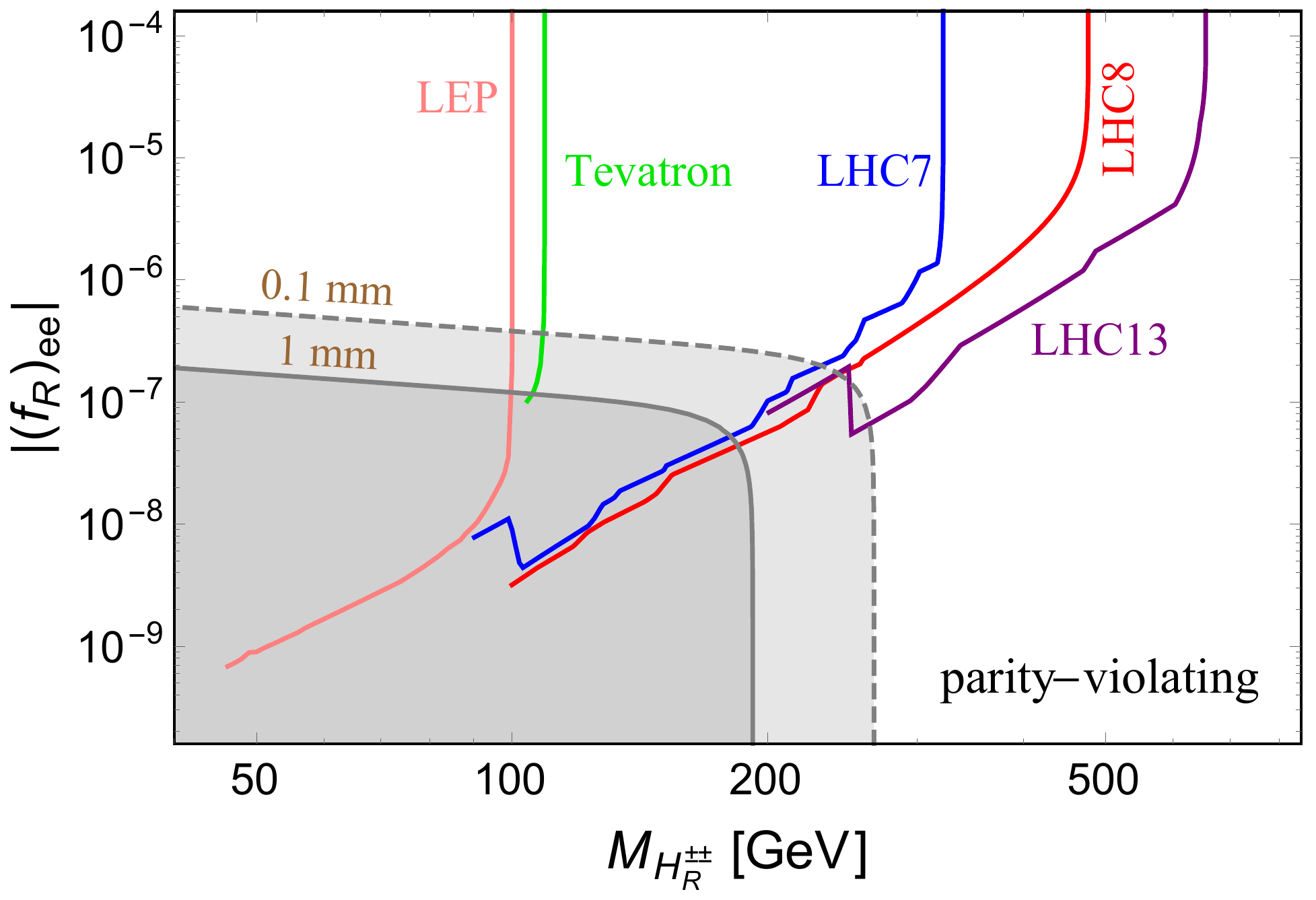}
  \caption{Lower limits on $M_{H_R^{\pm\pm}}$ in the parity-violating LRSM from the $e^\pm e^\pm$ data in Fig.~\ref{fig:dilepton:right:1}, as functions of the Yukawa coupling $|(f_R)_{ee}|$. All the regions above the curves are excluded. The darker and lighter gray regions correspond respectively to the lifetime $c\tau_0 (H_R^{\pm\pm}) > 1$ mm and 0.1 mm; within these regions the prompt dilepton limits are not applicable.}
  \label{fig:dilepton:PV}
\end{figure}

%In the benchmark scenarios in this section, $H_R^{\pm\pm}$ could decay into a pair of same-sign electrons/positrons $H_R^{\pm\pm} \to e^\pm e^\pm$ or into a pair of (off-shell) heavy $W_R$ bosons $H_R^{\pm\pm} \to W_R^{\pm\, (\ast)} W_R^{\pm\, (\ast)}$.
The $H_R^{\pm\pm}\to e^\pm e^\pm$ limits from LEP, Tevatron and LHC are the same as that in the upper left panel of Fig.~\ref{fig:dilepton:right:1} for the parity-conserving case. As in Sections~\ref{sec:left} and \ref{sec:lrsm}, for sufficiently small $|(f_R)_{ee}|$, $H_R^{\pm\pm}$ decays into $W_R$ boson with a sizable branching fraction. Thus the $e^\pm e^\pm$ limits from LEP and LHC can be used to set an upper bound on the Yukawa coupling $|(f_R)_{ee}|$ as functions of the mass of $H_R^{\pm\pm}$, as shown in Fig.~\ref{fig:dilepton:PV}. Following Figs.~\ref{fig:dilepton:left:3} and \ref{fig:dilepton:right:4}, within the darker and lighter gray regions the lifetime $c\tau_0 (H_R^{\pm\pm}) > 1$ mm and 0.1 mm respectively, and the dilepton limits are not applicable. The Yukawa coupling $(f_R)_{ee}$ here is not directly related to the active neutrino data, thus the limits in Fig.~\ref{fig:dilepton:PV} are free from the neutrino oscillation data uncertainties in Table~\ref{tab:neutriodata}.

\begin{figure}[t!]
  \centering
  \includegraphics[width=0.55\textwidth]{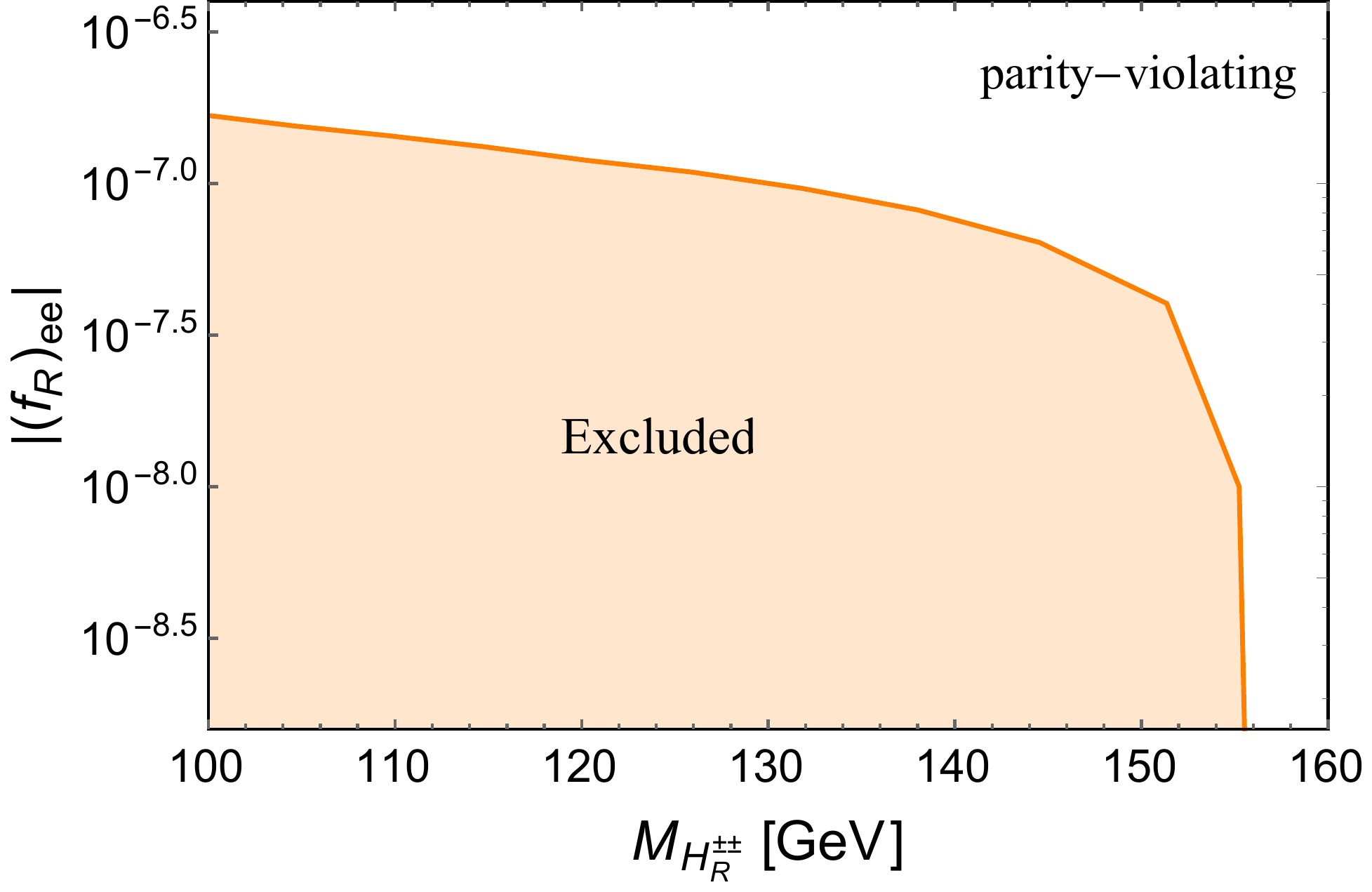}
  \caption{Limits on $M_{H_R^{\pm\pm}}$ in parity-violating LRSM and the Yukawa coupling $|(f_R)_{ee}|$ from searches of doubly-charged HSCPs by the CMS group~\cite{CMS:2016ybj}. The orange region is excluded. The RH scale $v_R = 5\sqrt2$ TeV and $g_R = g_L$.}
  \label{fig:HSCP3}
\end{figure}

Analogous to the parity-conserving case in Section~\ref{sec:lrsm}, the HSCP limits from Ref.~\cite{CMS:2016ybj} can be applied to $H_R^{\pm\pm}$ in the parity-violating LRSM, which is shown in Fig.~\ref{fig:HSCP3}.
For concreteness, we adopt again the RH scale $v_R = 5\sqrt2$ TeV, the gauge coupling $g_R = g_L$ and use conservatively only the  ``tracker-only'' data to set the limit. As $H_R^{\pm\pm}$ here decays only into $e^\pm e^\pm$ and the $W_R$ bosons, $H_R^{\pm\pm}$ is a little longer-lived than in the parity-conserving case and the HSCP limit in Fig.~\ref{fig:HSCP3} is slightly stronger than that in Fig.~\ref{fig:HSCP2}.

\subsubsection{Neutrinoless double beta decay}

In the parity-violating LRSM, the contribution of RHNs $N_i$ and $H_R^{\pm\pm}$ are roughly the same as in Eqs.~(\ref{eqn:etaN}) and (\ref{eqn:etaDCSR}). However, without the parity relation $f_L = f_R$, the RHN mixing matrix $U_R$ is in general different from the PMNS matrix $U$ in Eq.~(\ref{eqn:PMNS}), and the contribution of $H_R^{\pm\pm}$ to $0\nu\beta\beta$ can be re-written as explicit function of the coupling $(f_R)_{ee}$, i.e.
\begin{eqnarray}
\label{eqn:etaN2}
\eta_N^{\rm (PV)} & \ = \ & m_p
\left( \frac{g_R}{g_L} \right)^4
\left( \frac{m_W}{M_{W_R}} \right)^4
\sum_i \frac{(U_R)_{ei}^2}{M_{N_i}}  \ = \
\frac{m_p}{4}
\left( \frac{v_{\rm EW}}{v_{R}} \right)^4
\sum_i \frac{(U_R)_{ei}^2}{M_{N_i}} \,, \\
\label{eqn:etaDCSR2}
\eta_{\rm DCS}^{R \, ({\rm PV})} & \ = \ & m_p
\left( \frac{g_R}{g_L} \right)^4
\left( \frac{m_W}{M_{W_R}} \right)^4
\frac{ \sqrt2 (f_R)_{ee} v_R }{M_{H_R^{\pm\pm}}^2} \ = \
\frac{m_p}{2\sqrt2}
\left( \frac{v_{\rm EW}}{v_{R}} \right)^4
\frac{ (f_R)_{ee} v_R }{M_{H_R^{\pm\pm}}^2} \,.
\end{eqnarray}
Comparing the two terms above, with $(f_R)_{ee} v_R \sim M_{N_i}$, we get the ratio $\eta_N / \eta_{\rm DCS}^R \sim M_{H_R^{\pm\pm}}^2 / M_{N_i}^2$, which means that the $H_R^{\pm\pm}$ contribution is expected to be larger than that from the RHNs if $H_R^{\pm\pm}$ is lighter, i.e. $M_{H_R^{\pm\pm}} \lesssim M_{N_i}$. To set $0\nu\beta\beta$ limits on $H_R^{\pm\pm}$, we need to compare further the $H_R^{\pm\pm}$ term in Eq.~(\ref{eqn:etaDCSR2}) with the canonical term $\eta_\nu$ in Eq.~(\ref{eqn:0nubetabeta}):
\begin{eqnarray}
\label{eqn:ratio}
\frac{\eta_{\rm DCS}^R}{\eta_\nu} \ = \ \frac{1}{2\sqrt2}
\left( \frac{v_{\rm EW}}{v_{R}} \right)^4
\left( \frac{(m_\nu)_{ee}}{(f_R)_{ee} v_R} \right)^{-1}
\left( \frac{m_e m_p}{M_{H_R^{\pm\pm}}^2} \right) \,.
\end{eqnarray}
As in the parity-conserving case in Section~\ref{sec:lrsm:DBD}, if the doubly-charged scalar mass $M_{H_R^{\pm\pm}} \sim {\rm TeV}$ and the Yukawa coupling $(f_R)_{ee} \sim {\cal O} (1)$, the contribution from $H_R^{\pm\pm}$ could be comparable to the $\eta_\nu$ term and thus get constrained by the limits from KamLAND-Zen~\cite{KamLAND-Zen:2016pfg} and  GERDA~\cite{Agostini:2018tnm}.

\begin{figure}[t!]
  \centering
  \includegraphics[width=0.48\textwidth]{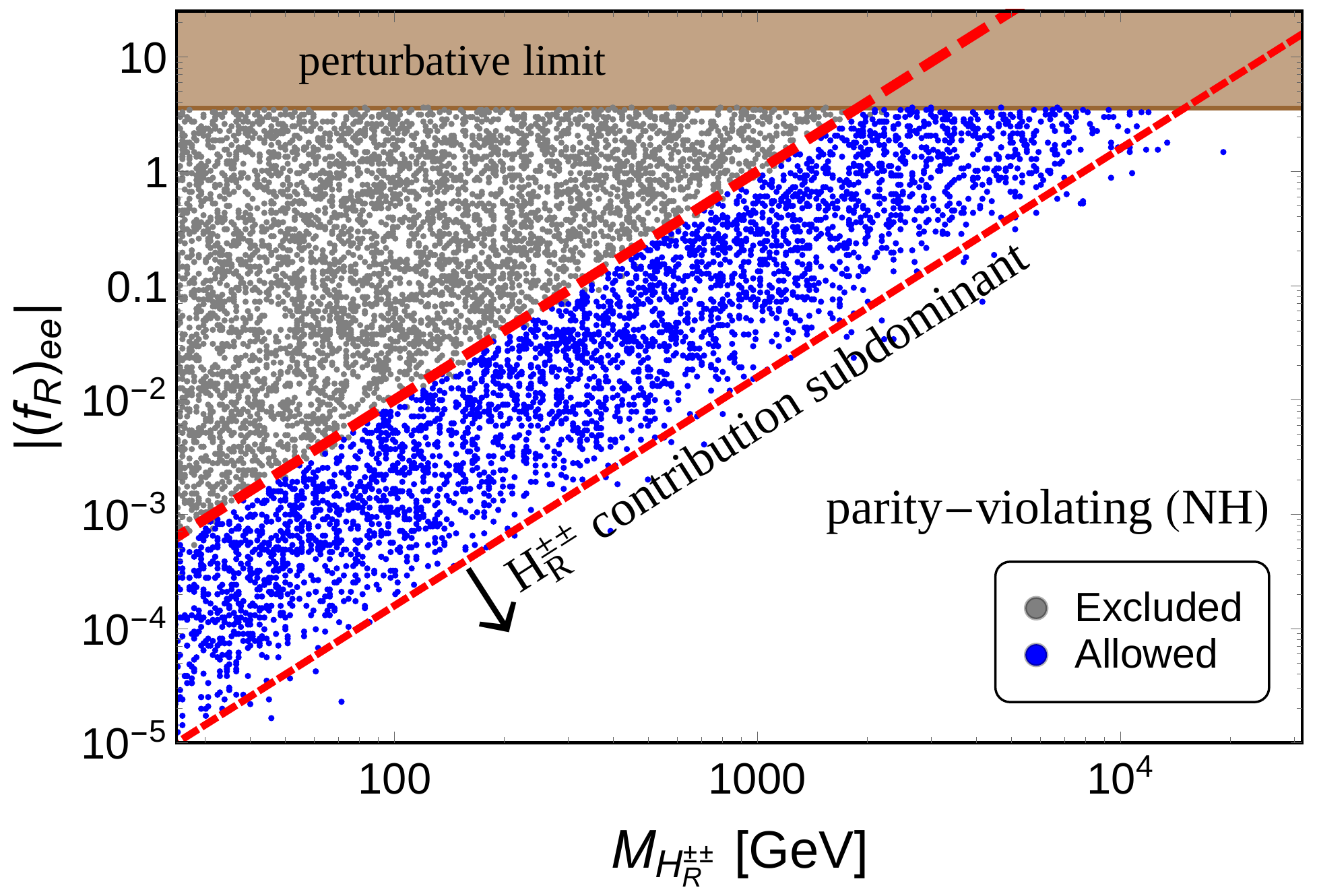}
  \includegraphics[width=0.48\textwidth]{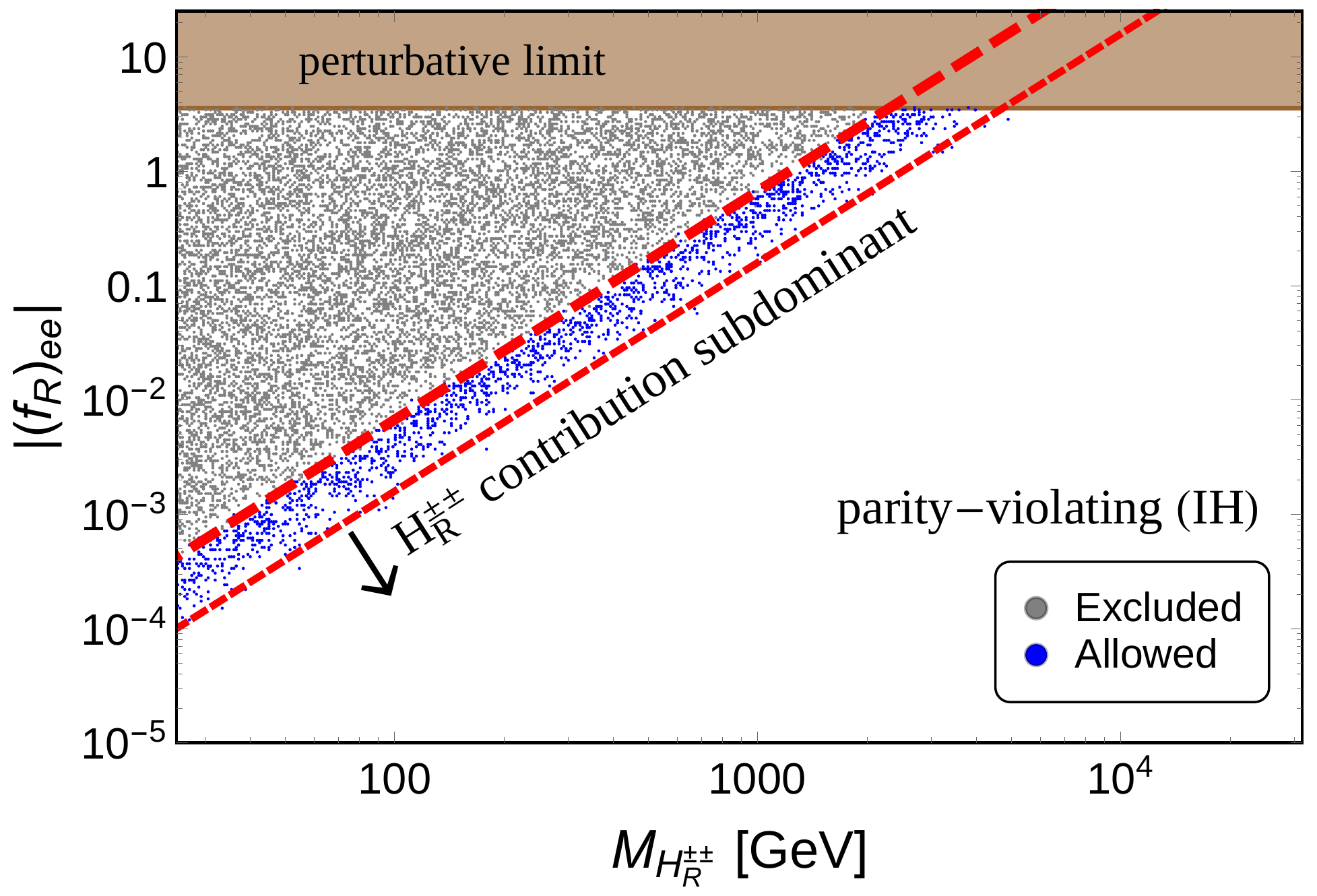}
  \caption{The same as in Fig.~\ref{fig:0nubetabeta}, but for $H_R^{\pm\pm}$ in the parity-violating LRSM.}
  \label{fig:0nubetabeta2}
\end{figure}

As in Section~\ref{sec:lrsm:DBD}, we vary the neutrino data in Table~\ref{tab:neutriodata} within their $3\sigma$ ranges and the lightest neutrino mass $m_0 \in [0,\, 0.05 \, {\rm eV}]$ and the RH scale $v_R = 5\sqrt2$ TeV, and the results are shown in Fig.~\ref{fig:0nubetabeta2} for both the NH (left) and IH (right) cases. As in Fig.~\ref{fig:0nubetabeta}, all the gray points are excluded by the current limits from KamLAND-Zen~\cite{KamLAND-Zen:2016pfg} and GERDA~\cite{Agostini:2018tnm} and the blue points are allowed. The $0\nu\beta\beta$ limits on the Yukawa coupling $(f_R)_{ee}$ turn out to be roughly the same as in the parity-conserving LRSM:
\begin{eqnarray}
\label{eqn:DBDlimit2}
\frac{|(f_R)_{ee}|} {M_{H_R^{\pm\pm}}^2} <
\begin{cases}
1.0 \times 10^{-6} \, {\rm GeV}^{-2} & \text{for NH , } \\
6.7 \times 10^{-7} \, {\rm GeV}^{-2} & \text{for IH .}
\end{cases} \text{ (parity-violating case)} \,,
\end{eqnarray}
as indicated by the long-dashed red curves in Fig.~\ref{fig:0nubetabeta2}. For heavier $H_R^{\pm\pm}$ and/or smaller coupling $|(f_R)_{ee}|$, the contribution of $H_R^{\pm\pm}$ is sub-dominant to the canonical light neutrino term $\eta_\nu$ in Eq.~(\ref{eqn:0nubetabeta}), and the KamLAND-Zen and  GERDA limits are no longer applicable to $H_R^{\pm\pm}$, as indicated by  the short-dashed red lines in Fig.~\ref{fig:0nubetabeta2}.

\subsection{Displaced vertex prospects}

\begin{figure}[t!]
  \centering
  \includegraphics[width=0.48\textwidth]{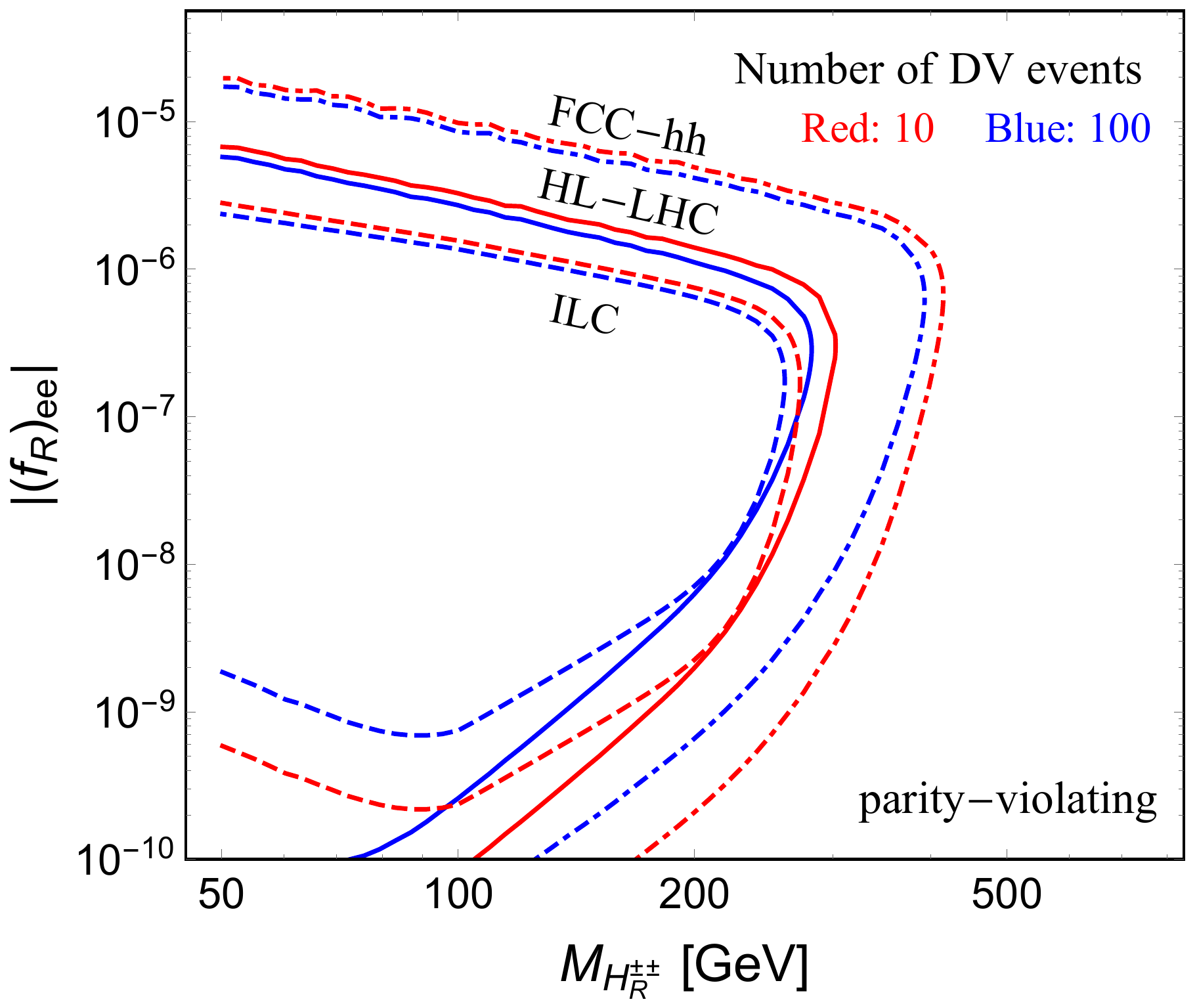}
  \caption{DV prospects from the decay of $H_R^{\pm\pm} \to e^\pm e^\pm$ in the parity-violating LRSM, at the HL-LHC 14 TeV with an integrated luminosity of 3000 fb$^{-1}$ (solid contours), the 100 TeV FCC-hh with a luminosity of 30 ab$^{-1}$ (dot-dashed contours) and ILC 1 TeV with 1 ab$^{-1}$ (dashed contours). The red and blue contours respectively correspond to 10 and 100 DV events, as functions of $M_{H_R^{\pm\pm}}$ and the Yukawa coupling $|(f_R)_{ee}|$.}
  \label{fig:DV:right2}
\end{figure}

With the total width in Eq.~(\ref{eqn:widthtotal3}), it is straightforward to estimate the number of displaced $e^\pm e^\pm$ events from $H_R^{\pm\pm}$ decay at the HL-LHC, ILC 1 TeV and 100 TeV FCC-hh. With the same setups for these colliders as in Section~\ref{sec:lrsm:DV} (including the benchmark values of $v_R = 5\sqrt2$ TeV and $g_R = g_L$) {and taking only the leptonic decays $H_R^{\pm\pm} \to e^\pm e^\pm$}, the prospects are collected in Fig.~\ref{fig:DV:right2}. Within the red and blue contours we can have respectively at least 10 and 100 DV events. Again, for extremely small couplings $|(f_R)_{ee}| \lesssim 10^{-10}$, the RHNs are expected to be lighter than the keV scale and get constrained by the cosmological data~\cite{Vincent:2014rja, Abazajian:2017tcc}. Therefore we set a lower cut of $10^{-10}$ on the Yukawa coupling $|(f_R)_{ee}|$ in Fig.~\ref{fig:DV:right2}. Though the dilepton partial width of $H_R^{\pm\pm}$ might be smaller than in the parity-conserving case, as we have only the $e^\pm e^\pm$ decay modes in the leptonic sector, the sensitivity contours in Fig.~\ref{fig:DV:right2} do not differ too much from those in Fig.~\ref{fig:DV:right}.

\begin{figure}[t!]
  \centering
  \includegraphics[height=0.38\textwidth]{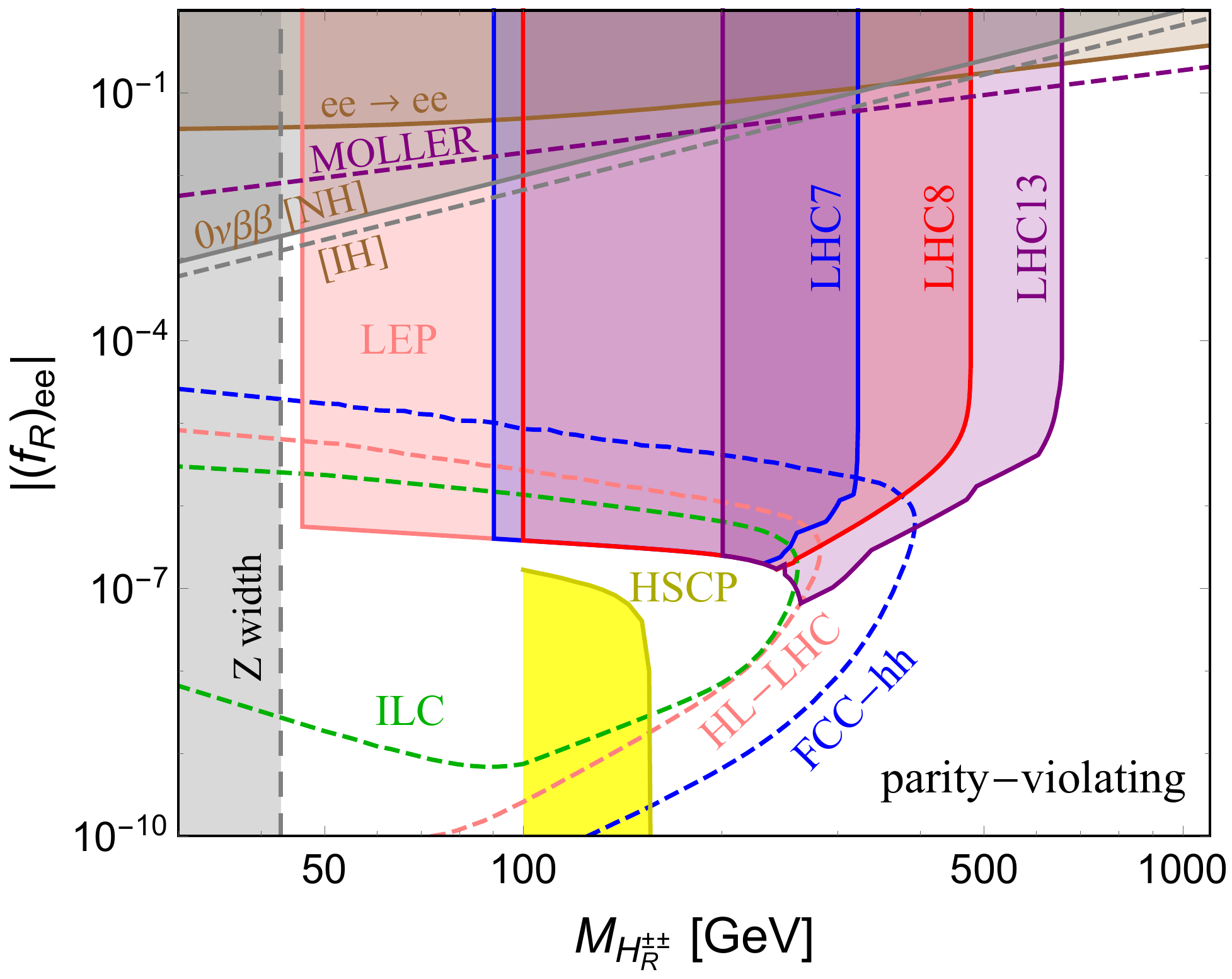}
  \caption{Summary of the most important constraints and sensitivities on $M_{H_R^{\pm\pm}}$ and the Yukawa coupling $|(f_R)_{ee}|$ in the parity-violating LRSM, extracted from \ref{fig:dilepton:PV}, \ref{fig:HSCP3}, \ref{fig:0nubetabeta2} and \ref{fig:DV:right2}. The DV prospects at ILC 1 TeV with a luminosity of 1 ab$^{-1}$ (dashed green), HL-LHC 14 TeV with 3000 fb$^{-1}$ (dashed pink) and FCC-hh 100 TeV with 30 ab$^{-1}$ (dashed blue) are shown assuming at least 100 signal events. All the shaded regions are excluded, which are derived from the $e^+e^- \to e^+e^-$ limit from LEP (orange), the $0\nu\beta\beta$ constraints for NH (brown) and IH (dashed brown) neutrino spectra, the MOLLER prospect (dashed purple), limit on $M_{H_R^{\pm\pm}}$ from $Z$ boson width (gray), the dilepton constraints from LEP (pink), LHC 7 TeV (blue), 8 TeV (red) and 13 TeV (purple), and the CMS HSCP limit (bright yellow). For all the dilepton limits at LEP and LHC, we have left out the regions with $c\tau_0 (H_R^{\pm\pm}) > 0.1$ mm (cf. Fig.~\ref{fig:dilepton:PV}).}
  \label{fig:complementarity3}
\end{figure}

The summary plot for the $H_R^{\pm\pm}$ in the parity-violating LRSM is shown in Fig.~\ref{fig:complementarity3}. As in the case of $H_L^{\pm\pm}$ in type-II seesaw in Section~\ref{sec:left} and $H_R^{\pm\pm}$ in the parity-conserving LRSM in Section~\ref{sec:lrsm}, with the coupling $(f_R)_{ee}$, the displaced $e^\pm e^\pm$ pair from $H_R^{\pm\pm}$ decay in the parity-violating LRSM at future HL-LHC (dashed pink), ILC (dashed green) and FCC-hh (dashed blue) are largely complementary to the prompt $e^\pm e^\pm$ limits from LEP (pink) and LHC (blue, red and purple) and the CMS HSCP data (bright yellow), as well as to the low-energy constraints from the LEP $e^+e^- \to e^+e^-$ data (brown) and $0\nu\beta\beta$ (solid and dashed gray). As in Fig.~\ref{fig:complementarity2}, we have spared the regions with $c\tau_0 (H_L^{\pm\pm}) > 0.1$ mm from all the dilepton limits from LEP and LHC (cf. Fig.~\ref{fig:dilepton:PV}). The future sensitivity of MOLLER is also shown in this plot (dashed purple), which exceeds the current $0\nu\beta\beta$ and LEP constraints for high-mass $H_R^{\pm\pm}$~\cite{Dev:2018sel}.

\section{Conclusion}
\label{sec:conclusion}

As one of the well-motivated solution to the tiny neutrino mass puzzle, the type-II seesaw and its left-right symmetric extensions with new scalar triplets offer a rich phenomenology at both low- and high-energy frontiers. As a class of almost background-free processes at high-energy colliders, the same-sign dilepton decays from the doubly-charged scalars $H_{L,R}^{\pm\pm} \to \ell_\alpha^\pm \ell_\beta^\pm$  undoubtedly provide a ``smoking-gun'' signal of triplet scalars beyond the SM. However, all the past searches at LEP, Tevatron and LHC 7 TeV, 8 TeV and 13 TeV have focused on the prompt $H_{L,R}^{\pm\pm}$ decays assuming relatively large Yukawa couplings $\gtrsim 10^{-7}$ to $10^{-6}$ (see e.g. Figs.~\ref{fig:dilepton:left:3}, \ref{fig:dilepton:left:4}, \ref{fig:dilepton:right:3}, \ref{fig:dilepton:right:4} and \ref{fig:dilepton:PV}).  We point out that in a large parameter space where the Yukawa couplings $f_{L,R}$ are small, the doubly-charged scalars $H^{\pm\pm}_{L,R}$ could be long-lived at the high-energy colliders, and decay into a pair of displaced same-sign leptons (and potentially other final states in the bosonic channels), which offer a complementary probe of the type-II seesaw.

We have estimated the DV prospects at the HL-LHC, 100 TeV FCC-hh and 1 TeV ILC for the LH doubly-charged scalar $H_L^{\pm\pm}$ in the type-II seesaw, and the RH doubly-charged scalar $H_R^{\pm\pm}$ in both parity-conserving and violating LRSM.  Our main results are summarized in Figs.~\ref{fig:complementarity}, \ref{fig:complementarity2} and \ref{fig:complementarity3}. For $H_L^{\pm\pm}$ in the pure type-II seesaw, within a broad region of the parameter space with $10^{-10} \lesssim |f_L| \lesssim 10^{-6}$ and $m_Z/2 < M_{H_L^{\pm\pm}} \lesssim 200$ (500) GeV, we can have at least 100 displaced same-sign dilepton events at the future colliders HL-LHC and ILC 1 TeV (FCC-hh).

In the LRSM, if we have the parity symmetry at the TeV scale, then the Yukawa couplings $f_L = f_R$, and the leptonic decays of $H_R^{\pm\pm}$ are the same as of $H_L^{\pm\pm}$ in type-II seesaw, dictated by the active neutrino data. With the RH gauge interaction, the decay $H_R^{\pm\pm} \to W_R^{\pm \ast} W_R^{\pm \ast}$ is possible, but highly suppressed by the $W_R$ mass. Therefore, for sufficiently small coupling $f_R$, $H^{\pm\pm}_R$ is long-lived at the high-energy colliders, just like $H_L^{\pm\pm}$. As shown in Fig.~\ref{fig:DV:right}, in a large parameter space with $10^{-10} \lesssim |f_R| \lesssim 10^{-6}$ and $m_Z/2 < M_{H_R^{\pm\pm}} \lesssim 200$ (400) GeV, we could detect the DV signals at the HL-LHC and ILC 1 TeV (FCC-hh). If parity is violated in the LRSM at the TeV scale, then $f_R$ could be different from $f_L$ and cannot be {\it directly} constrained by the light neutrino data. As a benchmark study of this case, we have investigated a sample texture of $f_R$ with only one non-vanishing element $(f_R)_{ee}$. Although the dileptonic decay width of $H_R^{\pm\pm}$ in the parity-violating case is smaller than in the parity-conserving case, the life-time of $H_R^{\pm\pm}$ and the DV sensitivity region do not change significantly, as shown in Fig.~\ref{fig:DV:right2}.

In both type-II seesaw and its LRSM extension, the low-energy, high-precision/intensity constraints, such as those from the LFV decays $\ell_\alpha \to \ell_\beta \ell_\gamma \ell_\delta$, $\ell_\alpha \to \ell_\beta \gamma$, electron and muon $g-2$, muonium oscillation, LEP $e^+e^- \to \ell^+\ell^-$ data, and $0\nu\beta\beta$ searches, set severe constraints on the (LFV) couplings $(f_{L,R})_{\alpha\beta}$ of the doubly-charged scalars. However, these limits only restrict $f_{L,R} \lesssim 10^{-3}$ for a 100 GeV doubly-charged scalar. The LFV constraints get weaker for heavier doubly-charged scalars (cf. Table~\ref{tab:limits} and Figs.~\ref{fig:flavor1}, \ref{fig:0nubetabeta}, \ref{fig:0nubetabeta2}). The displaced vertex signals,  as discussed here, are also largely complementary to these low-energy constraints, as clearly shown in Figs.~\ref{fig:complementarity}, \ref{fig:complementarity2} and \ref{fig:complementarity3}. Therefore, the displaced vertex signatures of doubly-charged scalars offer a new avenue to probe the origin of the tiny neutrino masses at future colliders.

\section*{Acknowledgments}

We gratefully acknowledge enlightening discussions with Frank Deppisch, Rabindra Mohapatra and Michael Ramsey-Musolf. BD thanks the organizers of the FPCP 2018 at the University of Hyderabad and IIT Hyderabad for the local hospitality, when this work was being finalized. YZ is grateful to the University of Maryland, College Park, University of Massachusetts, Amherst, and the Center for Future High Energy Physics, IHEP, CAS, for the hospitality and local support, where part of the work was done. This work is supported by the U.S. Department of Energy under Grant No. DE-SC0017987.

\appendix
\section{Calculation of the four-body decays of doubly-charged scalars }
\label{sec:decay}

In the type-II seesaw, if the doubly-charged scalar $H_L^{\pm\pm}$ from the scalar triplet $\Delta_L$ is light, which is directly relevant to the displaced vertex searches at future hadron and lepton colliders, it could decay into the SM quarks and leptons through two (off-shell) $W$-bosons, i.e.
\begin{eqnarray}
H^{\pm\pm} \ \to \ W^{\pm (\ast)} W^{\pm (\ast)} \ \to \ f \bar{f}^\prime f^{\prime\prime} \bar{f}^{\prime\prime\prime} \,,
\end{eqnarray}
with $f = q,\, \ell,\, \nu$ running over all the quark and lepton flavors except for the top and bottom quarks. For simplicity we neglect the small quark mixings among different generations. Note that for the cases with identical particles in the final states $f = f^{\prime\prime}$ and $f^\prime = f^{\prime\prime\prime}$, there are two Feynman diagrams, which correspond respectively to the processes $(W^{\pm (\ast)} \to f \bar{f}^\prime) (W^{\pm (\ast)} \to f^{\prime\prime} \bar{f}^{\prime\prime\prime})$ and $(W^{\pm (\ast)} \to f \bar{f}^{\prime\prime\prime}) (W^{\pm (\ast)} \to f^{\prime\prime} \bar{f}^{\prime})$. The two diagrams interfere with each other, which is important for a light $H_L^{\pm\pm}$~\cite{Kanemura:2014goa}. Take the leptonic decays $\ell \ell^\prime \nu \nu$ as an explicit example. Denoting the momenta of the charged leptons and neutrinos respectively as $p_1$ to $p_4$, we obtain the reduced amplitude squared
\begin{align}
|{\cal M}|^2 \ = \ (p_1 \cdot p_2) (p_3 \cdot p_4) \, ,
\end{align}
and the propagator factor
\begin{eqnarray}
\Delta \ = \
\Delta_{13} \Delta_{24} + \delta_{\ell\ell^\prime}
\Delta_{14} \Delta_{23} \,,
\end{eqnarray}
with $\Delta_{ij} = [(p_i+p_j)^2 - m_W^2 + i m_W \Gamma_W]^{-1}$ ($\Gamma_W$ being the $W$ boson width). The Lorentz invariant four-body phase space $\int {\rm d} \Pi_4$ has only five independent kinematic variables, which can be chosen as the Cabibbo variables~\cite{Cabibbo:1965zzb}, i.e. the effective mass squared $s_{12} = (p_1 + p_2)^2$ for the two charged leptons, the effective mass squared $s_{34} = (p_3 + p_4)^2$ for the two neutrinos, the angle $\theta_{12}$ of the momentum of the charged lepton $\ell$ in the dilepton rest frame with respect to the dilepton momentum in the rest frame of $H_L^{\pm\pm}$, the angle $\theta_{34}$ of the momentum of the neutrino $\nu$ in the dineutrino rest frame with respect to the dineutrino momentum in the rest frame of $H_L^{\pm\pm}$, and the angle $\phi$ between the planes defined by the dilepton and dineutrino momenta. The ranges for these variables are respectively,
\begin{align}
s_{12} & \ \in \  [(m_\ell + m_{\ell^\prime})^2,\, M^2] \,, \hspace{-60pt} &
s_{34} & \ \in \  [0,\, (M-\sqrt{s_{12}})^2] \,, \nonumber \\
\theta_{12,\,34} & \ \in \ [0,\,\pi] \,,  &
\phi_{} & \ \in \ [-\pi,\,\pi] \,,
\end{align}
where $M$ is the doubly-charged scalar mass.
Then the partial decay width
\begin{eqnarray}
\label{eqn:width}
&& \Gamma_{\ell\ell^\prime \nu \nu} \ = \
\frac{S g_L^8 v_{L}^2}{2^{12} \pi^6 M_{}^3}
\int \frac{{\rm d} s_{12}}{s_{12}} \int \frac{{\rm d} s_{34}}{s_{34}}
\int {\rm d} \cos\theta_{12} \int {\rm d} \cos\theta_{34} \int {\rm d} \phi
\, {\cal K} \, |\Delta|^2 |{\cal M}|^2 \,,
\end{eqnarray}
where $S = 1 \, (1/4)$ for $\ell \neq \ell^\prime$ ($\ell = \ell^\prime$) is the symmetry factor for identical charged leptons and neutrinos in the final state, and
\begin{eqnarray}
&&{\cal K} \  =  \
\lambda^{1/2} (M^2, s_{12},s_{34})
\lambda^{1/2} (s_{12},m_{\ell}^2, m_{\ell^\prime}^2) s_{34} \,, \\
\text{with}~&& \lambda (a,b,c) \ \equiv  \ a^2 + b^2 + c^2 - 2ab - 2ac - 2bc \,.
\end{eqnarray}
In the limit of $M \gg 2m_W$, to a good approximation,
\begin{eqnarray}
\Gamma_{\ell\ell^\prime \nu\bar\nu} & \ \simeq \ &
\Gamma (H^{\pm\pm} \to W^{\pm} W^\pm) \times {\rm BR} (W \to \ell\nu)
\times {\rm BR} (W \to \ell^\prime \bar\nu) \,.
\end{eqnarray}
In an analogous way, one can calculate the decay widths into four quarks $\Gamma (H_L^{\pm\pm} \to q\bar{q}^\prime q^{\prime\prime} \bar{q}^{\prime\prime\prime})$ or quarks plus leptons $\Gamma (H_L^{\pm\pm} \to q\bar{q}^\prime \ell\nu)$. For concreteness, we define the same flavor (SF) and different flavor (DF) partial width units~\cite{Kanemura:2014goa}:
\begin{eqnarray}
\Gamma_{\rm SF} & \ \equiv \ &
\Gamma (H_L^{\pm\pm} \to W^{\pm(\ast)} W^{\pm(\ast)} \to e^\pm e^\pm \nu_e \nu_e) \,, \nonumber \\
\Gamma_{\rm DF} & \ \equiv \ &
\Gamma (H_L^{\pm\pm} \to W^{\pm(\ast)} W^{\pm(\ast)} \to e^\pm \mu^\pm \nu_e \nu_\mu) \,.
\end{eqnarray}
Then the leptonic, semileptonic and hadronic decay widths are respectively
\begin{eqnarray}
\Gamma_{\rm lep} & \ \simeq \ &
3 \Gamma_{\rm SF} + 3 \Gamma_{\rm DF} \,, \nonumber \\
\Gamma_{\rm semilep} & \ \simeq \ &
2 N_C \Gamma_{\rm DF} \,, \nonumber \\
\Gamma_{\rm had} & \ \simeq \ &
2 N_C \Gamma_{\rm DF} + N_C (2N_C-1) \Gamma_{\rm DF} \,,
\end{eqnarray}
(where $N_C$ is the color factor) which sum up to the total width
\begin{eqnarray}
\Gamma_{\rm total}
(H_L^{\pm\pm} \to W^{\pm(\ast)} W^{\pm(\ast)} ) \ \simeq \
(3+2N_C) \Gamma_{\rm SF} + (3+5N_C+2N_C^2) \Gamma_{\rm DF} \,.
\end{eqnarray}

For the decay of RH doubly-charged scalar $H_R^{\pm\pm}$ in both the parity conserving and violating LRSMs:
\begin{eqnarray}
H_R^{\pm\pm} \to W_R^{\pm \, \ast} W_R^{\pm \, \ast}
\to f \bar{f}^\prime f^{\prime\prime} \bar{f}^{\prime\prime\prime} \,,
\end{eqnarray}
the calculation is almost the same as in the case of $H_L^{\pm\pm}$, with the SM $W$ boson replaced by the heavy $W_R$ boson. The calculations of the four-body decay here are done keeping in mind the displaced vertex searches at colliders; therefore, the Yukawa couplings $f_R$ are supposed to be very small. For the RH scale $v_R\sim$ few TeV, the RHN masses $m_N = 2 f_R v_R$ are expected to be much smaller than the masses of $W_R$ and $H_R^{\pm\pm}$. Hence, we include here all the decay modes of $W_R^{\pm \,\ast} \to q\bar{q}^\prime,\, \ell N$ (with $q,\, q' = u,\, d,\, s,\, c$) in the limit of $M_N / M_{H_R^{\pm\pm}} \ll 1$.

\section{Formulas for the LFV decays}
\label{sec:appendix:LFV}

The partial width for the tree level three-body decay $\ell_\alpha \to \ell_\beta \ell_\gamma \ell_\delta$ is~\cite{Akeroyd:2009nu, Dinh:2012bp}
\begin{eqnarray}
{\rm BR} (\ell_\alpha^- \to \ell_\beta^- \ell_\gamma^+ \ell_\delta^-) \ \simeq \
\frac{|f_{\alpha\gamma}|^2 |f_{\beta\delta}|^2}{2 (1+ \delta_{\beta\delta}) G_F^2 M_{{\pm\pm}}^4} \times
{\rm BR} (\ell_\alpha \to e\nu\bar\nu) \,,
\end{eqnarray}
with $M_{\pm\pm}$ the doubly-charged scalar mass, $f_{\alpha\beta}$ the Yukawa couplings of doubly-charged scalar to the charged leptons, and $G_F$ the Fermi constant. All the formulas in this appendix apply to both the left-handed doubly-charged scalar $H_L^{\pm\pm}$ in type-II seesaw and the RH doubly-charged scalar $H_R^{\pm\pm}$ in the LRSM (therefore, we drop the subscripts ``L" and ``R"). %As the doubly-charged scalar mass is much larger than the charged lepton masses, the constraints on $|f^\dagger f|/M_{\pm\pm}^2$ are almost constants, which correspond to an effective cut-off scale of $\Lambda \simeq M_{\pm\pm} / \sqrt{|f^\dagger f|}$.
At 1-loop level, the LFV couplings $f_{\alpha\beta}$ contribute to the two-body decays~\cite{Mohapatra:1992uu}
\begin{eqnarray}
{\rm BR} (\ell_\alpha \to \ell_\beta \gamma) \ \simeq \
\frac{\alpha_{\rm EM} |\sum_\gamma f_{\alpha\gamma}^\dagger f_{\beta\gamma}|^2}{3\pi G_F^2 M_{\pm\pm}^4} \times
{\rm BR} (\ell_\alpha \to e\nu\bar\nu) \,,
\end{eqnarray}
where $\alpha_{\rm EM}$ is the fine structure constant, and we have summed up all the diagrams involving a lepton $\ell_\gamma$ running in the loop. In the type-II seesaw, it is equivalent to doing the summation $\sum_\gamma (m_\nu)_{\alpha\gamma}^{\sf T} (m_\nu)_{\beta\gamma}$.
%The experimental data of $\mu \to e \gamma$, $\tau \to e\gamma$ and $\tau \to \mu \gamma$ could be used to set limits on the couplings $|\sum_k f_{ik}^\dagger f_{jk}|/M_{\pm\pm}^2$, which are also presented in Table~\ref{tab:limits}.

Similarly, we can calculate the contributions of the doubly-charged scalar loops to the anomalous magnetic moments of electron and muon (with $\alpha=e,\,\mu$)~\cite{Leveille:1977rc, Moore:1984eg, Gunion:1989in, Lindner:2016bgg}:
%The contribution of doubly-charged scalar loops to the electron $g-2$ is
\begin{eqnarray}
\label{eqn:g-2}
\Delta a_\alpha & \ \simeq \ &
- \frac{m_{\ell_\alpha}^2}{6\pi^2 \, M_{\pm\pm}^2} \sum_\beta |f_{\alpha\beta}|^2 \,,
\end{eqnarray}
where $m_{\ell_\alpha}$ is the charged lepton mass and we have summed up again the loops involving all the three lepton flavors $\beta = e,\, \mu,\, \tau$.
%The current $2\sigma$ experimental uncertainty $\Delta a_e = 5.2 \times 10^{-13}$~\cite{PDG} can be used to set limits on the couplings $\sum_j |f_{ej}|^2$ as function of the doubly-charged scalar mass.
%As the contributions from the doubly-charged scalar loops are always negative, the controversial theoretical and experimental discrepancy $\Delta a_\mu = (2.87 \pm 0.80) \times 10^{-9}$~\cite{PDG} can not be explained; we use instead the $5\sigma$ uncertainty of $5 \times 0.80 \times 10^{-9}$ to constrain the Yukawa couplings, as shown in Table~\ref{tab:limits} and Fig.~\ref{fig:flavor1}.
%Enhanced by the muon mass, the muon $g-2$ limit is one magnitude stronger than that from the electron $g-2$, as clearly shown in Table~\ref{tab:limits}.

The muonium-antimuonium oscillation, i.e. the LFV conversion of the bound states $(\mu^+ e^-) \leftrightarrow (\mu^- e^+)$, can be induced by the effective four-fermion Lagrangian via exchanging the doubly-charged scalar~\cite{HM}:
\begin{eqnarray}
{\cal L}_{M\overline{M}} \ = \
\frac{G_{M\overline{M}}}{\sqrt2} \Big[ \bar{\mu} \gamma_\alpha (1+\gamma_5) e \Big]
\Big[ \bar{\mu} \gamma^\alpha (1+\gamma_5) e \Big]
\end{eqnarray}
with the oscillation probability~\cite{Swartz:1989qz, Clark:2003tv}
\begin{eqnarray}
{\cal P} \ \simeq \
\frac{(\Delta M)^2}{2\Gamma_\mu^2 } \,,
\end{eqnarray}
where the mass splitting
\begin{eqnarray}
\Delta M  \ = \
2 \langle \overline{M} | {\cal L}_{M\overline{M}} | M \rangle \ = \
\frac{16 G_{M\overline{M}}}{\sqrt2 \pi a^3} \,,
\end{eqnarray}
with $a = (\alpha_{\rm EM} \mu)^{-1}$ and $\mu = m_e m_\mu / (m_e + m_\mu)$ the reduced mass of the muonium system. By performing a Fierz transformation, the effective coefficient is related to the couplings and doubly-charged scalar mass via
\begin{eqnarray}
G_{M\overline{M}} \ = \
\frac{f_{ee} f_{\mu\mu}^\dagger}{4\sqrt2 M_{\pm\pm}^2} \,.
\end{eqnarray}
%The MACS experiment~\cite{Willmann:1998gd} sets a 90\% C.L. upper bound of ${\cal P}< 8.2 \times 10^{-11}$, which requires that $|f_{ee}^\dagger f_{\mu\mu}| / M_{\pm\pm}^2 < 1.2 \times 10^{-7} \, {\rm GeV}^{-2}$, as shown in Table~\ref{tab:limits}.

%Regarding the constraints on $h^{ee}$ from the low-energy lepton flavor data, there exists only that limits from the electron $g-2$,

These formulas have been used in the derivation of the LFV bounds in Table~\ref{tab:limits}.


\begin{thebibliography}{99}

\bibitem{Mohapatra:2006gs}
  R.~N.~Mohapatra and A.~Y.~Smirnov,
  %``Neutrino Mass and New Physics,''
  Ann.\ Rev.\ Nucl.\ Part.\ Sci.\  {\bf 56}, 569 (2006)
 % doi:10.1146/annurev.nucl.56.080805.140534
  [hep-ph/0603118].

\bibitem{Konetschny:1977bn}
  W.~Konetschny and W.~Kummer,
  %``Nonconservation of Total Lepton Number with Scalar Bosons,''
  Phys.\ Lett.\  {\bf 70B}, 433 (1977).
 % doi:10.1016/0370-2693(77)90407-5

\bibitem{Magg:1980ut}
  M.~Magg and C.~Wetterich,
  %``Neutrino Mass Problem and Gauge Hierarchy,''
  Phys.\ Lett.\  {\bf 94B}, 61 (1980).

\bibitem{Schechter:1980gr}
  J.~Schechter and J.~W.~F.~Valle,
  %``Neutrino Masses in SU(2) x U(1) Theories,''
  Phys.\ Rev.\ D {\bf 22}, 2227 (1980).

\bibitem{Cheng:1980qt}
  T.~P.~Cheng and L.~F.~Li,
  %``Neutrino Masses, Mixings and Oscillations in SU(2) x U(1) Models of Electroweak Interactions,''
  Phys.\ Rev.\ D {\bf 22}, 2860 (1980).
%  doi:10.1103/PhysRevD.22.2860

\bibitem{Mohapatra:1980yp}
  R.~N.~Mohapatra and G.~Senjanovic,
  %``Neutrino Masses and Mixings in Gauge Models with Spontaneous Parity Violation,''
  Phys.\ Rev.\ D {\bf 23}, 165 (1981).

\bibitem{Lazarides:1980nt}
  G.~Lazarides, Q.~Shafi and C.~Wetterich,
  %``Proton Lifetime and Fermion Masses in an SO(10) Model,''
  Nucl.\ Phys.\ B {\bf 181}, 287 (1981).

%\cite{Perez:2008ha}
\bibitem{Perez:2008ha}
  P.~Fileviez Perez, T.~Han, G.~y.~Huang, T.~Li and K.~Wang,
  %``Neutrino Masses and the CERN LHC: Testing Type II Seesaw,''
  Phys.\ Rev.\ D {\bf 78}, 015018 (2008)
  %%doi:10.1103/PhysRevD.78.015018
  [arXiv:0805.3536 [hep-ph]].
  %%CITATION = %doi:10.1103/PhysRevD.78.015018;%%
  %269 citations counted in INSPIRE as of 03 May 2018

\bibitem{Melfo:2011nx}
  A.~Melfo, M.~Nemevsek, F.~Nesti, G.~Senjanovic and Y.~Zhang,
  %``Type II Seesaw at LHC: The Roadmap,''
  Phys.\ Rev.\ D {\bf 85}, 055018 (2012)
%  doi:10.1103/PhysRevD.85.055018
  [arXiv:1108.4416 [hep-ph]].

%\cite{Kanemura:2014goa}
\bibitem{Kanemura:2014goa}
  S.~Kanemura, M.~Kikuchi, K.~Yagyu and H.~Yokoya,
  %``Bounds on the mass of doubly-charged Higgs bosons in the same-sign diboson decay scenario,''
  Phys.\ Rev.\ D {\bf 90}, no. 11, 115018 (2014)
  %doi:10.1103/PhysRevD.90.115018
  [arXiv:1407.6547 [hep-ph]].
  %%CITATION = doi:10.1103/PhysRevD.90.115018;%%
  %46 citations counted in INSPIRE as of 14 Mar 2018



\bibitem{Apollinari:2017cqg}
  G.~Apollinari, O.~Brüning, T.~Nakamoto and L.~Rossi,
  %``High Luminosity Large Hadron Collider HL-LHC,''
  CERN Yellow Report, no. 5, 1 (2015)
%  doi:10.5170/CERN-2015-005.1
  [arXiv:1705.08830 [physics.acc-ph]].



\bibitem{fcc-hh} \url{https://fcc.web.cern.ch/Pages/default.aspx}


%\cite{Baer:2013cma}
\bibitem{Baer:2013cma}
  H.~Baer {\it et al.},
  %``The International Linear Collider Technical Design Report - Volume 2: Physics,''
  arXiv:1306.6352 [hep-ph].
  %%CITATION = ARXIV:1306.6352;%%
  %520 citations counted in INSPIRE as of 11 Nov 2017


%\cite{Tang:2015qga}
\bibitem{Tang:2015qga}
  J.~Tang {\it et al.},
  %``Concept for a Future Super Proton-Proton Collider,''
  arXiv:1507.03224 [physics.acc-ph].
  %%CITATION = ARXIV:1507.03224;%%
  %27 citations counted in INSPIRE as of 14 Apr 2018

%\cite{CEPC-SPPCStudyGroup:2015csa}
\bibitem{CEPC-SPPCStudyGroup:2015csa}
  CEPC-SPPC Study Group,
  %``CEPC-SPPC Preliminary Conceptual Design Report. 1. Physics and Detector,''
  \url{http://cepc.ihep.ac.cn/preCDR/main_preCDR.pdf}.
  %%CITATION = IHEP-CEPC-DR-2015-01, IHEP-TH-2015-01, IHEP-EP-2015-01;%%
  %50 citations counted in INSPIRE as of 15 Nov 2017


%\cite{Gomez-Ceballos:2013zzn}
\bibitem{Gomez-Ceballos:2013zzn}
  M.~Bicer {\it et al.} [TLEP Design Study Working Group],
  %``First Look at the Physics Case of TLEP,''
  JHEP {\bf 1401}, 164 (2014)
  %doi:10.1007/JHEP01(2014)164
  [arXiv:1308.6176 [hep-ex]].
  %%CITATION = doi:10.1007/JHEP01(2014)164;%%
  %354 citations counted in INSPIRE as of 15 Nov 2017

%\cite{Battaglia:2004mw}
\bibitem{Battaglia:2004mw}
  E.~Accomando {\it et al.} [CLIC Physics Working Group],
  %``Physics at the CLIC multi-TeV linear collider,''
  hep-ph/0412251.
  %%CITATION = HEP-PH/0412251;%%
  %288 citations counted in INSPIRE as of 15 Nov 2017

\bibitem{LR1} J. C. Pati and A. Salam, Phys. Rev. D {\bf 10}, 275 (1974).

\bibitem{LR2} R. N. Mohapatra and J. C. Pati, Phys. Rev. D {\bf 11} 2558 (1975).

\bibitem{LR3} G. Senjanovi\'{c} and R. N. Mohapatra, Phys. Rev. D {\bf 12} 1502 (1975).

%\cite{Gunion:1989in}
\bibitem{Gunion:1989in}
  J.~F.~Gunion, J.~Grifols, A.~Mendez, B.~Kayser and F.~I.~Olness,
  %``Higgs Bosons in Left-Right Symmetric Models,''
  Phys.\ Rev.\ D {\bf 40}, 1546 (1989).
  %doi:10.1103/PhysRevD.40.1546
  %%CITATION = doi:10.1103/PhysRevD.40.1546;%%
  %253 citations counted in INSPIRE as of 19 Dec 2017

%\cite{Dev:2016dja}
\bibitem{Dev:2016dja}
  P.~S.~B.~Dev, R.~N.~Mohapatra and Y.~Zhang,
  %``Probing the Higgs Sector of the Minimal Left-Right Symmetric Model at Future Hadron Colliders,''
  JHEP {\bf 1605}, 174 (2016)
  %doi:10.1007/JHEP05(2016)174
  [arXiv:1602.05947 [hep-ph]].
  %%CITATION = doi:10.1007/JHEP05(2016)174;%%
  %48 citations counted in INSPIRE as of 14 Apr 2018

\bibitem{Azuelos:2004mwa}
  G.~Azuelos, K.~Benslama and J.~Ferland,
  %``Prospects for the search for a doubly-charged Higgs in the left-right symmetric model with ATLAS,''
  J.\ Phys.\ G {\bf 32}, no. 2, 73 (2006)
%  doi:10.1088/0954-3899/32/2/002
  [hep-ph/0503096].

\bibitem{Han:2007bk}
  T.~Han, B.~Mukhopadhyaya, Z.~Si and K.~Wang,
  %``Pair production of doubly-charged scalars: Neutrino mass constraints and signals at the LHC,''
  Phys.\ Rev.\ D {\bf 76}, 075013 (2007)
%  doi:10.1103/PhysRevD.76.075013
  [arXiv:0706.0441 [hep-ph]].

\bibitem{delAguila:2008cj}
  F.~del Aguila and J.~A.~Aguilar-Saavedra,
  %``Distinguishing seesaw models at LHC with multi-lepton signals,''
  Nucl.\ Phys.\ B {\bf 813}, 22 (2009)
 % doi:10.1016/j.nuclphysb.2008.12.029
  [arXiv:0808.2468 [hep-ph]].

\bibitem{Akeroyd:2009hb}
  A.~G.~Akeroyd and C.~W.~Chiang,
  %``Doubly charged Higgs bosons and three-lepton signatures in the Higgs Triplet Model,''
  Phys.\ Rev.\ D {\bf 80}, 113010 (2009)
  %doi:10.1103/PhysRevD.80.113010
  [arXiv:0909.4419 [hep-ph]].
  %%CITATION = doi:10.1103/PhysRevD.80.113010;%%

\bibitem{Akeroyd:2010je}
  A.~G.~Akeroyd and C.~W.~Chiang,
  %``Phenomenology of Large Mixing for the CP-even Neutral Scalars of the Higgs Triplet Model,''
  Phys.\ Rev.\ D {\bf 81}, 115007 (2010)
  %doi:10.1103/PhysRevD.81.115007
  [arXiv:1003.3724 [hep-ph]].
  %%CITATION = doi:10.1103/PhysRevD.81.115007;%%

\bibitem{Akeroyd:2010ip}
  A.~G.~Akeroyd, C.~W.~Chiang and N.~Gaur,
  %``Leptonic signatures of doubly charged Higgs boson production at the LHC,''
  JHEP {\bf 1011}, 005 (2010)
  %doi:10.1007/JHEP11(2010)005
  [arXiv:1009.2780 [hep-ph]].
  %%CITATION = doi:10.1007/JHEP11(2010)005;%%



\bibitem{Alloul:2013raa}
  A.~Alloul, M.~Frank, B.~Fuks and M.~Rausch de Traubenberg,
  %``Doubly-charged particles at the Large Hadron Collider,''
  Phys.\ Rev.\ D {\bf 88}, 075004 (2013)
 % doi:10.1103/PhysRevD.88.075004
  [arXiv:1307.1711 [hep-ph]].

\bibitem{Chun:2013vma}
  E.~J.~Chun and P.~Sharma,
  %``Search for a doubly-charged boson in four lepton final states in type II seesaw,''
  Phys.\ Lett.\ B {\bf 728}, 256 (2014)
%  doi:10.1016/j.physletb.2013.11.056
  [arXiv:1309.6888 [hep-ph]].

\bibitem{delAguila:2013mia}
  F.~del Aguila and M.~Chala,
  %``LHC bounds on Lepton Number Violation mediated by doubly and singly-charged scalars,''
  JHEP {\bf 1403}, 027 (2014)
%  doi:10.1007/JHEP03(2014)027
  [arXiv:1311.1510 [hep-ph]].

\bibitem{Bambhaniya:2013wza}
  G.~Bambhaniya, J.~Chakrabortty, J.~Gluza, M.~Kordiaczyńska and R.~Szafron,
  %``Left-Right Symmetry and the Charged Higgs Bosons at the LHC,''
  JHEP {\bf 1405}, 033 (2014)
%  doi:10.1007/JHEP05(2014)033
  [arXiv:1311.4144 [hep-ph]].

\bibitem{Mitra:2016wpr}
  M.~Mitra, S.~Niyogi and M.~Spannowsky,
  %``Type-II Seesaw Model and Multilepton Signatures at Hadron Colliders,''
  Phys.\ Rev.\ D {\bf 95}, no. 3, 035042 (2017)
%  doi:10.1103/PhysRevD.95.035042
  [arXiv:1611.09594 [hep-ph]].

\bibitem{Babu:2016rcr}
  K.~S.~Babu and S.~Jana,
  %``Probing Doubly Charged Higgs Bosons at the LHC through Photon Initiated Processes,''
  Phys.\ Rev.\ D {\bf 95}, no. 5, 055020 (2017)
 % doi:10.1103/PhysRevD.95.055020
  [arXiv:1612.09224 [hep-ph]].

\bibitem{Ghosh:2017pxl}
  D.~K.~Ghosh, N.~Ghosh, I.~Saha and A.~Shaw,
  %``Revisiting the high-scale validity of the type II seesaw model with novel LHC signature,''
  Phys.\ Rev.\ D {\bf 97}, no. 11, 115022 (2018)
%  doi:10.1103/PhysRevD.97.115022
  [arXiv:1711.06062 [hep-ph]].

\bibitem{Agrawal:2018pci}
  P.~Agrawal, M.~Mitra, S.~Niyogi, S.~Shil and M.~Spannowsky,
  %``Probing the Type-II Seesaw Mechanism through the Production of Higgs Bosons at a Lepton Collider,''
  Phys.\ Rev.\ D {\bf 98}, no. 1, 015024 (2018)
%  doi:10.1103/PhysRevD.98.015024
  [arXiv:1803.00677 [hep-ph]].

%\cite{Dev:2018upe}
\bibitem{Dev:2018upe}
  P.~S.~B.~Dev, R.~N.~Mohapatra and Y.~Zhang,
  %``Probing TeV scale origin of neutrino mass at lepton colliders,''
  arXiv:1803.11167 [hep-ph].
  %%CITATION = ARXIV:1803.11167;%%

%\cite{Borah:2018yxd}
\bibitem{Borah:2018yxd}
  D.~Borah, B.~Fuks, D.~Goswami and P.~Poulose,
  %``Investigating the scalar sector of left-right symmetric models with leptonic probes,''
  Phys.\ Rev.\ D {\bf 98}, no. 3, 035008 (2018)
%  doi:10.1103/PhysRevD.98.035008
  [arXiv:1805.06910 [hep-ph]].

%\cite{Abbiendi:2001cr}
\bibitem{Abbiendi:2001cr}
  G.~Abbiendi {\it et al.} [OPAL Collaboration],
  %``Search for doubly charged Higgs bosons with the OPAL detector at LEP,''
  Phys.\ Lett.\ B {\bf 526}, 221 (2002)
  %doi:10.1016/S0370-2693(01)01474-5
  [hep-ex/0111059].
  %%CITATION = doi:10.1016/S0370-2693(01)01474-5;%%
  %90 citations counted in INSPIRE as of 19 Mar 2018

%\cite{Abdallah:2002qj}
\bibitem{Abdallah:2002qj}
  J.~Abdallah {\it et al.} [DELPHI Collaboration],
  %``Search for doubly charged Higgs bosons at LEP-2,''
  Phys.\ Lett.\ B {\bf 552}, 127 (2003)
  %doi:10.1016/S0370-2693(02)03125-8
  [hep-ex/0303026].
  %%CITATION = doi:10.1016/S0370-2693(02)03125-8;%%
  %92 citations counted in INSPIRE as of 19 Mar 2018

%\cite{Achard:2003mv}
\bibitem{Achard:2003mv}
  P.~Achard {\it et al.} [L3 Collaboration],
  %``Search for doubly charged Higgs bosons at LEP,''
  Phys.\ Lett.\ B {\bf 576}, 18 (2003)
  %doi:10.1016/j.physletb.2003.09.082
  [hep-ex/0309076].
  %%CITATION = doi:10.1016/j.physletb.2003.09.082;%%
  %82 citations counted in INSPIRE as of 13 Mar 2018




%\cite{Acosta:2004uj}
\bibitem{Acosta:2004uj}
  D.~Acosta {\it et al.} [CDF Collaboration],
  %``Search for doubly-charged Higgs bosons decaying to dileptons in $p\bar{p}$ collisions at $\sqrt{s} = 1.96$ TeV,''
  Phys.\ Rev.\ Lett.\  {\bf 93}, 221802 (2004)
  %doi:10.1103/PhysRevLett.93.221802
  [hep-ex/0406073].
  %%CITATION = doi:10.1103/PhysRevLett.93.221802;%%
  %128 citations counted in INSPIRE as of 12 Apr 2018

%\cite{Aaltonen:2008ip}
\bibitem{Aaltonen:2008ip}
  T.~Aaltonen {\it et al.} [CDF Collaboration],
  %``Search for Doubly Charged Higgs Bosons with Lepton-Flavor-Violating Decays involving Tau Leptons,''
  Phys.\ Rev.\ Lett.\  {\bf 101}, 121801 (2008)
  %doi:10.1103/PhysRevLett.101.121801
  [arXiv:0808.2161 [hep-ex]].
  %%CITATION = doi:10.1103/PhysRevLett.101.121801;%%
  %60 citations counted in INSPIRE as of 12 Apr 2018


%\cite{Abazov:2008ab}
\bibitem{Abazov:2008ab}
  V.~M.~Abazov {\it et al.} [D0 Collaboration],
  %``Search for pair production of doubly-charged Higgs bosons in the $H^{++} H^{--} \to \mu^{+} \mu^{+} \mu^{-} \mu^{-}$ final state at D0,''
  Phys.\ Rev.\ Lett.\  {\bf 101}, 071803 (2008)
  %doi:10.1103/PhysRevLett.101.071803
  [arXiv:0803.1534 [hep-ex]].
  %%CITATION = doi:10.1103/PhysRevLett.101.071803;%%
  %65 citations counted in INSPIRE as of 12 Apr 2018

%\cite{Abazov:2011xx}
\bibitem{Abazov:2011xx}
  V.~M.~Abazov {\it et al.} [D0 Collaboration],
  %``Search for doubly-charged Higgs boson pair production in $p\bar {p}$ collisions at $\sqrt{s} = 1.96$ TeV,''
  Phys.\ Rev.\ Lett.\  {\bf 108}, 021801 (2012)
  %doi:10.1103/PhysRevLett.108.021801
  [arXiv:1106.4250 [hep-ex]].
  %%CITATION = doi:10.1103/PhysRevLett.108.021801;%%
  %39 citations counted in INSPIRE as of 12 Apr 2018

%\cite{Aaboud:2017qph}
\bibitem{Aaboud:2017qph}
  M.~Aaboud {\it et al.} [ATLAS Collaboration],
  %``Search for doubly charged Higgs boson production in multi-lepton final states with the ATLAS detector using proton-proton collisions at $\sqrt{s}=13\,\text {TeV}$,''
  Eur.\ Phys.\ J.\ C {\bf 78}, no. 3, 199 (2018)
  %doi:10.1140/EPJC/S10052-018-5661-Z, 10.1140/epjc/s10052-018-5661-z
  [arXiv:1710.09748 [hep-ex]].
  %%CITATION = doi:10.1140/EPJC/S10052-018-5661-Z, 10.1140/epjc/s10052-018-5661-z;%%
  %8 citations counted in INSPIRE as of 12 Apr 2018

\bibitem{CMS:2017pet}
  CMS Collaboration,
  %``A search for doubly-charged Higgs boson production in three and four lepton final states at $\sqrt{s}=13~\mathrm{TeV}$,''
  CMS-PAS-HIG-16-036.

\bibitem{Pal:1983bf}
  P.~B.~Pal,
  %``Constraints on a Muon - Neutrino Mass Around 100-kev,''
  Nucl.\ Phys.\ B {\bf 227}, 237 (1983).
 % doi:10.1016/0550-3213(83)90021-4

\bibitem{Leontaris:1985qc}
  G.~K.~Leontaris, K.~Tamvakis and J.~D.~Vergados,
  %``Lepton and Family Number Violation From Exotic Scalars,''
  Phys.\ Lett.\  {\bf 162B}, 153 (1985).
 % doi:10.1016/0370-2693(85)91078-0

%\cite{Swartz:1989qz}
\bibitem{Swartz:1989qz}
  M.~L.~Swartz,
  %``Limits on Doubly Charged Higgs Bosons and Lepton Flavor Violation,''
  Phys.\ Rev.\ D {\bf 40}, 1521 (1989).
  %doi:10.1103/PhysRevD.40.1521
  %%CITATION = doi:10.1103/PhysRevD.40.1521;%%
  %139 citations counted in INSPIRE as of 19 Dec 2017

\bibitem{Mohapatra:1992uu}
  R.~N.~Mohapatra,
  %``Rare decays of the tau lepton as a probe of the left-right symmetric theories of weak interactions,''
  Phys.\ Rev.\ D {\bf 46}, 2990 (1992).
%  doi:10.1103/PhysRevD.46.2990

\bibitem{Cirigliano:2004mv}
  V.~Cirigliano, A.~Kurylov, M.~J.~Ramsey-Musolf and P.~Vogel,
  %``Lepton flavor violation without supersymmetry,''
  Phys.\ Rev.\ D {\bf 70}, 075007 (2004)
 % doi:10.1103/PhysRevD.70.075007
  [hep-ph/0404233].

%\cite{Cirigliano:2004tc}
\bibitem{Cirigliano:2004tc}
  V.~Cirigliano, A.~Kurylov, M.~J.~Ramsey-Musolf and P.~Vogel,
  %``Neutrinoless double beta decay and lepton flavor violation,''
  Phys.\ Rev.\ Lett.\  {\bf 93}, 231802 (2004)
  %doi:10.1103/PhysRevLett.93.231802
  [hep-ph/0406199].
  %%CITATION = doi:10.1103/PhysRevLett.93.231802;%%
  %68 citations counted in INSPIRE as of 24 May 2018

\bibitem{Akeroyd:2009nu}
  A.~G.~Akeroyd, M.~Aoki and H.~Sugiyama,
  %``Lepton Flavour Violating Decays tau ---> anti-l ll and mu ---> e gamma in the Higgs Triplet Model,''
  Phys.\ Rev.\ D {\bf 79}, 113010 (2009)
%  doi:10.1103/PhysRevD.79.113010
  [arXiv:0904.3640 [hep-ph]].

\bibitem{Tello:2010am}
  V.~Tello, M.~Nemevsek, F.~Nesti, G.~Senjanovic and F.~Vissani,
  %``Left-Right Symmetry: from LHC to Neutrinoless Double Beta Decay,''
  Phys.\ Rev.\ Lett.\  {\bf 106}, 151801 (2011)
 % doi:10.1103/PhysRevLett.106.151801
  [arXiv:1011.3522 [hep-ph]].

\bibitem{Chakrabortty:2012vp}
  J.~Chakrabortty, P.~Ghosh and W.~Rodejohann,
  %``Lower Limits on $\mu \to e \gamma$ from New Measurements on $U_{e3}$,''
  Phys.\ Rev.\ D {\bf 86}, 075020 (2012)
%  doi:10.1103/PhysRevD.86.075020
  [arXiv:1204.1000 [hep-ph]].

\bibitem{Barry:2013xxa}
  J.~Barry and W.~Rodejohann,
  %``Lepton number and flavour violation in TeV-scale left-right symmetric theories with large left-right mixing,''
  JHEP {\bf 1309}, 153 (2013)
%  doi:10.1007/JHEP09(2013)153
  [arXiv:1303.6324 [hep-ph]].



\bibitem{Bambhaniya:2015ipg}
  G.~Bambhaniya, P.~S.~B.~Dev, S.~Goswami and M.~Mitra,
  %``The Scalar Triplet Contribution to Lepton Flavour Violation and Neutrinoless Double Beta Decay in Left-Right Symmetric Model,''
  JHEP {\bf 1604}, 046 (2016)
%  doi:10.1007/JHEP04(2016)046
  [arXiv:1512.00440 [hep-ph]].

\bibitem{Chakrabortty:2015zpm}
  J.~Chakrabortty, P.~Ghosh, S.~Mondal and T.~Srivastava,
  %``Reconciling (g-2)$_μ$ and charged lepton flavor violating processes through a doubly charged scalar,''
  Phys.\ Rev.\ D {\bf 93}, no. 11, 115004 (2016)
%  doi:10.1103/PhysRevD.93.115004
  [arXiv:1512.03581 [hep-ph]].


%\cite{Borah:2016iqd}
\bibitem{Borah:2016iqd}
  D.~Borah and A.~Dasgupta,
  %``Charged lepton flavour violcxmation and neutrinoless double beta decay in left-right symmetric models with type I+II seesaw,''
  JHEP {\bf 1607}, 022 (2016)
  %doi:10.1007/JHEP07(2016)022
  [arXiv:1606.00378 [hep-ph]].
  %%CITATION = doi:10.1007/JHEP07(2016)022;%%
  %10 citations counted in INSPIRE as of 09 Mar 2018


%\cite{Lindner:2016bgg}
\bibitem{Lindner:2016bgg}
  M.~Lindner, M.~Platscher and F.~S.~Queiroz,
  %``A Call for New Physics : The Muon Anomalous Magnetic Moment and Lepton Flavor Violation,''
   Phys.\ Rept.\  {\bf 731}, 1 (2018)
  %doi:10.1016/j.physrep.2017.12.001
  [arXiv:1610.06587 [hep-ph]].

\bibitem{Bonilla:2016fqd}
  C.~Bonilla, M.~E.~Krauss, T.~Opferkuch and W.~Porod,
  %``Perspectives for Detecting Lepton Flavour Violation in Left-Right Symmetric Models,''
  JHEP {\bf 1703}, 027 (2017)
%  doi:10.1007/JHEP03(2017)027
  [arXiv:1611.07025 [hep-ph]].



%\cite{Borgohain:2017akh}
\bibitem{Borgohain:2017akh}
  H.~Borgohain and M.~K.~Das,
  %``Lepton number violation, lepton flavor violation, and baryogenesis in left-right symmetric model,''
  Phys.\ Rev.\ D {\bf 96}, no. 7, 075021 (2017)
  %doi:10.1103/PhysRevD.96.075021
  [arXiv:1709.09542 [hep-ph]].
  %%CITATION = doi:10.1103/PhysRevD.96.075021;%%

\bibitem{Crivellin:2018ahj}
  A.~Crivellin, M.~Ghezzi, L.~Panizzi, G.~M.~Pruna and A.~Signer,
  %``Low- and high-energy phenomenology of a doubly charged scalar,''
  arXiv:1807.10224 [hep-ph].


%\cite{Leveille:1977rc}
\bibitem{Leveille:1977rc}
  J.~P.~Leveille,
  %``The Second Order Weak Correction to (G-2) of the Muon in Arbitrary Gauge Models,''
  Nucl.\ Phys.\ B {\bf 137}, 63 (1978).
  %doi:10.1016/0550-3213(78)90051-2
  %%CITATION = doi:10.1016/0550-3213(78)90051-2;%%
  %165 citations counted in INSPIRE as of 09 Dec 2017

%\cite{Moore:1984eg}
\bibitem{Moore:1984eg}
  S.~R.~Moore, K.~Whisnant and B.~L.~Young,
  %``Second Order Corrections to the Muon Anomalous Magnetic Moment in Alternative Electroweak Models,''
  Phys.\ Rev.\ D {\bf 31}, 105 (1985).
  %doi:10.1103/PhysRevD.31.105
  %%CITATION = doi:10.1103/PhysRevD.31.105;%%
  %29 citations counted in INSPIRE as of 09 Dec 2017

\bibitem{Chang:1989uk}
  D.~Chang and W.~Y.~Keung,
  %``Constraints on Muonium - anti-Muonium Conversion,''
  Phys.\ Rev.\ Lett.\  {\bf 62}, 2583 (1989).
%  doi:10.1103/PhysRevLett.62.2583

\bibitem{HM} P.~Herczeg and R.~N.~Mohapatra,
  %``Muonium to anti-muonium conversion and the decay mu+ ---> e+ anti-neutrino neutrino in left-right symmetric models,''
  Phys.\ Rev.\ Lett.\  {\bf 69}, 2475 (1992).






%\cite{Clark:2003tv}
\bibitem{Clark:2003tv}
  T.~E.~Clark and S.~T.~Love,
  %``Muonium - anti-muonium oscillations and massive Majorana neutrinos,''
  Mod.\ Phys.\ Lett.\ A {\bf 19}, 297 (2004)
  %doi:10.1142/S0217732304013143
  [hep-ph/0307264].
  %%CITATION = doi:10.1142/S0217732304013143;%%



\bibitem{PDG}
  M. Tanabashi {\it et al.} (Particle Data Group), Phys. Rev. D {\bf 98}, 030001 (2018).



%\cite{Dev:2017ouk}
\bibitem{Dev:2017ouk}
  P.~S.~B.~Dev, C.~M.~Vila and W.~Rodejohann,
  %``Naturalness in testable type II seesaw scenarios,''
  Nucl.\ Phys.\ B {\bf 921}, 436 (2017)
  %doi:10.1016/j.nuclphysb.2017.06.007
  [arXiv:1703.00828 [hep-ph]].
  %%CITATION = doi:10.1016/j.nuclphysb.2017.06.007;%%
  %8 citations counted in INSPIRE as of 09 Mar 2018




\bibitem{Dev:2018sel}
  P.~S.~B.~Dev, M.~J.~Ramsey-Musolf and Y.~Zhang,
  %``Doubly-Charged Scalars in the Type-II Seesaw Mechanism: Fundamental Symmetry Tests and High-Energy Searches,''
  arXiv:1806.08499 [hep-ph].


\bibitem{Arhrib:2011uy}
  A.~Arhrib, R.~Benbrik, M.~Chabab, G.~Moultaka, M.~C.~Peyranere, L.~Rahili and J.~Ramadan,
  %``The Higgs Potential in the Type II Seesaw Model,''
  Phys.\ Rev.\ D {\bf 84}, 095005 (2011)
  %doi:10.1103/PhysRevD.84.095005
  [arXiv:1105.1925 [hep-ph]].

\bibitem{DGOS}
  P.~S.~B.~Dev, D.~K.~Ghosh, N.~Okada and I.~Saha,
  %``125 GeV Higgs Boson and the Type-II Seesaw Model,''
  JHEP {\bf 1303}, 150 (2013)
  [Erratum-ibid.\  {\bf 1305}, 049 (2013)]
  [arXiv:1301.3453 [hep-ph]].

\bibitem{Chabab:2015nel}
  M.~Chabab, M.~C.~Peyranre and L.~Rahili,
  %``Naturalness in a type II seesaw model and implications for physical scalars,''
  Phys.\ Rev.\ D {\bf 93}, no. 11, 115021 (2016)
  %doi:10.1103/PhysRevD.93.115021
  [arXiv:1512.07280 [hep-ph]].



\bibitem{delAguila:2008ks}
  F.~del Aguila, J.~A.~Aguilar-Saavedra, J.~de Blas and M.~Perez-Victoria,
  %``Electroweak constraints on see-saw messengers and their implications for LHC,''
  arXiv:0806.1023 [hep-ph].
  %%CITATION = ARXIV:0806.1023;%%
  %26 citations counted in INSPIRE as of 13 Mar 2018

%\cite{Chun:2012jw}
\bibitem{Chun:2012jw}
  E.~J.~Chun, H.~M.~Lee and P.~Sharma,
  %``Vacuum Stability, Perturbativity, EWPD and Higgs-to-diphoton rate in Type II Seesaw Models,''
  JHEP {\bf 1211}, 106 (2012)
  %doi:10.1007/JHEP11(2012)106
  [arXiv:1209.1303 [hep-ph]].
  %%CITATION = doi:10.1007/JHEP11(2012)106;%%
  %87 citations counted in INSPIRE as of 14 Apr 2018

%\cite{Aoki:2012jj}
\bibitem{Aoki:2012jj}
  M.~Aoki, S.~Kanemura, M.~Kikuchi and K.~Yagyu,
  %``Radiative corrections to the Higgs boson couplings in the triplet model,''
  Phys.\ Rev.\ D {\bf 87}, no. 1, 015012 (2013)
  %doi:10.1103/PhysRevD.87.015012
  [arXiv:1211.6029 [hep-ph]].
  %%CITATION = doi:10.1103/PhysRevD.87.015012;%%
  %56 citations counted in INSPIRE as of 14 Mar 2018

\bibitem{Esteban:2016qun}
  I.~Esteban, M.~C.~Gonzalez-Garcia, M.~Maltoni, I.~Martinez-Soler and T.~Schwetz,
  %``Updated fit to three neutrino mixing: exploring the accelerator-reactor complementarity,''
  JHEP {\bf 1701}, 087 (2017)
 % doi:10.1007/JHEP01(2017)087
  [arXiv:1611.01514 [hep-ph]].

\bibitem{nufit} \url{http://www.nu-fit.org/}




\bibitem{Abe:2017vif}
  K.~Abe {\it et al.} [T2K Collaboration],
  %``Measurement of neutrino and antineutrino oscillations by the T2K experiment including a new additional sample of $\nu_e$ interactions at the far detector,''
  Phys.\ Rev.\ D {\bf 96}, no. 9, 092006 (2017)
 % doi:10.1103/PhysRevD.96.092006
  [arXiv:1707.01048 [hep-ex]].

\bibitem{NOvA:2018gge}
  M.~A.~Acero {\it et al.} [NOvA Collaboration],
  %``New constraints on oscillation parameters from $\nu_e$ appearance and $\nu_\mu$ disappearance in the NOvA experiment,''
  arXiv:1806.00096 [hep-ex].


















%\cite{Abbiendi:2003pr}
\bibitem{Abbiendi:2003pr}
  G.~Abbiendi {\it et al.} [OPAL Collaboration],
  %``Search for the single production of doubly charged Higgs bosons and constraints on their couplings from Bhabha scattering,''
  Phys.\ Lett.\ B {\bf 577}, 93 (2003)
  %doi:10.1016/j.physletb.2003.10.034
  [hep-ex/0308052].
  %%CITATION = doi:10.1016/j.physletb.2003.10.034;%%
  %59 citations counted in INSPIRE as of 13 Mar 2018

%\cite{Abdallah:2005ph}
\bibitem{Abdallah:2005ph}
  J.~Abdallah {\it et al.} [DELPHI Collaboration],
  %``Measurement and interpretation of fermion-pair production at LEP energies above the Z resonance,''
  Eur.\ Phys.\ J.\ C {\bf 45}, 589 (2006)
  %doi:10.1140/epjc/s2005-02461-0
  [hep-ex/0512012].
  %%CITATION = doi:10.1140/epjc/s2005-02461-0;%%
  %65 citations counted in INSPIRE as of 13 Mar 2018




\bibitem{Benesch:2014bas}
  J.~Benesch {\it et al.} [MOLLER Collaboration],
  %``The MOLLER Experiment: An Ultra-Precise Measurement of the Weak Mixing Angle Using M{\o}ller Scattering,''
  arXiv:1411.4088 [nucl-ex].

\bibitem{Moller}
  \url{http://hallaweb.jlab.org/12GeV/Moller/}


\bibitem{Amhis:2016xyh}
  Y.~Amhis {\it et al.} [HFLAV Collaboration],
  %``Averages of $b$-hadron, $c$-hadron, and $\tau$-lepton properties as of summer 2016,''
  Eur.\ Phys.\ J.\ C {\bf 77}, no. 12, 895 (2017)
 % doi:10.1140/epjc/s10052-017-5058-4
  [arXiv:1612.07233 [hep-ex]].

\bibitem{Hanneke:2008tm}
  D.~Hanneke, S.~Fogwell and G.~Gabrielse,
  %``New Measurement of the Electron Magnetic Moment and the Fine Structure Constant,''
  Phys.\ Rev.\ Lett.\  {\bf 100}, 120801 (2008)
%  doi:10.1103/PhysRevLett.100.120801
  [arXiv:0801.1134 [physics.atom-ph]].

\bibitem{Bennett:2006fi}
  G.~W.~Bennett {\it et al.} [Muon g-2 Collaboration],
  %``Final Report of the Muon E821 Anomalous Magnetic Moment Measurement at BNL,''
  Phys.\ Rev.\ D {\bf 73}, 072003 (2006)
%  doi:10.1103/PhysRevD.73.072003
  [hep-ex/0602035].

\bibitem{Willmann:1998gd}
  L.~Willmann {\it et al.},
  %``New bounds from searching for muonium to anti-muonium conversion,''
  Phys.\ Rev.\ Lett.\  {\bf 82}, 49 (1999)
 % doi:10.1103/PhysRevLett.82.49
  [hep-ex/9807011].

%\cite{Dev:2017ftk}
\bibitem{Dev:2017ftk}
  P.~S.~B.~Dev, R.~N.~Mohapatra and Y.~Zhang,
  %``Lepton Flavor Violation Induced by a Neutral Scalar at Future Lepton Colliders,''
  arXiv:1711.08430 [hep-ph].
  %%CITATION = ARXIV:1711.08430;%%
  %2 citations counted in INSPIRE as of 03 May 2018

%\cite{ATLAS:2011rha}
\bibitem{ATLAS:2011rha}
  ATLAS Collaboration,
  %``Search for Doubly Charged Higgs Boson Production in Like-sign Muon Pairs in pp Collisions at \A1\CCs=7 TeV,''
  ATLAS-CONF-2011-127.
  %%CITATION = ATLAS-CONF-2011-127;%%
  %9 citations counted in INSPIRE as of 11 Apr 2018

%\cite{CMS:2011sqa}
\bibitem{CMS:2011sqa}
  CMS Collaboration,
  %``Inclusive search for doubly charged higgs in leptonic final states at sqrt s=7 TeV,''
  CMS-PAS-HIG-11-007.
  %%CITATION = CMS-PAS-HIG-11-007;%%
  %5 citations counted in INSPIRE as of 12 Apr 2018

%\cite{ATLAS:2014kca}
\bibitem{ATLAS:2014kca}
  G.~Aad {\it et al.} [ATLAS Collaboration],
  %``Search for anomalous production of prompt same-sign lepton pairs and pair-produced doubly charged Higgs bosons with $ \sqrt{s}=8 $ TeV $pp$ collisions using the ATLAS detector,''
  JHEP {\bf 1503}, 041 (2015)
  %doi:10.1007/JHEP03(2015)041
  [arXiv:1412.0237 [hep-ex]].
  %%CITATION = doi:10.1007/JHEP03(2015)041;%%
  %109 citations counted in INSPIRE as of 12 Apr 2018

%\cite{CMS:2016cpz}
\bibitem{CMS:2016cpz}
  CMS Collaboration,
  %``Search for a doubly-charged Higgs boson with $\sqrt{s}=8~\mathrm{TeV}$ $pp$ collisions at the CMS experiment,''
  CMS-PAS-HIG-14-039.
  %%CITATION = CMS-PAS-HIG-14-039;%%
  %19 citations counted in INSPIRE as of 12 Apr 2018







%\cite{Arkani-Hamed:2015vfh}
\bibitem{Arkani-Hamed:2015vfh}
  N.~Arkani-Hamed, T.~Han, M.~Mangano and L.~T.~Wang,
  %``Physics opportunities of a 100 TeV proton-proton collider,''
  Phys.\ Rept.\  {\bf 652}, 1 (2016)
  %doi:10.1016/j.physrep.2016.07.004
  [arXiv:1511.06495 [hep-ph]].
  %%CITATION = doi:10.1016/j.physrep.2016.07.004;%%
  %128 citations counted in INSPIRE as of 14 Apr 2018


%\cite{Contino:2016spe}
\bibitem{Contino:2016spe}
  R.~Contino {\it et al.},
  %``Physics at a 100 TeV pp collider: Higgs and EW symmetry breaking studies,''
  CERN Yellow Report, no. 3, 255 (2017)
  %doi:10.23731/CYRM-2017-003.255
  [arXiv:1606.09408 [hep-ph]].
  %%CITATION = doi:10.23731/CYRM-2017-003.255;%%
  %81 citations counted in INSPIRE as of 14 Apr 2018


\bibitem{Aktas:2006nu}
  A.~Aktas {\it et al.} [H1 Collaboration],
  %``Search for doubly-charged Higgs boson production at HERA,''
  Phys.\ Lett.\ B {\bf 638}, 432 (2006)
 % doi:10.1016/j.physletb.2006.05.061
  [hep-ex/0604027].


\bibitem{Ade:2015xua}
  P.~A.~R.~Ade {\it et al.} [Planck Collaboration],
  %``Planck 2015 results. XIII. Cosmological parameters,''
  Astron.\ Astrophys.\  {\bf 594}, A13 (2016)
 % doi:10.1051/0004-6361/201525830
  [arXiv:1502.01589 [astro-ph.CO]].


%\cite{Aad:2014yea}
\bibitem{Aad:2014yea}
  G.~Aad {\it et al.} [ATLAS Collaboration],
  %``Search for long-lived neutral particles decaying into lepton jets in proton-proton collisions at $ \sqrt{s}=8 $ TeV with the ATLAS detector,''
  JHEP {\bf 1411}, 088 (2014)
%  doi:10.1007/JHEP11(2014)088
  [arXiv:1409.0746 [hep-ex]].
  %%CITATION = doi:10.1007/JHEP11(2014)088;%%
  %96 citations counted in INSPIRE as of 10 Oct 2018



%\cite{Kanemura:2014ipa}
\bibitem{Kanemura:2014ipa}
  S.~Kanemura, M.~Kikuchi, H.~Yokoya and K.~Yagyu,
  %``LHC Run-I constraint on the mass of doubly charged Higgs bosons in the same-sign diboson decay scenario,''
  PTEP {\bf 2015}, 051B02 (2015)
  %doi:10.1093/ptep/ptv071
  [arXiv:1412.7603 [hep-ph]].
  %%CITATION = doi:10.1093/ptep/ptv071;%%
  %33 citations counted in INSPIRE as of 14 Apr 2018

%\cite{CMS:2016ybj}
\bibitem{CMS:2016ybj}
  CMS Collaboration,
  %``Search for heavy stable charged particles with $12.9~\mathrm{fb}^{-1}$ of 2016 data,''
  CMS-PAS-EXO-16-036.
  %%CITATION = CMS-PAS-EXO-16-036;%%
  %27 citations counted in INSPIRE as of 14 Apr 2018

%\cite{Chatrchyan:2008aa}
\bibitem{Chatrchyan:2008aa}
  S.~Chatrchyan {\it et al.} [CMS Collaboration],
  %``The CMS Experiment at the CERN LHC,''
  JINST {\bf 3}, S08004 (2008).
  %doi:10.1088/1748-0221/3/08/S08004
  %%CITATION = doi:10.1088/1748-0221/3/08/S08004;%%
  %5182 citations counted in INSPIRE as of 14 Apr 2018

\bibitem{Cerri:2018rkm}
  O.~Cerri, S.~Xie, C.~Pena and M.~Spiropulu,
  %``Identification of Long-lived Charged Particles using Time-Of-Flight Systems at the Upgraded LHC detectors,''
  arXiv:1807.05453 [hep-ex].

\bibitem{Aaboud:2018jbr}
  M.~Aaboud {\it et al.} [ATLAS Collaboration],
  %``Search for long-lived particles in final states with displaced dimuon vertices in $pp$ collisions at $\sqrt{s}=$ 13 TeV with the ATLAS detector,''
  arXiv:1808.03057 [hep-ex].

%\cite{Englert:2016ktc}
\bibitem{Englert:2016ktc}
  C.~Englert, P.~Schichtel and M.~Spannowsky,
  %``Same-sign W pair production in composite Higgs models,''
  Phys.\ Rev.\ D {\bf 95}, no. 5, 055002 (2017)
 % doi:10.1103/PhysRevD.95.055002
  [arXiv:1610.07354 [hep-ph]].
  %%CITATION = doi:10.1103/PhysRevD.95.055002;%%
  %6 citations counted in INSPIRE as of 10 Oct 2018

%\cite{Alcaide:2017dcx}
\bibitem{Alcaide:2017dcx}
  J.~Alcaide, M.~Chala and A.~Santamaria,
  %``LHC signals of radiatively-induced neutrino masses and implications for the Zee–Babu model,''
  Phys.\ Lett.\ B {\bf 779}, 107 (2018)
%  doi:10.1016/j.physletb.2018.02.001
  [arXiv:1710.05885 [hep-ph]].
  %%CITATION = doi:10.1016/j.physletb.2018.02.001;%%


\bibitem{ATLAS:2013hta}
  [ATLAS Collaboration],
  %``Physics at a High-Luminosity LHC with ATLAS,''
  arXiv:1307.7292 [hep-ex].

%\cite{Muhlleitner:2003me}
\bibitem{Muhlleitner:2003me}
  M.~Muhlleitner and M.~Spira,
  %``A Note on doubly charged Higgs pair production at hadron colliders,''
  Phys.\ Rev.\ D {\bf 68}, 117701 (2003)
  %doi:10.1103/PhysRevD.68.117701
  [hep-ph/0305288].
  %%CITATION = doi:10.1103/PhysRevD.68.117701;%%
  %140 citations counted in INSPIRE as of 31 May 2018


%\cite{Belyaev:2012qa}
\bibitem{Belyaev:2012qa}
  A.~Belyaev, N.~D.~Christensen and A.~Pukhov,
  %``CalcHEP 3.4 for collider physics within and beyond the Standard Model,''
  Comput.\ Phys.\ Commun.\  {\bf 184}, 1729 (2013)
  %doi:10.1016/j.cpc.2013.01.014
  [arXiv:1207.6082 [hep-ph]].
  %%CITATION = doi:10.1016/j.cpc.2013.01.014;%%
  %522 citations counted in INSPIRE as of 05 May 2018

\bibitem{seesaw1} P. Minkowski, Phys. Lett. B {\bf 67}, 421 (1977).

\bibitem{seesaw2}  R. N. Mohapatra and G. Senjanovi\'{c}, Phys. Rev. Lett. {\bf 44}, 912 (1980).

\bibitem{seesaw3}  T. Yanagida, Conf.  Proc.  C {\bf 7902131},  95  (1979).

\bibitem{seesaw4} M. Gell-Mann, P. Ramond and R. Slansky, Conf. Proc. C {\bf 790927}, 315 (1979) [arXiv:1306.4669 [hep-th]].

\bibitem{seesaw5} S.~L.~Glashow, NATO Sci. Ser. B {\bf 61}, 687 (1980).

\bibitem{Dev:2013vxa}
  P.~S.~B.~Dev, S.~Goswami, M.~Mitra and W.~Rodejohann,
  %``Constraining Neutrino Mass from Neutrinoless Double Beta Decay,''
  Phys.\ Rev.\ D {\bf 88}, 091301 (2013)
%  doi:10.1103/PhysRevD.88.091301
  [arXiv:1305.0056 [hep-ph]].

\bibitem{Awasthi:2015ota}
  R.~L.~Awasthi, P.~S.~B.~Dev and M.~Mitra,
  %``Implications of the Diboson Excess for Neutrinoless Double Beta Decay and Lepton Flavor Violation in TeV Scale Left Right Symmetric Model,''
  Phys.\ Rev.\ D {\bf 93}, no. 1, 011701 (2016)
 % doi:10.1103/PhysRevD.93.011701
  [arXiv:1509.05387 [hep-ph]].



%\cite{Pritimita:2016fgr}
\bibitem{Pritimita:2016fgr}
  P.~Pritimita, N.~Dash and S.~Patra,
  %``Neutrinoless Double Beta Decay in LRSM with Natural Type-II seesaw Dominance,''
  JHEP {\bf 1610}, 147 (2016)
  %doi:10.1007/JHEP10(2016)147
  [arXiv:1607.07655 [hep-ph]].
  %%CITATION = doi:10.1007/JHEP10(2016)147;%%
  %6 citations counted in INSPIRE as of 09 Mar 2018



%\cite{Zhang:2007da}
\bibitem{Zhang:2007da}
  Y.~Zhang, H.~An, X.~Ji and R.~N.~Mohapatra,
  %``General CP Violation in Minimal Left-Right Symmetric Model and Constraints on the Right-Handed Scale,''
  Nucl.\ Phys.\ B {\bf 802}, 247 (2008)
  %doi:10.1016/j.nuclphysb.2008.05.019
  [arXiv:0712.4218 [hep-ph]].
  %%CITATION = doi:10.1016/j.nuclphysb.2008.05.019;%%
  %165 citations counted in INSPIRE as of 18 Apr 2018

%\cite{Bertolini:2014sua}
\bibitem{Bertolini:2014sua}
  S.~Bertolini, A.~Maiezza and F.~Nesti,
  %``Present and Future K and B Meson Mixing Constraints on TeV Scale Left-Right Symmetry,''
  Phys.\ Rev.\ D {\bf 89}, no. 9, 095028 (2014)
  %doi:10.1103/PhysRevD.89.095028
  [arXiv:1403.7112 [hep-ph]].
  %%CITATION = doi:10.1103/PhysRevD.89.095028;%%
  %91 citations counted in INSPIRE as of 18 Apr 2018


%\cite{Dev:2016vle}
\bibitem{Dev:2016vle}
  P.~S.~Bhupal Dev, R.~N.~Mohapatra and Y.~Zhang,
  %``Displaced photon signal from a possible light scalar in minimal left-right seesaw model,''
  Phys.\ Rev.\ D {\bf 95}, no. 11, 115001 (2017)
 % doi:10.1103/PhysRevD.95.115001
  [arXiv:1612.09587 [hep-ph]].
  %%CITATION = doi:10.1103/PhysRevD.95.115001;%%
  %23 citations counted in INSPIRE as of 21 Aug 2018

%\cite{Dev:2017dui}
\bibitem{Dev:2017dui}
  P.~S.~B.~Dev, R.~N.~Mohapatra and Y.~Zhang,
  %``Long Lived Light Scalars as Probe of Low Scale Seesaw Models,''
  Nucl.\ Phys.\ B {\bf 923}, 179 (2017)
%  doi:10.1016/j.nuclphysb.2017.07.021
  [arXiv:1703.02471 [hep-ph]].
  %%CITATION = doi:10.1016/j.nuclphysb.2017.07.021;%%
  %24 citations counted in INSPIRE as of 21 Aug 2018


%\cite{Dev:2017ozg}
\bibitem{Dev:2017ozg}
  P.~S.~B.~Dev, R.~N.~Mohapatra and Y.~Zhang,
  %``Long-lived Light Scalars at the LHC,''
  Acta Phys.\ Polon.\ B {\bf 48}, 969 (2017).
 % doi:10.5506/APhysPolB.48.969
  %%CITATION = doi:10.5506/APhysPolB.48.969;%%
  %2 citations counted in INSPIRE as of 21 Aug 2018





\bibitem{Keung:1983uu}
  W.~Y.~Keung and G.~Senjanovic,
  %``Majorana Neutrinos and the Production of the Right-handed Charged Gauge Boson,''
  Phys.\ Rev.\ Lett.\  {\bf 50}, 1427 (1983).
 % doi:10.1103/PhysRevLett.50.1427

\bibitem{Sirunyan:2018pom}
  A.~M.~Sirunyan {\it et al.} [CMS Collaboration],
  %``Search for a heavy right-handed W boson and a heavy neutrino in events with two same-flavor leptons and two jets at $\sqrt{s}=$ 13 TeV,''
  JHEP {\bf 1805}, no. 05, 148 (2018)
%  doi:10.1007/JHEP05(2018)148
  [arXiv:1803.11116 [hep-ex]].

\bibitem{Mohapatra:1981pm}
  R.~N.~Mohapatra and J.~D.~Vergados,
  %``A New Contribution to Neutrinoless Double Beta Decay in Gauge Models,''
  Phys.\ Rev.\ Lett.\  {\bf 47}, 1713 (1981).
 % doi:10.1103/PhysRevLett.47.1713

\bibitem{Hirsch:1996qw}
  M.~Hirsch, H.~V.~Klapdor-Kleingrothaus and O.~Panella,
  %``Double beta decay in left-right symmetric models,''
  Phys.\ Lett.\ B {\bf 374}, 7 (1996)
 % doi:10.1016/0370-2693(96)00185-2
  [hep-ph/9602306].

%\cite{Chakrabortty:2012mh}
\bibitem{Chakrabortty:2012mh}
  J.~Chakrabortty, H.~Z.~Devi, S.~Goswami and S.~Patra,
  %``Neutrinoless double-$\beta$ decay in TeV scale Left-Right symmetric models,''
  JHEP {\bf 1208}, 008 (2012)
  %doi:10.1007/JHEP08(2012)008
  [arXiv:1204.2527 [hep-ph]].
  %%CITATION = doi:10.1007/JHEP08(2012)008;%%
  %56 citations counted in INSPIRE as of 09 Mar 2018

%\cite{Borah:2015ufa}
\bibitem{Borah:2015ufa}
  D.~Borah and A.~Dasgupta,
  %``Neutrinoless Double Beta Decay in Type I+II Seesaw Models,''
  JHEP {\bf 1511}, 208 (2015)
  %doi:10.1007/JHEP11(2015)208
  [arXiv:1509.01800 [hep-ph]].
  %%CITATION = doi:10.1007/JHEP11(2015)208;%%
  %12 citations counted in INSPIRE as of 09 Mar 2018



%\cite{Deppisch:2017vne}
\bibitem{Deppisch:2017vne}
  F.~F.~Deppisch, C.~Hati, S.~Patra, P.~Pritimita and U.~Sarkar,
  %``Neutrinoless double beta decay in left-right symmetric models with a universal seesaw mechanism,''
  Phys.\ Rev.\ D {\bf 97}, no. 3, 035005 (2018)
  %doi:10.1103/PhysRevD.97.035005
  [arXiv:1701.02107 [hep-ph]].
  %%CITATION = doi:10.1103/PhysRevD.97.035005;%%
  %7 citations counted in INSPIRE as of 09 Mar 2018




%\cite{Ge:2015yqa}
\bibitem{Ge:2015yqa}
  S.~F.~Ge, M.~Lindner and S.~Patra,
  %``New physics effects on neutrinoless double beta decay from right-handed current,''
  JHEP {\bf 1510}, 077 (2015)
  %doi:10.1007/JHEP10(2015)077
  [arXiv:1508.07286 [hep-ph]].
  %%CITATION = doi:10.1007/JHEP10(2015)077;%%
  %27 citations counted in INSPIRE as of 14 Mar 2018



%\cite{Prezeau:2003xn}
\bibitem{Prezeau:2003xn}
  G.~Prezeau, M.~Ramsey-Musolf and P.~Vogel,
  %``Neutrinoless double beta decay and effective field theory,''
  Phys.\ Rev.\ D {\bf 68}, 034016 (2003)
  %doi:10.1103/PhysRevD.68.034016
  [hep-ph/0303205].
  %%CITATION = doi:10.1103/PhysRevD.68.034016;%%
  %78 citations counted in INSPIRE as of 24 May 2018

\bibitem{Albert:2017owj}
  J.~B.~Albert {\it et al.} [EXO Collaboration],
  %``Search for Neutrinoless Double-Beta Decay with the Upgraded EXO-200 Detector,''
  Phys.\ Rev.\ Lett.\  {\bf 120}, no. 7, 072701 (2018)
%  doi:10.1103/PhysRevLett.120.072701
  [arXiv:1707.08707 [hep-ex]].


%\cite{KamLAND-Zen:2016pfg}
\bibitem{KamLAND-Zen:2016pfg}
  A.~Gando {\it et al.} [KamLAND-Zen Collaboration],
  %``Search for Majorana Neutrinos near the Inverted Mass Hierarchy Region with KamLAND-Zen,''
  Phys.\ Rev.\ Lett.\  {\bf 117}, no. 8, 082503 (2016)
  Addendum: [Phys.\ Rev.\ Lett.\  {\bf 117}, no. 10, 109903 (2016)]
  %doi:10.1103/PhysRevLett.117.109903, 10.1103/PhysRevLett.117.082503
  [arXiv:1605.02889 [hep-ex]].
  %%CITATION = doi:10.1103/PhysRevLett.117.109903, 10.1103/PhysRevLett.117.082503;%%
  %228 citations counted in INSPIRE as of 27 Feb 2018

%\cite{Agostini:2013mzu}
%\bibitem{Agostini:2013mzu}
%  M.~Agostini {\it et al.} [GERDA Collaboration],
  %``Results on Neutrinoless Double-$\beta$ Decay of $^{76}$Ge from Phase I of the GERDA Experiment,''
%  Phys.\ Rev.\ Lett.\  {\bf 111}, no. 12, 122503 (2013)
  %doi:10.1103/PhysRevLett.111.122503
%  [arXiv:1307.4720 [nucl-ex]].
  %%CITATION = doi:10.1103/PhysRevLett.111.122503;%%
  %509 citations counted in INSPIRE as of 27 Feb 2018

%\cite{Agostini:2017iyd}
%\bibitem{Agostini:2017iyd}
%  M.~Agostini {\it et al.},
  %``Background-free search for neutrinoless double-$\beta$ decay of $^{76}$Ge with GERDA,''
%  Nature {\bf 544}, 47 (2017)
  %doi:10.1038/nature21717
%  [arXiv:1703.00570 [nucl-ex]].
  %%CITATION = doi:10.1038/nature21717;%%
  %99 citations counted in INSPIRE as of 27 May 2018

\bibitem{Agostini:2018tnm}
  M.~Agostini {\it et al.} [GERDA Collaboration],
  %``Improved Limit on Neutrinoless Double-$\beta$ Decay of $^{76}$Ge from GERDA Phase II,''
  Phys.\ Rev.\ Lett.\  {\bf 120}, no. 13, 132503 (2018)
%  doi:10.1103/PhysRevLett.120.132503
  [arXiv:1803.11100 [nucl-ex]].

\bibitem{Aalseth:2017btx}
  C.~E.~Aalseth {\it et al.} [Majorana Collaboration],
  %``Search for Neutrinoless Double-β Decay in $^{76}$Ge with the Majorana Demonstrator,''
  Phys.\ Rev.\ Lett.\  {\bf 120}, no. 13, 132502 (2018)
%  doi:10.1103/PhysRevLett.120.132502
  [arXiv:1710.11608 [nucl-ex]].


%\cite{Alfonso:2015wka}
%\bibitem{Alfonso:2015wka}
%  K.~Alfonso {\it et al.} [CUORE Collaboration],
  %``Search for Neutrinoless Double-Beta Decay of $^{130}$Te with CUORE-0,''
%  Phys.\ Rev.\ Lett.\  {\bf 115}, no. 10, 102502 (2015)
%  doi:10.1103/PhysRevLett.115.102502
%  [arXiv:1504.02454 [nucl-ex]].
  %%CITATION = doi:10.1103/PhysRevLett.115.102502;%%
  %151 citations counted in INSPIRE as of 27 May 2018s

%\cite{Alduino:2017ehq}
\bibitem{Alduino:2017ehq}
  C.~Alduino {\it et al.} [CUORE Collaboration],
  %``First Results from CUORE: A Search for Lepton Number Violation via $0\nu\beta\beta$ Decay of $^{130}$Te,''
  Phys.\ Rev.\ Lett.\  {\bf 120}, no. 13, 132501 (2018)
  %doi:10.1103/PhysRevLett.120.132501
  [arXiv:1710.07988 [nucl-ex]].
  %%CITATION = doi:10.1103/PhysRevLett.120.132501;%%
  %30 citations counted in INSPIRE as of 27 May 2018


\bibitem{Arnold:2018tmo}
  R.~Arnold {\it et al.} [NEMO-3 Collaboration],
  %``Final results on $^\textbf{82}$Se double beta decay to the ground state of $^\textbf{82}$Kr from the NEMO-3 experiment,''
  arXiv:1806.05553 [hep-ex].


%\cite{Meroni:2012qf}
\bibitem{Meroni:2012qf}
  A.~Meroni, S.~T.~Petcov and F.~Simkovic,
  %``Multiple CP non-conserving mechanisms of $(\beta\beta)_{0\nu}$-decay and nuclei with largely different nuclear matrix elements,''
  JHEP {\bf 1302}, 025 (2013)
  %doi:10.1007/JHEP02(2013)025
  [arXiv:1212.1331 [hep-ph]].
  %%CITATION = doi:10.1007/JHEP02(2013)025;%%
  %52 citations counted in INSPIRE as of 20 Apr 2018

%\cite{Vincent:2014rja}
\bibitem{Vincent:2014rja}
  A.~C.~Vincent, E.~F.~Martinez, P.~Hernandez, M.~Lattanzi and O.~Mena,
  %``Revisiting cosmological bounds on sterile neutrinos,''
  JCAP {\bf 1504}, no. 04, 006 (2015)
  %doi:10.1088/1475-7516/2015/04/006
  [arXiv:1408.1956 [astro-ph.CO]].
  %%CITATION = doi:10.1088/1475-7516/2015/04/006;%%
  %24 citations counted in INSPIRE as of 15 May 2018

%\cite{Abazajian:2017tcc}
\bibitem{Abazajian:2017tcc}
  K.~N.~Abazajian,
  %``Sterile neutrinos in cosmology,''
  Phys.\ Rept.\  {\bf 711-712}, 1 (2017)
  %doi:10.1016/j.physrep.2017.10.003
  [arXiv:1705.01837 [hep-ph]].
  %%CITATION = doi:10.1016/j.physrep.2017.10.003;%%
  %26 citations counted in INSPIRE as of 15 May 2018

\bibitem{CMP}
  D.~Chang, R.~N.~Mohapatra and M.~K.~Parida,
  %``Decoupling Parity and SU(2)-R Breaking Scales: A New Approach to Left-Right Symmetric Models,''
  Phys.\ Rev.\ Lett.\  {\bf 52}, 1072 (1984).


%\cite{Cabibbo:1965zzb}
\bibitem{Cabibbo:1965zzb}
  N.~Cabibbo and A.~Maksymowicz,
  %``Angular Correlations in Ke-4 Decays and Determination of Low-Energy pi-pi Phase Shifts,''
  Phys.\ Rev.\  {\bf 137}, B438 (1965)
  Erratum: [Phys.\ Rev.\  {\bf 168}, 1926 (1968)].
  %doi:10.1103/PhysRev.137.B438, 10.1103/PhysRev.168.1926
  %%CITATION = doi:10.1103/PhysRev.137.B438, 10.1103/PhysRev.168.1926;%%
  %194 citations counted in INSPIRE as of 14 Mar 2018



%\cite{Dinh:2012bp}
\bibitem{Dinh:2012bp}
  D.~N.~Dinh, A.~Ibarra, E.~Molinaro and S.~T.~Petcov,
  %``The $\mu - e$ Conversion in Nuclei, $\mu \to e \gamma, \mu \to 3e$ Decays and TeV Scale See-Saw Scenarios of Neutrino Mass Generation,''
  JHEP {\bf 1208}, 125 (2012)
  Erratum: [JHEP {\bf 1309}, 023 (2013)]
  %doi:10.1007/JHEP09(2013)023, 10.1007/JHEP08(2012)125
  [arXiv:1205.4671 [hep-ph]].
  %%CITATION = doi:10.1007/JHEP09(2013)023, 10.1007/JHEP08(2012)125;%%
  %96 citations counted in INSPIRE as of 09 Dec 2017












\end{thebibliography}
\end{document}